%% file: main.tex
\documentclass[a4paper,11pt]{article}
\usepackage{jheppub}
\usepackage{lineno}
\usepackage{subfiles}
\usepackage{xcolor}
\usepackage{subcaption} 
\usepackage{tikz} 
\usepackage{array}
\usepackage{xspace}
\usepackage[normalem]{ulem}

\preprint{
\begin{flushright}
CERN-TH-2026-101 \\
MCNET-26-09
\end{flushright}}

\definecolor{CS}{rgb}{0.847, 0.102, 0.098}
\definecolor{PG}{rgb}{0.392, 0.561, 1.0}
\definecolor{FHP}{rgb}{0.471, 0.369, 0.941}
\definecolor{AO}{rgb}{0.996, 0.380, 0.0}

\newcommand{\PG}{PG\textsubscript{0}\xspace}
\newcommand{\PGAOB}{\ensuremath{\mathrm{PG}_{0}^{\mathrm{AOB}}}\xspace}
\newcommand{\PGAOBbold}{\ensuremath{\mathbf{PG}_{\mathbf{0}}^{\mathbf{AOB}}}\xspace}
\newcommand{\FHPAOB}{FHP\textsuperscript{AOB}\xspace}
\newcommand{\FHPAOBbold}{\ensuremath{\mathbf{FHP}^{\mathbf{AOB}}}\xspace}

\newcommand{\labeledfig}[2]{%
\begin{tikzpicture}
\node[anchor=south west, inner sep=0] (img) {\includegraphics[width=0.24\linewidth]{#1}};
\node[anchor=north west, xshift=13pt, yshift=-3pt, fill=white, fill opacity=0.75, text opacity=1, inner sep=2pt] at (img.north west) {\tiny{#2}};
\end{tikzpicture}%
}

\title{Studying the Infrared Behaviour of Improved Logarithmic Accuracy Parton Showers with Herwig}

\author[a]{Jack Holguin}
\author[b,c]{Simon Pl\"atzer}
\author[a,d]{Michael H. Seymour}
\author[a]{Siddharth Sule}
\affiliation[a]{Department of Physics and Astronomy, University of Manchester, Manchester M13 9PL, United Kingdom}
\affiliation[b]{Institute of Physics, NAWI Graz, University of Graz, Universit\"atsplatz 5, A-8010 Graz, Austria}
\affiliation[c]{Particle Physics, Faculty of Physics, University of Vienna, Boltzmanngasse 5, A-1090 Wien, Austria}
\affiliation[d]{Theoretical Physics Department, CERN, CH-1211 Geneva 23, Switzerland}
\emailAdd{jack.holguin@manchester.ac.uk}
\emailAdd{simon.plaetzer@uni-graz.at}
\emailAdd{michael.seymour@manchester.ac.uk}
\emailAdd{siddharth.sule@manchester.ac.uk}

\abstract{
We have implemented two recently proposed dipole shower algorithms that have next-to-leading-logarithmic accuracy at leading colour in the Herwig event generator. We study their properties and compare them to Herwig's existing dipole and angular ordered parton shower algorithms. In addition to their improved properties in the logarithmic regime, we find important roles for their extrapolations into the hard regime, where we perform NLO matching, and into the infrared regime, where we perform cluster hadronization. We emphasise the importance of this infrared regime and the precise definition of the infrared cutoff used by each shower as the initial state for Herwig's hadronization model. Studying the results at the hadron level, we find important consequences of this infrared cutoff difference and propose it as a starting point for further study of the interplay between parton showers and hadronization models. We conclude by studying the models' tunability and identifying the best-fit parameters for each.
}

\begin{document}

\maketitle
\flushbottom

\input{sections/intro}

\input{sections/showers}

\input{sections/cluster}

\input{sections/tune}

\input{sections/conc}

\acknowledgments
We thank the members of the Herwig Collaboration for their useful discussions during the development of the codebase and for their comments on this paper.
We acknowledge the contributions of E. Simpson Dore and C. B. Duncan to earlier iterations of the Herwig NLL dipole shower codebase.
We also thank the members of the PanScales and Sherpa Collaborations for their fruitful insights and look forward to the collaborative development of these tools.
Notably, we thank S. Ferrario Ravasio for assistance with setting up the updated PanGlobal shower. 
SS thanks A. Buckley for assistance with using Professor for the analyses.

We are grateful to Keith Hamilton and Gavin Salam for suggesting that there was an error in the appendix of the original version of this paper.

\paragraph{Funding Information}
The numerical results presented in this paper have been obtained on the computing clusters of the Particle Physics Groups of Universit\"at Wien and The University of Manchester. 
We are grateful for the opportunity to have used these facilities. 
SS would like to thank the UK Science and Technology Facilities Council (STFC) for the award of a studentship. 
MS also acknowledges STFC's support through grants ST/T001038/1 and ST/X00077X/1.
This research was supported by the Munich Institute for Astro-, Particle and BioPhysics (MIAPbP), which is funded by the Deutsche Forschungsgemeinschaft (DFG, German Research Foundation) under Germany's Excellence Strategy – EXC-2094 – 390783311.
JH is supported by the Leverhulme Trust as an Early Career Fellow.

\appendix
\input{sections/appndx}

\bibliographystyle{JHEP}
\bibliography{biblio.bib}

\end{document}

%% file: sections/intro.tex
\vspace*{-2ex}
\section{Introduction}\label{sec:intro}

Monte Carlo event generators have played a fundamental role in the success of particle physics for over forty years, simulating high-energy collisions at LEP, SLAC, HERA, the Tevatron, and the LHC \cite{Buckley:2011ms}. By translating theoretical predictions into simulated data as observed by detectors, event generators enable comparison of experimental results with theory. Indeed, event generators such as Herwig \cite{Bewick:2023tfi, Bellm:2025pcw}, Pythia \cite{Bierlich:2022pfr}, and Sherpa \cite{Sherpa:2024mfk} are extensively used across the majority of analyses performed with the LHC data. The backbone of event generators is their simulation of radiation produced from quantum chromodynamics (QCD). This starts with the sampling of parton configurations from low-multiplicity matrix elements for the high-energy hard processes which underlie a collision. These configurations are then dressed with large multiplicities of QCD radiation by the application of a parton shower algorithm. This produces a cascade of quarks and gluons which ultimately must form the hadrons observed in experimental measurements. Large multiplicities of partonic radiation are produced because the matrix elements for the emission of soft and collinear partons diverge when the emitted parton becomes soft or collinear relative to its parent(s). These divergences leave their trace in large logarithms in the cross sections for QCD observables. The past decade has seen drastic advances in the formulation of parton shower algorithms by matching the spectrum of the radiation produced against the formal accuracy with which the logarithmic structures in observable cross sections are computed.

The cumulative cross sections of simple exponentiated global observables have logarithmic structures that can be written in the form
\begin{align}
    \Sigma(V) = C(\alpha_s)\exp(L \, g_{1} (\alpha_s L) + g_{2} (\alpha_s L) + \alpha_s \, g_{3}(\alpha_s L) + \dots) \, , \label{eq:simpleExp}
\end{align}
where $V \ll 1$ is the value of an observable and $L = \ln (1/V)$, up to corrections that are suppressed by powers of $V$. As we typically consider the limit where $L \gg 1$, it is conventional to define accuracy by considering the limit $\alpha_s L \sim 1$. Consequently, each of the functions $g_{i}(\alpha_s L) \sim 1$ and sums terms of order $\alpha_s^NL^N$ to all orders in $N$. A complete calculation of $g_i$ for $i \leq n+1$ will include corrections of up to $\mathcal{O}(\alpha_s^n)$ in the exponent, relative to the leading $L \, g_1$ term. Traditionally, the logarithmic accuracy of a calculation is defined by analogy to Eq.~\eqref{eq:simpleExp} with N$^n$LL accuracy achieving theoretical control over the region where $\alpha_s L \sim 1$ up to and including corrections $\mathcal{O}(\alpha_s^n)$: i.e. fully computing $g_i$ for $i \leq n+1$ for these simple observables. The paradigm shift achieved by the PanScales collaboration \cite{Dasgupta:2020fwr,Hamilton:2020rcu,Karlberg:2021kwr,Hamilton:2021dyz,vanBeekveld:2022zhl} and concurrent developments \cite{Forshaw:2020wrq,Holguin:2020joq,Herren:2022jej,Hoche:2024dee} has been to map the criteria on N$^n$LL accuracy onto criteria for the probability distribution against which partonic configurations are sampled to create a parton shower algorithm. The success of this approach has seen parton showers transition from rarely making statements on formal accuracy\footnote{A notable exception is the Herwig angular-ordered shower, which achieves NLL accuracy for many global event-shape observables \cite{Bewick:2019rbu}, including massive quark effects \cite{Hoang:2018zrp}, by faithfully producing a coherent branching parton cascade \cite{Catani:1990rr}.}, to NLL accuracy becoming the new standard for shower development, and with N$^2$LL accuracy achieved for some processes \cite{FerrarioRavasio:2023kyg,vanBeekveld:2024wws}.

However, there is a remaining link in the chain which has not yet been subject to this paradigm shift -- hadronization. Hadronization modelling is the final, detector-independent step in event generation. It aims to sample a population of hadrons, upon which measurements can be made, from the population of quarks and gluons as produced by the parton shower. Systematising and improving implementations of hadronization is an extremely challenging problem which has seen relatively limited attention in recent years \cite{Platzer:2022jny, Masouminia:2023zhb, Hoang:2024nqi, Altmann:2024kwx, Altmann:2024odn, Gieseke:2025mcy}. This is in part because of the wide scope of the physics a hadronization model must describe, from the power corrections to infrared and collinear safe observables where some analytical insight exists \cite{Webber:1994cp,Dokshitzer:1995zt,Dokshitzer:1995qm,Salam:2001bd,Mateu:2012nk,Farren-Colloty:2025amh}, to hadron spectroscopy where the only viable ground-up approach is lattice QCD \cite{Bulava:2022ovd} and secondary decays of the produced hadrons, where often experimental results \cite{ParticleDataGroup:2024cfk} are the dominant input. Nevertheless, hadronization modelling is of crucial importance to the realistic simulation of collider processes, particularly as the precision of experimental results continues to improve \cite{ZurbanoFernandez:2020cco, Amoroso:2020lgh} and the accuracy of parton showers and hadronization becomes an increasingly significant limitation on overall uncertainties. It is therefore important to achieve a detailed understanding of how hadronization interplays with the recent advances in parton shower development, even when formal statements cannot yet be made.

To this end, in this paper we present the implementation of two previously introduced, distinct NLL leading-colour accurate dipole-type parton showers within the Herwig event generator \cite{Bewick:2023tfi, Bellm:2025pcw} for processes where the initial state forms a colour singlet: namely simplified implementations of the $\beta=0$ PanGlobal shower \cite{Dasgupta:2020fwr, FerrarioRavasio:2023kyg} and the Forshaw-Holguin-Pl\"atzer shower \cite{Forshaw:2020wrq, Holguin:2020joq}~\footnote{For a previous implementation of the FHP inspired shower scheme by the PanScales Collaboration, see the appendices of \cite{Luisoni:2020efy}. For a discussion of power corrections to the FHP scheme, see \cite{Caola:2022vea}.}$\,$\footnote{We simplify our implementation of PanGlobal by omitting subleading $N_c$ corrections~\cite{Hamilton:2020rcu} and spin correlations from interleaved soft and collinear radiation~\cite{Hamilton:2021dyz} and use a simplified dipole partitioning.}. These are in addition to the Herwig angular-ordered shower, which already achieves NLL accuracy across many event-shape and global observables \cite{Bewick:2019rbu} and the Catani-Seymour dipole shower~\cite{Platzer:2009jq}, which is known to be only LL accurate. Since these showers are implemented into a complete generator, they benefit from the pre-existing functionality of the generator, facilitating matching them to NLO hard processes with the MC@NLO framework \cite{Platzer:2011bc, Frixione:2002ik, Bellm:2025pcw} and hadronizing the partonic output with the Herwig cluster hadronization model \cite{Bahr:2008pv}. It is one of our goals to document the breadth of the predictions which these showers make, providing some indication of the uncertainty inherent in NLL parton showering. Building on this, we study their infrared behaviour, leading towards their properties under cluster hadronization. 

Ultimately, we aim to highlight the directions for future progress needed for phenomenology with logarithmically accurate event generators, particularly the need for a detailed understanding of the infrared behaviour of the shower: how it fills the partonic phase space above the shower cutoff, and how this influences the hadronization modelling. Crucially, we find that the shape of the shower cutoff boundary, can have considerable effects on the quality of the hadronization modelling by the Herwig cluster model. We find that our implementations of both the FHP and PanGlobal showers are improved at the hadron level by modifying their natural shower cutoffs (given directly by the ordering variable) to match the cutoff introduced by the Herwig angular-ordered shower. These observations suggest more broadly that phenomenology affecting the IR phase space will be important for hadronization and practical event generation, such as the inclusion of heavy quark masses within the framework of these new parton showers and the complete population of fixed-order phase spaces. Parton mass effects were recently considered in the Alaric shower framework~\cite{Assi:2023rbu}.

This paper is structured as follows. In Section~\ref{sec:showers}, we outline the Catani-Seymour and the new implementations of the PanGlobal and Forshaw-Holguin-Pl\"atzer shower schemes. In this section, we will also discuss results from these showers at the parton level. Following this, in Section~\ref{sec:cluster}, we provide an overview of the cluster hadronization model, detailing each step and highlighting the key parameters. Lastly, in Section~\ref{sec:trials}, we present tuned results for $e^+e^- \to q\bar{q}$ at NLO and compare the three showers, highlighting the physical properties of these showers which influence their behaviour under tuning. All histograms plotted in this paper were made using Rivet \cite{Bierlich:2024vqo} and Yoda \cite{Buckley:2023xqh}.

%% file: sections/showers.tex
\section{Overview of the Showers}\label{sec:showers}

In this section, we begin by introducing the different shower algorithms we will consider, translating each into the evolution and splitting variables defined in the Catani-Seymour dipole shower framework in Herwig \cite{Platzer:2011bc}, which allows us to address all the showers in a flexible fashion. This involves defining the kinematics, splitting kernels, and the choice of eikonal partitioning.
Building on these foundations, we re-derive the splitting rates for these new shower algorithms, including the Jacobian obtained by factorising the matrix element and phase space. While formally a subleading contribution giving rise to only non-logarithmic terms in the exponent, this Jacobian is inevitable for the hard process matching of the first emission(s), but also affects the population of the higher-multiplicity non-singular partonic phase space. To our knowledge, explicit Jacobian contributions are currently included in the Catani-Seymour-based dipole showers \cite{Schumann:2007mg, Dinsdale:2007mf, Platzer:2009jq}. We also anticipate that its effects will be central for multi-jet merging with the Herwig dipole shower \cite{Bellm:2017ktr} and the filling of the tails in the event-shape and multiplicity distributions, which we study in Section~\ref{sec:hadronlevel}. We also discuss the collinear limits of the different showers, the NLO matching for $e^+ e^- \to q \bar{q}$ using the MC@NLO method or subtractive matching \cite{Frixione:2002ik, Bellm:2025pcw}, and analyse first parton-level results. The showers which we implement assume massless QCD, and we leave the generalisation to massive partons to future work.

\subsection{Kinematics and Kernels}
\label{sec:Kinematics}

To use a single framework for all three showers, we work throughout in the Catani-Seymour dipole shower variables introduced in \cite{Catani:1996vz}, $p_T^2 = -k_\perp \cdot k_\perp$, $z$, and $\phi$, and adapt the descriptions of the other showers accordingly. In this convention, an emission is defined using the pre- and post-emission momenta,
\begin{align}
    p_i + p_j \to q_i + q + q_j \, ,
\end{align}
where $p_i$ and $q_i$ denote the emitter parton, $p_j$ and $q_j$ the spectator parton, and $q$ the emitted parton. For all three showers, the showers evolve according to an ordered cascade in $p_T^2$. The sampling procedures implemented in the Herwig dipole shower, specifically the use of the ExSample library \cite{Platzer:2011dr}, allow us to handle a generic, $p_T$-type evolution variable which has a common logarithmic behaviour to all splitting kernels we study. Differences between various showers are then implemented at the level of boundaries for the other variables of the splitting kernels, partitionings, and different Jacobian weights. The generic ordering variable gains its physical significance at the point where the kinematics are constructed in terms of the full four-momenta of the branching. This only happens at the point when a certain dipole and splitting channel have been selected. 

\subsubsection{The Catani-Seymour Shower}

Starting with the Catani-Seymour shower, as prescribed in \cite{Nagy:2006kb, Dinsdale:2007mf, Schumann:2007mg} and adapted for Herwig in \cite{Platzer:2009jq, Platzer:2011bc}, which we describe as the CS scheme from here on, the kinematic mapping for a final-final dipole is defined by
\begin{align}
    q_i &= z p_i +  y (1 - z) p_j + k_\perp \, , \nonumber \\
    q   &= (1 - z) p_i + y z p_j - k_\perp \, , \nonumber \\
    q_j &= (1 - y) p_j \, .
\end{align}
For convenience, we define
\begin{align}
    y = \frac{p_T^2}{2 p_i \cdot p_j} \frac{1}{z(1-z)} \, ,
\end{align}
and
\begin{align}
    p_T^2 = -k_\perp \cdot k_\perp \, ,
\end{align}
with
\begin{align}
    k_\perp \cdot p_i = k_\perp \cdot p_j = 0 \, .
\end{align}
This is a local recoil prescription, in which only the partons belonging to the emitting dipole are affected by the emission.
The splitting kernels are defined by the dipoles from Catani-Seymour subtraction \cite{Catani:1996vz},
\begin{align}
P_{q \to qg}^{\text{CS}} &= C_F \left( \frac{2}{1-z(1-y)} - (1+z)\right) \, , \nonumber \\
P_{g \to gg}^{\text{CS}} &= \frac{C_A}{2} \left( \frac{2}{1-z(1-y)} - 2 + z(1-z)\right) \, .
\end{align}
In the collinear limit, they are equal to the standard DGLAP splitting functions, but they \emph{partition} the soft radiation, which is singular in both dipole collinear limits, into parts that are singular only in one collinear limit or the other,
\begin{align}
\frac{p_i\!\cdot\!p_j}{(q_i\!\cdot\!q)(q_j\!\cdot\!q)}=
\frac{p_i\!\cdot\!p_j}{(q_i\!\cdot\!q)(q_i\!\cdot\!q+q_j\!\cdot\!q)}+
\frac{p_i\!\cdot\!p_j}{(q_j\!\cdot\!q)(q_i\!\cdot\!q+q_j\!\cdot\!q)}=
\frac1{q_i\!\cdot\!q} \, \frac1{1-z(1-y)} + (i \leftrightarrow j).
\end{align}
The $g\rightarrow q \bar q$ kernel is the standard DGLAP splitting function.

The kinematic limits are given by
$z,y \in [0,1]$. At fixed $p_{T}$, this corresponds to
\begin{align}
    z_- < z < z_+ \, ,
    \quad
    z_\pm = \frac12\left(1\pm\sqrt{1-\frac{4p_{T}^2}{s_{ij}}}\right),
\end{align}
with a maximum allowed transverse momentum of $p_{T,\max}^2 = {s_{ij}}/{4}$.

\subsubsection{The PanGlobal Shower}
The PanScales collaboration have defined a family of recoil schemes \cite{Dasgupta:2020fwr,FerrarioRavasio:2023kyg}, which incorporate the 
total momentum of the collision $Q$.
We use their $\beta=0$ variant, as defined in \cite{FerrarioRavasio:2023kyg}, and describe it as the \PG\ scheme from here on.
The mapping from $\{p\}$ to $\{q\}$ is defined in several steps.
The initial kinematic mapping is given by
\begin{align}
q_i'' &= z p_i \, , \nonumber \\
q''   &= (1 - z) p_i + z y p_j - k_\perp \, , \nonumber \\
q_j'' &= (1 - z y) p_j \, .
\label{eq:pg-mapping}
\end{align}
At this stage, energy and momentum are not conserved. The momenta $q_i''$, $q''$, and $q_j''$ are then rescaled by a factor
\begin{align}
\alpha = \frac{-q_m\cdot p_m + \sqrt{\left( q_m \cdot p_m \right)^2 + q_m^2(Q^2 - p_m^2)}}{q_m^2} \ ,
\label{eq:alpha_rescaling}
\end{align}
where $q_{m} = q_i'' + q'' + q_j''$ and $p_{m} = Q - p_i - p_j$. This rescaling acts to ensure the total invariant mass of the event is preserved before and after the emission.
Following this, \textit{all} particles in the event undergo a Lorentz boost
\begin{align}
\Lambda(Q, Q'): p \to p + \frac{2 Q' \cdot  p}{Q^2} Q - \frac{2 (Q + Q')\cdot p}{(Q +Q')^2} (Q + Q') \, ,
\end{align}
where $Q' = p_m + \alpha q_m$. This transformation can be understood as realigning the centre-of-mass frame after the emission with the centre-of-mass frame before the emission.

The full \PG\ momentum mapping is then given by
\begin{align}
q_i &= \Lambda q_i'= \Lambda \alpha q_i'' \, , \nonumber \\
q &= \Lambda q' = \Lambda \alpha q'' \, , \nonumber \\
q_j &= \Lambda q_j' = \Lambda \alpha q_j'' \, , \nonumber \\
q_k &= \Lambda p_k \, ,
\end{align}
where $q_k$, $k\ne i,j$, represents the momenta of the remaining partons in the event.

The soft radiation pattern is partitioned by defining a generalised lab-frame rapidity
\begin{align}
\bar{\eta} &= \frac{1}{2} \ln \left( \frac{p_j \cdot q''}{p_i \cdot q''} \right) - \frac{1}{2} \ln \left( \frac{p_j \cdot Q}{p_i \cdot Q} \right) \, , \label{eq:etabar} \\
g^{\text{PS}} &= \frac{1}{1 + e^{-2 \bar{\eta}}} = \frac{1-z}{1 - z \left(1 - y\frac{p_j \cdot Q}{p_i \cdot Q} \right)} = \frac{1-z}{1 - z \left(1 - \tilde{y} \right)} \, ,
\end{align}
where we have defined $\tilde{y} = y({p_j \cdot Q}/{p_i \cdot Q})$ for convenience.
It is worth pointing out that although $\bar{\eta}$ is not preserved by the recoil (i.e.\ it is not equal to the same expression with $q''$ replaced by $q$), the momentum mapping is symmetric under exchange of the role of emitter and spectator, which is sufficient to preserve the desirable property that the sum of $g^{\text{PS}}$ and its `mirror image' is equal to~1.

The emissions are generated according to DGLAP kernels\footnote{For the $g \to gg$ kernel, one may choose either a symmetric or asymmetric assignment of $z$ between the two outgoing gluons; we adopt the symmetric choice here.} multiplied by this partitioning factor,
\begin{align}
P_{q \to qg}^{\text{PG}} &= C_F \left(  \frac{2}{1 - z} - (1 + z)\right) \left( \frac{1 - z}{1 - z(1 - \tilde{y})} \right) = C_F \frac{1 + z^2}{1 - z(1 - \tilde{y})} \, , \nonumber \\
P_{g \to gg}^{\text{PG}} &= \frac{C_A}{2} \left(  \frac{2}{1 - z} - 2 + z(1 - z)\right) \left( \frac{1 - z}{1 - z(1 - \tilde{y})} \right) = \frac{C_A}{2} \frac{z (3 - 2 z + z^2)}{1 - z(1 - \tilde{y})} \, .
\end{align}
In this algorithm, the kinematic limits are given by $z, zy \in [0,1]$. At fixed $p_T$, this corresponds to
\begin{align}
    0 < z < 1-\frac{p_T^2}{s_{ij}} \, ,
\end{align}
with a maximum allowed transverse momentum of $p_{T,\max}^2 = s_{ij}$.

\subsubsection{The Forshaw-Holguin-Pl\"atzer Shower}

Ref.~\cite{Forshaw:2020wrq} reconsidered the basis of parton shower algorithms and proposed a scheme that has some similarities to \PG, but with important differences both in the recoil scheme and in the partitioning. The original Forshaw-Holguin-Pl\"atzer shower employed a partitioning that is continuous and populates the complete phase space. However, this partitioning is not strictly positive. Under averaging over the azimuthal degrees of freedom, it reduces to a $\Theta$-function constraint.

We have implemented this shower within Herwig and describe it as the FHP scheme from here on.
The initial kinematic mapping is given by
\begin{align}
q_i'' &= z p_i \, , \nonumber \\
q''   &= (1 - z) p_i + z y p_j - k_\perp \, , \nonumber \\
q_j'' &= p_j \, ,
\label{eq:fhp-mapping}
\end{align}
with the same boost and rescaling applied as in the \PG\ case. However, in contrast to PG$_{0}$, since the light-cone momentum in the direction of the spectator parton is also not conserved, the boost and rescaling now additionally recoil that light-cone momentum against the rest of the event, as well as the momentum transverse to the dipole.

FHP partitions the soft radiation pattern in a way that is related to \PG's, but with a $\Theta$~function rather than a smooth suppression, which reduces the effect of a tail of emissions that are generated in a very backward direction.
One is initially tempted to implement a $\Theta$~function on $\bar{\eta}$, as defined in Eq.~(\ref{eq:etabar}), but since the FHP scheme absorbs the recoil in the emitter direction locally, and in the spectator direction globally, it does not preserve the symmetry of the partition function observed for \PG.
However, preserving the logarithmic properties of the shower requires only that the rapidity of the partition be fixed in the soft limit, and one is free to choose an extrapolation away from that limit, i.e.\ to include additional factors that tend to~1 in the limits $y\to0$ or $z\to1$.
We choose to implement a function $\Theta(\tilde{\eta})$, with
\begin{align}
\tilde{\eta} &= \frac{1}{2} \ln \left( \frac{q''_j \cdot q''}{q''_i \cdot q''} \right) - \frac{1}{2} \ln \left( \frac{p_j \cdot Q}{p_i \cdot Q} \right)
\nonumber \\
&= \frac{1}{2} \ln \left( \frac{1-z}{z^2y} \right) - \frac{1}{2} \ln R \, .
\end{align}
Within the region allowed by this $\Theta$~function, the standard DGLAP splitting functions are used.

Populating the region allowed by this $\Theta$~function constrains the allowed values of $z$,
\begin{align}
    0 < z < z_+ \, ,
    \quad
    z_+ = 1-\sqrt{\frac{p_{T}^2}{s_{ij}}R+\frac{p_{T}^4}{4s_{ij}^2}R^2}+\frac{p_{T}^2}{2s_{ij}}R.
\end{align}
Within this scheme, there is no kinematic limit, but requiring the final momenta to stay within the region where the spectator parton remains harder than the emitter and
emitted partons give a physically reasonable bound,
$p_{T,\max}^2 = s_{ij}/2$.

Finally, we note that we have not implemented the additional colour-factor corrections introduced for the NLL PanScales showers \cite{Hamilton:2020rcu} or the FHP shower \cite{Holguin:2020joq}, which limits the accuracy of these showers to leading-colour NLL for some global observables.\footnote{For non-global observables, or observables with $\geq 4$ coloured partons in the Born hard process, the shower accuracy is similarly limited to leading colour within the dipole shower approach.} Collinear spin correlations can be accommodated via the pre-existing implementation of the Collins--Knowles algorithm \cite{Collins:1987cp,Knowles:1988hu,Knowles:1988vs} within the Herwig dipole shower framework \cite{Richardson:2018pvo,Bellm:2019zci}, although to keep the analysis and comparisons in this paper simple, we leave this option switched off.

\subsection{The Shower Splitting Rate}
\label{sec:showersplitting}

For complete integration into an event generator, which is subject to matching and merging, we must account for the effect of the kinematic mapping on the emission matrix element and phase-space.
The kinematic mapping introduces an additional Jacobian when factorising the emission phase-space measure from that of the remaining particles in the event. This Jacobian is power suppressed, scaling as\footnote{The Jacobian factor for the \PG and FHP algorithms, derived in Appendix~\ref{sec:Jacobian}, has a leading correction for emissions after the first $\sim p_T/\sqrt{s_{ij}}$ that is odd in azimuth so averages to zero, leaving a correction $\sim p_T^2/s_{ij}$.} $1 + \mathcal{O}(p_T^2/s_{ij})$, and is therefore strictly subleading in tests of logarithmic accuracy. Nevertheless, it is an integral part of matching. In the PanScales implementation of PG$_{0}$, the Jacobian is included only for the first emission \cite{Hamilton:2023dwb,vanBeekveld:2025lpz}, which is sufficient for NLO matching. In contrast, the Herwig implementation of the CS shower includes the Jacobian for all emissions, amounting to a suppression of hard emissions that is essentially given by $1-y$ for the final-final dipole considered here. This Jacobian is included for all emissions with the aim of achieving completely flexible event generation.

Including Jacobian corrections that are formally of order $p_T^2/s_{ij}$, even though we neglect corrections of the same order in the matrix elements, might seem arbitrary or inconsistent. However, the approach we take is that the phase space mapping is under our control, and should be implemented exactly, i.e.\ we should include Jacobian factors in the distributions we generate. But power corrections to the matrix elements are out of our control, in the sense that they are not universal (process-independent). That is, neglecting the corrections in Eq.~(\ref{eq:factorisation}) below is a fundamental limitation in our accuracy, and we may wish to explore the implications of that limitation, whereas neglecting the terms in the Jacobian factors would be just an arbitrary choice of our particular shower scheme. We want to make our predictions as independent as possible of our arbitrary choices, and we can do so by retaining the full result for the phase space Jacobian factors. In some cases, these become zero or (integrably) divergent at phase space edges, and it becomes mandatory to include them to avoid wildly behaving results.

To derive the Jacobian for the PG$_{0}$ and FHP showers, we follow a procedure similar to that of \cite[ch.~5]{Ellis:1996mzs}, studying the effect of the kinematic mapping on both the factorisation of the squared matrix element and the partonic phase space (see also the related discussion in \cite{Loschner:2021keu}).
First, consider the matrix element squared.
When a pair of partons is much more collinear than all other pairs in the event, the $(n+1)$-parton squared matrix element $\left|\mathcal{M}_{n+1}\right|^2$ factorises onto the $n$-parton squared matrix element $\left|\mathcal{M}_{n}\right|^2$ as
\begin{align}
\Bigl|\mathcal{M}_{n+1} (q_1, \dots , q_n , q)\Bigr|^2 = \Bigl|\mathcal{M}_n(p_1, \dots , p_n) \Bigr|^2 \frac{8\pi\alpha_s}{(q_i + q)^2} P(z) \left(1+\mathcal{O}\left(\frac{p_T^2}{s_{ij}}\right)\right),
\label{eq:factorisation}
\end{align}
where $q_i$ and $q$ are the momenta of the collinear partons such that $q_i + q \approx p_i$ and $P(z)$ is the splitting kernel.
$p_i$ and $q$ are identified with the emitter and emission momenta defined in the kinematic mappings of the previous sections. For both \PG\ and FHP showers, this gives
\begin{align}
(q_i + q)^2  = 2q_i \cdot q = 2 \alpha^2 q_i'' \cdot q'' = 2\alpha^2yz^2 p_i \cdot p_j = \alpha^2yz^2 s_{ij} \, ,
\end{align}
and therefore
\begin{align}
\Bigl|\mathcal{M}_{n+1} (q_1, \cdots , q_n , q)\Bigr|^2 = \Bigl|\mathcal{M}_n(p_1, \cdots , p_n) \Bigr|^2 \frac{8\pi\alpha_s}{\alpha^2 yz^2s_{ij}} P(z) 
\left(1+\mathcal{O}\left(\frac{p_T^2}{s_{ij}}\right)\right).
\end{align}
The CS shower gives the same expression, but with $\alpha^2yz$ replaced by just $y$.

Different recoil schemes will give different mappings between the post- and pre-emission momenta, $\{q_1, \cdots , q_n , q\}$ and $\{p_1, \cdots , p_n\}$, together with the emission variables, $\{p_T, z, \phi\}$. The only absolute requirements are firstly that both should represent \emph{physical} events, i.e.\ with all momenta on-shell and with conservation of four-momentum, and secondly that in the soft and collinear limits ($p_T\to0$), they should represent kinematically indistinguishable events. One might wonder whether this arbitrariness can also be removed with a suitable Jacobian factor, but in general, this depends on the detailed properties of the matrix elements, which again are process dependent. Therefore, our approach in developing different shower schemes is to include the Jacobian factors to remove the scheme dependence in the phase space distribution that they cover, but different schemes will still differ in how well they match the dynamics of the underlying matrix elements. In particular, as was pointed out in Ref.~\cite{Dasgupta:2018nvj}, an emission that is, in itself, generated correctly according to its phase space and matrix elements, can push the remainder of the event into a region in which the approximation that was used to generate it is no longer valid, and thus undermine the logarithmic accuracy of the shower.

To derive these Jacobian factors, we consider the $n+1$ parton phase-space measure associated with $\bigl|\mathcal{M}_{n+1} (q_1, \cdots , q_n , q)\bigr|^2$, $\Phi_{n+1}$.
As we show in Appendix~\ref{sec:Jacobian}, this can be written in a factorised form,
\begin{equation}
\int \text{d}\Phi_{n+1} = \int \text{d}\Phi_{n} \int \frac{\mathrm{d}^4q''}{(2\pi)^4}(2\pi)\delta(q''^2) \alpha^6 z(1-zy){X}
\end{equation}
for the \PG\ scheme, the same expression without the $(1-zy)$ factor for the FHP scheme and with $\alpha^6z(1-zy){X}$ replaced by $(1-y)/z$ for the CS scheme.
The factor $X=p_{\text{dip}}\cdot Q/q_{\text{dip}}\cdot Q$ is the ratio of the dipole energy in the centre-of-mass frame, $Q$, before and after emission, and is equal to~1 for the first emission.
From here, the integral $\text{d}^4 q''$ can be replaced with the emission variables,
\begin{equation}
\frac{\text{d}^4q''}{(2\pi)^4}(2\pi)\delta(q''^2) =
\frac{1}{16\pi^2}{\text{d}p_T^2} \frac{d\phi}{2\pi}\frac{\text{d}z}{1-z}\,.
\end{equation}

We combine the matrix element and phase space to derive the differential cross section and, with it, the emission rate.
The final result is
\begin{equation}
\begin{aligned}
\text{d} \mathcal{P}^{\text{CS}} &= \frac{\text{d}p_T^2}{p_T^2} \text{d}z \frac{\text{d} \phi}{2\pi} \cdot \frac{\alpha_s^{\rm CMW}(p_T^2)}{2\pi} \cdot P(z,y) \cdot (1-y) \, , \\
\text{d} \mathcal{P}^{\text{PG}} &= \frac{\text{d}p_T^2}{p_T^2} \text{d}z \frac{\text{d} \phi}{2\pi} \cdot \frac{\alpha_s^{\rm CMW}(p_T^2)}{2\pi}\cdot  g^{\text{PS}}P(z) \cdot \alpha^4 (1-zy) {\cdot X} \, , \\
\text{d} \mathcal{P}^{\text{FHP}} &=  \frac{\text{d}p_T^2}{p_T^2} \text{d}z \frac{\text{d} \phi}{2\pi} \cdot \frac{\alpha_s^{\rm CMW}(p_T^2)}{2\pi}\cdot \Theta^{\text{FHP}} P(z)\cdot \alpha^4 {\cdot X} \, .
\end{aligned}
\end{equation}
All three showers use the CMW running coupling \cite{Catani:1990rr} with the scale evaluated at the shower $p_T$ ordering variable:
\begin{align}
    \alpha_s^{\rm CMW} (p_T^2)= \alpha_s(p_T^2)\left(1 + \frac{\alpha_s(p_T^2)}{2\pi}{K_{\rm CMW}}\right),
    \quad
    {K_{\rm CMW}=\left(\frac{67}{18} - \frac{\pi^{2}}{6}\right)C_{A} -\frac{5}{9}n_{f}} \,,
\end{align} 
where $\alpha_s(\mu^{2})$ is the 2-loop $\overline{\rm{MS}}$ coupling.

The general expressions for the rescaling factor $\alpha$ in the two schemes are quite complicated, encoding information about the boost of the dipole relative to the centre-of-mass frame, but for the first emission, in which the dipole frame is the centre-of-mass frame, they are given by
\begin{align}
    \alpha^2_{\text{PG}} &= \frac{Q^2}{Q^2-p_T^2} \, , \\
    \alpha^2_{\text{FHP}} &= \frac{Q^2}{Q^2+p_T^2\frac{z}{1-z}} \, .
\end{align}
Note that $\alpha^2_{\text{PG}}$ is divergent at the phase space boundary, $p_T^2=Q^2$, whereas $\alpha^2_{\text{FHP}}$ always lies between~1 and~$\frac23$.

\subsection{Collinear Limit~-- Running Coupling and Infrared Cutoff}
\label{sec:cutoff}

In the \PG\ and FHP schemes, to avoid transverse recoil effects generating spurious correlations between well-separated soft gluons, the local dipole recoil does not account for transverse momentum conservation. Instead, it is absorbed by the event as a whole. As we will discuss in this section, this has important consequences for the hard collinear limit of the showers. In this discussion, the \PG\ and FHP schemes have the same properties and are different from the CS scheme, which, in fact, has the same properties as Herwig's angular ordered parton shower (AO). Moreover, since we are discussing the collinear limit, $p_T^2\ll s_{ij}$, we can ignore the rescaling and boosting and treat the double-primed and un-primed momenta as locally equivalent.

To recap, the \PG\ and FHP schemes define momenta
\begin{align}
    q_{i\text{PG,FHP}} &=
    zp_i \, , \\
    q_{\text{PG,FHP}} &=
    (1-z)p_i + \frac{p_{T\text{PG,FHP}}^2}{2p_i\!\cdot\!p_j(1-z)}p_j
    -k_{\perp\text{PG,FHP}} \, ,
\end{align}
and the CS (and AO) schemes define momenta
\begin{align}
    \label{eq:CSAO1}
    q_{i\text{CS,AO}} &=
    zp_i + \frac{p_{T\text{CS,AO}}^2}{2p_i\!\cdot\!p_j\,z}p_j
    +k_{\perp\text{CS,AO}} \, , \\
    q_{\text{CS,AO}} &=
    (1-z)p_i + \frac{p_{T\text{CS,AO}}^2}{2p_i\!\cdot\!p_j(1-z)}p_j
    -k_{\perp\text{CS,AO}} \, .
    \label{eq:CSAO2}
\end{align}
In order to compare them, it is helpful to define a massless momentum in the direction of their common axis, $\sim q_i+q$, and a momentum transverse to this direction. In the \PG\ and FHP schemes, these are
\begin{align}
    \tilde{p}_i &= p_i + \frac{p_{T\text{PG,FHP}}^2}{2p_i\!\cdot\!p_j}p_j
    -k_{\perp\text{PG,FHP}} \, , \\
    \tilde{k}_\perp &= k_{\perp\text{PG,FHP}} - \frac{p_{T\text{PG,FHP}}^2}{p_i\!\cdot\!p_j}p_j \, ,
\end{align}
and their magnitudes are unchanged, $\tilde{p}_i\!\cdot\!p_j=p_i\!\cdot\!p_j$ and $\tilde{k}_{\perp\text{PG,FHP}}\!\cdot\!\tilde{k}_{\perp\text{PG,FHP}}=k_{\perp\text{PG,FHP}}\!\cdot\!k_{\perp\text{PG,FHP}}$. In terms of these momenta, the \PG\ and FHP momenta are given by
\begin{align}
    \label{eq:PGFHP1}
    q_{i\text{PG,FHP}} &=
    z\tilde{p}_i + \frac{z\,p_{T\text{PG,FHP}}^2}{2\tilde{p}_i\!\cdot\!p_j}p_j
    +z\tilde{k}_{\perp\text{PG,FHP}} \, , \\
    q_{\text{PG,FHP}} &=
    (1-z)\tilde{p}_i + \frac{z^2p_{T\text{PG,FHP}}^2}{2\tilde{p}_i\!\cdot\!p_j(1-z)}p_j
    -z\tilde{k}_{\perp\text{PG,FHP}} \, .
    \label{eq:PGFHP2}
\end{align}
For the CS and AO schemes, $p_i$ is already the common axis, and trivially, $\tilde{p}_i=p_i$ and $\tilde{k}_\perp=k_{\perp\text{CS,AO}}$, so Eqs.~(\ref{eq:CSAO1},\ref{eq:CSAO2}) are already relative to their common axis.

Thus, finally, we have to compare Eqs.~(\ref{eq:CSAO1},\ref{eq:CSAO2}) with Eqs.~(\ref{eq:PGFHP1},\ref{eq:PGFHP2}). We conclude that if both schemes populate the same phase space point, they describe it with different emission variables, $(z,p_{T\text{PG,FHP}})$ and $(z,p_{T\text{CS,AO}})$,
with
\begin{equation}
    p_{T\text{CS,AO}} =
    z \, p_{T\text{PG,FHP}} \, .
\end{equation}
In the soft and soft collinear limits, $1\!-\!z\ll1$, the two definitions are equivalent and the difference is insignificant. However, in the hard collinear limit, $z\sim1\!-\!z$, the difference is potentially important, and we have to understand its consequences.

It is natural (although, as we now discuss, not mandatory) to use, in a given scheme, its definition of $p_T$ as the scale of the running coupling, $\alpha_s(p_T^2)$, and as the variable on which the infrared cutoff is defined, $p_T>p_{T,\min}$, and all of the schemes we are discussing do this by default.

In order to discuss the running coupling effect, it is helpful to recall the argument for choosing a particular scale. The NLO collinear $q \to q$ splitting function was studied by Dasgupta and El-Menoufi~\cite{Dasgupta:2021hbh}, who showed that it can be written in the form
\begin{align}
    \label{eq:Dasgupta1}
    \frac1{\sigma_0}\mathrm{d}\sigma &= 
    C_F\frac{\alpha_s(\mu_R^2)}{2\pi}\,
    \frac{\mathrm{d}\theta_g^2}{\theta_g^2}\,
    \mathrm{d}z
    \Biggl[
    \frac{1+z^2}{1-z}
    \Biggl\{
    1+\frac{\alpha_s}{2\pi}
    \Biggl(
    -b_0\ln\frac{(1-z)^2\theta_g^2E^2}{\mu_R^2}
    +K_{\rm CMW}
    \nonumber\\&\hspace{13em}
    -b_0\ln(z)
    \Biggr)
    \Biggr\}
    +\frac{\alpha_s}{2\pi}b_0(1-z)
    +\frac{\alpha_s}{2\pi}R^{\text{nab.}}(z)
    \Biggr],
\end{align}
where $\theta_g$ is the angle between the quark and total momentum of the emitted parton(s), and $R^{\text{nab.}}(z)$ is a smooth function proportional to $C_A$, whose exact form is not needed here.
The different terms are written to distinguish their role in logarithmic accuracy. The $b_0$ term on the first line is clearly related to the scale of the running coupling. In fact, we can identify the numerator of its logarithm as the \PG and FHP definition of the transverse momentum of the gluon, relative to the direction of the quark, $p_{T\text{PG,FHP}}^2=(1-z)^2\theta_g^2E^2$. This term can be absorbed by choosing to evaluate the running coupling at a renormalization scale set equal to this transverse momentum. At the same time, we can also rewrite $\mathrm{d}\theta_g^2/\theta_g^2=\mathrm{d}p_{T\text{PG,FHP}}^2/p_{T\text{PG,FHP}}^2$, to give
\begin{align}
    \label{eq:Dasgupta2}
    \frac1{\sigma_0}\mathrm{d}\sigma &= 
    C_F\frac{\alpha_s(p_{T\text{PG,FHP}}^2)}{2\pi}\,
    \frac{\mathrm{d}p_{T\text{PG,FHP}}^2}{p_{T\text{PG,FHP}}^2}\,
    \mathrm{d}z
    \Biggl[
    \frac{1+z^2}{1-z}
    \Biggl\{
    1+\frac{\alpha_s}{2\pi}
    \Biggl(
    K_{\rm CMW}
    \nonumber\\&\hspace{13em}
    -b_0\ln(z)
    \Biggr)
    \Biggr\}
    +\frac{\alpha_s}{2\pi}b_0(1-z)
    +\frac{\alpha_s}{2\pi}R^{\text{nab.}}(z)
    \Biggr].
\end{align}
With this choice, integrating the leading order splitting function sums the tower of LLs, $\alpha_s^n\log^{n+1}(p_{T,\max}^2/p_{T,\min}^2)$, to all orders, as well as a tower of NLLs. Next, the $K_{\rm CMW}$ term, together with the soft part of the splitting function, gives another tower of NLLs. In early implementations of the CMW factor \cite{Catani:1990rr,Marchesini:1991ch}, it was conventional to consider it a change of scheme, i.e.~to absorb $K_{\rm CMW}$ into the denominator of the solution of the running coupling equation, or, equivalently, to evaluate the running coupling at a scale $p_T^2\exp(-K_{\rm CMW}/b_0)\approx0.4p_T^2$. Instead, more recent implementations treat $K_{\rm CMW}$ as a universal correction, rather than a scheme, leaving it in the numerator. To the NLL accuracy at which $K_{\rm CMW}$ is first relevant, either implementation is equally valid, and they only differ by higher order corrections. This illustrates a general point, that at a given order, one is free to trade corrections between those that are summed into the denominator, and those that are left in the numerator.

The terms in the second line of Eq.~(\ref{eq:Dasgupta2}) all contribute at NNLL and, as such, can be neglected in our NLL shower implementations. Nevertheless, seeing the structure of the first term, $b_0$ times a $z$-dependent logarithm, it is tempting to absorb this into the scale of the coupling and evaluate $\alpha_s$ at $z(1-z)^2\theta_g^2E^2=zp_{T\text{PG,FHP}}^2$. Interestingly, this is the scale used in the first dipole shower algorithm, ARIADNE~\cite{Lonnblad:1992tz}. Although this choice absorbs the $-b_0\ln(z)$ term into the running coupling, the remaining terms in the second line (or at least their $z$ integral~\cite{vanBeekveld:2024wws}) must still be included as a correction for NNLL accuracy. From this point of view, one could equally well choose $z^2p_{T\text{PG,FHP}}^2=p_{T\text{CS,AO}}^2$, as long as one supplies a correction $\mathbf{+}b_0\ln(z)$.

In summary, we see that in a NLL-accurate shower, we can choose to use $p_{T\text{PG,FHP}}^2$, $zp_{T\text{PG,FHP}}^2$ or $z^2p_{T\text{PG,FHP}}^2=p_{T\text{CS,AO}}^2$ as equally-valid scales. Even in an NNLL-accurate shower, we can choose any of these scales, provided the correct term is added back into the NNLL correction factor to compensate. The $g \to g$ splitting function has not been analysed to the same level of detail, so the effect of these scale choices for this case cannot be established at present.

The infrared cutoff of a parton shower plays two roles. Firstly, it prevents the parton shower from entering non-perturbative regions where the running coupling blows up. For this reason, it is sensible to cut off the distributions using the same definition of $p_T$ as is used for the running coupling. Without this, one may have to add some non-perturbative physics to the running coupling, e.g.\ a freezing scale. Secondly, one can view the infrared cutoff as a matching scale between two components of the full event generation model, the parton shower and hadronization \cite{Hoang:2018zrp,Platzer:2022jny,Hoang:2024nqi}. If one understood hadronization well enough, different choices of the form of the cutoff, as well as different values, would be compensated between the two components, in the same way that the factorisation scheme and scale dependence cancel between perturbative calculations and parton distribution or fragmentation functions. In the absence of this level of understanding, it again becomes a practical question, which form of cutoff is best matched to a particular hadronization model.

\begin{figure}[t]
\centering
\begin{tikzpicture}
\draw (0, 0) node[inner sep=0] {\includegraphics[width=0.30\linewidth]{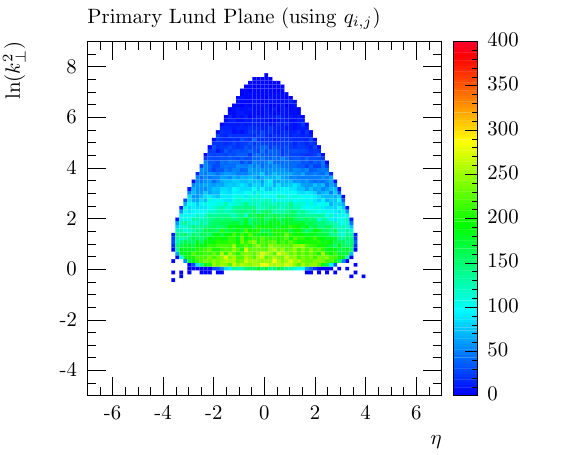}};
\draw (-0.2,-0.9) node {\small CS};
\end{tikzpicture}
\begin{tikzpicture}
\draw (0, 0) node[inner sep=0] {\includegraphics[width=0.30\linewidth]{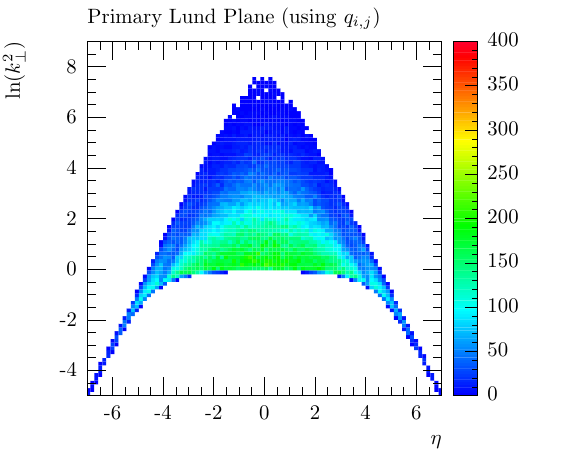}};
\draw (-0.14,-0.9) node {\small \PG};
\end{tikzpicture}
\begin{tikzpicture}
\draw (0, 0) node[inner sep=0] {\includegraphics[width=0.30\linewidth]{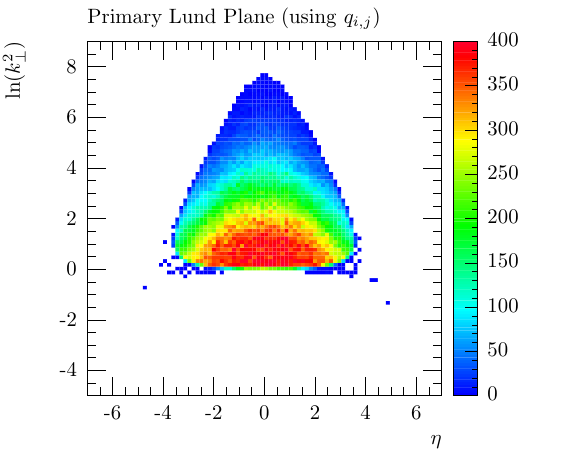}};
\draw (-0.1,-0.9) node {\small \PGAOB};
\end{tikzpicture}
\begin{tikzpicture}
\draw (0, 0) node[inner sep=0] {\includegraphics[width=0.30\linewidth]{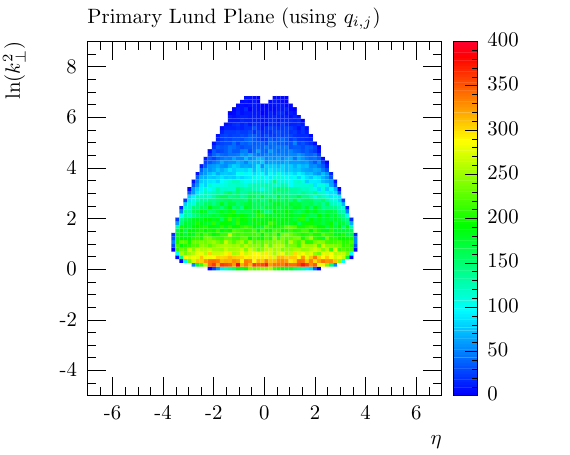}};
\draw (-0.2,-0.9) node {\small AO};
\end{tikzpicture}
\begin{tikzpicture}
\draw (0, 0) node[inner sep=0] {\includegraphics[width=0.30\linewidth]{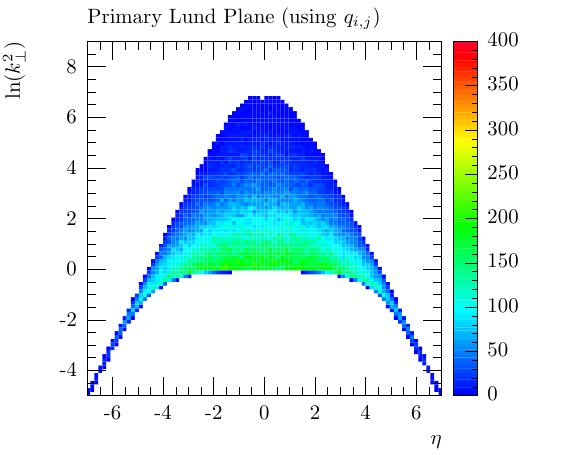}};
\draw (-0.17,-0.9) node {\small FHP};
\end{tikzpicture}
\begin{tikzpicture}
\draw (0, 0) node[inner sep=0] {\includegraphics[width=0.30\linewidth]{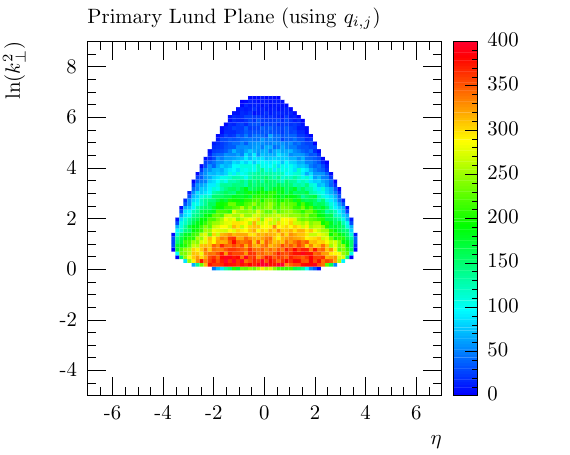}};
\draw (-0.1,-0.9) node {\small \FHPAOB};
\end{tikzpicture}
\caption{Lund planes for events with a single emission. The CS and AO showers share the same lower boundary, $p_{T\text{CS,AO}}^2>p_{T,\min}^2\Rightarrow k_\perp^2>p_{T,\min}^2/z$, as do the \PG\ and FHP showers, $p_{T\text{PG,FHP}}^2>p_{T,\min}^2\Rightarrow k_\perp^2>p_{T,\min}^2\times z$. The CS shower has a small scatter of events beyond this boundary, due to hard anti-collinear emissions, which are power-suppressed by the partitioning but nonetheless present. Also shown are the results of supplementing the \PG\ and FHP showers with an angular-ordered-like shower boundary (``AOB''), $p_{T\text{PG,FHP}}^2>p_{T,\min}^2/z^2\Rightarrow k_\perp^2>p_{T,\min}^2/z$. We see that, indeed, the kinematic limit becomes like the AO one. We also see that the \PG\ shower gives a small scatter of hard backward events like the CS shower (these are also present in the standard \PG\ events, but overshadowed by the larger number of hard forward events). The strict $\Theta$-function partitioning of both the FHP and AO schemes does not allow such backward emission. The ``temperatures'' of the heat maps are not directly comparable, because the distributions are normalised to the fraction of single emission events, which is not the same in the different schemes.}
\label{fig:Lund}
\end{figure}
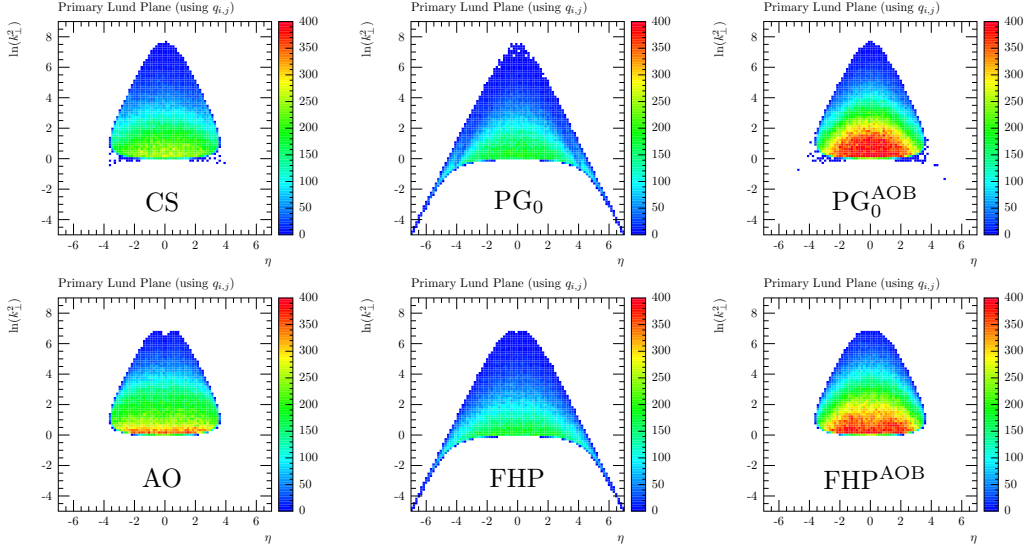

To illustrate this point, we show the distribution in the Lund plane \cite{Andersson:1988gp,Andersson:1990dp,Dreyer:2018nbf} of events for which there is exactly one emission, in each of the schemes we consider. Various, slightly different, definitions of the Lund plane variables are in use; we use the original ones of \cite{Andersson:1990dp},
\begin{equation}
    k_\perp^2 = \frac{2q_i\!\cdot\!q\,2q_j\!\cdot\!q}
    {2p_i\!\cdot\!p_j}\,, \qquad
    \eta = \frac12\ln\frac{2q_j\!\cdot\!q}{2q_i\!\cdot\!q}\,,
\end{equation}
with the event orientated so that the quark is $i$ (positive rapidity) and the antiquark is $j$ (negative). We can easily check that this definition of $k_\perp$ is actually the intermediate between $p_{T\text{CS,AO}}$ and $p_{T\text{PG,FHP}}$ discussed above. The results are shown in Figure~\ref{fig:Lund}, for $e^+ e^- \to q\bar{q}$ events with $Q=91.2~\text{GeV}$ and $p_{T,\min}^2=1~\text{GeV}^2$.

We see clearly the very different boundaries that cuts on the shower ordering variables dictate. As we have argued, we cannot say which of these cutoffs is better. But, given that we will be interfacing these showers with Herwig's cluster hadronization model, which has been developed over several decades together with Herwig's angular-ordered shower to describe hadron-level data well, one might wonder whether the angular-ordered-like boundary of the CS shower would match better with this hadronization model than the very different \PG\ and FHP boundary. In particular, the latter will produce many more events for the model to hadronize with an extremely hard gluon and extremely soft quark.

To test this idea, whether the hadron-level results are improved by imposing an angular-ordered-like boundary on the \PG\ and FHP showers, we also introduce modified versions, which we call the AOB versions, in which emissions with $p_{T\text{PG,FHP}}<p_{T,\min}/z$ are vetoed. We see in Figure~\ref{fig:Lund} that this veto has had the desired effect on the generated distributions. We shall study below whether this has an impact on the hadron-level results.

One might also wonder whether this additional veto, requiring $p_{T\text{PG,FHP}}>p_{T,\min}/z$ also affects the hard region. We can see from the Lund plane distributions that it primarily affects a small region in the bottom corner (whose area remains constant as $p_{T,\min}\to0$). The region it affects becomes a vanishingly small strip for $k_\perp^2\gtrsim10p_{T,\min}^2$. Viewed in the Dalitz plane for \PG, this strip corresponds to the region of extremely hard gluon emission, $x_q+x_{\bar{q}}\sim1$. For example, at the point at which a hard gluon recoils from an equal energy quark-antiquark pair, it limits the gluon energy to be smaller than $Q/2-p_{T,\min}/4$. Thus the region of extremely high $p_T\to Q$, where the Jacobian factor diverges, is vetoed, with the maximum allowed value of $p_T$ being $Q-p_{T,\min}/2$ and hence the maximum possible value of $\alpha^2$ being $Q/p_{T,\min}$. For FHP, since this strip is in the region in which the spectator is not the hardest parton, which is anyway vetoed by the partition function, the additional veto does not affect high-$p_T$ emission.

\subsection{NLO Matching}

In this section, we detail how we match the different dipole showers to an NLO calculation in order to facilitate meaningful comparisons to LEP data.
While the CS and \PG\ schemes allow the phase space to be filled, the FHP shower in particular requires NLO matching to ensure that the hard-emission region is filled. We use the
MC@NLO method~\cite{Frixione:2002ik}, building on the infrastructure of Herwig's Matchbox module \cite{Platzer:2011bc}. Matchbox is flexible enough in regards to the shower algorithms, and allows the inclusion of general kinematic mappings through according remapping factors from the underlying subtraction term kinematics (in this case, by default, the CS subtraction \cite{Catani:1996vz}). More details of this approach can be found in \cite[ch. 5]{Bellm:2025pcw}.

The general result for what we call the H~events, with real-emission kinematics from an $n$-parton process, is
\begin{equation}
    \sigma^H[\text{PS}(\phi_{n+1})] =
    \int \left(
    \text{d}\sigma^{R}(\phi_{n+1})
    -\sum_i \text{d}\sigma^{PS}_{(i)}(\phi_{n+1})
    \right)\text{PS}(\phi_{n+1}),
\end{equation}
where $\text{PS}(\phi_{n+1})$ represents the parton shower algorithm starting from the $n\!+\!1$-parton configuration. The general result for the S~events, with Born kinematics, contains shower-related terms
\begin{equation}
    \sigma^S[\text{PS}(\phi_n)] \ni
    \sum_i
    \int \left(
    \text{d}\sigma^{PS}_{(i)}(\phi_n,r)
    -\text{d}\sigma^{A}_{(i)}(\phi_n,r)
    \right)\text{PS}(\phi_n),
\end{equation}
where $r$ are emission variables and $\text{d}\sigma^{A}_{(i)}$ is the NLO subtraction term, calculated in the Catani-Seymour algorithm. In both these expressions, $\text{d}\sigma^{PS}_{(i)}(\phi_{n+1})=\text{d}\sigma^{PS}_{(i)}(\phi_n,r)$ is the analytical expression of the distribution generated by a parton shower emission.

In order to construct $\text{d}\sigma^{PS}_{(i)}(\phi_n,r)$, we need to address the mismatch between the shower distribution and the QCD subtraction terms. To this end, we consider two {\it distinct} phase space factorisations from the three-particle phase space to the two-particle phase space. Generally, these phase space mappings satisfy
\begin{equation}
  {\cal M}(\phi_3) {\rm d}\phi_3 = {\cal M}(\psi_3(\phi_2,r)) 
  {\cal J}_\psi(\phi_2,r) {\rm d}\phi_2 {\rm d}r
\end{equation}
where $r$ are the radiation variables and $\phi_n$ are the phase space points. ${\cal J}$ is the Jacobian of interest. The mappings are also invertible in the sense that
\begin{equation}
\psi_3(\psi_2(\phi_3),R_\psi(\phi_3)) = \phi_3 \qquad
\psi_2(\psi_3(\phi_2,r)) = \phi_2\qquad R_\psi(\psi_3(\phi_2,r)) = r \ .
\end{equation}
We now face two such mappings: a mapping $\chi$ from the matching subtraction, which is the standard CS one which covers all of phase space ({\it i.e.}, there is no point where ${\cal J}_\chi$ vanishes except on boundaries of phase space), and $\psi$, which implements the shower emission.

Our particular concern is with functions that depend on an underlying Born phase space point of a different nature,
\begin{equation}
    {\cal M}(\phi_3){\rm d}\phi_3 = {\cal F}(\phi_3,\chi_2(\phi_3),R_\chi(\phi_3)){\rm d}\phi_3 \ ,
\end{equation}
which in particular applies to the virtual and real shower subtractions, using a shower mapping $\chi$ which we aim to replace by a subtraction term mapping $\psi$. In the case of the virtual shower subtraction, the change in mapping can be accompanied by a ratio of Jacobians
\begin{equation}
    {\cal F}(\phi_3,\chi_2(\phi_3),R_\chi(\phi_3)){\rm d}\phi_3 = 
    {\cal F}(\chi_3(\psi_2,R_\chi),\psi_2,R_\chi)\frac{{\cal J}_\chi}{{\cal J}_\psi}\left|\frac{\partial R_\chi}{\partial R_\psi}\right|{\rm d}\phi_3 \ ,
\end{equation}
where all terms are now functions of the real emission phase space point. We can therefore evaluate virtual shower subtractions as a re-weighted class of subtraction terms, which will generate the same underlying Born kinematics, since there is no explicit dependence on the remapped three particle phase space. For evaluating the underlying Born matrix element for the real shower subtraction, we required the inverse shower mapping. This can be found for the PG${}_0$ and FHP showers, in the case of a three-particle final state, as follows:
\begin{eqnarray}
    p_i &=& \frac{1}{\alpha\ z}\Lambda^{-1} q_i\\\nonumber
    p_j & = & \frac{1}{\alpha\ \kappa}\Lambda^{-1} q_j
\end{eqnarray}
with $\kappa = 1-yz$ for PG${}_0$ and $\kappa = 1$ for FHP. The Lorentz transform is now required to map
\begin{equation}
    \Lambda Q = \frac{1}{\alpha}\left(Q-q +\frac{1-z}{z}q_i+\frac{1-\kappa}{\kappa}q_j\right)
\end{equation}
giving us
\begin{equation}
    \alpha^2 = \frac{1}{Q^2}\left(\frac{1}{z}q_i+\frac{1}{\kappa}q_j\right)^2 \ .
\end{equation}

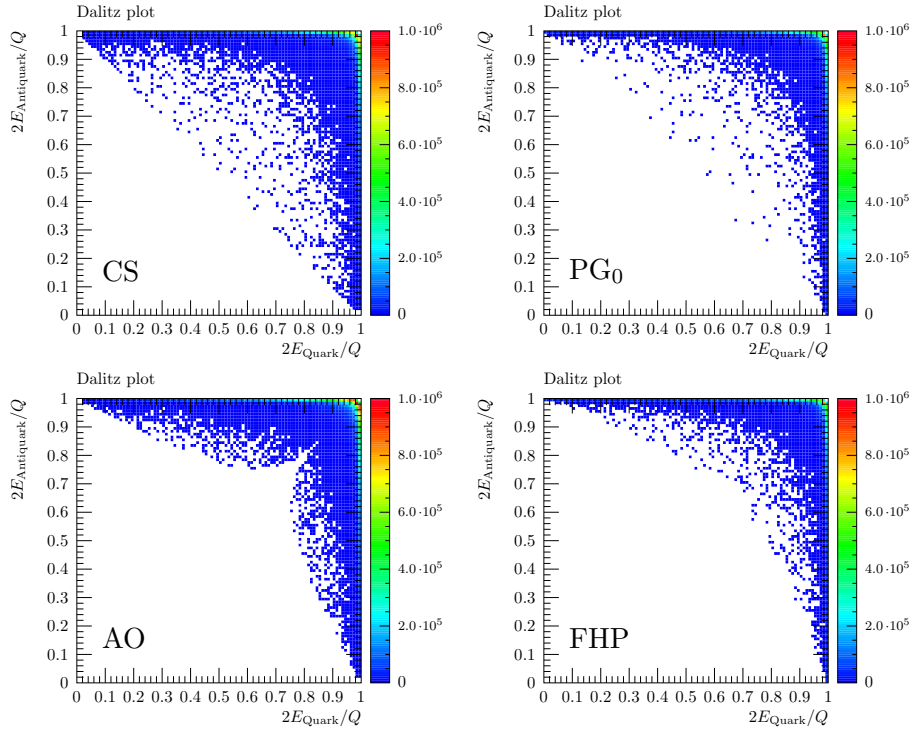
\begin{figure}[!htbp]
\centering

\begin{tikzpicture}
\draw (0, 0) node[inner sep=0] {\includegraphics[width=0.40\linewidth]{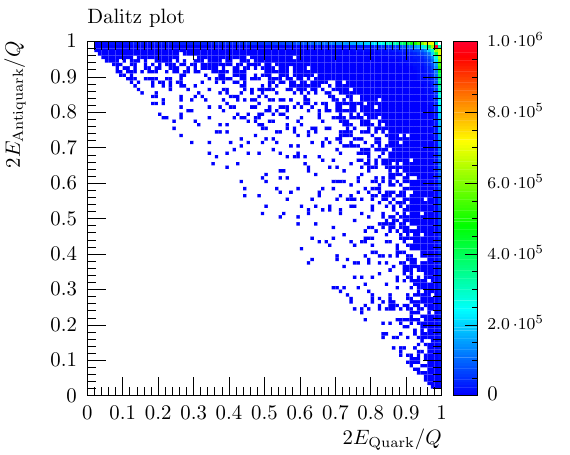}};
\draw (-1.9,-1.2) node[anchor=west] {CS};
\end{tikzpicture}
\begin{tikzpicture}
\draw (0, 0) node[inner sep=0] {\includegraphics[width=0.40\linewidth]{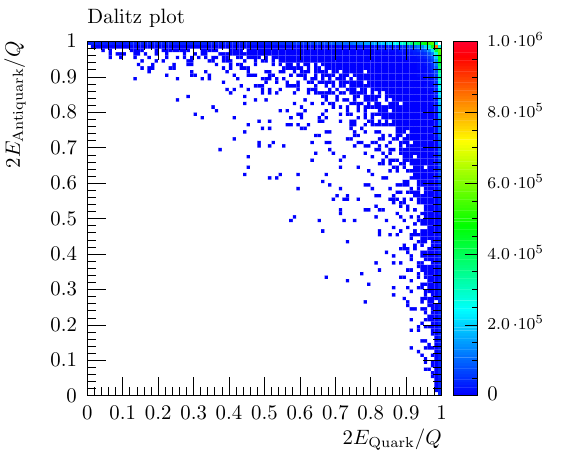}};
\draw (-1.9,-1.2) node[anchor=west] {\PG};
\end{tikzpicture}
\begin{tikzpicture}
\draw (0, 0) node[inner sep=0] {\includegraphics[width=0.40\linewidth]{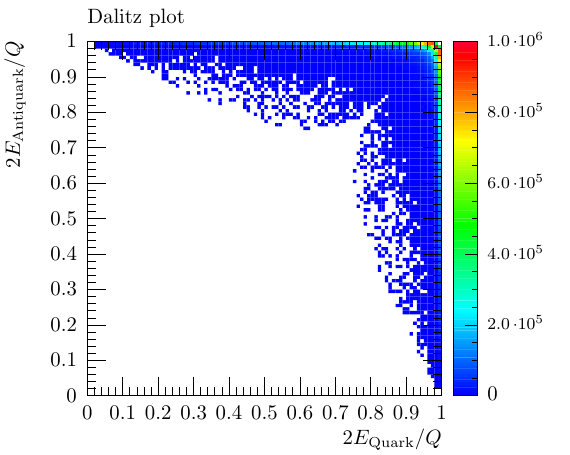}};
\draw (-1.9,-1.2) node[anchor=west] {AO};
\end{tikzpicture}
\begin{tikzpicture}
\draw (0, 0) node[inner sep=0] {\includegraphics[width=0.40\linewidth]{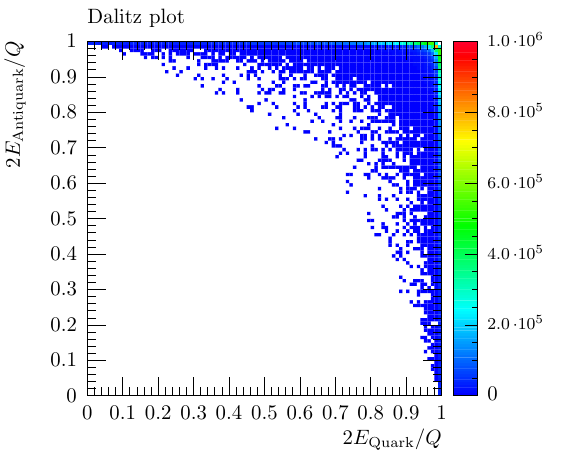}};
\draw (-1.9,-1.2) node[anchor=west] {FHP};
\end{tikzpicture}

\label{fig:dalitz-dalitz}

\caption{Dalitz plots illustrating the three-jet phase-space population from the first shower emission off a $q\bar{q}$ dipole (without NLO matching). Although the Dalitz plot density differs, both CS and \PG\ occupy the same phase space. In contrast, the AO and FHP showers do not fill the entire first-emission phase space. The AO shower's Dalitz plot forms the well-known butterfly pattern, which carves out the region populated by very hard emissions. The FHP shower similarly does not completely fill the region populated by very hard emissions, instead filling up to the symmetric Mercedes star point.}
\label{fig:dalitz}
\end{figure}

In this way, we can construct the shower subtraction terms in either direction, $\text{d}\sigma^{PS}_{(i)}(\phi_3)$ or $\text{d}\sigma^{PS}_{(i)}(\phi_2,r)$, for each of the shower schemes. It is useful to compare them with the real emission distribution, $\text{d}\sigma^{R}(\phi_3)$. We find that all three are good approximations in the phase space region that they cover. For example, near the `Mercedes star' point, i.e.~$T=2/3$, at which all three final-state partons have the same energy, the CS shower is $\frac{9}{8}=1.125$ times the real matrix element, while the \PG\ and FHP showers are both $\frac{15}{16}=0.938$ times it. However, by construction, the FHP shower only covers the region $p_T^2<Q^2/2$ and does not produce events in which the emitted gluon is harder than both the emitter and the spectator, whereas the other two showers both fill the full phase space. This is illustrated in Figure~\ref{fig:dalitz}, where the Dalitz plot for the first gluon emission from the initial $q \bar{q}$ dipole is shown. This is produced by simulating at parton-level an $e^+ e^- \to q\bar{q}$ process and selecting the events with only one emission. We do not normalise the plots, so one count is one event. Also shown is Herwig's angular-ordered shower, AO, for comparison. We see that CS and \PG\ fill the whole of phase space, whereas FHP and AO do not fill the region of very hard emission. The NLO matching correction will consequently be much more significant for them.

\begin{figure}[!htbp]
\centering
\begin{tikzpicture}
\draw (0, 0) node[inner sep=0] {\includegraphics[width=0.30\linewidth]{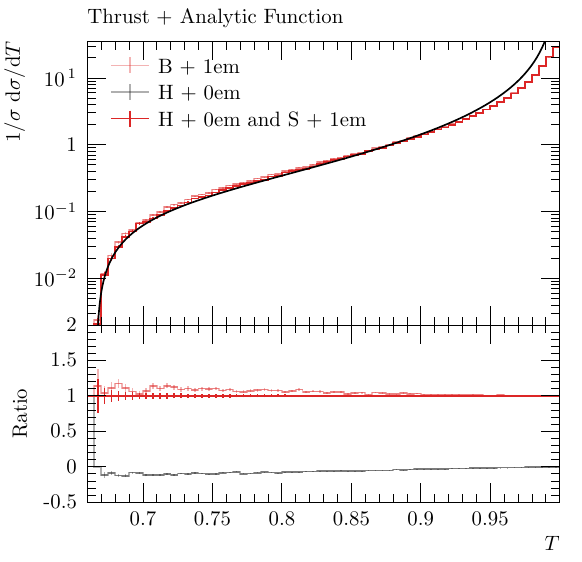}};
\draw (2.,0.) node[anchor=east] {\small CS};
\end{tikzpicture}
\begin{tikzpicture}
\draw (0, 0) node[inner sep=0] {\includegraphics[width=0.30\linewidth]{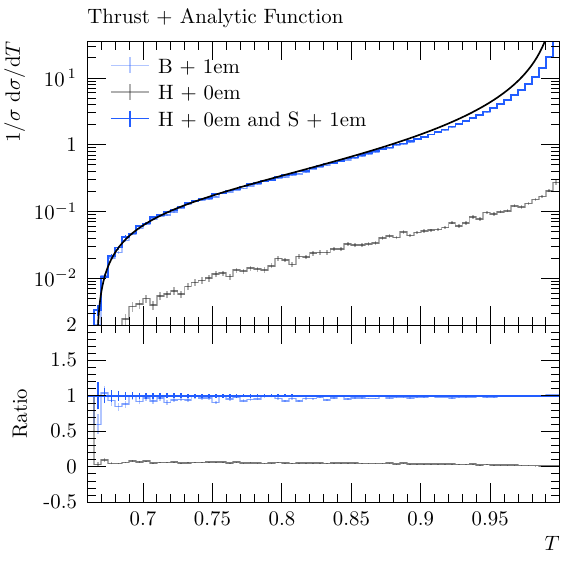}};
\draw (2.,0.) node[anchor=east] {\small \PG};
\end{tikzpicture}
\begin{tikzpicture}
\draw (0, 0) node[inner sep=0] {\includegraphics[width=0.30\linewidth]{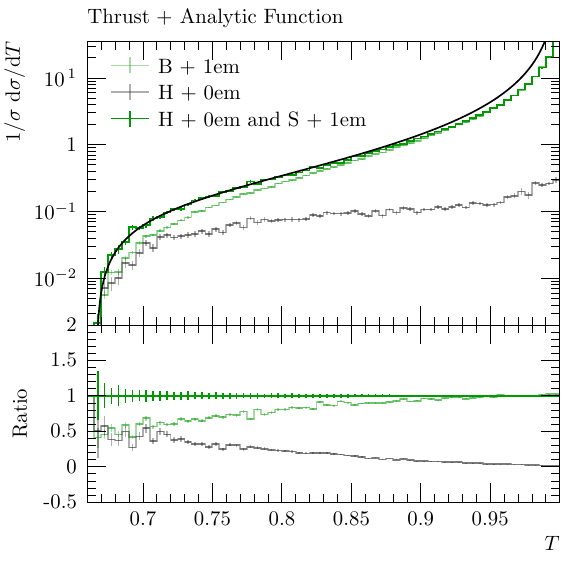}};
\draw (2.,0.) node[anchor=east] {\small FHP};
\end{tikzpicture}
\caption{\label{fig:matchingclosure} Thrust distributions for the matching closure tests. The simulations were all run with fixed $\alpha_s$. The smooth curve on each is the analytic function for the thrust distribution.}
\end{figure}
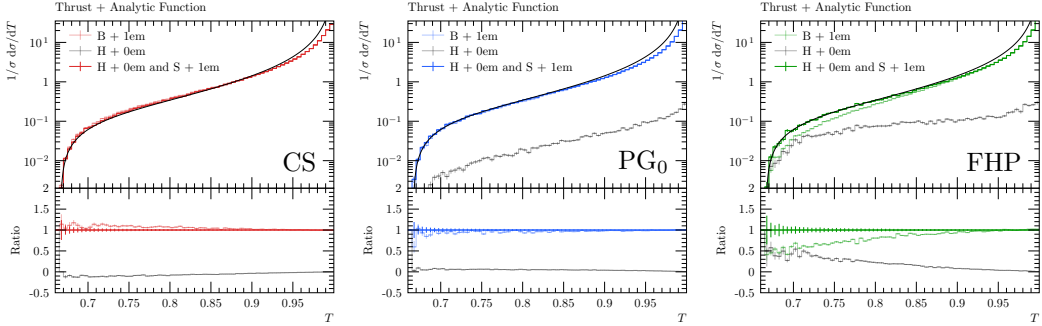

As a test of the construction of $\text{d}\sigma^{PS}_{(i)}$, and hence $\text{d}\sigma^H$, we show in Figure~\ref{fig:matchingclosure} the thrust distribution produced by S events with a single shower emission, and H events without emission, both with a fixed coupling $\alpha_s(m_Z)$. The expectation is that at small $T$, the sum of the two should reproduce the real matrix element, while at large $T$, they should start to differ due to the Sudakov suppression of the shower when required to produce only one emission. To test this, we compare them with the analytic thrust formula at this order,
\begin{equation}
\left. \frac{1}{\sigma} \frac{\text{d} \sigma}{\text{d} \tau} \right|_{\text{NLO}} = {\delta(1-T)+}
C_F \frac{\alpha_s}{2\pi} \left( \frac{2(3T^2 - 3T + 2)}{T(1-T)} \ln \frac{2T -1}{1-T} -  \frac{3(3T-2)(2-T)}{1-T} \right)_{+} \, ,
\label{eq:thrust-nlo-calc}
\end{equation}
{shown as the black curves in Figure~\ref{fig:matchingclosure}.}
We see that this is the case for all three and that, as anticipated, the H events are a small contribution for CS (where they are negative) and \PG\ (positive), whereas they are an essential contribution for FHP, reaching a factor of~2 at the Mercedes star point.

\subsection{Parton-Level Results}

\begin{figure}[!htbp]
\centering

\begin{subfigure}{\linewidth}
\centering
\includegraphics[width=0.30\linewidth]{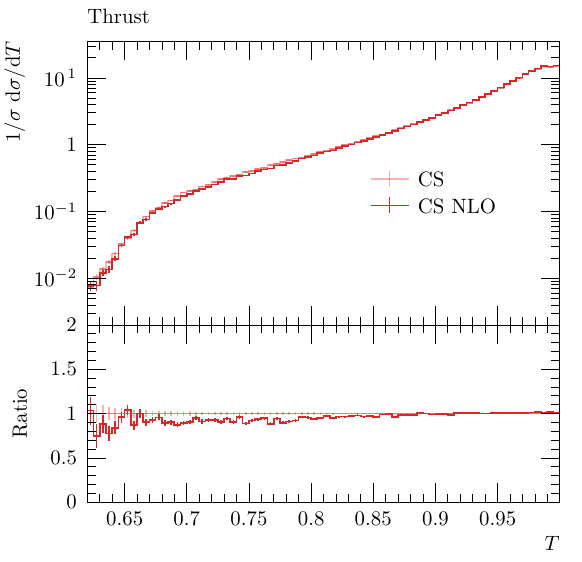}
\includegraphics[width=0.30\linewidth]{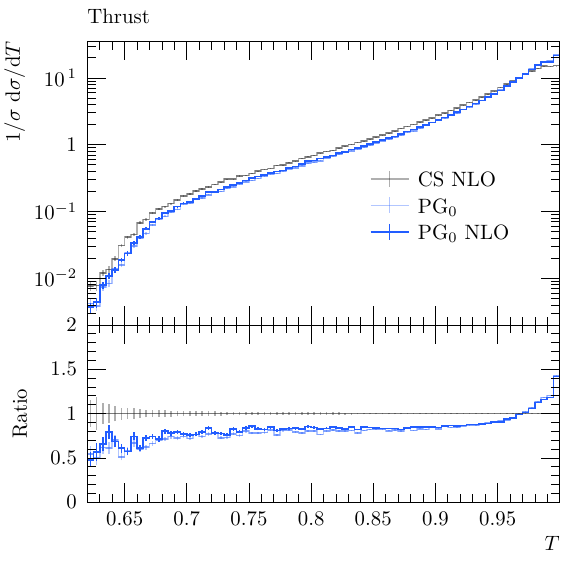}
\includegraphics[width=0.30\linewidth]{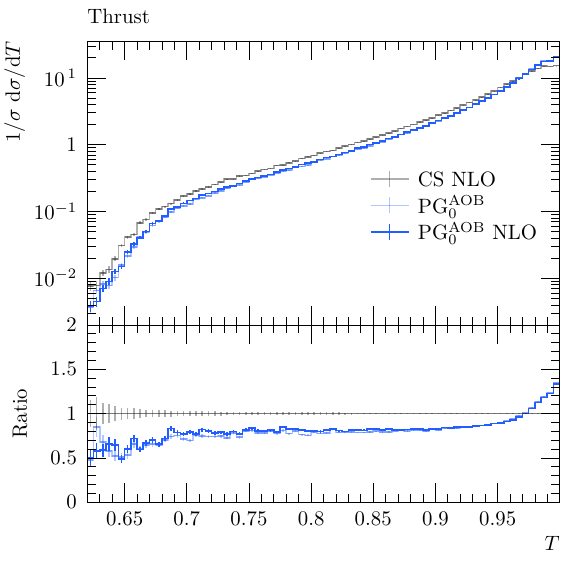}
\includegraphics[width=0.30\linewidth]{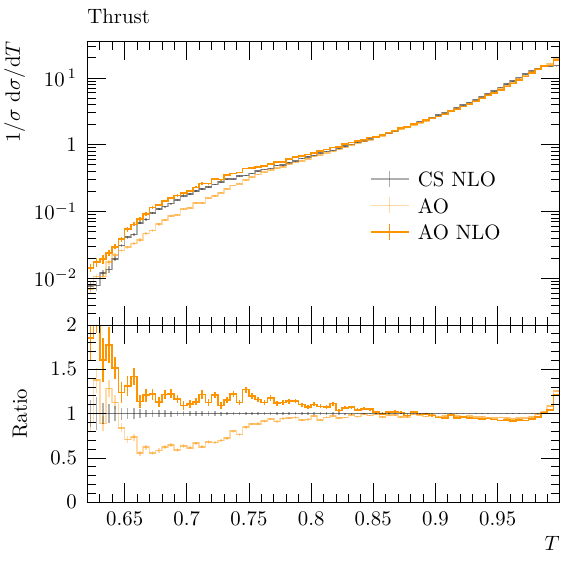}
\includegraphics[width=0.30\linewidth]{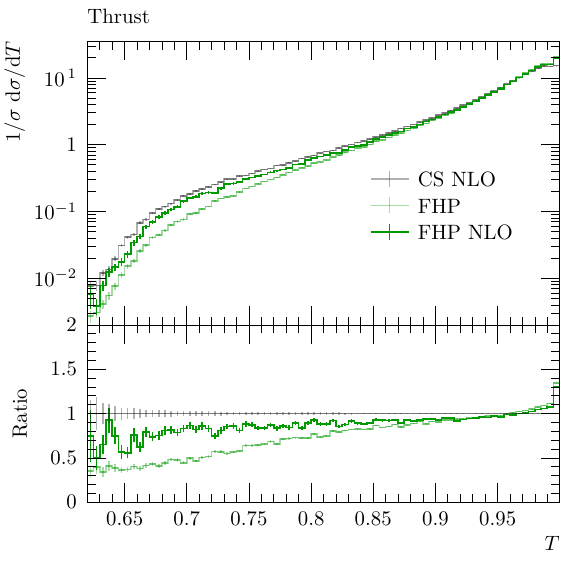}
\includegraphics[width=0.30\linewidth]{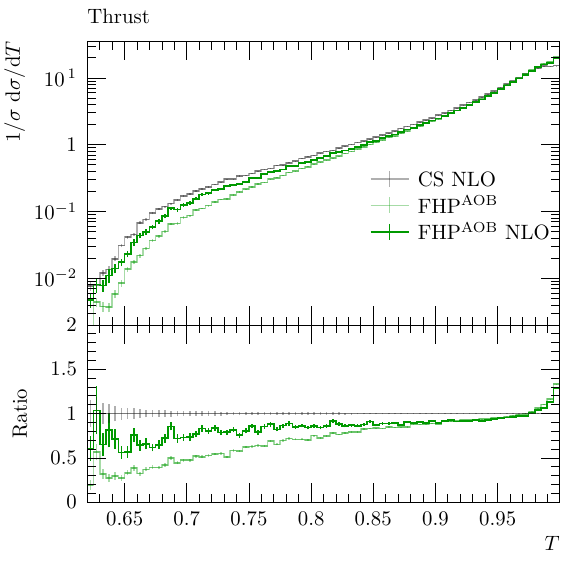}
\caption{Thrust, $T$}
\label{fig:first-look-thrust}
\end{subfigure}

\vspace{1em}

\begin{subfigure}{\linewidth}
\centering
\includegraphics[width=0.30\linewidth]{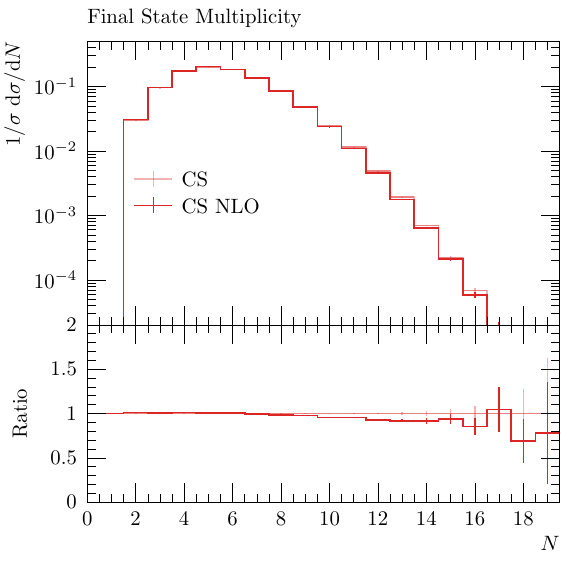}
\includegraphics[width=0.30\linewidth]{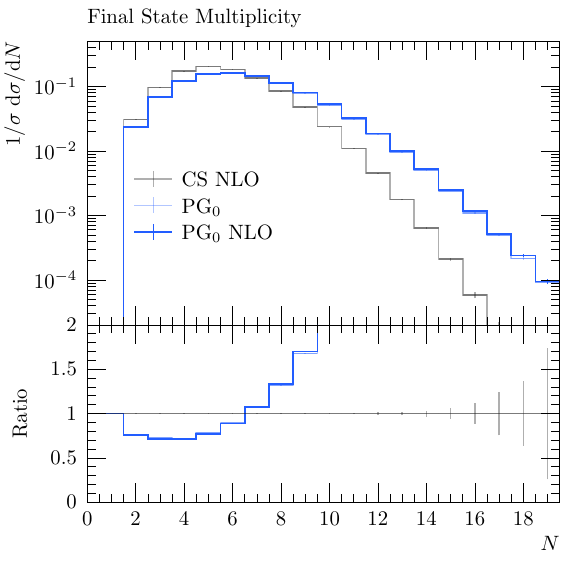}
\includegraphics[width=0.30\linewidth]{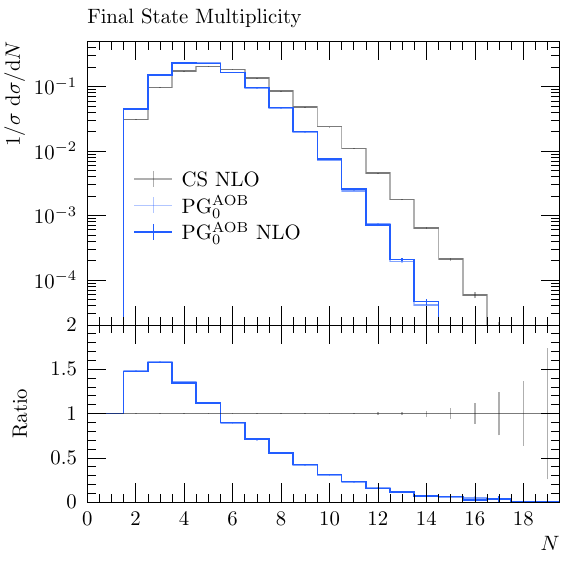}
\includegraphics[width=0.30\linewidth]{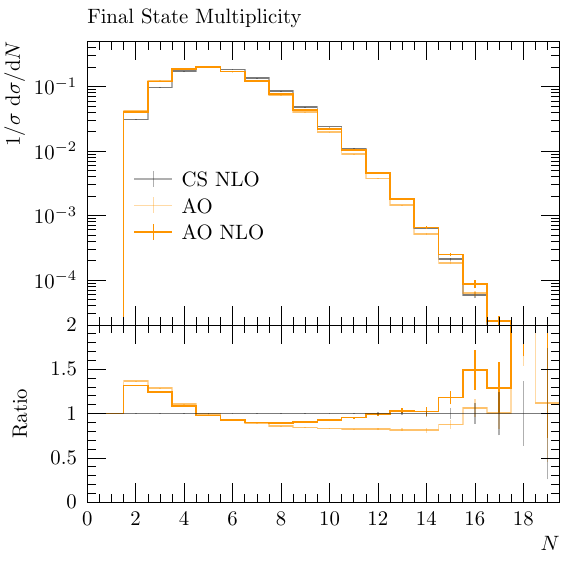}
\includegraphics[width=0.30\linewidth]{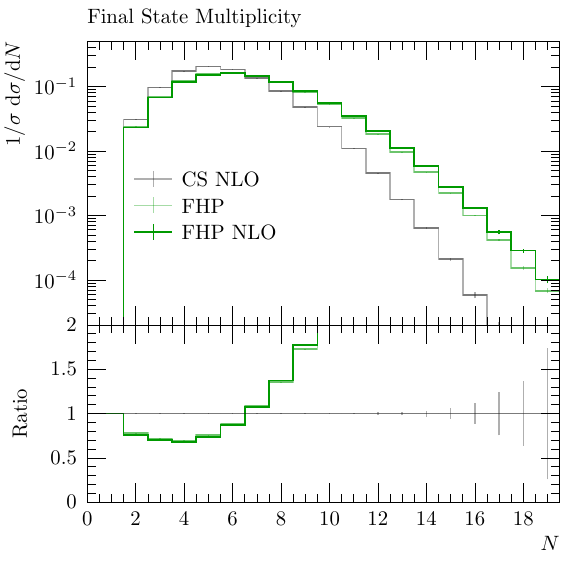}
\includegraphics[width=0.30\linewidth]{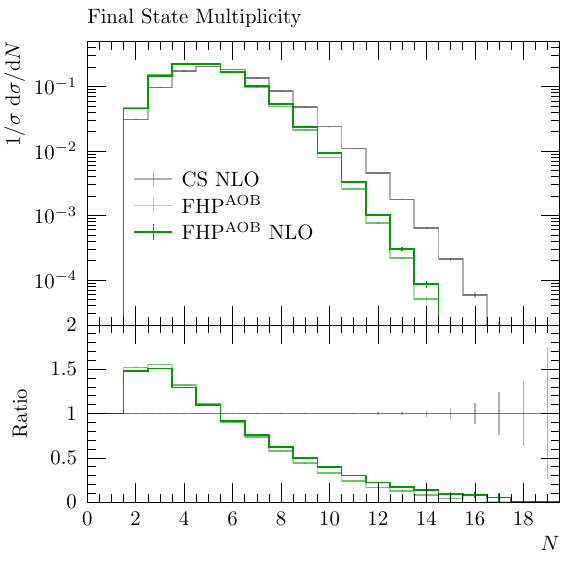}
\caption{Final State Multiplicity}
\label{fig:first-look-nump}
\end{subfigure}

\caption{Thrust and final-state multiplicity distributions for the four showers (CS, AO, \PG, FHP), and for \PG\ and FHP with the angular-ordered-like infrared cutoff boundary. For the CS shower, NLO matching reduces the tails of the distribution, whereas for \PG, adjustments occur as mild increases throughout the tails of the distributions. For FHP, NLO matching has a similar effect to \PG, but with a larger contribution to the high-$p_T$ tail of the distributions. NLO matching is necessary to fill the high-$p_T$ tails of the AO shower distributions. \PG\ and FHP with the angular-ordered-like boundary both exhibit dampening in the deeply infrared region.
}
\label{fig:first-look-thrust-chargedmult}
\end{figure}

\begin{figure}[!htbp]
\centering

\begin{subfigure}{\linewidth}
\centering
\includegraphics[width=0.30\linewidth]{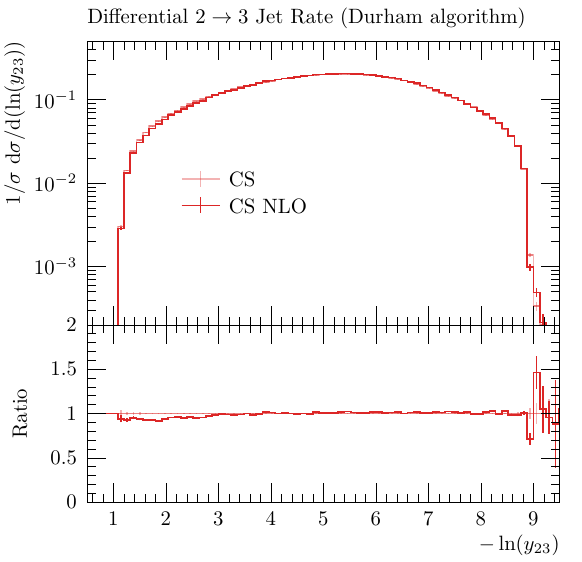}
\includegraphics[width=0.30\linewidth]{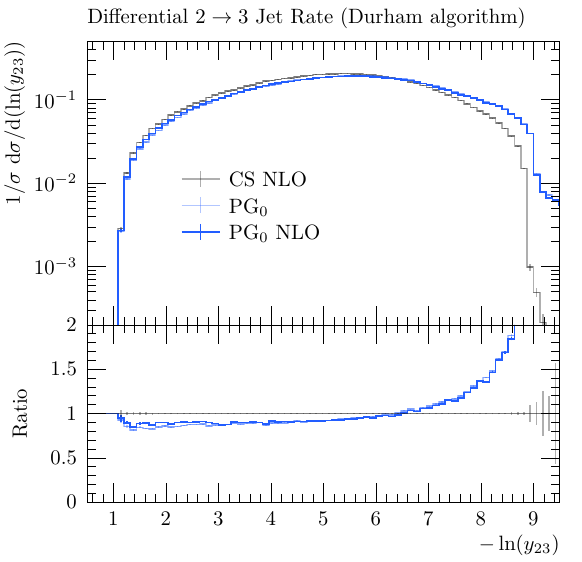}
\includegraphics[width=0.30\linewidth]{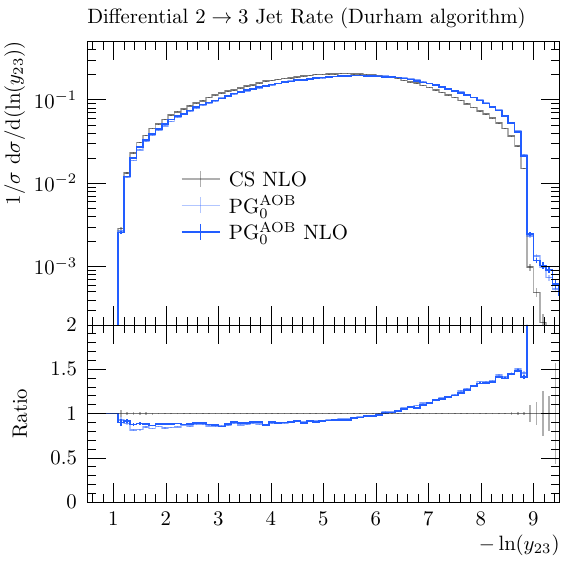}
\includegraphics[width=0.30\linewidth]{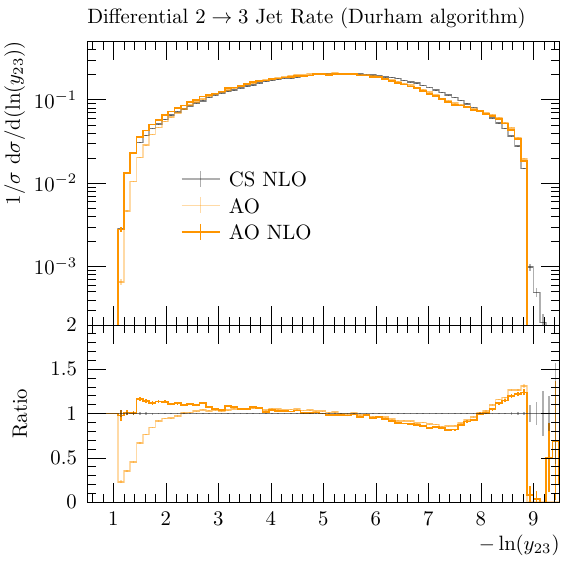}
\includegraphics[width=0.30\linewidth]{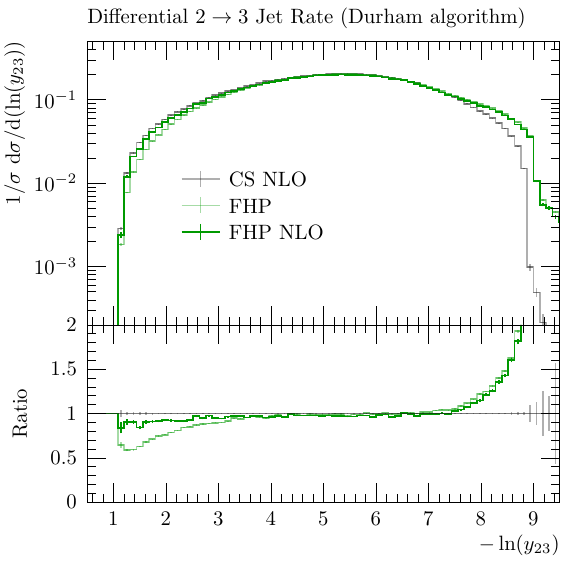}
\includegraphics[width=0.30\linewidth]{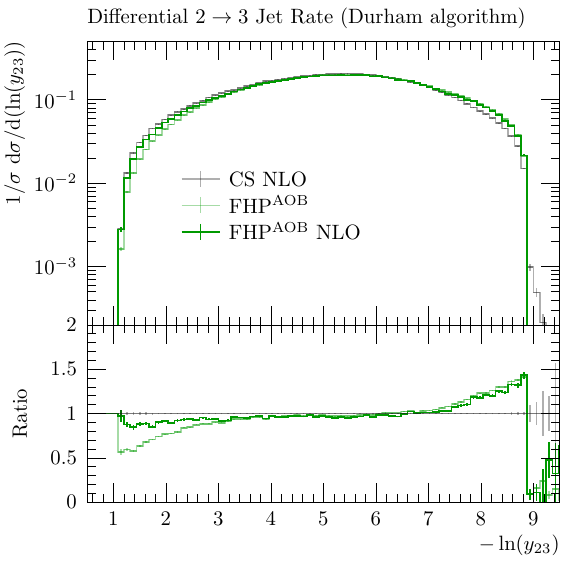}
\caption{$-\ln y_{23}$}
\label{fig:first-look-y23}
\end{subfigure}

\vspace{1em}

\begin{subfigure}{\linewidth}
\centering
\includegraphics[width=0.30\linewidth]{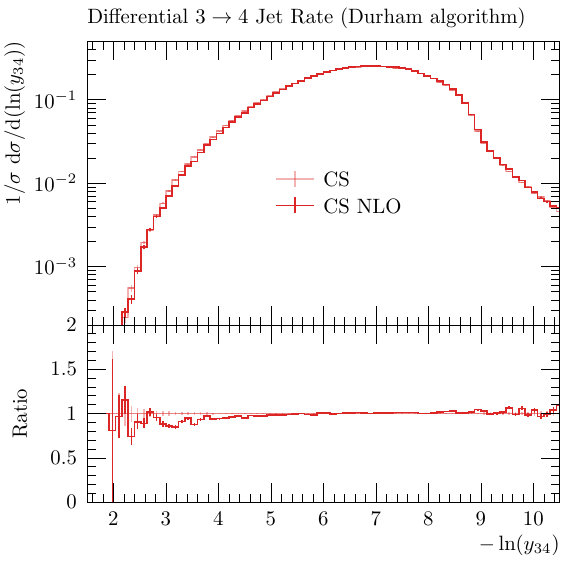}
\includegraphics[width=0.30\linewidth]{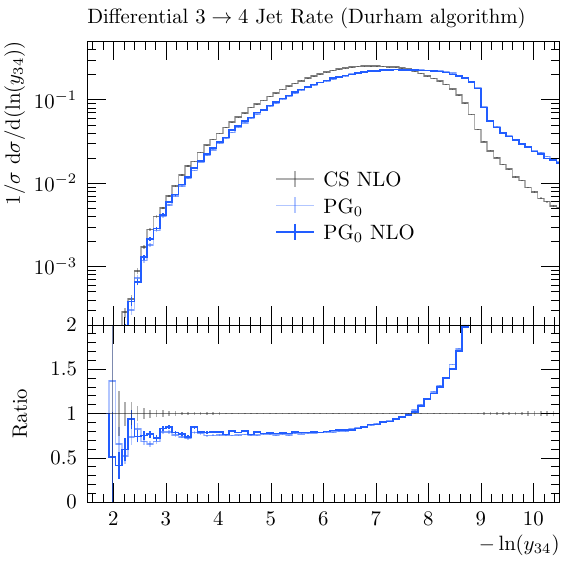}
\includegraphics[width=0.30\linewidth]{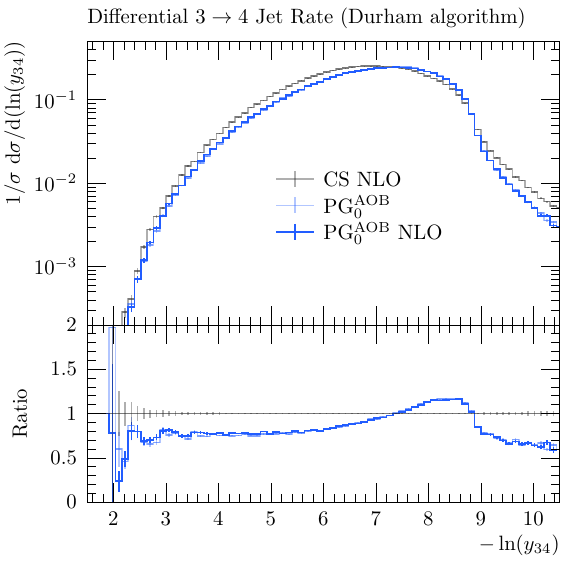}
\includegraphics[width=0.30\linewidth]{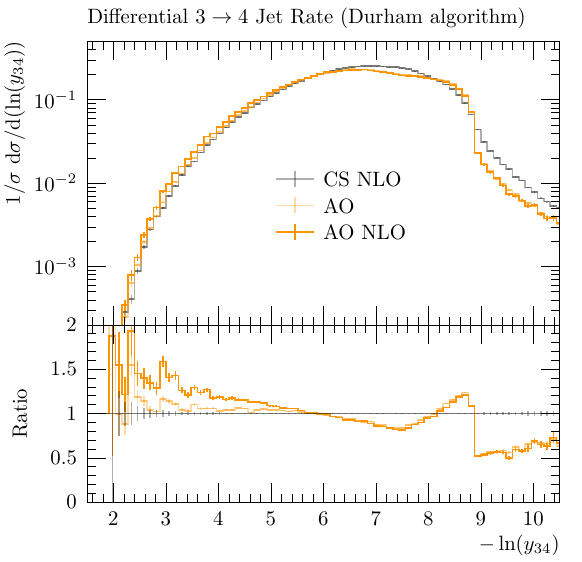}
\includegraphics[width=0.30\linewidth]{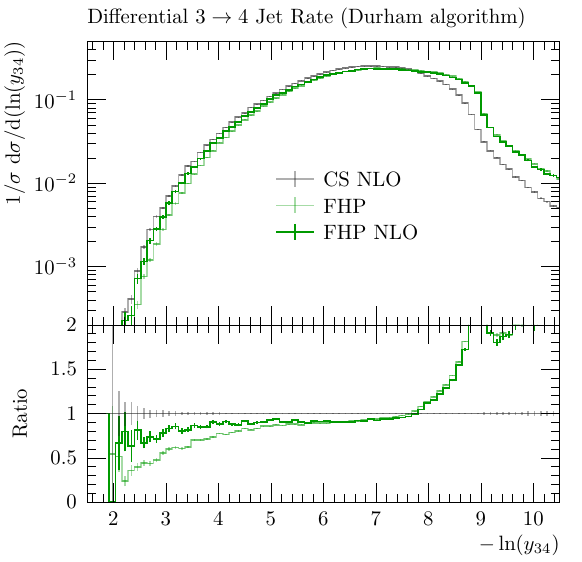}
\includegraphics[width=0.30\linewidth]{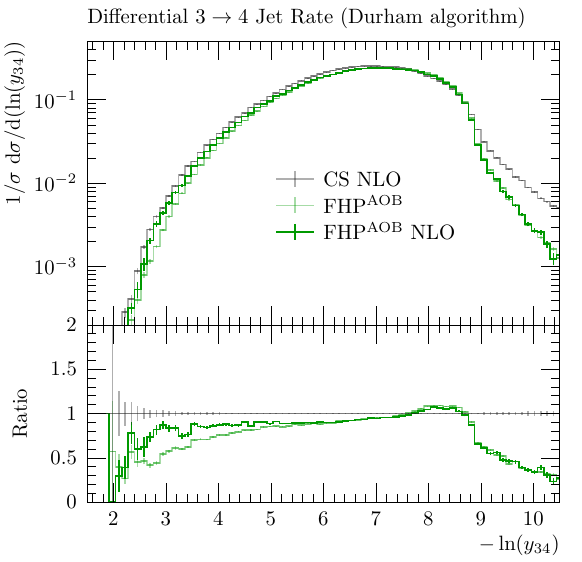}
\caption{$-\ln y_{34}$}
\label{fig:first-look-y34}
\end{subfigure}

\caption{The $-\ln y_{23}$ and $-\ln y_{34}$ distributions for the four showers (CS, AO, \PG, FHP), and for \PG\ and FHP with the angular-ordered-like infrared cutoff boundary. For the CS shower, NLO matching has a minimal effect. For \PG\ and FHP, adjustments lead to mild increases throughout the tails of the distributions, as in Figure~\ref{fig:first-look-nump}. NLO matching is necessary to fill the high-$p_T$ tails of the AO shower distributions. \PG\ and FHP with the angular-ordered-like boundary both exhibit dampening in the deeply infrared region.}
\label{fig:first-look-y23-y34}
\end{figure}

Having discussed the formulation of these three showers, we now consider their parton-level output in order to examine the breadth of their predictions from the infrared limit to the effects of NLO matching. In addition to the `out-of-the-box' FHP and \PG\ showers, we also present results for these showers where we enforce a shower cutoff boundary that matches the angular-ordered shower boundary. To illustrate the output of these showers, we consider exemplary infrared and collinear-safe distributions: Thrust, and the $y_{23}$ and $y_{34}$ distributions, displayed in Figures~\ref{fig:first-look-thrust} and~\ref{fig:first-look-y23-y34}, as well as the infrared and collinear-unsafe multiplicity distribution in Figure~\ref{fig:first-look-nump}. For each case, we use the CS shower as a baseline and additionally show the Herwig angular-ordered shower for further reference.
For all showers, we generated $2 \times10^6$ events using the canonical parameter values $\alpha_s(m_Z) = 0.118$ and $p_{T,\min}^2 = 1 \text{ GeV}^2$.

Both the \PG\ and FHP shower schemes under-populate the fixed-order regions relative to the CS shower. There are several reasons for this. Firstly, as mentioned in the previous section, the effective matrix elements they embody have different extrapolations away from the soft and collinear limits. In particular, at the Mercedes star point, CS is 20\% above the other two. Secondly, all the schemes use their $p_T$ value as the scale of $\alpha_s$ in the shower, but $m_Z$ in the H event. Not only does this mean that the smaller the H-event contribution (and CS's is negative) the larger the rate, but also the different $p_T$ definitions have different extrapolations away from the soft and collinear limit: at the Mercedes star point, the CS scale is $p_T=m_Z/\sqrt{12}$, giving $\alpha_s(p_T)\approx1.22\alpha_s(m_Z)$, the \PG\ scale is $p_T=m_Z/2$, giving $\alpha_s(p_T)\approx1.11\alpha_s(m_Z)$ and the FHP scale is $p_T=m_Z/\sqrt{2}$, giving $\alpha_s(p_T)\approx1.05\alpha_s(m_Z)$. Moreover, all three schemes use the CMW correction in the shower but not the H event, giving a further approximately $0.06\alpha_s(m_Z)$ relative to the fixed-coupling results shown in the previous section. 
Finally, FHP has a dead region: only half of the region surrounding the Mercedes star point is covered.
Putting together all these factors, we expect their rates near $T=\frac23$ to be factors
\begin{eqnarray}
    \mbox{CS:}&&
    \phantom{\stackrel{\text{dead}}{\textstyle\frac12}
    \times{}}
    \stackrel{\text{ME}}{1.125}
    \times
    \;(\stackrel{\alpha_s(p_T)}{1.22}
    +
    \stackrel{\text{CMW}}{0.06})\;
    -
    \stackrel{\text{H event}}{0.125}
    =1.32
    \nonumber\\
    \mbox{\PG:}&&
    \phantom{\stackrel{\text{dead}}{\textstyle\frac12}
    \times{}}
    \stackrel{\text{ME}}{0.938}
    \times
    \;(\stackrel{\alpha_s(p_T)}{1.11}
    +
    \stackrel{\text{CMW}}{0.06})\;
    +
    \stackrel{\text{H event}}{0.062}
    =1.16
    \\
    \mbox{FHP:}&&
    \stackrel{\text{dead}}{\textstyle\frac12}
    \times
    \stackrel{\text{ME}}{0.938}
    \times
    \;(\stackrel{\alpha_s(p_T)}{1.05}
    +
    \stackrel{\text{CMW}}{0.06})\;
    +
    \stackrel{\text{H event}}{0.531}
    =1.05
    \nonumber
\end{eqnarray}
times larger than the fixed-order result. Or, normalising to CS, \PG\ should be 12\% lower and FHP 20\% lower. Similar effects are seen at the hard end of the jet rate distributions.
Therefore, for the same value of $\alpha_s$, \PG and FHP can be expected to be below CS, or, conversely, to describe the same data, they will require a larger $\alpha_s$ value.

Turning to the angular-ordered-like boundary (AOB) versions, we see a significant effect at the low Durham-$k_T$ end of the jet rates. With the default boundary, the rate of very low Durham-$k_T$ emission is significantly higher than in CS, while the AOB version is similar to CS, although still a little above. The final state multiplicity shows very large differences, with \PG\ and FHP showing significantly more emissions than CS and their AOB versions significantly fewer.

The hard region of thrust and $y_{23}$ are improved by the NLO matching and this feeds to some extent into the hard region of $y_{34}$, which is determined by two hard emissions. Further improvement to this region would require matching to the 4-parton tree-level matrix elements, as was first explored for a dipole shower in~\cite{Andersson:1991he}.

%% file: sections/cluster.tex
\section{Cluster Hadronization and Cluster Masses}\label{sec:cluster}

At the end of the parton shower, the event arrives at a state with many partons with energies of order the infrared parton shower cutoff ({\it i.e.} around $1\ {\rm GeV}$), where the transition to hadrons begins.
Perturbative QCD fails to describe how such partons transition into hadrons. First-principles theoretical approaches to describe this regime are limited to reactions with a very small multiplicity of hadrons, and consequently, very exclusive measurements, \cite{Bauer:2000yr,Beneke:2003zv,Brambilla:2004jw,Bulava:2022ovd} or simple bulk properties of the hadronic energy-flow \cite{Webber:1994cp,Dokshitzer:1995zt,Dokshitzer:1995qm,Salam:2001bd,Mateu:2012nk,Farren-Colloty:2025amh}.
More complex final states will need to be described by phenomenological models, which also scale to larger multiplicities. Novel developments in this area have only recently taken place, within both the major paradigms of the string and cluster hadronization models.

The focus of this paper, and the model we use, is the cluster hadronization model.
Experimental and phenomenological studies have demonstrated that spectra of hadrons contain comparable properties as the spectra of partons after parton showering \cite{Azimov:1984np}, motivating theories of ``preconfinement'' properties of QCD \cite{Amati:1979fg, Marchesini:1980cr, Mazzanti:1981nn}, where partons at this stage are organised in phase space such that neighbouring partons together form colour singlets with finite masses, independent of the hard interaction, addressing potential issues of maintaining colour conservation when the original $q\bar{q}$ pair is so far apart.
These ideas developed into the theory of \textit{Local Parton-Hadron Duality}, which implies a local transition in phase space, where nearby partons form hadrons and should not be affected by the rest of the system \cite[ch. 7]{Dokshitzer:1991wu}.
The resulting hadron system in this picture moves along the direction of the partons.
This idea is used to devise an algorithm that locally converts pairs of partons into colour-singlet pseudoparticles, called ``clusters'', which decay into hadrons.
This model is designed to complement the parton shower and relies on the showers' (physics) accuracy to predict hadrons.
Although many further developments have been incorporated in the Herwig cluster hadronization model \cite{Gieseke:2017clv, Hoang:2018zrp, Gieseke:2018gff, Gieseke:2019wcl, Masouminia:2023zhb, Hoang:2024nqi, Gieseke:2025mcy}, summarised in \cite[ch. 7]{Bellm:2025pcw}, for our study, we used the simplest form of the algorithm: we disable intrinsic transverse momentum and colour reconnection, in line with the earlier Herwig++ releases \cite{Bahr:2008pv}.
This choice of a rather minimal hadronization model does not make any further assumptions about the factorisation between partons and hadrons, and, as shown by the charged multiplicity in Figure~\ref{fig:first-look-thrust-chargedmult}, yields markedly different results for the showers.
In this section, we first provide an overview of the cluster hadronization model implemented in Herwig and specify the relevant parameters.
We then simulate the different showers with cluster hadronization and generate mass spectra of clusters formed both in the final state and after cluster fission.

\subsection{Cluster formation and cluster fission}

\begin{figure}[t]
\centering
\begin{tikzpicture}
\draw (0, 0) node[inner sep=0] {\includegraphics[width=0.30\linewidth, trim={3cm 1.5cm 2cm 0.5cm}, clip]{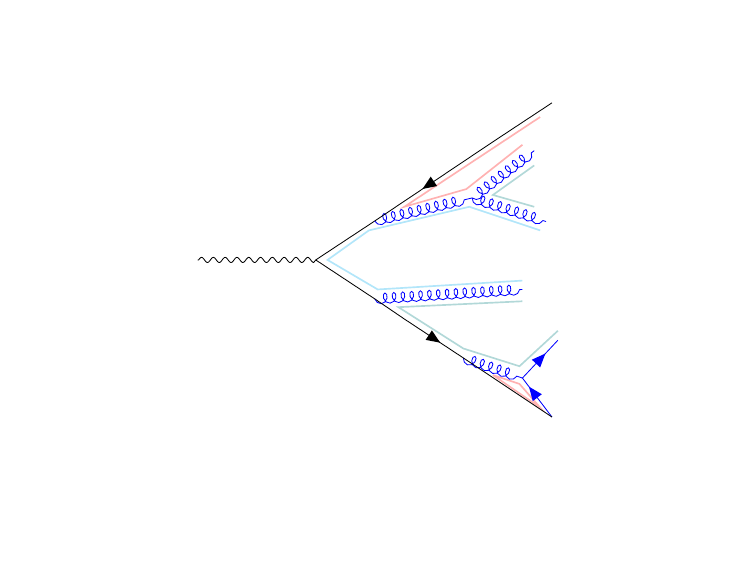}};
\draw (-1.8, 1.1) node {\small (a)};
\end{tikzpicture}\hfill
\begin{tikzpicture}
\draw (0, 0) node[inner sep=0] {\includegraphics[width=0.30\linewidth, trim={3cm 1.5cm 2cm 0.5cm}, clip]{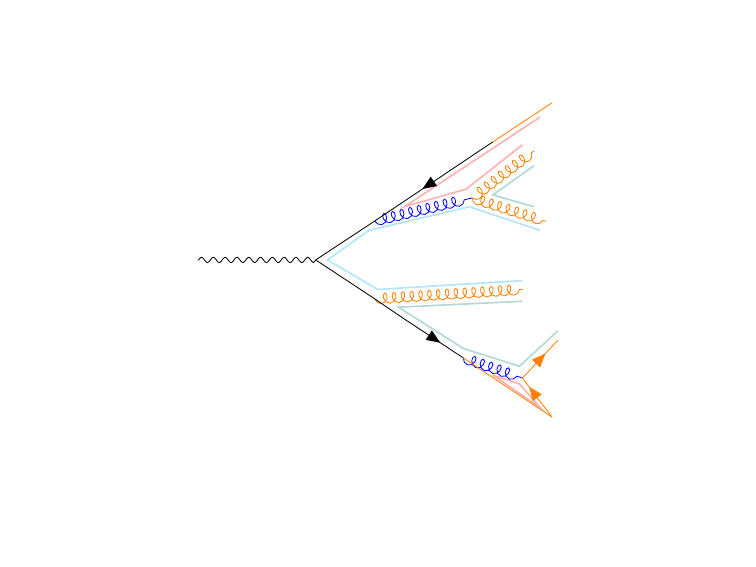}};
\draw (-1.8, 1.1) node {\small (b)};
\end{tikzpicture}\hfill
\begin{tikzpicture}
\draw (0, 0) node[inner sep=0] {\includegraphics[width=0.30\linewidth, trim={3cm 1.5cm 2cm 0.5cm}, clip]{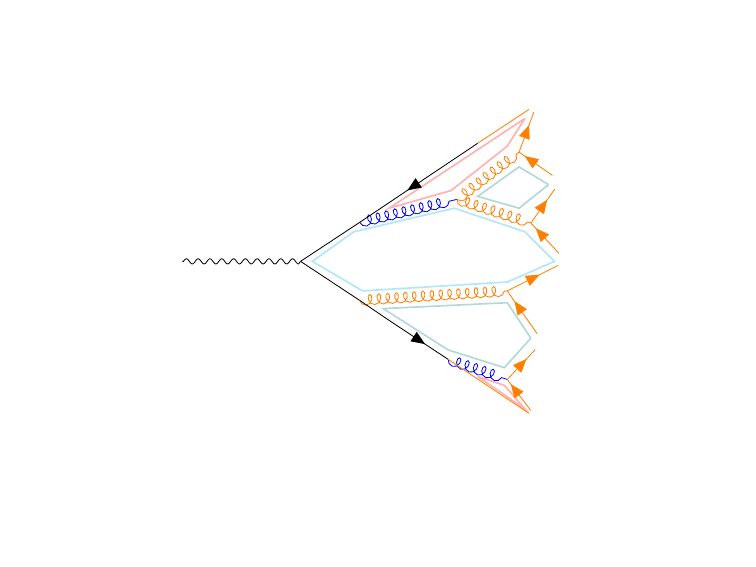}};
\draw (-1.8, 1.1) node {\small (c)};
\end{tikzpicture}\\[0.5em]
\begin{tikzpicture}
\draw (0, 0) node[inner sep=0] {\includegraphics[width=0.30\linewidth, trim={3cm 1.5cm 2cm 0.5cm}, clip]{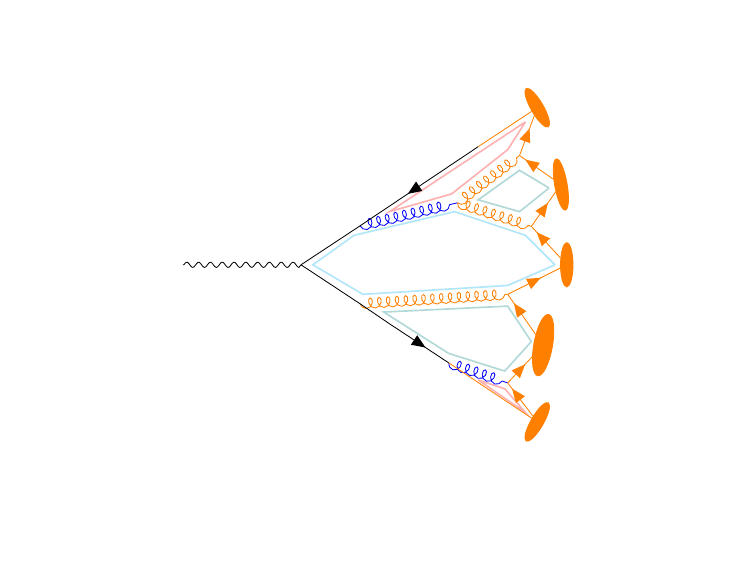}};
\draw (-1.8, 1.1) node {\small (d)};
\end{tikzpicture}\hfill
\begin{tikzpicture}
\draw (0, 0) node[inner sep=0] {\includegraphics[width=0.30\linewidth, trim={3cm 1.5cm 2cm 0.5cm}, clip]{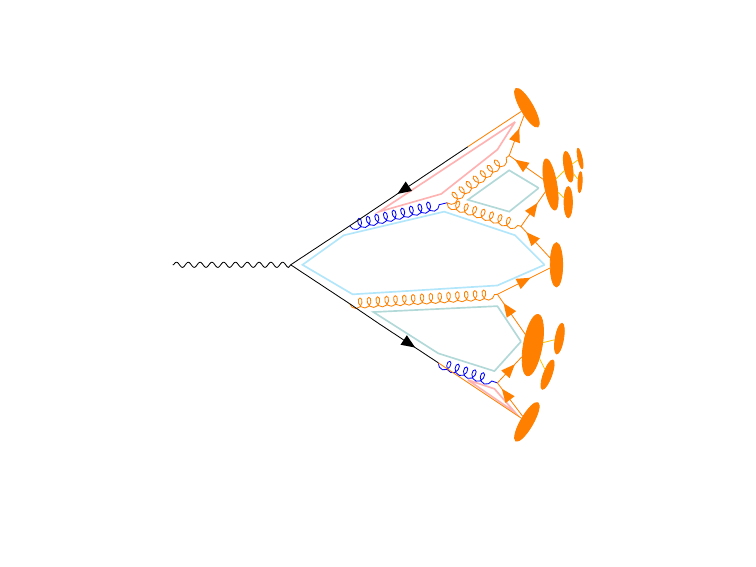}};
\draw (-1.8, 1.1) node {\small (e)};
\end{tikzpicture}\hfill
\begin{tikzpicture}
\draw (0, 0) node[inner sep=0] {\includegraphics[width=0.30\linewidth, trim={3cm 1.5cm 2cm 0.5cm}, clip]{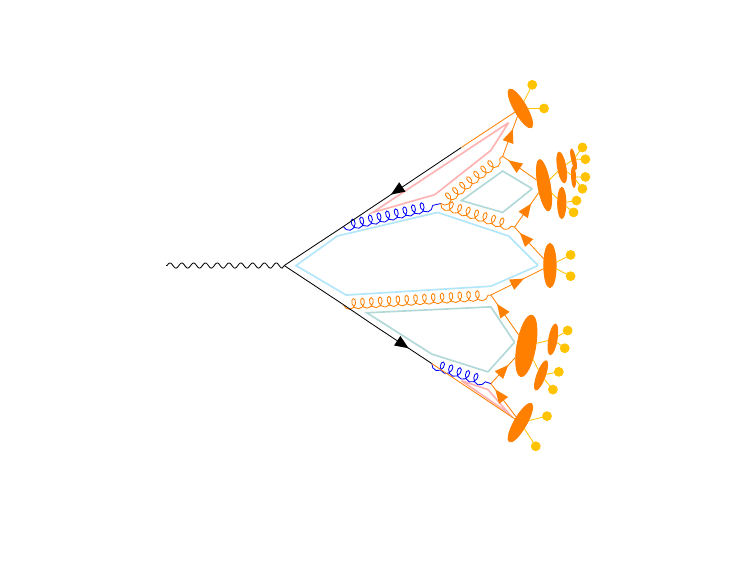}};
\draw (-1.8, 1.1) node {\small (f)};
\end{tikzpicture}
\caption{
An illustration of cluster hadronization.
First, the momenta of the partonic final state (a) are reshuffled such that each parton has its constituent mass (b). Next, all the gluons decay isotropically into $q\bar{q}$ pairs (c). Clusters are then formed (d).
After the heavy clusters undergo fission (e), all clusters decay into a pair
of hadrons (f).
}
\label{fig:clusterhad}
\end{figure}
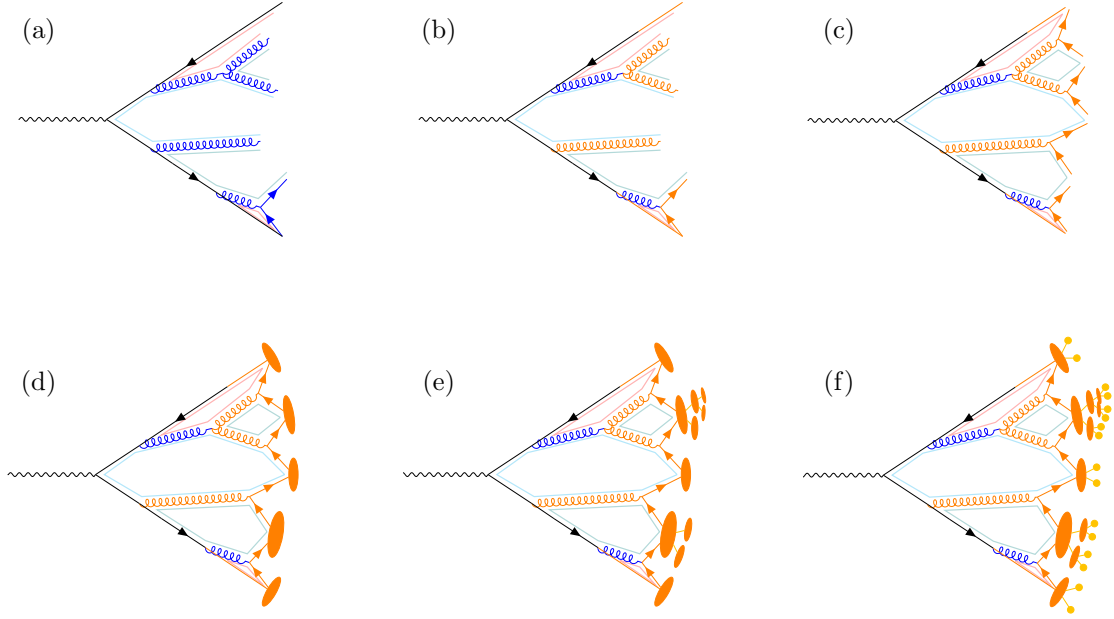

At the end of the parton shower, the final state consists of multiple massless quarks and gluons.
The first step involves reshuffling the momenta by a common factor $K$ such that all partons have their constituent mass,
\begin{equation}
\sum \sqrt{\mathbf{p}_i^2} = \sum\sqrt{K^2\mathbf{p}_i^2+m_i^2} \quad \Rightarrow \,\,\, E_i \to \sqrt{K^2\mathbf{p}_i^2 + m_i^2}, \quad \mathbf{p}_i \to K\mathbf{p}_i \, .
\end{equation}
The quarks are reshuffled to their constituent masses, while the gluons are assigned a \textit{gluon mass}.
The gluons then undergo massive isotropic decay into a quark-antiquark pair, where the momentum is shared equally between them.
If the gluons decay into light quarks, they travel in the gluon's path.
These decays leave the final state composed solely of quarks and antiquarks.
Following preconfinement, all such $q\bar{q}$ pair are combined into colour-singlet pseudoparticles called \textit{clusters}, with a mass $M^2 = (p_{q} + p_{\bar{q}})^2 = (p_{1} + p_{2})^2$.
Whilst most clusters are ``light'' ($M \leq 5$ GeV), there is often a small fraction that are too heavy to decay into hadrons. 
A notable case of this is the interaction $e^+e^- \to q \bar{q}$, where some events may evolve without emissions, resulting in a cluster of mass $Q$.
In these cases, these large clusters undergo \textit{cluster fission} if
\begin{equation}
    M^{a} \geq {M_{\text{max}}}^a + (m_1 + m_2)^a \, ,
\end{equation}
where $M_{\text{max}} = \texttt{Clmax}$ is a tunable condition for fission to occur and $a = \texttt{CLpow}$ adjusts the impact of \texttt{Clmax} and $(m_1 + m_2)$.
In this process, an additional $q\bar{q} \in \{u\bar{u}, d\bar{d}, s\bar{s}\}$ pair is generated with tunable weights $\texttt{Pwt}_{\texttt{light}}, \texttt{Pwt}_{\texttt{s}}$, and two clusters are generated, with one of the original quarks in each.
The masses of these clusters are sampled using a power law,
\begin{equation}
\begin{aligned}
M_1 &= m_1 + (M-m_1-m_{q, \text{new}})\,\mathcal{R}_1^{1/P} \, ,\\
M_2 &= m_2 + (M-m_2-m_{q, \text{new}})\,\mathcal{R}_2^{1/P} \, ,
\end{aligned}
\end{equation}
where $P = \texttt{Psplit}$.
The masses of the children are smaller than the mass of the parent cluster, but larger than the masses of their constituent quarks.
The fission is applied iteratively to any product clusters that are above the threshold.
Once all clusters are light, they are treated as excited hadron resonances and decay into stable hadrons.
For every cluster, a $q\bar{q}$ pair is generated (using different weights for all possible pairs of hadrons, together with a phase space factor that suppresses heavier hadrons relative to lighter hadrons with the same quantum numbers), and a pair of hadrons is formed.
An illustration is provided in Figure~\ref{fig:clusterhad}.

\subsection{Hadron-Level Results with Default Parameters}
\label{sec:hadronlevel}

The default cluster mass parameters in Herwig are based on a combination of phenomenological ideas and data-driven insights \cite[ch. 7]{Bellm:2025pcw}.
These are
\begin{equation}
\begin{gathered}
\alpha_s(m_Z) = 0.118 \, , \quad
p_{T,\min}^2 = 1.02\text{ GeV}^2 \, , \quad
m_{g, \text{const}} = 0.95\text{ GeV} \, , \\[0.5ex]
\texttt{Clmax}_{\texttt{Light}} = 3.53 \text{ GeV}\, , \quad
\texttt{Clpow}_{\texttt{Light}} = 1.85 \, , \quad
\texttt{PSplit}_{\texttt{Light}} = 0.91 \, , \\[0.5ex]
\texttt{Clmax}_{\texttt{Charm}} = 3.95 \text{ GeV}\, , \quad
\texttt{Clpow}_{\texttt{Charm}} = 2.56 \, , \quad
\texttt{PSplit}_{\texttt{Charm}} = 0.99 \, , \\[0.5ex]
\texttt{Clmax}_{\texttt{Bottom}} = 3.76 \text{ GeV}\, , \quad
\texttt{Clpow}_{\texttt{Bottom}} = 0.55 \, , \quad
\texttt{PSplit}_{\texttt{Bottom}} = 0.63 \, . \\[0.5ex]
\end{gathered}
\label{eq:default-params}
\end{equation}
The shower parameters $\alpha_s(m_Z)$ and $p_{T,\min}^2$ are set to typical values.
The gluon mass is required to be large enough to decay into $u \bar{u}$, $d \bar{d}$ and $s \bar{s}$ pairs; working with the strange quark constituent mass, $m_{s, \text{const}} = 0.45 \text{ GeV}$, this needs to be above $0.9 \text{ GeV}$.
The values of \texttt{Clmax} are similar for the three quark flavours.
The values around $3.5 \text{ GeV}$ stem from the observation that, when \texttt{Clmax} ranges between 3 and 4 GeV, the bulk of the events' features are insensitive to cluster fission.
On the other hand, \texttt{Clpow} is quite different.
For clusters with light quarks, $m_1 + m_2$ is much smaller than \texttt{Clmax} and for $\texttt{Clpow} = 2$, the fission is dependent on \texttt{Clmax}.
For clusters with one or two charm quarks, $m_1 + m_2$ approaches \texttt{Clmax}; a slightly higher \texttt{Clpow} increases the difference between them.
Lastly, clusters with even a single bottom quark result in $m_1 + m_2 > \texttt{Clmax}$, and the reduced \texttt{Clpow} reduced the difference instead.
This reduction in \texttt{Clpow} increases the $b$-baryon to $b$-hadron ratio.
The splitting parameter \texttt{PSplit} being close to 1 has a negligible impact and is designed to be tuned to better describe $b$ hadrons \cite{Corcella:2000bw}.
The take-home message from these parameters is that, unlike the shower parameters and the gluon mass, they need to be handled flavour-by-flavour, at least for the heavy flavours.

\begin{figure}[!htbp]
\centering

\begin{subfigure}{\linewidth}
\centering
\includegraphics[width=0.30\linewidth]{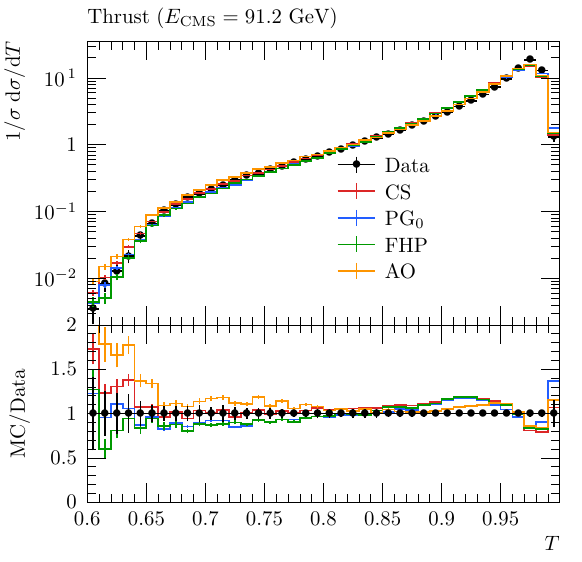}
\includegraphics[width=0.30\linewidth]{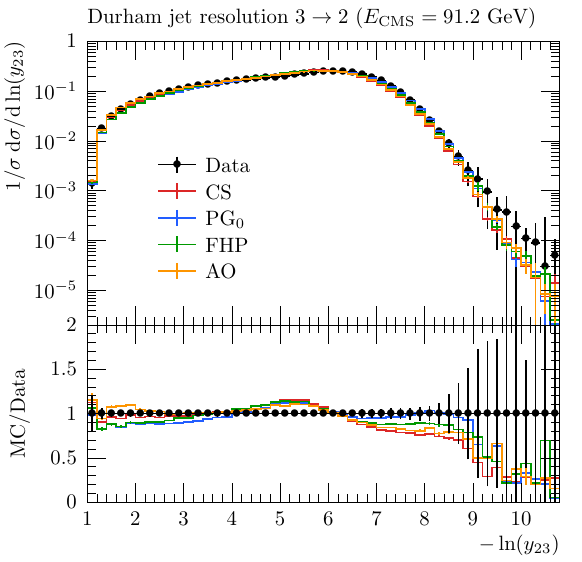}
\includegraphics[width=0.30\linewidth]{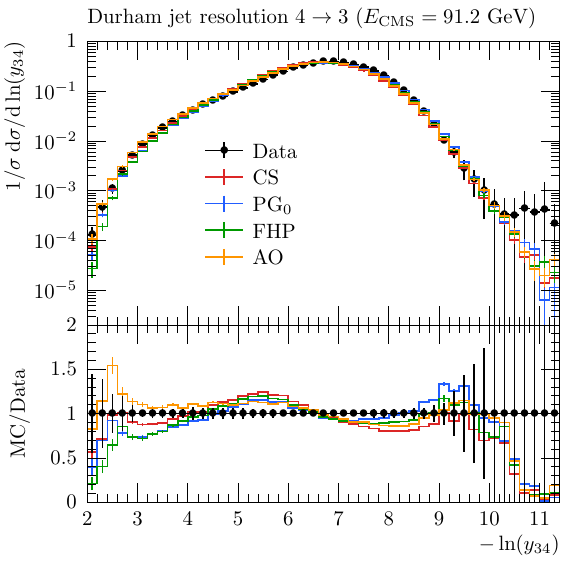}
\includegraphics[width=0.30\linewidth]{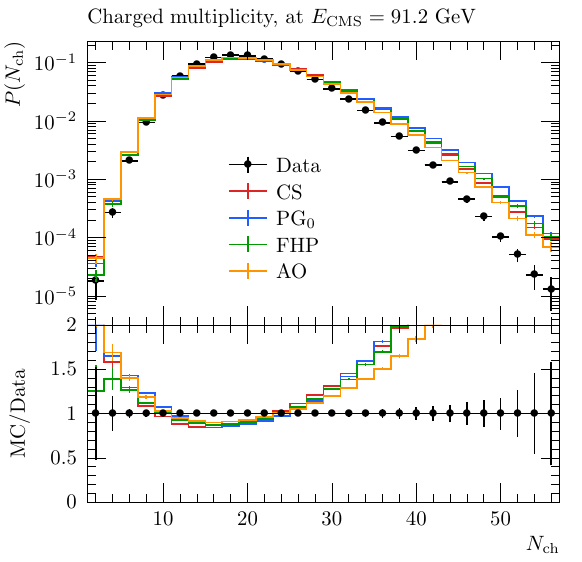}
\includegraphics[width=0.30\linewidth]{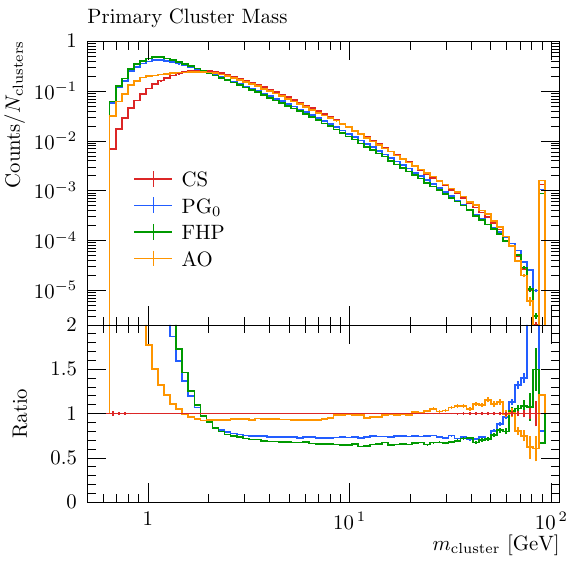}
\includegraphics[width=0.30\linewidth]{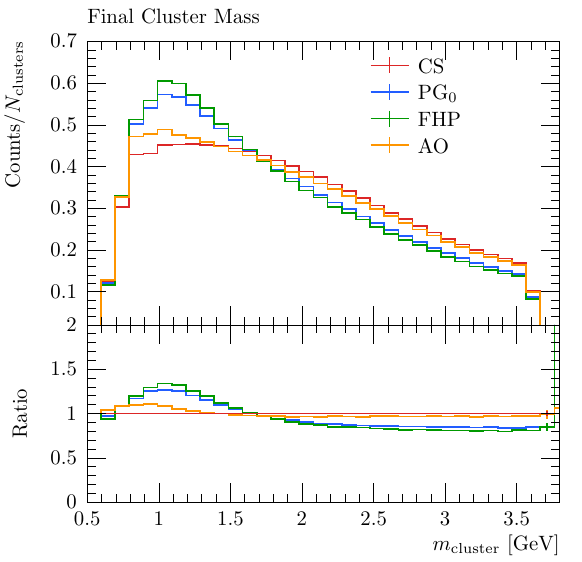}

\caption{Without AO-like Boundary}
\label{fig:second-look-no-aob}
\end{subfigure}

\vspace{1em}

\begin{subfigure}{\linewidth}
\centering
\includegraphics[width=0.30\linewidth]{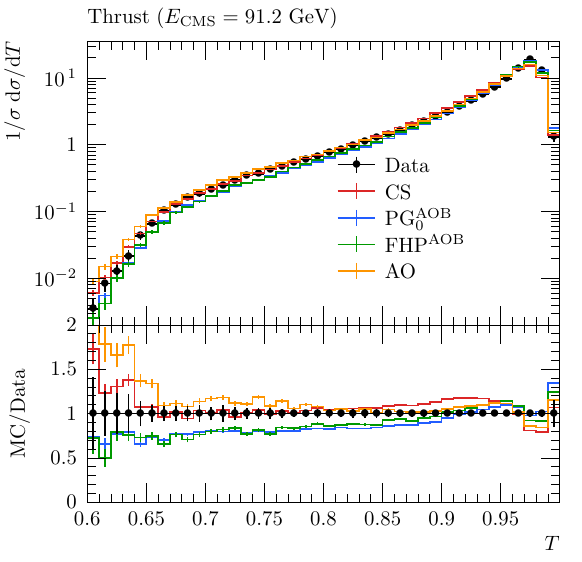}
\includegraphics[width=0.30\linewidth]{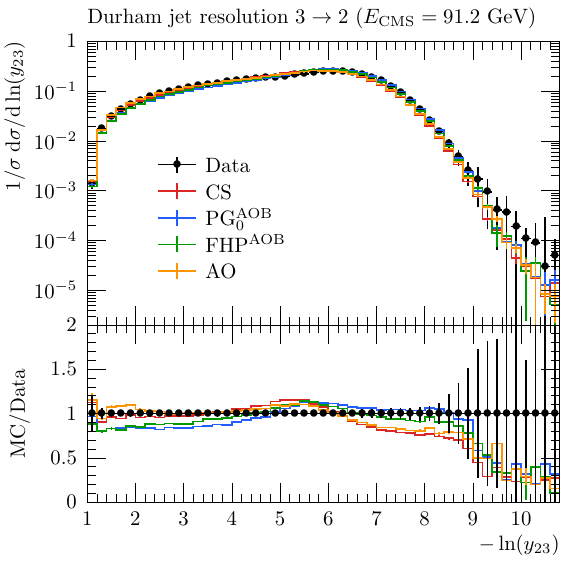}
\includegraphics[width=0.30\linewidth]{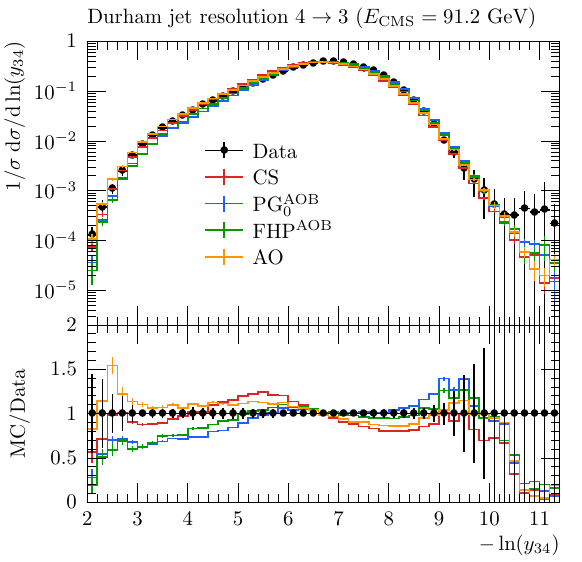}
\includegraphics[width=0.30\linewidth]{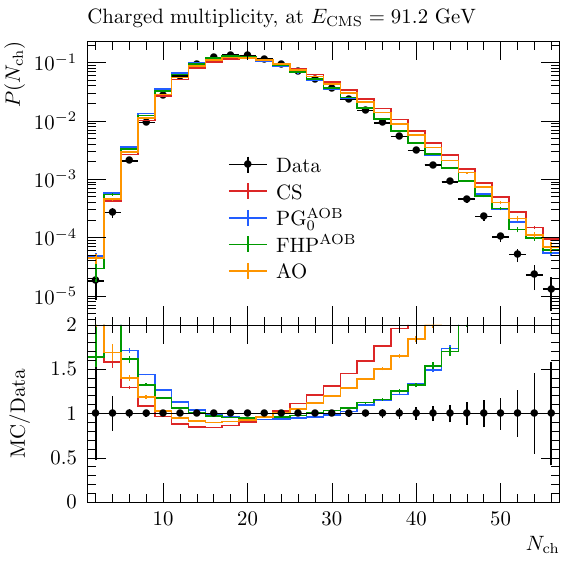}
\includegraphics[width=0.30\linewidth]{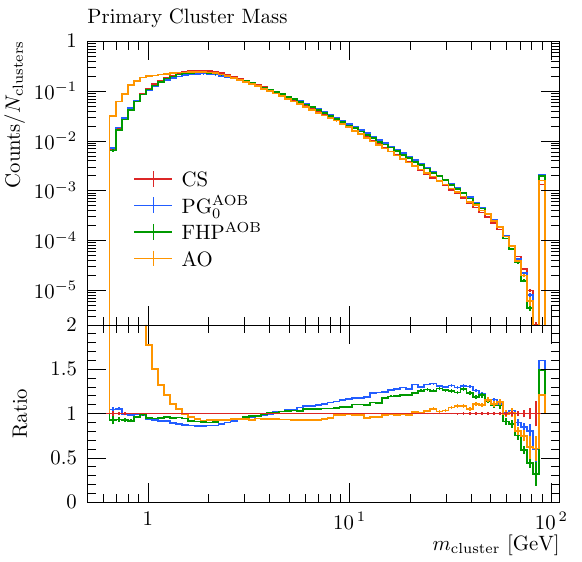}
\includegraphics[width=0.30\linewidth]{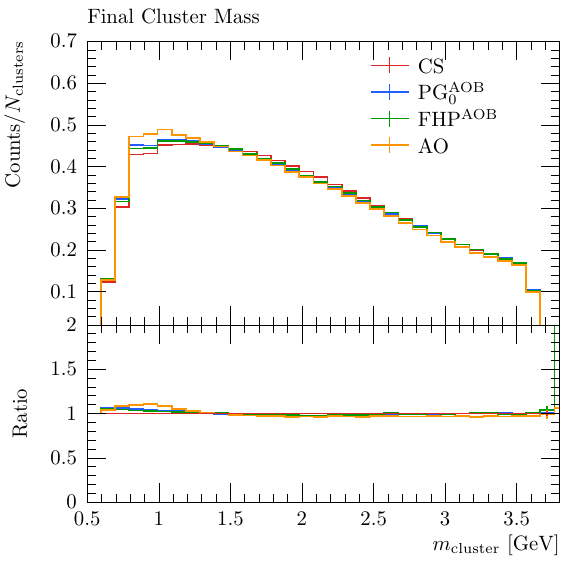}

\caption{With AO-like Boundary}
\label{fig:second-look-aob}
\end{subfigure}

\caption{
Comparison of hadron-level distributions without (a) and with (b) the AO-like Boundary applied.
As in the previous section, \PG and FHP produce very similar results, and the same is observed for \PGAOB and \FHPAOB.
}
\label{fig:second-look}
\end{figure}

We simulated $2 \times 10^6$ NLO events, including cluster hadronization, using the default parameter values above.
These simulations are compared to event shapes and jet rates from ALEPH \cite{ALEPH:2003obs}, as well as the multiplicity distribution from L3 \cite{L3:2004cdh}.
Additionally, to better study the effect of the shower on hadronization, we can plot spectra of cluster masses at formation and just before decay.
Event generators store the complete history of each event to create ``snapshots'' of the final state at these stages.
Any event with $c$ or $b$ quarks in the hard process was removed to study only light-quark processes throughout event generation.
A selection of plots is presented in Figure~\ref{fig:second-look}.
Here, the first subfigure represents the new showers without the AO-like boundary, and the second represents those with it.
Without the AO-like boundary, the new showers predict event shapes well, but predict the charged multiplicity very poorly.
Their behaviour, compared with CS and AO, in the cluster mass spectra is also very different, with the distribution peaking at a smaller mass.
These findings are swapped in the second row, as the showers fail to predict the hard tail, but offer a significant improvement in the charged multiplicity.
The results for the event shapes are also consistent with those observed by PanScales using the Pythia string hadronization model \cite{vanBeekveld:2024wws}.
The AO-like boundary also brings the cluster mass spectra much closer to AO and CS, implying that the hard-collinear emissions in the tail are the cause of the change in the peak.

%% file: sections/tune.tex
\section{Hadronization and Tuning} \label{sec:trials}

The previous two sections focused on strict comparisons of the shower schemes, without varying the hadronization modelling. In this section, we study the response of the Herwig cluster model to tuning with each of the different shower schemes providing the partonic input.

\subsection{The Professor Tuning Method}

We implement the tuning with Professor \cite{Buckley:2009bj}, which we briefly summarise here.
Professor models the response to parameter changes bin-by-bin for each observable.
Consider a vector of parameters $\vec{p}$.
The ``true'' event generator response for a bin $b$, $\text{G}_b$, is modelled by a series expansion for small variations in $\vec{p}$:
\begin{equation}
\text{G}_b(\vec{p} \, ) \approx f^{(b)}(\vec{p} \, ) = \alpha_0^{(b)} + \sum_i \beta_i^{(b)} p_i' + \sum_{i \leq j} \gamma_{ij}^{(b)} p_i' p_j' + \sum_{i \leq j \leq k} \delta_{ijk}^{(b)} p_i' p_j' p_k' \, ,
\end{equation}
where $p_i' = (\vec{p} - \vec{p}_0)_i$, with $\vec{p}_0$ being a starting point, usually chosen as the midpoint of all parameters.
Professor fits the coefficients $\alpha, \beta_i, \gamma_{ij}, \delta_{ijk}$ using generator samples that span the desired range for $\vec{p}$.
A goodness-of-fit measure is introduced, against which the optimal $\vec{p}$ is found. Here, the $\chi^2$ function is used:
\begin{equation}
\chi^2(\vec{p} \, ) = \sum_{O} \sum_{b\in O} w_{b} \frac{\left( f^{(b)}(\vec{p} \, ) - y_b \right)^2}{\Delta^2_b} \, ,
\end{equation}
where, for an observable $O$, $w_b$ is a weight assigned to a given bin, and $y_b, \Delta_b$ are the experimentally measured value and uncertainty for that bin.
The choice of weights has been a topic of discussion and has motivated successor programmes to Professor with algorithmic weight definition \cite{Bellm:2019owc, Krishnamoorthy:2021nwv}.
The reduced $\chi^2$ statistic can be obtained by dividing $\chi^2$ by the number of degrees of freedom:
\begin{equation}
N_{\text{df}} = \sum_{O,\,b \in O} w_b \, , \quad \chi^2_{\text{reduced}} = \chi^2 / N_{\text{df}} \, .
\end{equation}
Tunes typically employ an adjustment when utilising the experimental uncertainties, so as not to be dominated by the relatively few bins with large cross sections and consequently small statistical uncertainties.
We implement this as
\begin{equation}
\Delta_b \to \max \left( \Delta_b, 0.05\cdot y_b \right) , 
\end{equation}
matching the approach taken in \cite{Reichelt:2017hts}.

\subsection{Methodology for Tuning}

The showers discussed in this paper are not yet formulated for entirely realistic event generation. In particular, they do not yet include non-zero quark masses. Several of the parameters in the Herwig model directly relate to the spectroscopy of $b$ and $c$ hadrons, and so a complete tune of the Herwig cluster model is not possible. Concurrently, many experimental measurements of hadron spectra are also sensitive to the production rates of $b$ and $c$ hadrons. We therefore must be careful to tune only parameters relating to light quarks and to choose a subset of experimental measurements which will support this more limited approach to tuning. We first provide an overview of the parameters we tune and then the data we use for fitting these parameters.

The Herwig cluster model parameters are discussed in Section~\ref{sec:cluster}, of which six are not explicitly linked to $b$ and $c$ hadrons: $\alpha_s(m_Z)$, $p^2_{T,{\rm min}}$, $m_{g,{\rm const}}$, $\texttt{Clmax}_{\texttt{Light}}$, $\texttt{Clpow}_{\texttt{Light}}$, $\texttt{PSplit}_{\texttt{Light}}$. However, the parameters \texttt{Clpow} and \texttt{PSplit} have minimal effect on light-flavour observables and therefore represent degenerate directions in our tuning parameter space, $\vec{p}$. 
The value of the gluon mass, $m_{g,\text{const}} \sim 0.95 \text{ GeV}$, is not critical in determining infrared- and collinear-safe observables and, in common with \cite{Reichelt:2017hts}, we do not tune it.
For this simplified tune, we additionally fix $\texttt{Clmax}_{\texttt{Light}}= \texttt{Clmax}_{\texttt{Charm}} = \texttt{Clmax}_{\texttt{Bottom}}{}\equiv \texttt{Clmax}$ such that we tune one parameter \texttt{Clmax}, relying only on \texttt{Clpow} and \texttt{PSplit} to model the production of heavy hadrons.
We therefore keep \texttt{Clpow}, \texttt{PSplit} and $m_{g,\text{const}}$ at their default values from Eq.~(\ref{eq:default-params}), and only tune $\alpha_s(m_Z)$, $p^2_{T,{\rm min}}$ and $\texttt{Clmax}$.

Each of the three parameters is tuned within a physically reasonable range, with the default values at, or, for $\texttt{Clmax}$, near, their centres.
These are
\begin{equation}
\label{eq:tuneranges}
\alpha_s(m_Z) \in [0.106,0.130] \, , \quad p_{T,\min}^2 \in [0.5,1.5]\text{ GeV}^2  \, , \quad \texttt{Clmax} \in [2.5, 4.5] \, .
\end{equation}

We select the experimental measurements used in the tune so as to prioritise the accurate reproduction of the perturbatively dominated radiation pattern, where the improved logarithmic accuracy of PG$_0$ and FHP will be most significant. Consequently, our target observables are jet rates and event shapes, particularly those sensitive to parton showers and matching: thrust, $C$ parameter, heavy jet mass and mass difference, wide and total jet broadening, and the $y_{23}$ jet rate.
In addition, we also include the jet rates $y_{n,n+1}$ for $n = 3,4,5$ and thrust major/minor, oblateness, sphericity and aplanarity. The cleanest experimental $e^+e^-$ measurements are those of the LEP experiments at the $Z$-pole. We choose to use jet rate and event shape results from a single experiment, ALEPH \cite{ALEPH:2003obs}, so as to avoid tensions between measurements. Finally, we also include the infrared- and collinear-\textit{un}safe charged multiplicity from L3 \cite{L3:2004cdh} in order to further break degeneracies in the tuning.
The total number of observables used in the tune was 16.

\subsection{General Tune}\label{sec:general-tune}

\begin{figure}[!htbp]
\centering

\begin{subfigure}{\linewidth}
\centering
\includegraphics[width=0.24\linewidth]{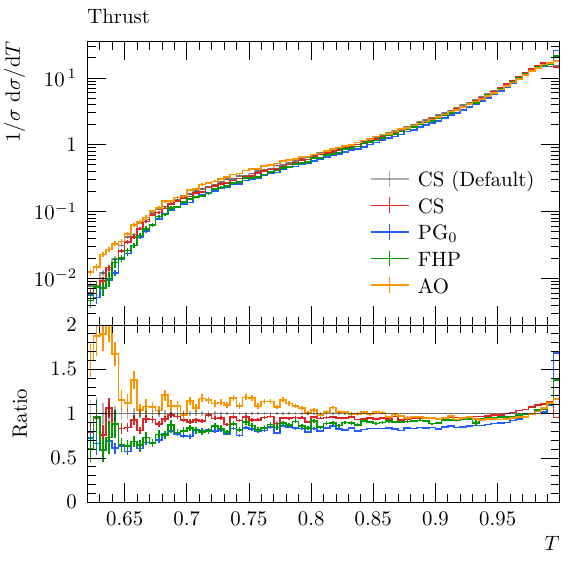}
\includegraphics[width=0.24\linewidth]{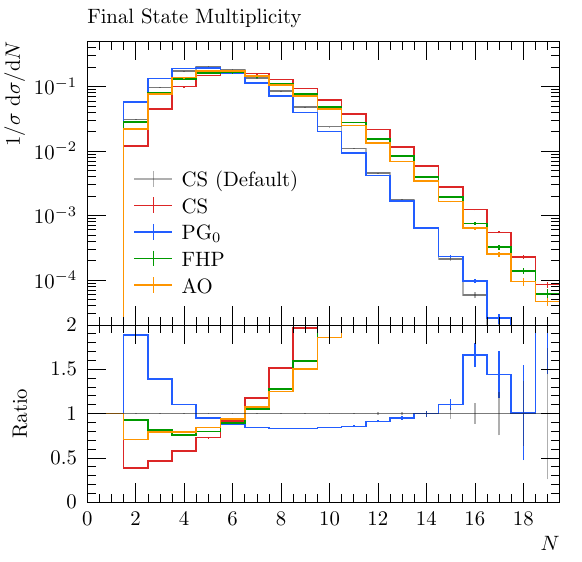}
\includegraphics[width=0.24\linewidth]{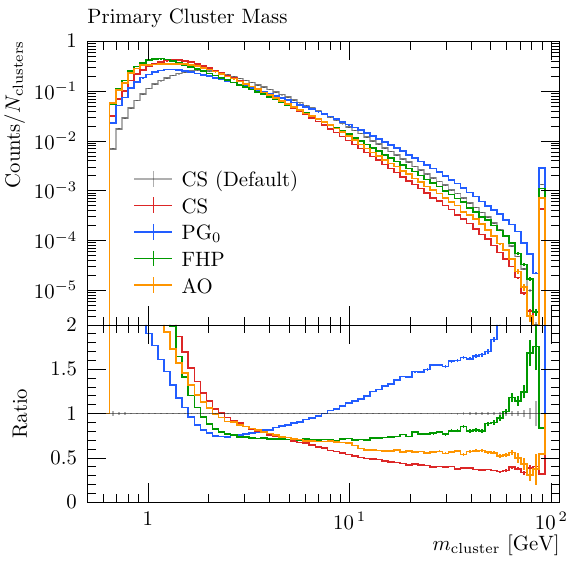}
\includegraphics[width=0.24\linewidth]{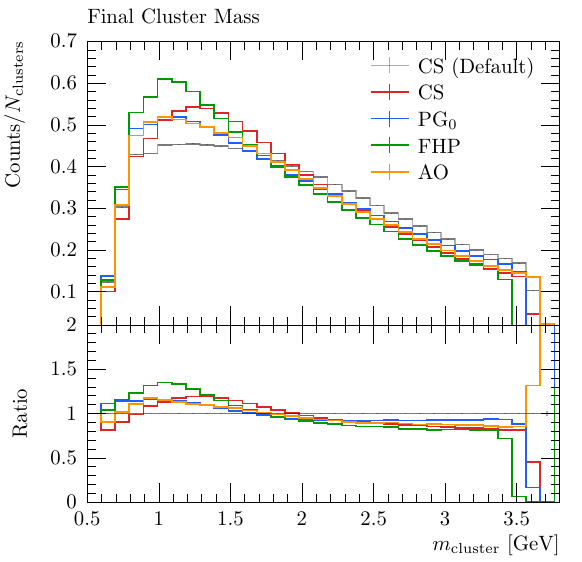}
\caption{Without AO-like Boundary}
\label{fig:tuned-clustermass-no-aob}
\end{subfigure}

\vspace{1em}

\begin{subfigure}{\linewidth}
\centering
\includegraphics[width=0.24\linewidth]{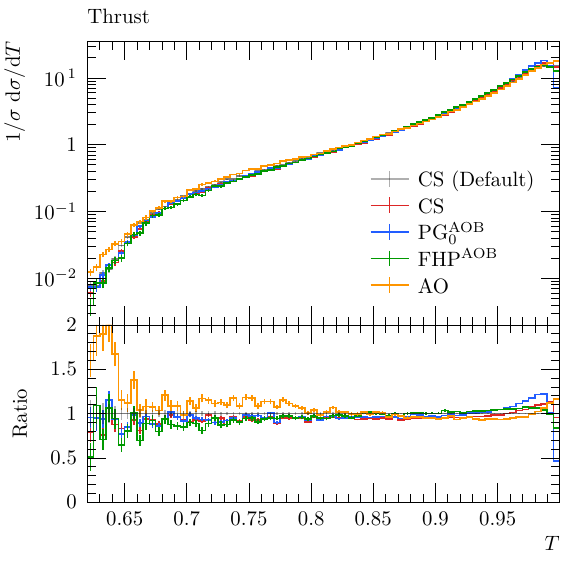}
\includegraphics[width=0.24\linewidth]{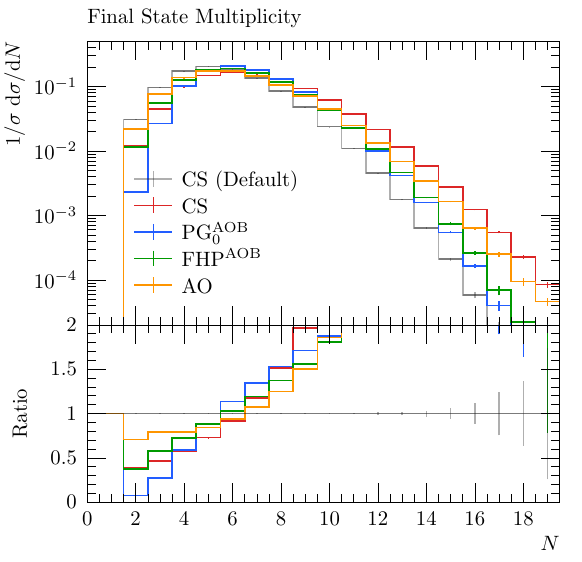}
\includegraphics[width=0.24\linewidth]{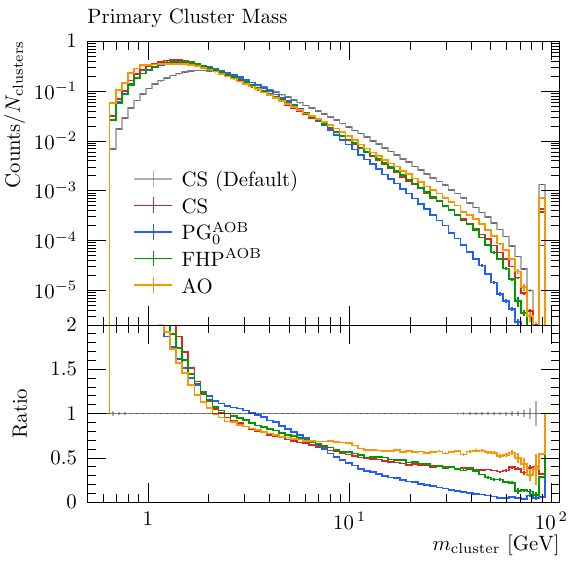}
\includegraphics[width=0.24\linewidth]{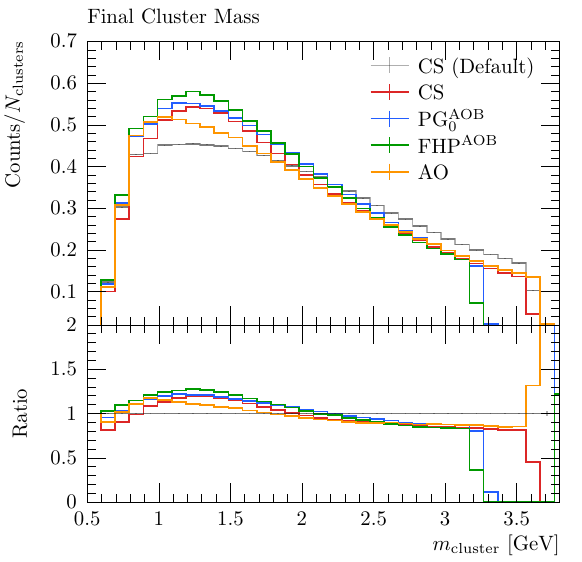}
\caption{With AO-like Boundary}
\label{fig:tuned-clustermass-aob}
\end{subfigure}

\caption{
Parton-level thrust and final-state multiplicity, together with the primary and final cluster mass spectra, generated using the tuned parameters of Table~\ref{tab:shower-comparison}. Rows~(a) and~(b) correspond to the showers without and with the AO-like boundary, respectively; the untuned CS prediction is shown in each row for reference.
Without the AO-like boundary, the final-state multiplicity of \PG approaches the untuned CS prediction while FHP does not, marking the first appreciable difference between the two schemes; the primary cluster mass spectrum also shows a mild shift due to tuning for \PG but not for FHP.
Once the AO-like boundary is imposed, \PGAOB and \FHPAOB align with each other and with AO and CS across all the shown distributions.
}
\label{fig:tuned-clustermass}
\end{figure}

We generated 100 parameter samples per shower across the ranges of Eq.~(\ref{eq:tuneranges}), each with 500{,}000 events, from which Professor constructs the parametric response of every observable and determines the best-fit point by minimising the total $\chi^2$. The resulting tuned parameters, together with the breakdown of the total $\chi^2$ into jet-rate, event-shape and charged-multiplicity contributions, are collected in Table~\ref{tab:shower-comparison}. Default and tuned predictions for the thrust and charged multiplicity are shown in Figure~\ref{fig:tuned-thrust-chargedmult}, and the complete set of tuned observables is provided in Appendix~\ref{sec:full-tuning}. To probe how each tune reshapes the partonic input to hadronization, we additionally show the tuned parton-level thrust and final-state multiplicity, together with the primary and final cluster mass spectra, in Figure~\ref{fig:tuned-clustermass}.
\begin{table}[htbp]
\vspace{5ex}
\centering
\begin{tabular}{l|c|c|c|c|c|c|c}
\hline
\textbf{Parameter} & \textbf{Range} & \textbf{CS} & \textbf{AO} & \textbf{\PG} & \textbf{FHP} & \PGAOBbold & \FHPAOBbold  \\ \hline
$\alpha_s(m_Z)$ & 0.106 - 0.130 & 0.114 & 0.115 & 0.119 & 0.117 & 0.129 & 0.124 \\
$p_{T,\min}^2/\text{GeV}^2$ & 0.5 - 1.5 & 0.645 & 0.764 & 1.403 & 1.051 & 0.890 & 0.777 \\
\texttt{Clmax} & 2.5 - 4.5 & 3.499 & 3.582 & 3.470 & 3.359 & 3.164 & 3.101 \\
\hline \hline
\multicolumn{2}{l|}{$\chi^2$ (Ev. Shapes)} & 3073 & 1947 & 4207 & 4061 & 2000 & 2726 \\
\multicolumn{2}{l|}{$\chi^2$ (Jet Rates)} & 1164 & 590 & 1137 & 1062 & 504 & 698 \\
\multicolumn{2}{l|}{$\chi^2$ (Charged Mult.)} & 1575 & 763 & 2790 & 1554 & 325 & 275 \\
\multicolumn{2}{l|}{Total $\chi^2$} & 5813 & 3300 & 8134 & 6676 & 2829 & 3698 \\
\hline
\multicolumn{2}{l|}{$\chi^2/N_{\text{df}} \, \, (N_{\text{df}} = 666)$} & 8.73 & 4.96 & 12.21 & 10.02 & 4.25 & 5.55 \\
\hline
\end{tabular}
\caption{Comparison of tuned Herwig cluster model parameters and the $\chi^2$ results for each tune for the CS, AO, \PG, FHP, \PGAOB, and \FHPAOB shower schemes. 
}
\label{tab:shower-comparison}
\end{table}

\begin{figure}[htbp]
\centering

\begin{subfigure}{\linewidth}
\centering
\includegraphics[width=0.30\linewidth]{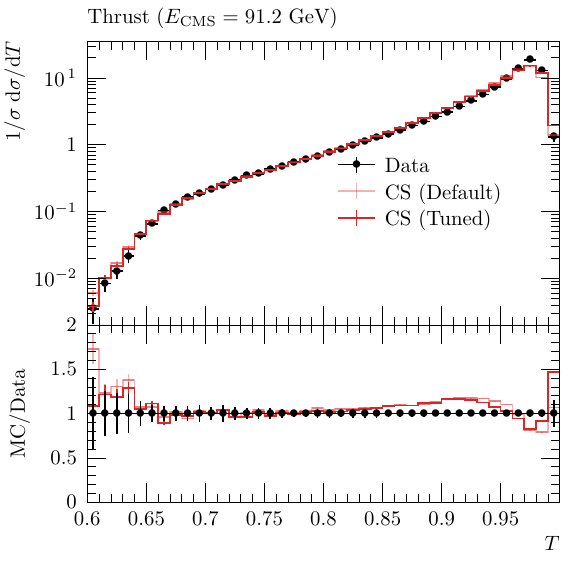}
\includegraphics[width=0.30\linewidth]{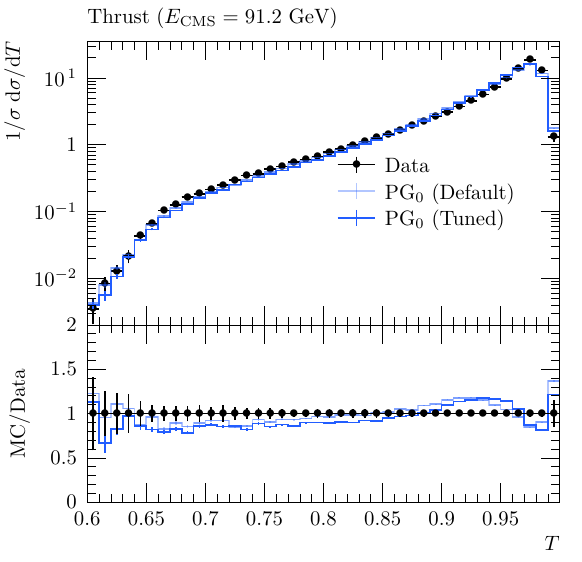}
\includegraphics[width=0.30\linewidth]{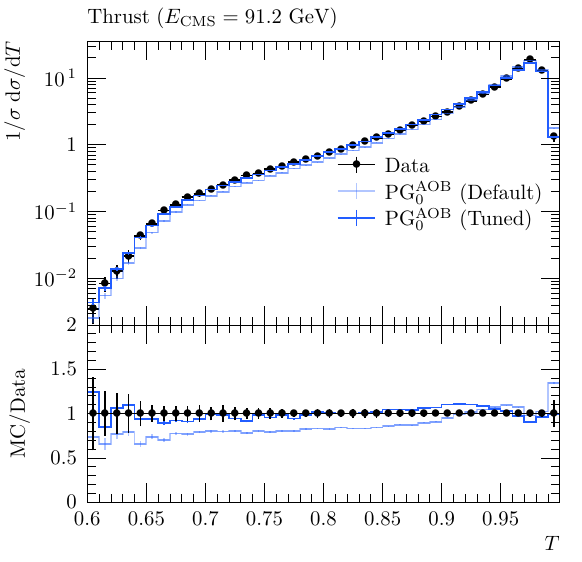}
\includegraphics[width=0.30\linewidth]{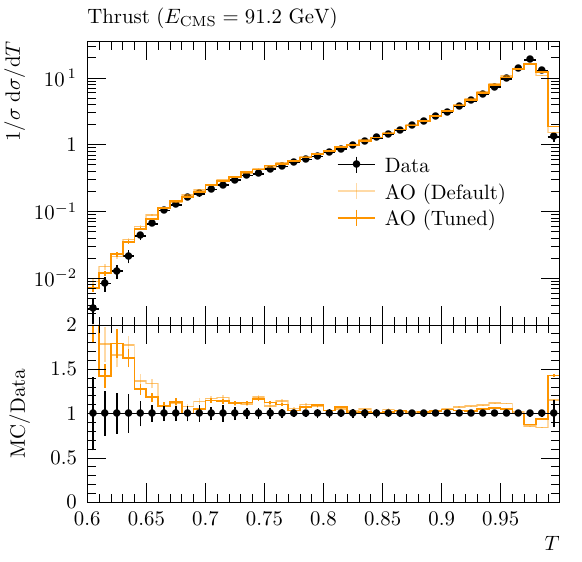}
\includegraphics[width=0.30\linewidth]{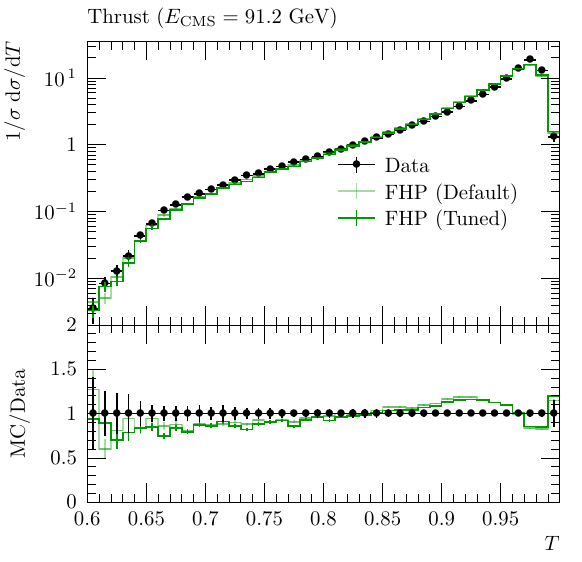}
\includegraphics[width=0.30\linewidth]{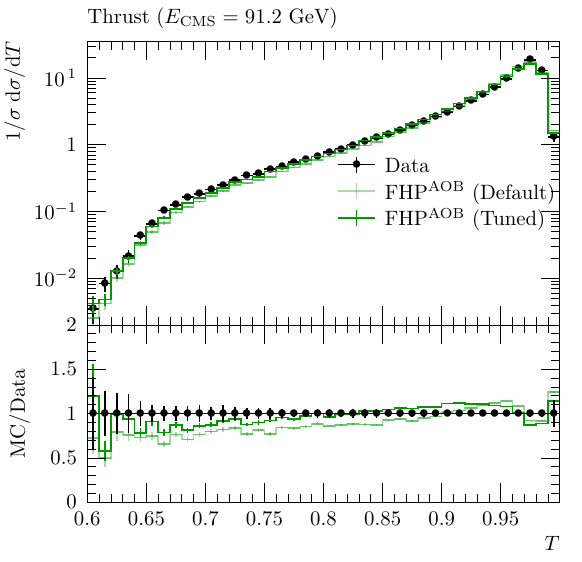}
\caption{Thrust, $T$}
\label{fig:tuned-thrust}
\end{subfigure}

\vspace{1em}

\begin{subfigure}{\linewidth}
\centering
\includegraphics[width=0.30\linewidth]{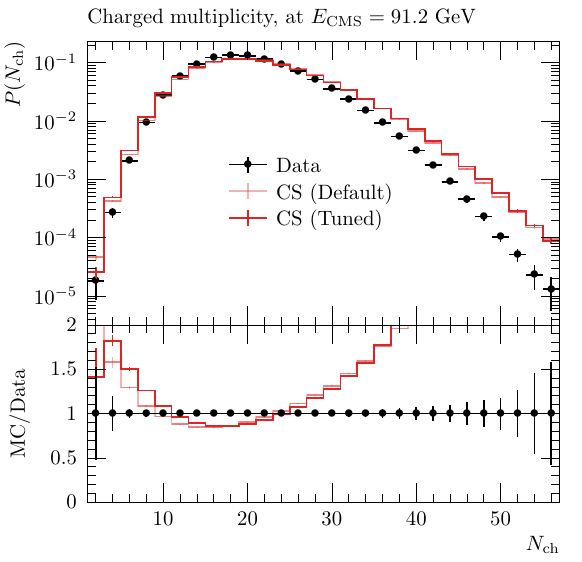}
\includegraphics[width=0.30\linewidth]{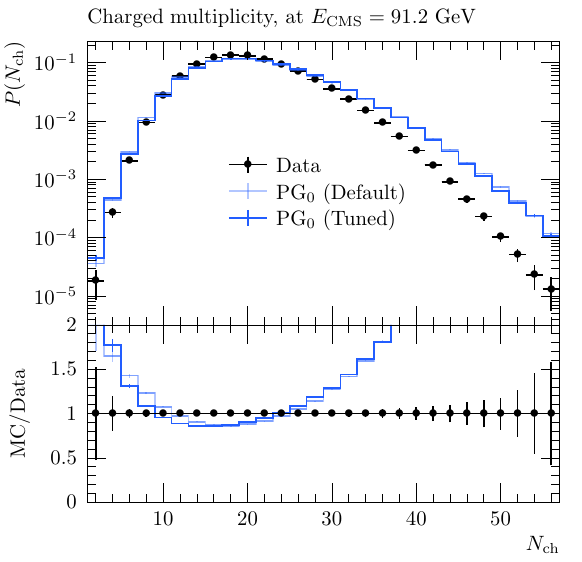}
\includegraphics[width=0.30\linewidth]{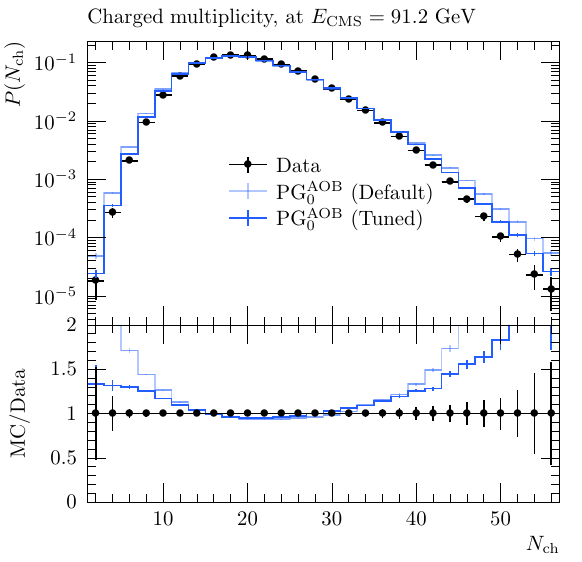}
\includegraphics[width=0.30\linewidth]{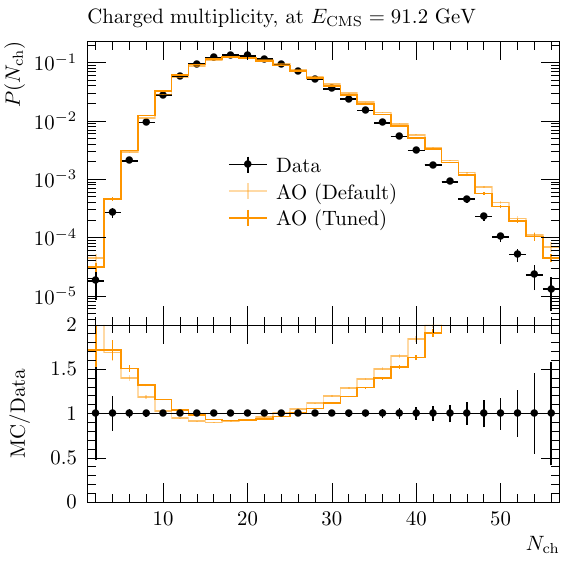}
\includegraphics[width=0.30\linewidth]{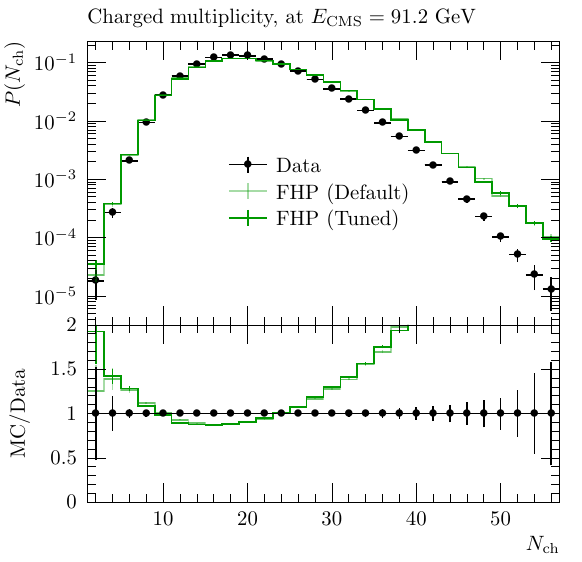}
\includegraphics[width=0.30\linewidth]{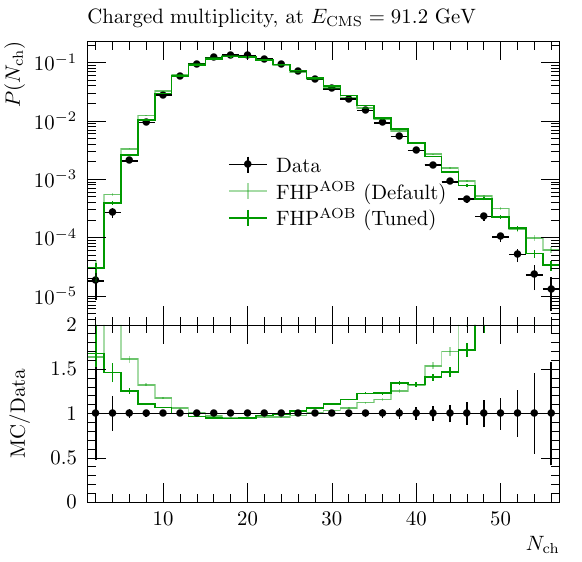}
\caption{Charged Multiplicity}
\label{fig:tuned-nump}
\end{subfigure}

\caption{
Default and tuned predictions for the thrust (a) and charged multiplicity (b), for each of the six shower schemes, compared to the ALEPH~\cite{ALEPH:2003obs} and L3~\cite{L3:2004cdh} data, respectively.
After tuning, all showers provide a good description of the event shapes; for \PGAOB and \FHPAOB, the charged multiplicity distribution is also improved substantially, approaching the ALEPH and L3 data. This improvement is achieved by increasing $\alpha_s(m_Z)$ and reducing $p_{T,\min}^2$ from their default values, so as to produce more emissions, correlated with a reduction of \texttt{Clmax}. The corresponding tuned parameter values and $\chi^2$ are collected in Table~\ref{tab:shower-comparison}.
}
\label{fig:tuned-thrust-chargedmult}
\end{figure}

The tunes organise themselves naturally into the three shower pairs identified in Section~\ref{sec:cutoff}, distinguished by the shape of their IR cutoff boundary and logarithmic accuracy. The convex-boundary showers, CS and AO, remain close to the default value of $\texttt{Clmax}\approx3.5$,
prefer a slightly reduced $\alpha_s(m_Z)$ value, $~0.114\!-\!5$, but noticeably reduced $p_{T,\min}^2$ relative to the default, which improves the description of observables sensitive to multiple emissions, in particular the higher jet rates and the charged multiplicity. The \PG\ and FHP schemes respond in a qualitatively different way: both favour $\alpha_s(m_Z)$ values close to the default and only marginally shift \texttt{Clmax} relative to the default. However, both increase $p_{T,\min}^2$ relative to the default; \PG substantially and FHP only moderately. Qualitatively, this has the effect of reducing the number of emissions in each scheme; however, this constitutes the first significant point of quantitative difference between the two. The resulting $\chi^2$ values in Table~\ref{tab:shower-comparison} are noticeably larger for FHP and \PG than those of CS and AO, and, as Figure~\ref{fig:tuned-thrust} shows, the tune induces only minor changes in the simulation itself. This is a direct consequence of the observation in Section~\ref{sec:cutoff}: a cut on the dipole-$p_T$ variable defines a concave physical cutoff boundary for \PG and FHP, declaring a larger volume of phase space perturbative than the convex AO boundary around which the cluster model was historically developed, leaving the tune little room to manoeuvre.

Enforcing the AO-like boundary on the two showers changes the picture substantially. Both \PGAOB and \FHPAOB raise $\alpha_s(m_Z)$ significantly relative to the default, towards the upper edge of the tune range, and reduce both $p_{T,\min}^2$ and \texttt{Clmax} below their defaults. These tuned values produce both more shower emissions and more cluster fissions, making the hadronic spectrum softer. The increase in $\alpha_s(m_Z)$ lifts the tails of the event-shape and jet-rate distributions into agreement with the data, a direction of tune movement consistent with the string-hadronization study of Ref.~\cite{vanBeekveld:2024wws}, while the combined reduction of $p_{T,\min}^2$ and \texttt{Clmax} improves the description of the charged multiplicity. The resulting total $\chi^2/N_{\text{df}}$ is drastically lower for \PGAOB and \FHPAOB, with the largest individual improvement coming from the charged multiplicity, where \PGAOB and \FHPAOB outperform even the AO shower by a substantial margin.

Despite this superior performance, we do not consider that the AOB versions of the algorithms are better than the default versions, merely better suited to the current version of the cluster hadronization model. Without developing a new hadronization model, adapting the shower algorithms to better match the hadronization model is the only way of gauging the importance of this physics.

The parton-level response to each tune, shown in Figure~\ref{fig:tuned-clustermass}, makes the direction of these adjustments transparent. Of particular note are the primary cluster mass spectra, which represent the shower-hadronization interface most directly. Without the AO-like boundary, each shower populates a distinct distribution of primary cluster masses, with AO and CS closest to one another. Once the AO-like boundary is imposed on the showers, the \PGAOB and \FHPAOB primary spectra collapse onto those of CS and AO, demonstrating that it is the shape of the cutoff boundary, rather than the details of the shower kinematics, that dominantly sets the cluster-mass scale fed into hadronization. The final-state cluster mass spectrum, by contrast, shows reduced sensitivity to both the shower choice and the cutoff boundary. Nevertheless, it retains the core features present in the low-mass end of the primary cluster mass spectrum. This low-mass region undergoes fewer fissions and so is seeded directly by the parton showers. The commonality of the high-mass tail suggests that cluster fission and decay provide enough flexibility, through tuning \texttt{Clmax}, to absorb differences in the initial cluster mass distribution into a common final-state spectrum, provided enough cluster fissions take place.

\subsection{Subtunes and Tuning Heatmaps}\label{sec:heatmaps}

\begin{figure}[t]
\centering
\begin{tikzpicture}
\node[anchor=south west, inner sep=0] (img) {\includegraphics[width=0.45\linewidth]{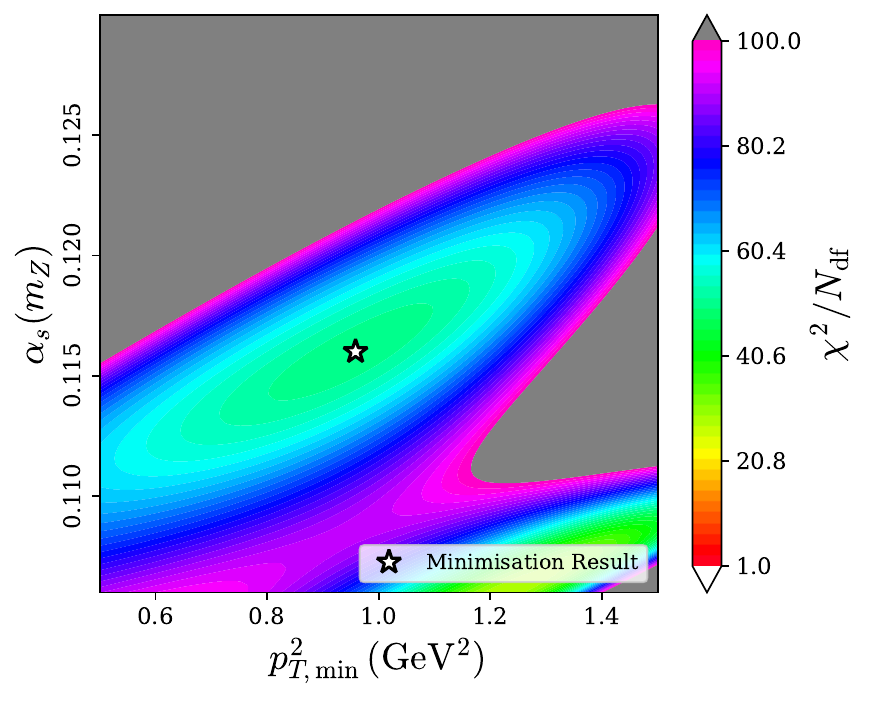}};
\node[anchor=north west, xshift=25pt, yshift=-5pt, fill=white, fill opacity=0.75, text opacity=1, inner sep=2pt] at (img.north west) {\large{FHP}};
\end{tikzpicture}
\begin{tikzpicture}
\node[anchor=south west, inner sep=0] (img) {\includegraphics[width=0.45\linewidth]{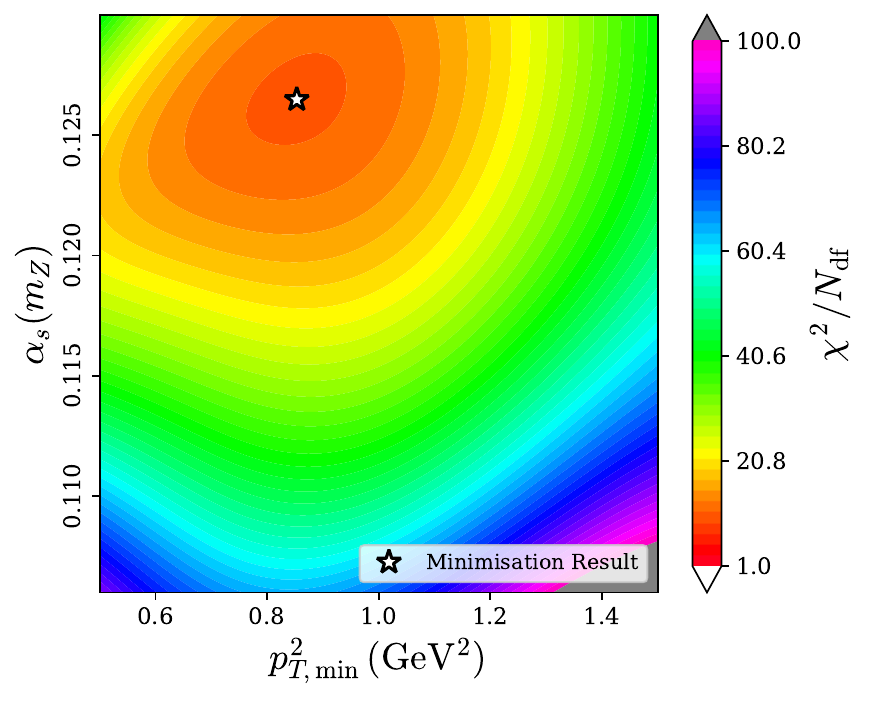}};
\node[anchor=north west, xshift=25pt, yshift=-5pt, fill=white, fill opacity=0.75, text opacity=1, inner sep=2pt] at (img.north west) {\large{\FHPAOB}};
\end{tikzpicture}
\caption{
$\chi^2/N_{\text{df}}$ heatmaps of $\alpha_s(m_Z)$ and $p_{T,\min}^2$ for the charged multiplicity observable, with the remaining tune parameter $\texttt{Clmax}$ fixed to the centre of its range. The left panel shows the FHP shower and the right panel the \FHPAOB shower.
The grey region indicates $\chi^2/N_{\text{df}}>100$.
For FHP, the only sub-100 region of the parameter space is a narrow sliver along the edge and a narrow valley, giving a strong correlation between the parameters, so the tune has very little flexibility in this observable. Imposing the AO-like boundary opens a broad, well-defined minimum and allows \FHPAOB to fit the charged multiplicity considerably better. Even without tuning, the default parameters, in the centre of the plot, give a significantly better description of data with AOB ($\chi^2/N_{\text{df}}\sim20$ compared to $\sim50$ without).}
\label{fig:placeholder}
\end{figure}

The global minimisation of Section~\ref{sec:general-tune} returns a single best-fit point per shower, but reveals little about the shape of the $\chi^2$ landscape or about the relative sensitivity of the different observable classes to each parameter. To complement the global tune, we therefore examine $\chi^2/N_{\text{df}}$ as a function of each of the three parameter pairs $\{\alpha_s(m_Z), p_{T,\min}^2\}$, $\{p_{T,\min}^2, \texttt{Clmax}\}$ and $\{\alpha_s(m_Z), \texttt{Clmax}\}$, with the remaining parameter fixed to the centre of its range. For each pair, we generate 50 parameter samples per shower and compute the resulting $\chi^2/N_{\text{df}}$ landscape separately for the event shapes, the jet rates and the charged multiplicity, as well as for the combined fit. The complete set of heatmaps, covering all five dipole showers, is collected in Appendix~\ref{sec:full-tuning}; here we summarise the main physical conclusions, illustrated in Figure~\ref{fig:placeholder} by the charged multiplicity heatmap of FHP and \FHPAOB.

The most striking feature of the subtunes is the sharp difference in the charged multiplicity channel between the showers with and without the AO-like boundary. For \PG and FHP, only very narrow regions of the $\{\alpha_s,p_{T,\min}^2\}$ plane reach $\chi^2/N_{\text{df}}<100$: the best fit sits in a shallow, narrow band, leaving the tune almost no flexibility. Imposing the AO-like boundary opens a broad, well-defined basin of low $\chi^2/N_{\text{df}}$ within which the tune can be readily optimised. The same qualitative behaviour, though less pronounced, is observed in the event-shape heatmaps. This is direct numerical evidence for the statement of Section~\ref{sec:general-tune}: the cluster hadronization model performs best when it receives the convex AO boundary partonic input it was developed to handle.

In general, the event-shape and jet-rate landscapes are comparatively flat; the $\chi^2/N_{\text{df}}$ typically varies by $\sim 10$ across the $\{p_{T,\min}^2,\texttt{Clmax}\}$ plane, and the variation in the $\{\alpha_s(m_Z),$ $\texttt{Clmax}\}$ plane is of similar magnitude while being mostly dominated by $\alpha_s(m_Z)$. This indicates that, for these IRC-safe observables, it is the parton shower rather than the cluster model that sets the overall quality of the fit, with $\alpha_s(m_Z)$ playing the dominant role. The charged multiplicity, on the other hand, constrains the tune much more tightly, and its heatmap shape is the principal driver of the combined-fit landscape.

Finally, we examine how the $\{ \alpha_s(m_Z), \texttt{Clmax} \}$ heatmaps respond to changes in the value of the cutoff, repeating the scan with $p_{T,\min}^2$ fixed to $0.75\,\text{GeV}^2$ and $1.25\,\text{GeV}^2$ (Figures~\ref{fig:heatmaps-as-cl-pt2min0p75} and~\ref{fig:heatmaps-as-cl-pt2min1p25}). The event-shape and jet-rate heatmaps are broadly unaffected, consistent with the IRC safety of these observables. The charged multiplicity heatmaps of \PG and FHP, in contrast, change dramatically between the two cutoff values, reflecting a strong sensitivity of the cluster model to the position of the concave cutoff. For \PGAOB and \FHPAOB, the effect is reduced to a smooth stretching and compression of the optimal region. These subtunes demonstrate that it is not only the value of the shower cutoff but also the shape of the boundary it defines that seeds the cluster hadronization model and has a significant phenomenological impact.

%% file: sections/conc.tex
\section{Discussion, Concluding Remarks and Outlook}
\label{sec:conc}

In this study, we implemented simplified versions of the PanGlobal $\beta = 0$ (PG$_0$) and Forshaw-Holguin-Pl\"atzer (FHP) massless final-state parton shower schemes in the Herwig event generator. Our implementations are sufficient for next-to-leading logarithmic and leading colour accuracy within the Herwig generator framework, complementing the default Angular Ordered (AO) and Catani-Seymour dipole-shower (CS) parton shower schemes. Benefiting from the Herwig framework, these showers are matched at NLO using the Matchbox module \cite{Platzer:2011bc} and include collinear spin correlations via implementations of the Collins--Knowles algorithm \cite{Collins:1987cp, Knowles:1988hu, Knowles:1988vs} within the Herwig dipole shower framework \cite{Richardson:2018pvo, Bellm:2019zci}. The goal of this study was to understand the impact of these new parton shower schemes on realistic event generation, particularly how the infrared limit of these showers seeds hadronization. To this end, we present the first implementation of the FHP and PG$_0$ shower schemes with the non-logarithmic Jacobians for phase-space mappings included at all orders.

The CS, FHP and PG$_0$ schemes are each dipole showers, ordered in a dipole-$p_T$ ordering variable, but differ in the mechanisms they use to conserve momentum. Consequently, the dipole-$p_T$ ordering variables define different physical transverse momenta after momentum conservation. We identify that this leads to a noticeable difference in the infrared limits of the showers, where the shower cutoff defines a different contour in the emission phase space for the FHP and PG$_0$ schemes than it does in the CS and AO schemes. In particular, a cut on the dipole-$p_T$ ordering variable for the FHP and PG$_0$ schemes defines a physical boundary that is concave from below when drawn in terms of the Lund phase space variables, in contrast to the convex-from-below CS and AO boundaries. Consequently, for a fixed value of the cutoff, a larger volume of phase space is accessible to the FHP and PG$_0$ shower schemes, and so is treated as perturbative.

The distinction between the convex and concave boundaries lies beyond discussions of NLL (or even NNLL) accuracy and so must instead be treated as a modelling choice. A truly faithful recreation of the dynamics of hadronization would be independent of such choices, and so a sufficiently flexible hadronization model should be able to accommodate either. However, the Herwig cluster hadronization model has historically been developed for the Herwig angular-ordered parton shower, whose shower cutoff scale defines the more conservative convex boundary. In this paper, we have tested the flexibility of the cluster model by additionally implementing the convex AO-like boundary as an optional constraint on the FHP and PG$_0$ parton shower schemes, which we label as FHP$^{\rm AOB}$ and PG$_0^{\rm AOB}$.

We consistently find that event generation with the FHP and PG$_0$ parton shower schemes performs similarly. However, we find that event generation with the FHP$^{\rm AOB}$ and PG$_0^{\rm AOB}$ parton shower schemes gives drastically better hadron-level results than the out-of-the-box FHP and PG$_0$ schemes, both with and without retuning the cluster model. Event generation with the FHP and PG$_0$ schemes underperforms relative to both the AO and CS parton shower schemes. However, once the AO-like boundary constraint is enforced, event generation with the FHP$^{\rm AOB}$ and PG$_0^{\rm AOB}$ schemes noticeably outperforms the CS dipole parton shower scheme. Relative to the Herwig AO shower, FHP$^{\rm AOB}$ and PG$_0^{\rm AOB}$ perform similarly with respect to event shape distributions and jet rates. This is expected since the AO parton shower scheme also achieves NLL accuracy for most of these observables. Strikingly, FHP$^{\rm AOB}$ and PG$_0^{\rm AOB}$ dramatically outperform all other parton shower schemes when considering the infrared- and collinear-\textit{un}safe hadron multiplicity spectra.

We wish to emphasise again that we do not consider that the AOB versions of the algorithms are \emph{better} than the default versions. In fact, ideally, we wish to describe as large a phase space region as possible perturbatively, and so would prefer the default versions. This would have the dual benefit of increasing the phase space over which analytical control is achieved, while also decreasing the number of events with no (or few) parton shower emissions. Such events are particularly challenging for the cluster hadronization model, since larger modifications from hadronization are required to produce locally colour-singlet hadrons, which can in turn lead to spurious results. We implemented the AOB as the simplest way to explore the physics of the interface between the perturbative and non-perturbative components of full event generation. It is remarkable that, without further modifications or optimisations, these versions, together with Herwig's cluster model, give us the best description of hadron-level results of any of the dipole showers we have considered.

One of the challenges in realising a reliable hadronization model is the tail of events with very large cluster masses, particularly those that, with probability given by the Sudakov form factor, produce no radiation at all. While such events are already known to be problematic in Herwig, its historical developments have mitigated the problem by the AO shower producing fewer of these events. We have seen that by adding a source of very hard, very collinear gluons, effectively splitting the very heavy $q\bar{q}$ cluster into an almost-as-heavy $g\bar{q}$ cluster and a light $qg$ cluster, without other changes to the model, the \PG and FHP schemes have not solved the problem; in fact, they have worsened it. This motivates us to propose that the study of parton shower accuracy should not be separated from that of hadronization models, and that improving parton shower accuracy necessitates a reconsideration and refinement of hadronization model accuracy. This perspective aligns with recent developments \cite{Hoang:2018zrp, Platzer:2022jny, Gieseke:2025mcy}, which highlight the importance of infrared sensitivity and the role of the shower cutoff in hadronization modelling, and which have been partly explored in the context of the angular-ordered parton shower in \cite{Hoang:2024nqi}. As a first step, it is reasonable to speculate that if the cluster model were adapted to handle the concave boundary, the minimisation of no-emission events could lead to the best modelling of hadronization by allowing it to remain a local effect. We hope our work motivates the development of newer hadronization algorithms to complement these logarithmically improved parton shower schemes.

More generally, our results motivate the study of the infrared limit of parton shower algorithms as an important aspect for reliable interfacing with hadronization and phenomenology. This suggests that multiple, as yet unexplored, directions for improvement will be needed for event generation with logarithmically accurate parton shower schemes. In particular, the inclusion of massive quarks also affects this infrared phase space (via the dead-cone effect) and so may be relevant beyond just the spectroscopy of heavy hadrons. Similarly, the complete and accurate population of the shower phase space may be significant (improvable by merging or by achieving yet higher perturbative accuracy \cite{vanBeekveld:2024wws}). We hope that parallel, collaborative development of these components can prepare us for future LHC runs, as well as for future colliders.

%% file: sections/appndx.tex
\section{Derivation of Shower Emission Jacobian Factors}
\label{sec:Jacobian}

In this appendix, we briefly recap the main steps to calculate the Jacobian factors quoted in Section~\ref{sec:showersplitting}. We do this explicitly for the \PG\ scheme and only mention the points where the other schemes differ.

We start from the expression for $m\!+\!1$-body phase space in terms of $m$-body phase space with an emission momentum $q''$ and an unknown Jacobian factor, which it is our aim to calculate:
\begin{align}
    \mathrm{d}\Phi_{m+1} =\;&
    \mathrm{d}\Phi_m \,
    \frac{\mathrm{d}^4q''}{(2\pi)^4}\,
    (2\pi)\delta(q^{\prime\prime2}) \,
    J_{\mathrm{emit}}
    \\
    =\;&
    \frac{\mathrm{d}^4p_i}{(2\pi)^4}\,
    (2\pi)\delta(p_i^2) \,
    \frac{\mathrm{d}^4p_j}{(2\pi)^4}\,
    (2\pi)\delta(p_j^2) \,
    \prod_k
    \left(
    \frac{\mathrm{d}^4p_k}{(2\pi)^4}\,
    (2\pi)\delta(p_k^2)
    \right)
    \nonumber\\&\;
    (2\pi)^4\delta^{(4)}\left(Q-p_i-p_j-\sum_kp_k\right)
    \times \frac{\mathrm{d}^4q''}{(2\pi)^4}\,
    (2\pi)\delta(q^{\prime\prime2}) \,
    J_{\mathrm{emit}}
    \,.
\end{align}
We introduce the first post-emission momenta $q''_i=zp_i$ and $q''_j=\kappa p_j$ by multiplying by
\begin{equation}
    \int \mathrm{d}z \,
    \delta\left(
    z-1+\frac{q''\!\cdot\!p_j}{p_i\!\cdot\!p_j}
    \right)
    \int \mathrm{d}^4q''_i \, \delta^{(4)}(q''_i-zp_i) \,,
\end{equation}
and similarly for $\kappa$. Integrating out $p_i$ on this delta-function, and replacing $p_i$ by $q''_i/z$ in its mass-shell delta-function, gives a factor of $1/z^2$ and a remaining integral of
\begin{equation}
    \int \mathrm{d}z \,
    \delta\left(
    z-1+z\frac{q''\!\cdot\!p_j}{q''_i\!\cdot\!p_j}
    \right)
    =\frac1{1+\frac{q''\cdot p_j}{q''_i\cdot p_j}}
    =\frac1{1+\frac{q''\cdot q''_j}{q''_i\cdot q''_j}}
    \equiv z \,.
\end{equation}
Thus, so far, we have
\begin{align}
    \mathrm{d}\Phi_{m+1} =\;&
    \frac{\mathrm{d}^4q''_i}{(2\pi)^4}\,
    (2\pi)\delta({q''_i}^2) \,
    \frac{\mathrm{d}^4q''_j}{(2\pi)^4}\,
    (2\pi)\delta({q''_j}^2) \,
    \prod_k
    \left(
    \frac{\mathrm{d}^4p_k}{(2\pi)^4}\,
    (2\pi)\delta(p_k^2)
    \right)
    \nonumber\\&\; \times
    (2\pi)^4\delta^{(4)}\left(Q
    -q''_i\left(1+\frac{q''\!\cdot\!q''_j}{q''_i\!\cdot\!q''_j}\right)
    -q''_j\left(1+\frac{q''\!\cdot\!q''_i}{q''_i\!\cdot\!q''_j}\right)
    -\sum_kp_k\right)
    \nonumber\\&\; \times
    \frac{\mathrm{d}^4q''}{(2\pi)^4}\,
    (2\pi)\delta(q^{\prime\prime2}) \,
    \frac{J_{\mathrm{emit}}}{z\kappa}
    \,.
\end{align}
Now we introduce the rescaling, $\alpha$, by multiplying by
\begin{equation}
    \int_0 \mathrm{d}\alpha \,
    2\bigl(q''_i+q''+q''_j\bigr)\!\cdot\!\bigl(p_m+\alpha\bigl(q''_i+q''+q''_j\bigr)\bigr) \,
    \delta\Bigl(\bigl(p_m+\alpha\bigl(q''_i+q''+q''_j\bigr)\bigr)^2-Q^2\Bigr) \,,
\end{equation}
where we have explicitly chosen the positive root for $\alpha$, and where $p_m$ is a shorthand for $\sum_kp_k$ (not a formal change of variables). At the same time, we also introduce the total momentum after emission and rescaling, $Q'$:
\begin{equation}
    \int \mathrm{d}^4Q' \,
    \delta^{(4)}\Bigl(Q'-p_m-\alpha\bigl(q''_i+q''+q''_j\bigr)\Bigr).
\end{equation}

Now we introduce rescaled versions of the emission momenta, $q'_i=\alpha q''_i$, etc. We obtain Jacobian factors of $1/\alpha^2$ from each of the integrals, and replace $q''_i$ by $q'_i/\alpha$, etc. We also trivially rename $p_k$ as $q'_k$:
\begin{align}
    \mathrm{d}\Phi_{m+1} =&\;
    \frac{\mathrm{d}^4q'_i}{(2\pi)^4}\,
    (2\pi)\delta({q'_i}^2) \,
    \frac{\mathrm{d}^4q'}{(2\pi)^4}\,
    (2\pi)\delta(q^{\prime2}) \,
    \frac{\mathrm{d}^4q'_j}{(2\pi)^4}\,
    (2\pi)\delta({q'_j}^2) \,
    \prod_k
    \left(
    \frac{\mathrm{d}^4q'_k}{(2\pi)^4}\,
    (2\pi)\delta({q'_k}^2)
    \right)
    \nonumber\\&\; \times
    (2\pi)^4\delta^{(4)}\left(Q
    -\frac1\alpha q'_i\left(1+\frac{q'\!\cdot\!q'_j}{q'_i\!\cdot\!q'_j}\right)
    -\frac1\alpha q'_j\left(1+\frac{q'\!\cdot\!q'_i}{q'_i\!\cdot\!q'_j}\right)
    -p_m\right)
    \frac{J_{\mathrm{emit}}}{z\kappa\alpha^6}
    \nonumber\\&\; \times
    \int_0 \frac{\mathrm{d}\alpha}{\alpha} \,
    2\bigl(q'_i+q'+q'_j\bigr)\!\cdot\!Q' \,
    \delta\Bigl(Q^{\prime2}-Q^2\Bigr)
    \int \mathrm{d}^4Q' \,
    \delta^{(4)}\Bigl(Q'-p_m-\bigl(q'_i+q'+q'_j\bigr)\Bigr) \,.
\end{align}
The factors in brackets in the second delta-function are $1/z$ and $1/\kappa$ respectively. Since these have uniform expressions in terms of unprimed, single-primed and double-primed momenta, they could safely be left as $1/z$ and $1/\kappa$, but we prefer to consistently only write delta-function arguments in terms of other integration variables, so we leave them in this clumsier form.

Now, since $Q$ is fixed by input, and $Q'$ is fixed as an integration variable, the Lorentz transformation from the rest-frame of $Q'$ to that of $Q$, $\Lambda(Q,Q')$ is a fixed transformation with determinant~1, so we can replace all primed integration measures by unprimed, $q_l=\Lambda(Q,Q')q'_l$, and replace $q'_l$ by the inverse transform, $\Lambda(Q',Q)q_l$:
\begin{align}
    \mathrm{d}\Phi_{m+1} =&\;
    \frac{\mathrm{d}^4q_i}{(2\pi)^4}\,
    (2\pi)\delta(q_i^2) \,
    \frac{\mathrm{d}^4q}{(2\pi)^4}\,
    (2\pi)\delta(q^2) \,
    \frac{\mathrm{d}^4q_j}{(2\pi)^4}\,
    (2\pi)\delta(q_j^2) \,
    \prod_k
    \left(
    \frac{\mathrm{d}^4q_k}{(2\pi)^4}\,
    (2\pi)\delta(q_k^2)
    \right)
    \nonumber\\&\; \times
    (2\pi)^4\delta^{(4)}\Biggl(Q
    -\frac1\alpha\Lambda(Q',Q)q_i
    \left(1+\frac{q\!\cdot\!q_j}{q_i\!\cdot\!q_j}\right)
    -\frac1\alpha\Lambda(Q',Q)q_j
    \left(1+\frac{q\!\cdot\!q_i}{q_i\!\cdot\!q_j}\right)
    \nonumber\\&\hspace{25em}
    -\Lambda(Q',Q)\sum_kq_k\Biggr)
    \nonumber\\&\; \times
    \frac{J_{\mathrm{emit}}}{z\kappa\alpha^6}
    \int_0 \frac{\mathrm{d}\alpha}{\alpha} \,
    2\bigl(q_i+q+q_j\bigr)\!\cdot\!Q \,
    \delta\Bigl(Q^{\prime2}-Q^2\Bigr)
    \int \mathrm{d}^4Q' \,
    \delta^{(4)}\Bigl(Q'-\Lambda(Q',Q)\sum_lq_l\Bigr) \,.
\end{align}
In the final delta-function, we can apply $\Lambda(Q,Q')$ to both terms, and obtain the post-emission four-momentum conserving delta-function. In the first four-momentum conserving delta-function, we can do the same, but have to evaluate
\begin{equation}
    \Lambda(Q,Q')Q=
    \frac{2 Q\!\cdot\!Q'}{Q^2}Q - Q' \,.
\end{equation}
We have succeeded in isolating the $m\!+\!1$-body phase space, and just have to evaluate the remaining factors:
\begin{align}
    \mathrm{d}\Phi_{m+1} =&\;
    \frac{\mathrm{d}^4q_i}{(2\pi)^4}\,
    (2\pi)\delta(q_i^2) \,
    \frac{\mathrm{d}^4q}{(2\pi)^4}\,
    (2\pi)\delta(q^2) \,
    \frac{\mathrm{d}^4q_j}{(2\pi)^4}\,
    (2\pi)\delta(q_j^2) \,
    \prod_k
    \left(
    \frac{\mathrm{d}^4q_k}{(2\pi)^4}\,
    (2\pi)\delta(q_k^2)
    \right)
    \nonumber\\&\; \times
    (2\pi)^4\delta^{(4)}\Bigl(Q-\sum_lq_l\Bigr)
    \times
    \frac{J_{\mathrm{emit}}}{z\kappa\alpha^6}
    \nonumber\\&\; \times
    2\bigl(q_i+q+q_j\bigr)\!\cdot\!Q
    \int_0 \frac{\mathrm{d}\alpha}{\alpha} \,
    \delta\Bigl(Q^{\prime2}-Q^2\Bigr)
    \nonumber\\&\; \times
    \int \mathrm{d}^4Q' \,
    \delta^{(4)}\left(\frac{2 Q\!\cdot\!Q'}{Q^2}Q - Q'
    -\frac1\alpha q_i
    \left(1+\frac{q\!\cdot\!q_j}{q_i\!\cdot\!q_j}\right)
    -\frac1\alpha q_j
    \left(1+\frac{q\!\cdot\!q_i}{q_i\!\cdot\!q_j}\right)
    -\sum_kq_k\right) . \!
\end{align}
To simplify the final delta-function, we can note that $2(Q\cdot Q'/Q^2)Q-Q'$ is the result of reflecting the spatial components of $Q'$ in the $Q$ rest-frame, an operation we write as $\mathcal{R}_QQ'$. $\mathcal{R}_Q$ has the properties that (a)~it is self-inverse, $\mathcal{R}_Q\mathcal{R}_Q=1$, (b)~it has determinant~$-1$, so $\mathrm{d}^4a=\mathrm{d}^4a\times \int\mathrm{d}^4b\,\delta^{(4)}(b-\mathcal{R}_Qa)=|\det \mathcal{R}_Q|^{-1}\mathrm{d}^4b=\mathrm{d}^4b$, and (c)~it preserves dot products, $(\mathcal{R}_Qa)\!\cdot\!(\mathcal{R}_Qb)=a\!\cdot\!b$. Properties (a) and (b) can be combined to make the $Q'$ integral trivial. Property (c) then means the result can be fed into the delta-function on the previous line, and we are left with
\begin{align}
    &
    2\bigl(q_i+q+q_j\bigr)\!\cdot\!Q
    \int_0 \frac{\mathrm{d}\alpha}{\alpha} \,
    \delta\left(\left[\frac1\alpha q_i
    \left(1+\frac{q\!\cdot\!q_j}{q_i\!\cdot\!q_j}\right)
    +\frac1\alpha q_j
    \left(1+\frac{q\!\cdot\!q_i}{q_i\!\cdot\!q_j}\right)
    +\sum_kq_k\right]^2-Q^2\right)
    \nonumber\\&\hspace{25em}
    =\frac{\bigl(q_i+q+q_j\bigr)\!\cdot\!Q}
    {\bigl(p_i+p_j\bigr)\!\cdot\!Q} \,.
\end{align}
That is, our final result for the Jacobian factor for the PG${}_0$ scheme is
\begin{equation}
    J_{\mathrm{emit}}=z(1-zy)\alpha^6X \, ,
    \qquad 
    X=\frac{\bigl(p_i+p_j\bigr)\!\cdot\!Q}
    {\bigl(q_i+q+q_j\bigr)\!\cdot\!Q} \,.
\end{equation}
$X$ is the ratio of the total dipole energy in the $Q$ rest-frame before and after  the emission (including the emitted particle in the latter), which is~1 for the first emission.

FHP gives the same expression without the $(1-zy)$ factor. The derivation for CS is similar, but there is no boost or rescaling and the expression for $p_i$ depends on $q''$ and $q''_j$ as well as $q''_i$. The end result is $(1-y)/z$.

\clearpage

\section{Complete Set of Tuning Plots\label{sec:full-tuning}}

\subsection*{Tuned Results -- Without AO-like Boundary}

\begin{figure}[!htbp]
\centering
\includegraphics[width=0.24\linewidth]{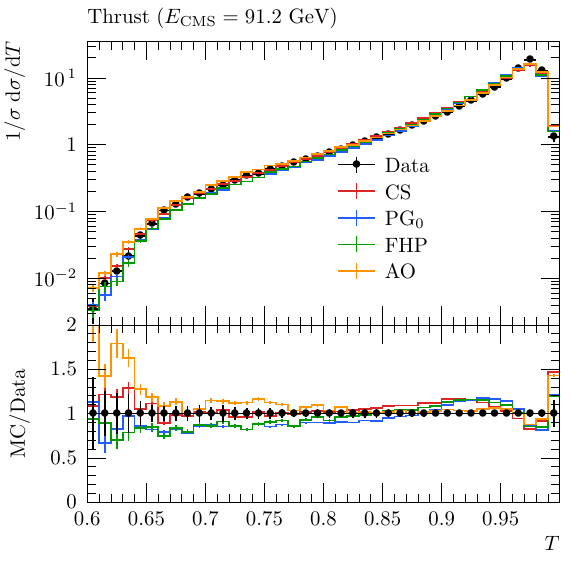}
\includegraphics[width=0.24\linewidth]{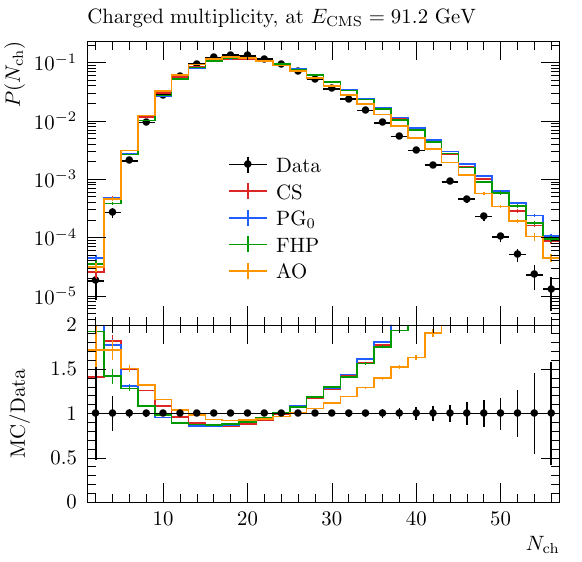}
\includegraphics[width=0.24\linewidth]{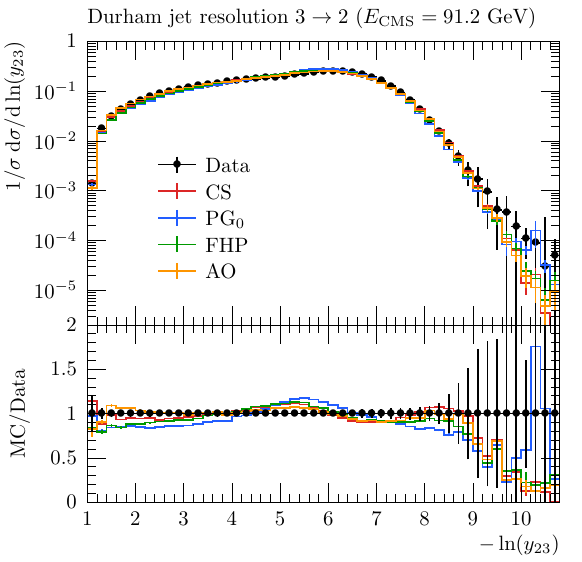}
\includegraphics[width=0.24\linewidth]{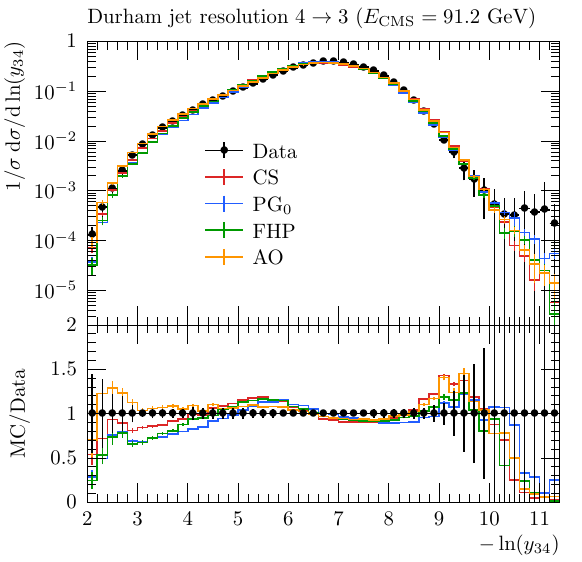}
\includegraphics[width=0.24\linewidth]{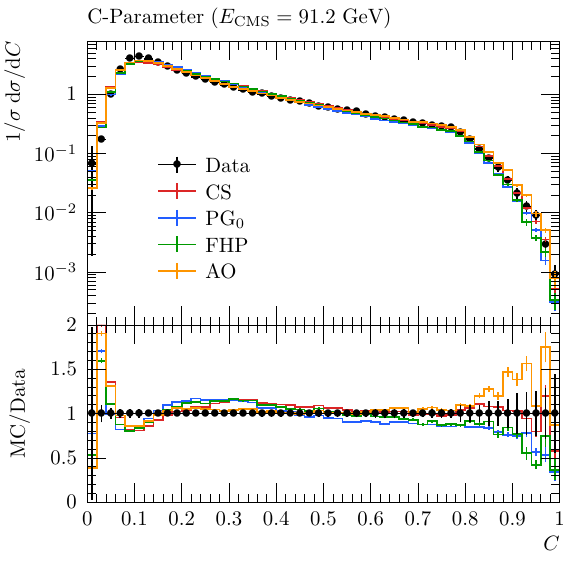}
\includegraphics[width=0.24\linewidth]{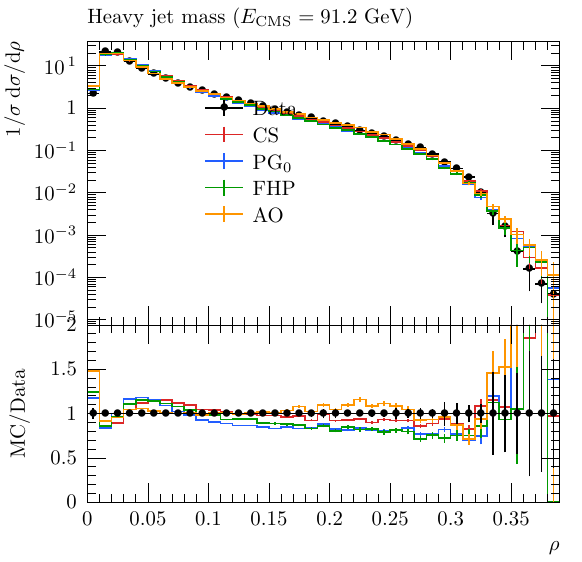}
\includegraphics[width=0.24\linewidth]{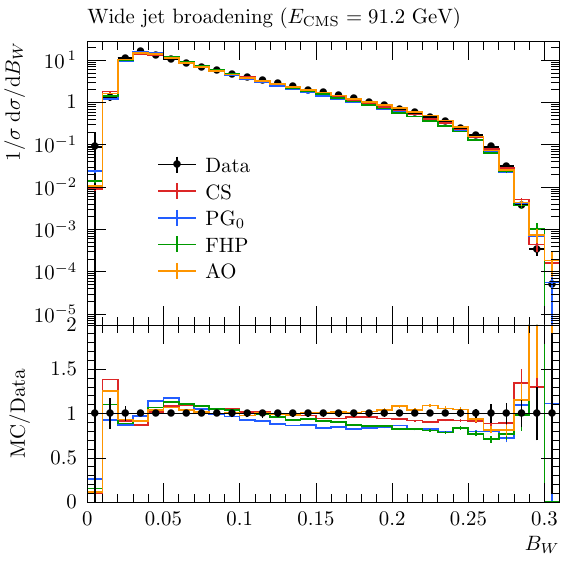}
\includegraphics[width=0.24\linewidth]{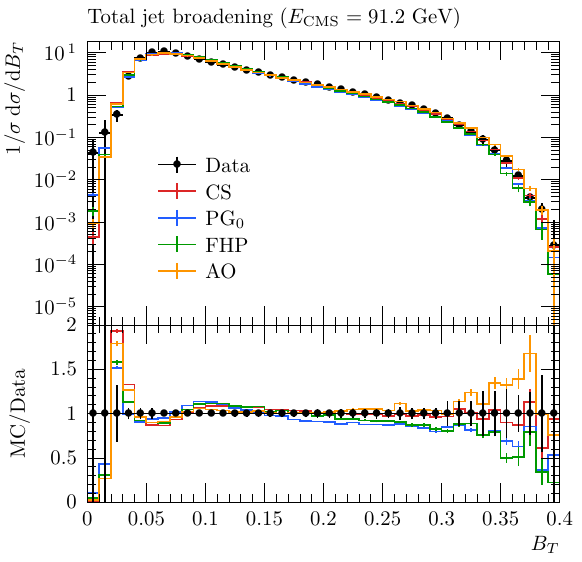}
\includegraphics[width=0.24\linewidth]{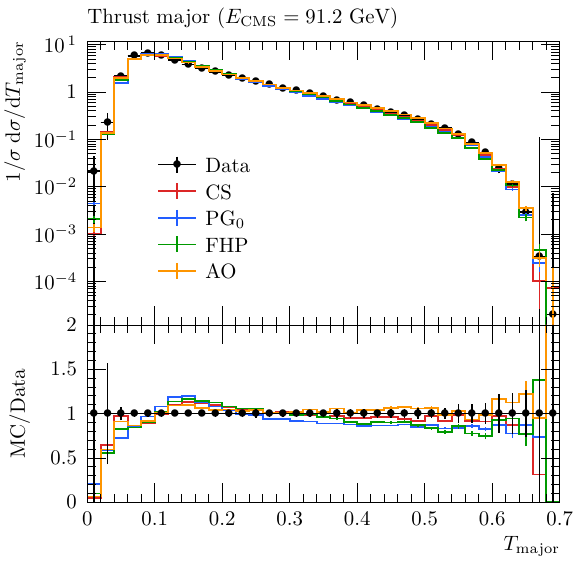}
\includegraphics[width=0.24\linewidth]{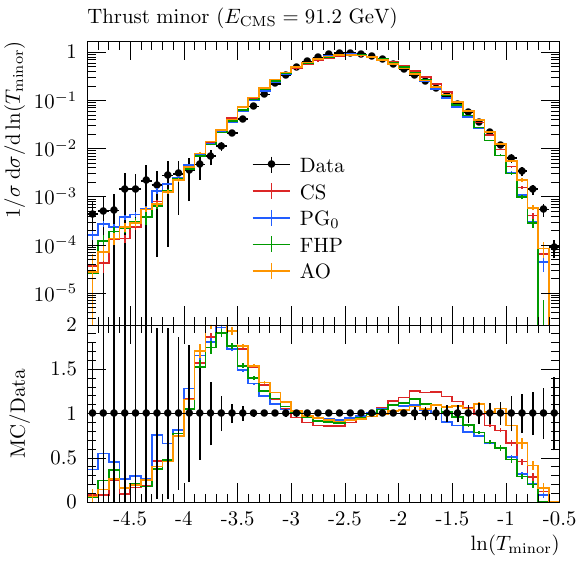}
\includegraphics[width=0.24\linewidth]{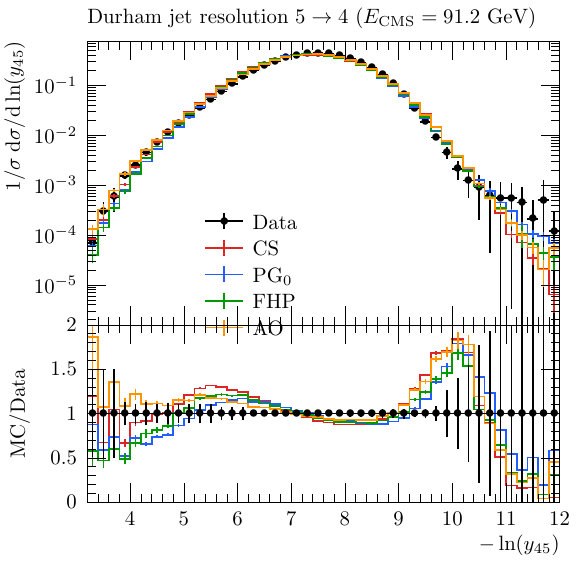}
\includegraphics[width=0.24\linewidth]{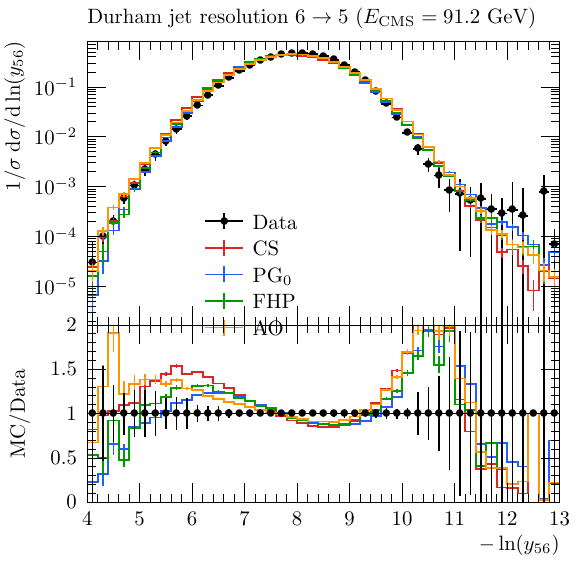}
\includegraphics[width=0.24\linewidth]{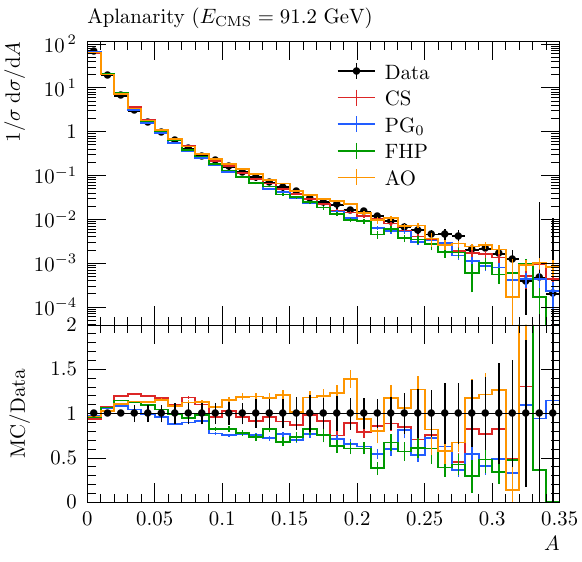}
\includegraphics[width=0.24\linewidth]{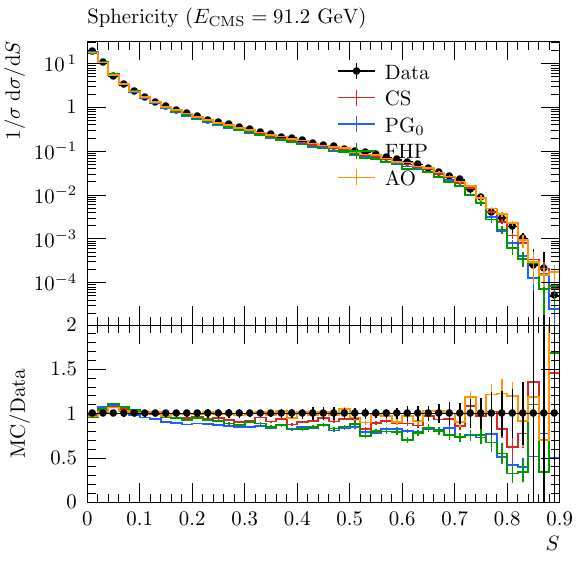}
\includegraphics[width=0.24\linewidth]{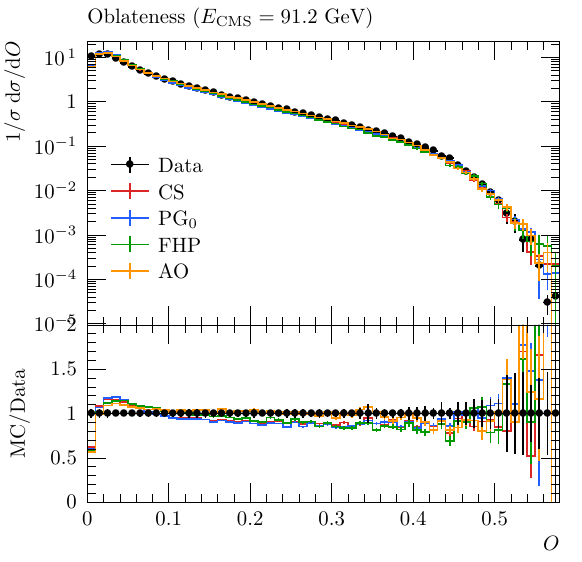}
\includegraphics[width=0.24\linewidth]{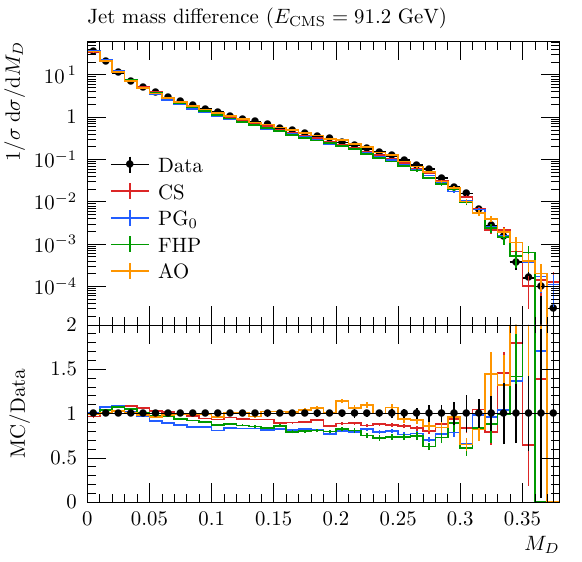}
\caption{%
Tuned results for CS, AO, \PG  and FHP.
CS still performs better in the tails of the event shapes, but the three showers share a similar behaviour near the peaks.
Some improvement is observed at higher jet rates, such as $y_{56}$.
None of the showers predicts the charged multiplicity well, with AO producing a slightly better result.
}
\label{fig:full-tune-no-AOB}
\end{figure}
\clearpage

\subsection*{Tuned Results -- With AO-like Boundary}

\begin{figure}[!htbp]
\centering
\includegraphics[width=0.24\linewidth]{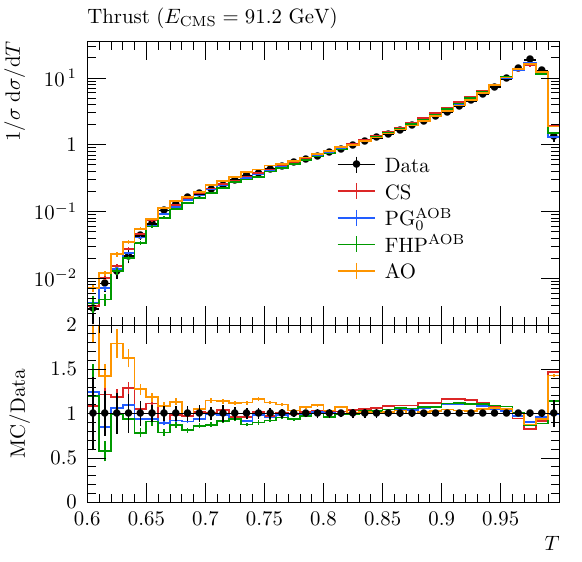}
\includegraphics[width=0.24\linewidth]{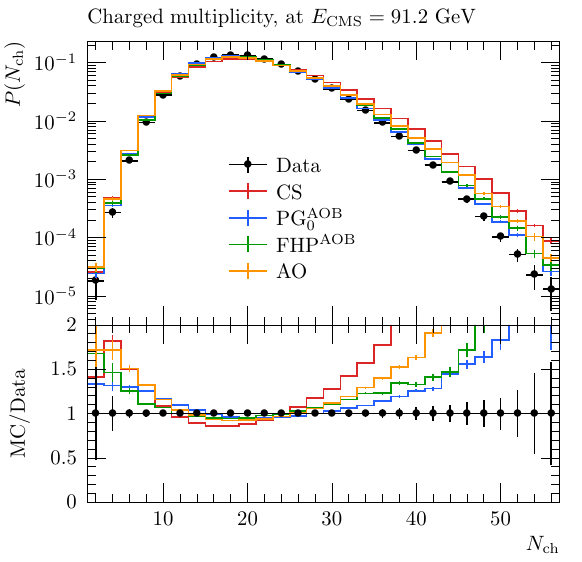}
\includegraphics[width=0.24\linewidth]{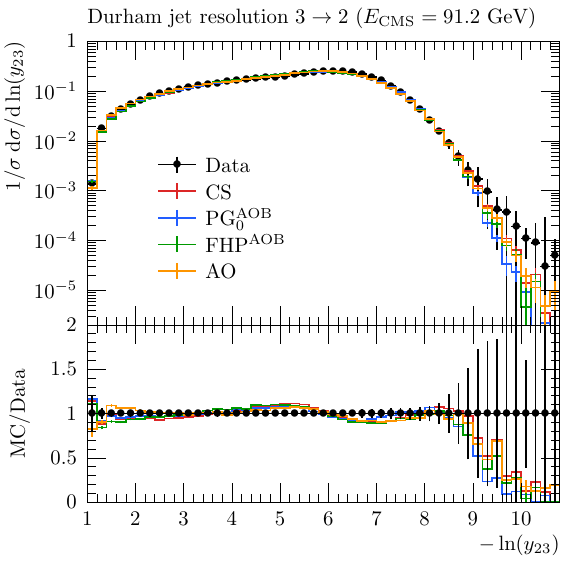}
\includegraphics[width=0.24\linewidth]{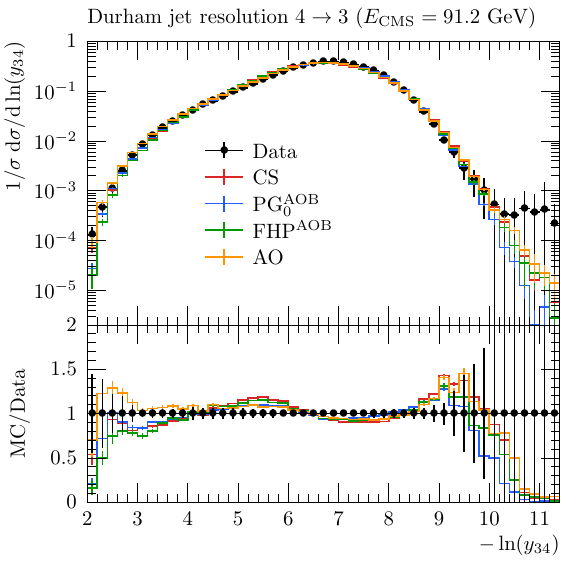}
\includegraphics[width=0.24\linewidth]{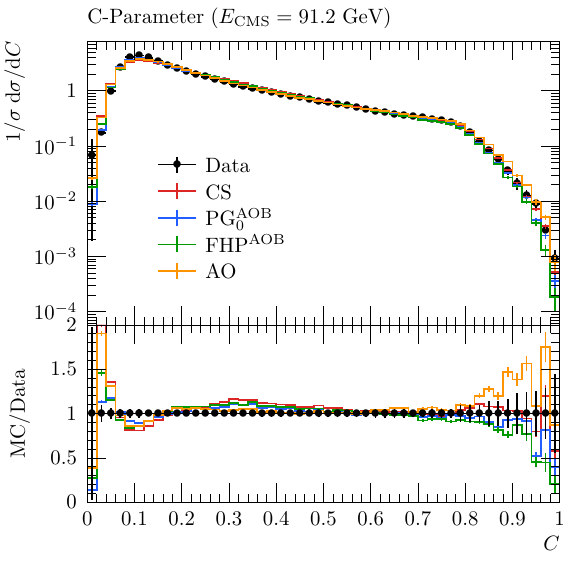}
\includegraphics[width=0.24\linewidth]{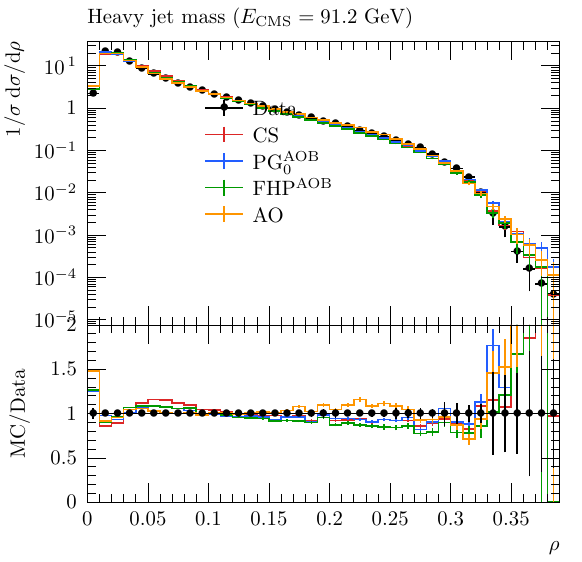}
\includegraphics[width=0.24\linewidth]{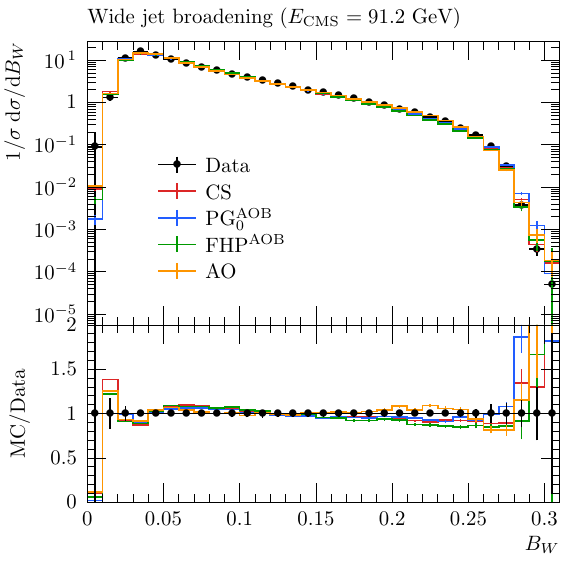}
\includegraphics[width=0.24\linewidth]{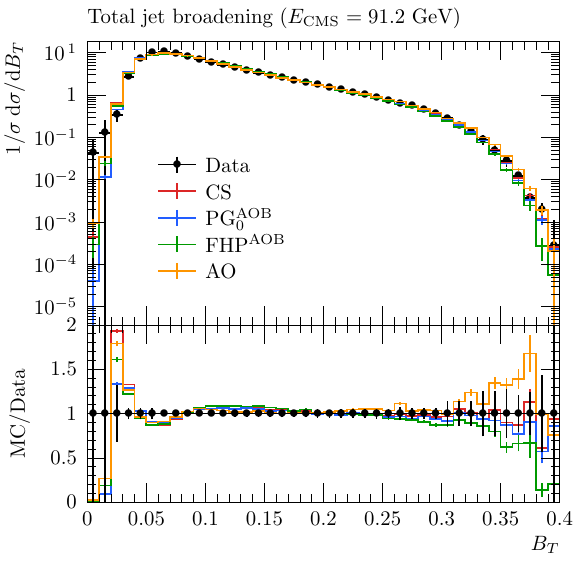}
\includegraphics[width=0.24\linewidth]{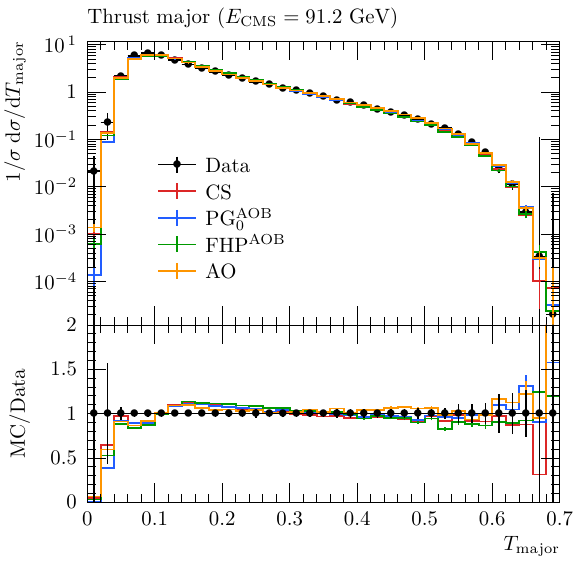}
\includegraphics[width=0.24\linewidth]{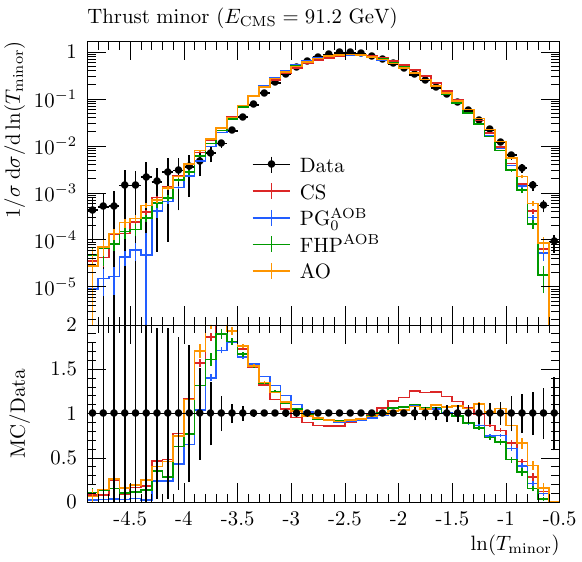}
\includegraphics[width=0.24\linewidth]{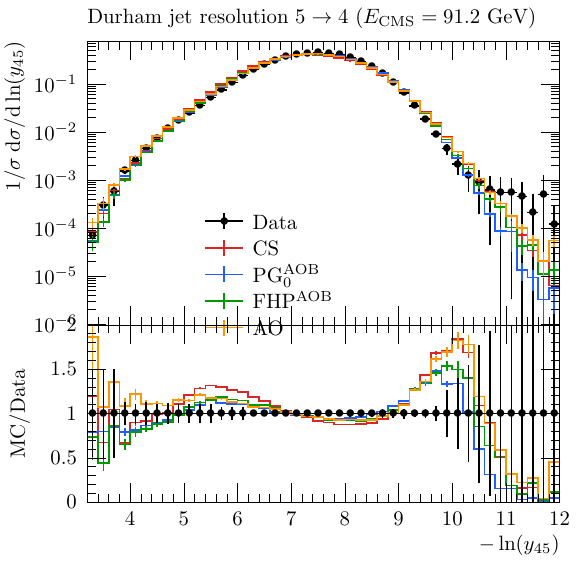}
\includegraphics[width=0.24\linewidth]{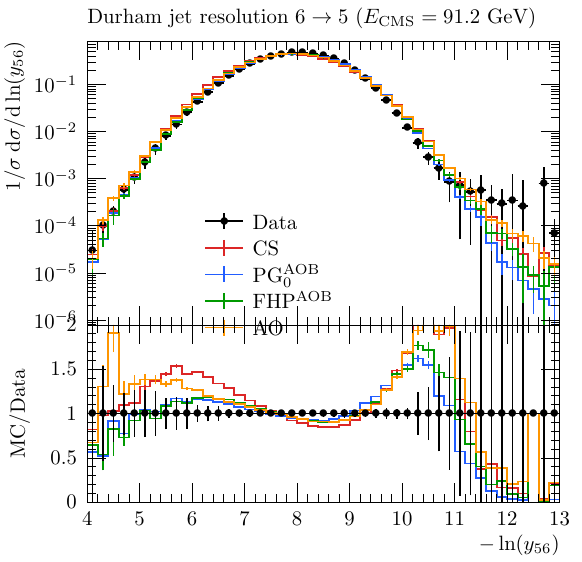}
\includegraphics[width=0.24\linewidth]{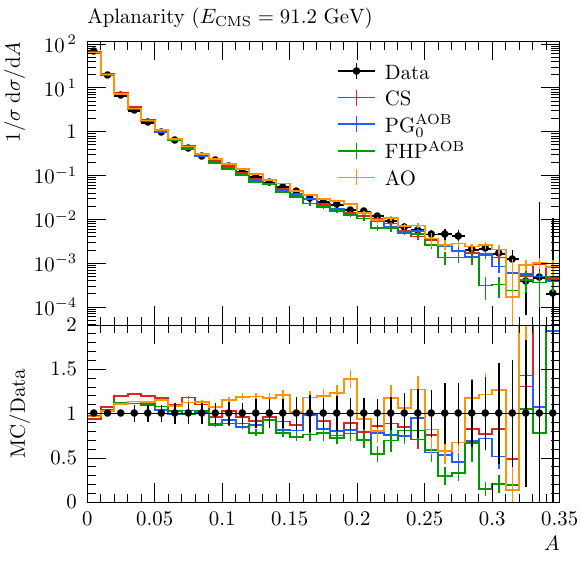}
\includegraphics[width=0.24\linewidth]{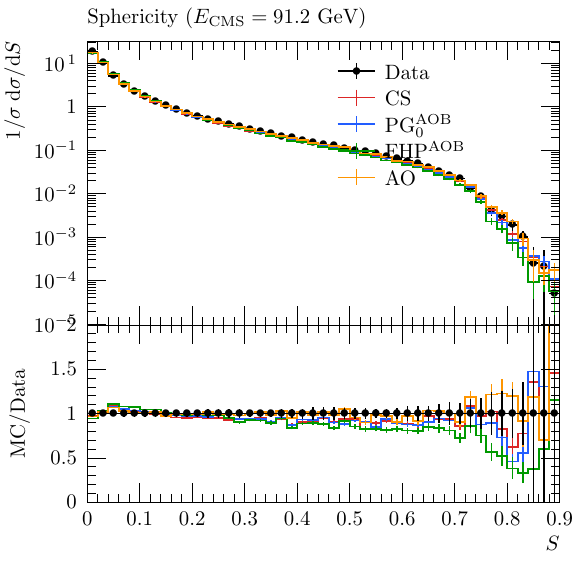}
\includegraphics[width=0.24\linewidth]{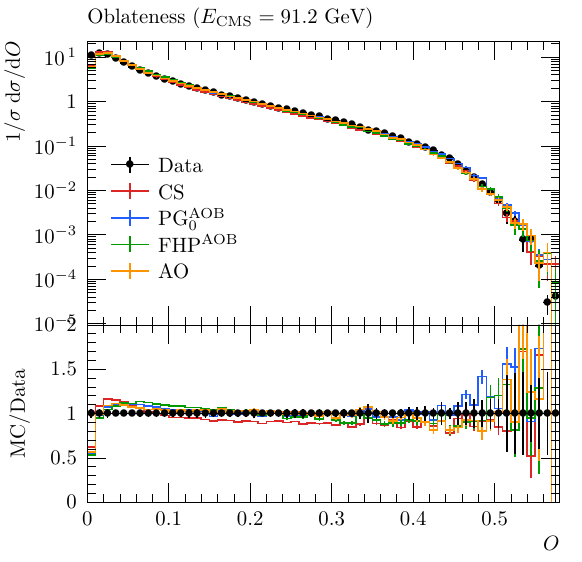}
\includegraphics[width=0.24\linewidth]{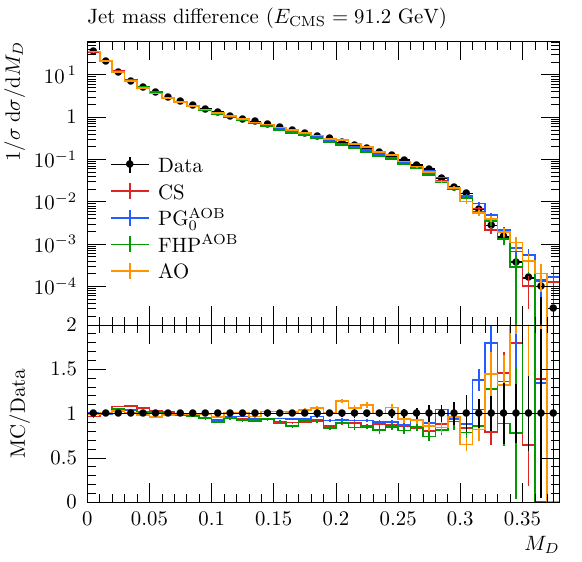}
\caption{
Tuned results for CS, \PGAOB and \FHPAOB.
The increase in $\alpha_s(m_Z)$ for the two NLL showers ensures they achieve a good fit in the tails.
Additionally, their improved description of the peak region significantly reduces the total $\chi^2$.
A further improvement is evident in the higher jet rates compared with the previous figure.
Here, CS performs very poorly in comparison.
As discussed before, the charged multiplicity is also significantly improved.
This improvement is not seen in AO.
}
\label{fig:full-tune-AOB}
\end{figure}
\clearpage

\subsection*{$\chi^2/N_{\text{df}}$ heatmaps -- $\alpha_s(m_Z)$ and $p_{T,\min}^2$}
\begin{figure}[!htbp]
\centering
\labeledfig{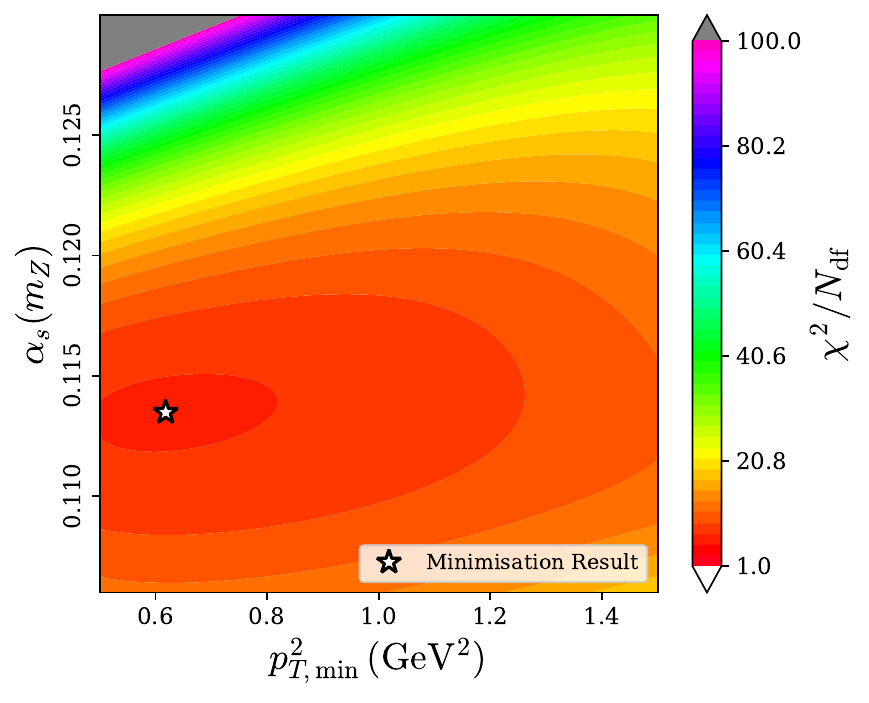}{CS}
\labeledfig{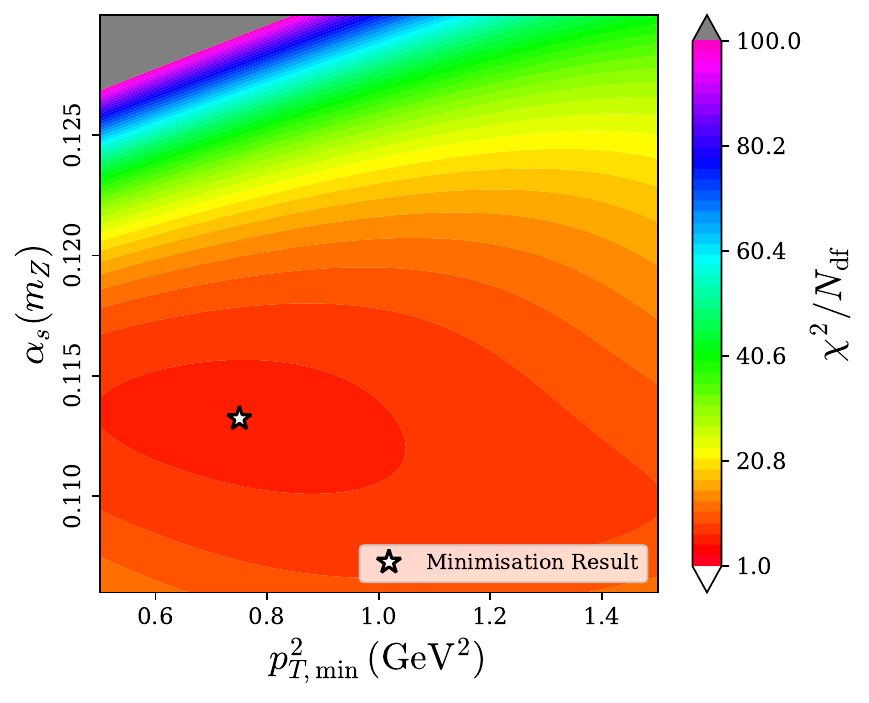}{CS}
\labeledfig{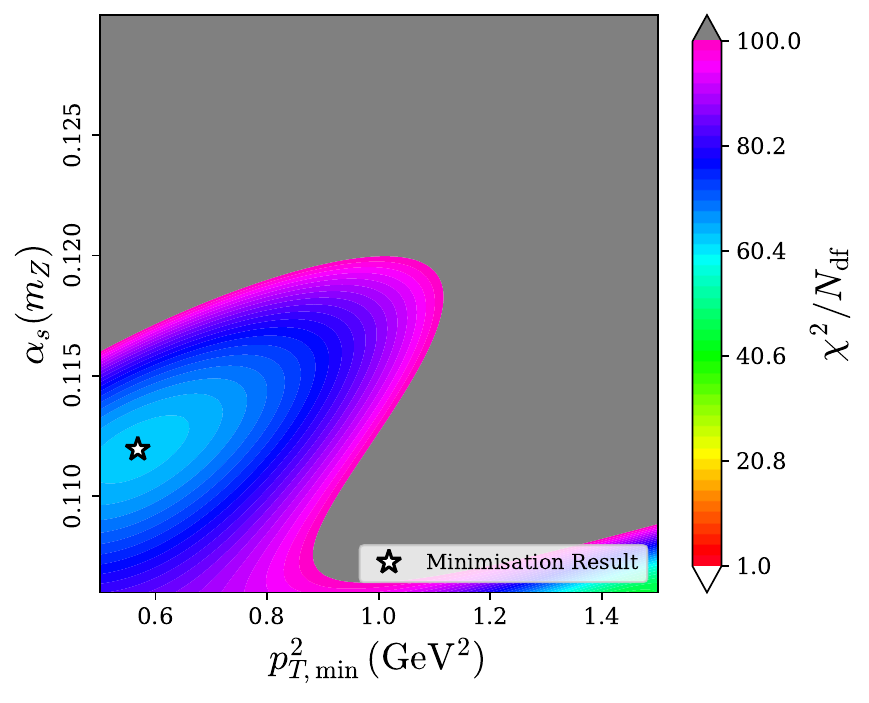}{CS}
\labeledfig{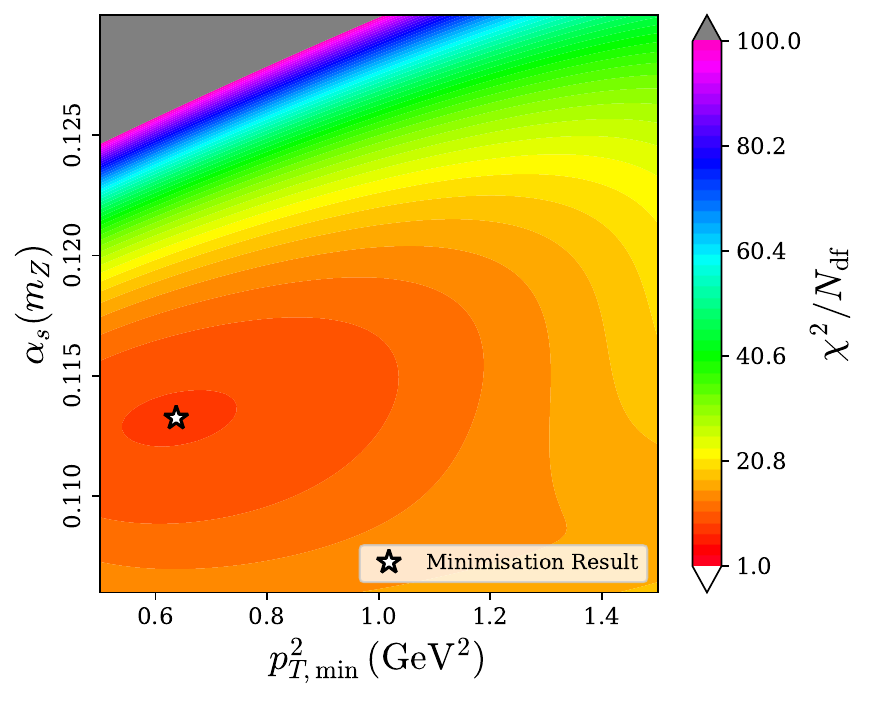}{CS}
\labeledfig{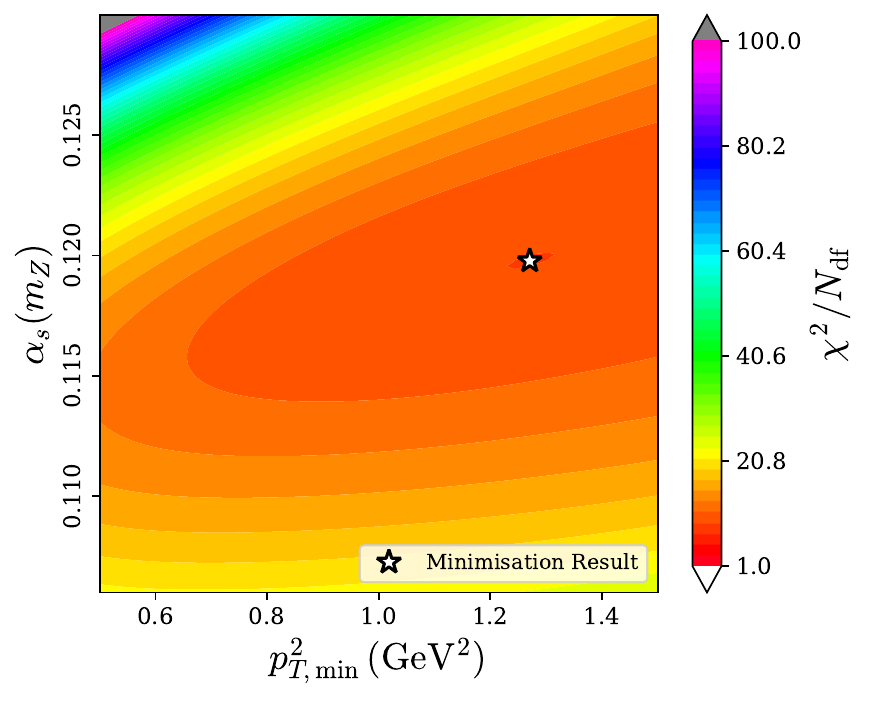}{\PG}
\labeledfig{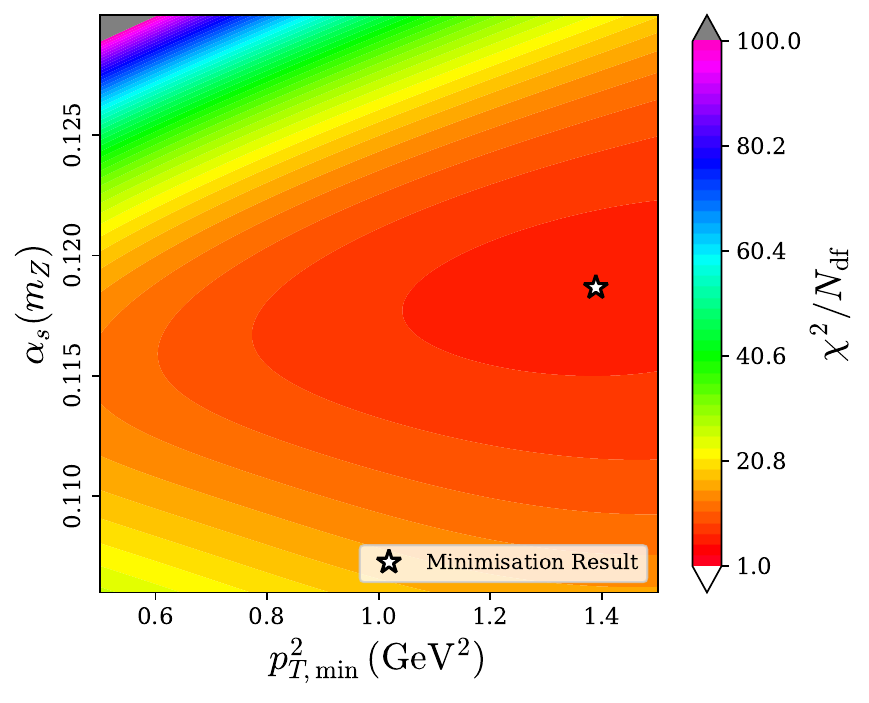}{\PG}
\labeledfig{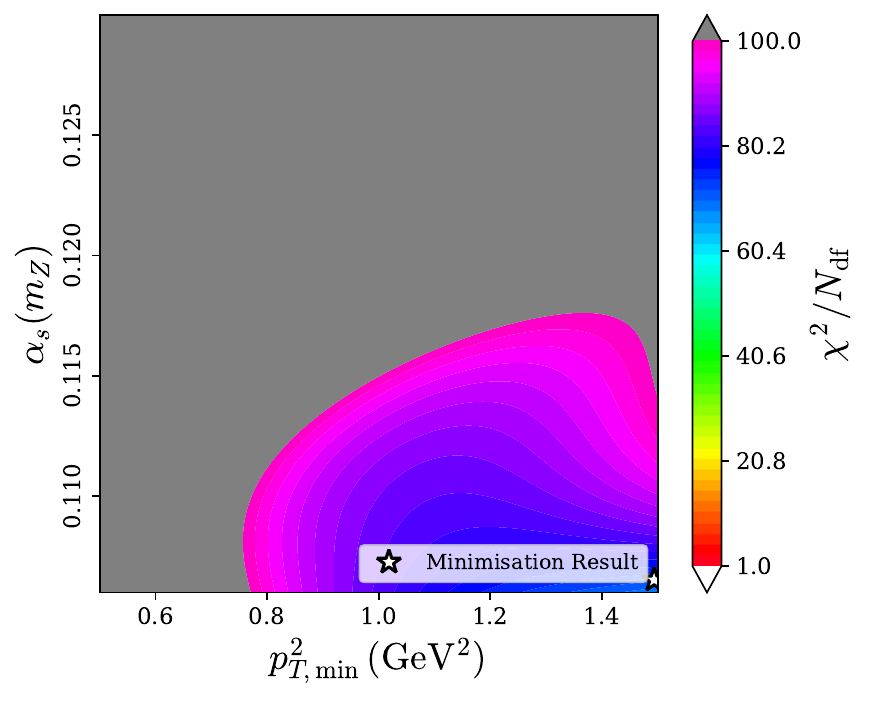}{\PG}
\labeledfig{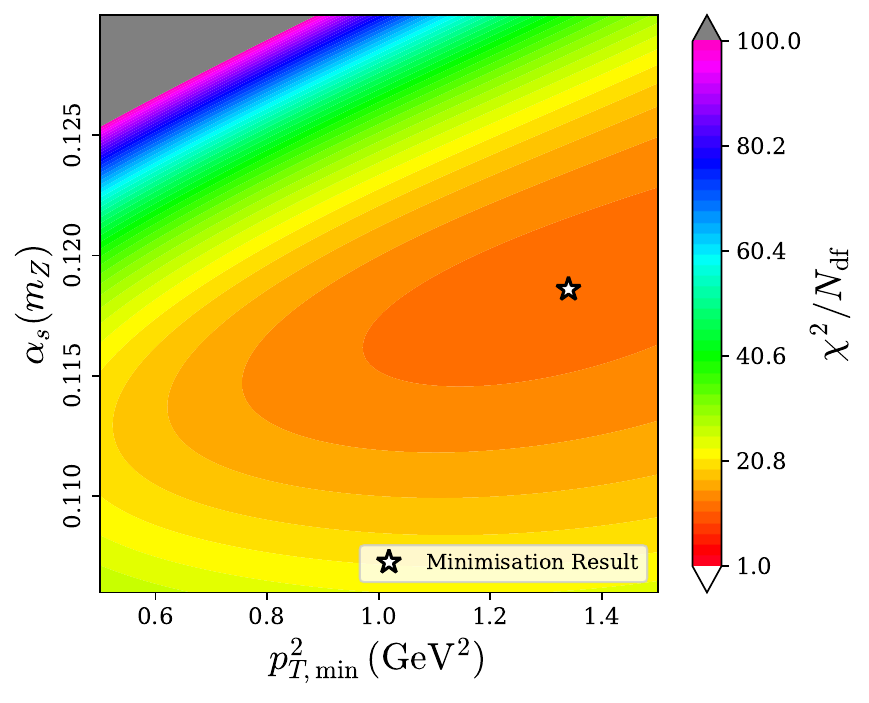}{\PG}
\labeledfig{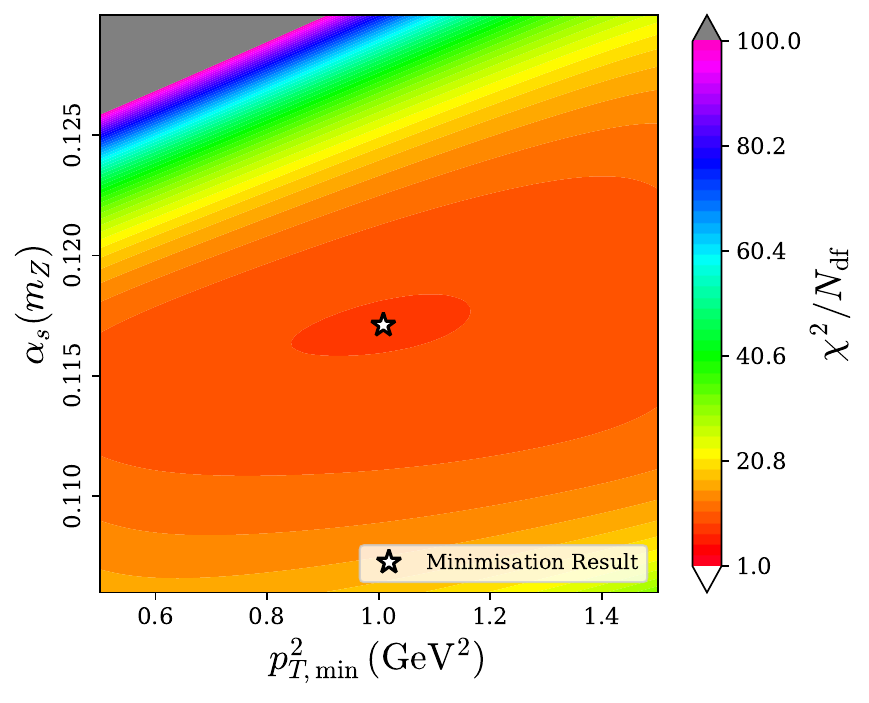}{FHP}
\labeledfig{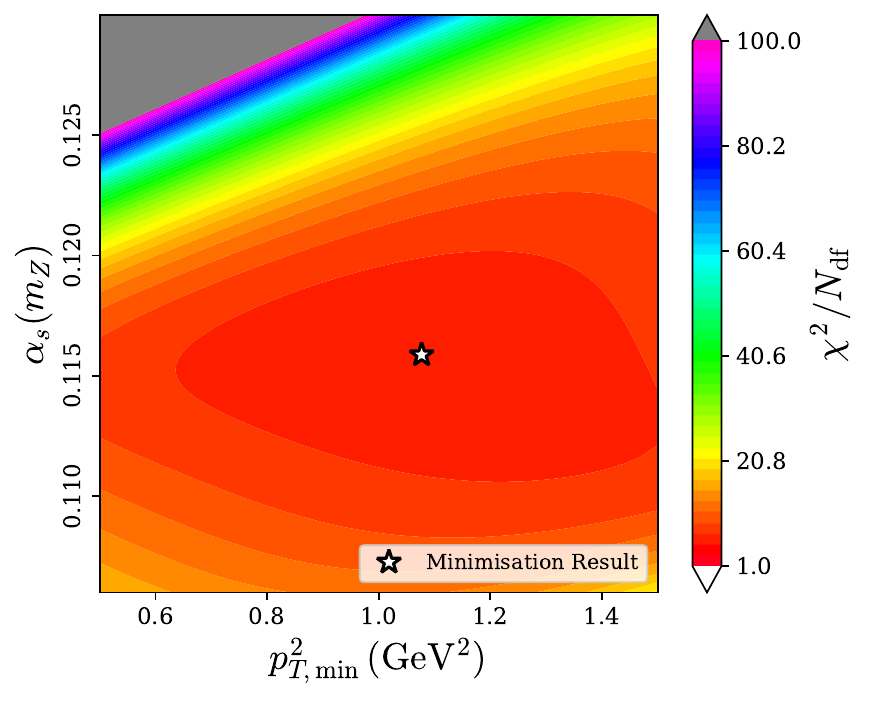}{FHP}
\labeledfig{figures/heatmaps/ChargedMult-as-pt-FHP.pdf}{FHP}
\labeledfig{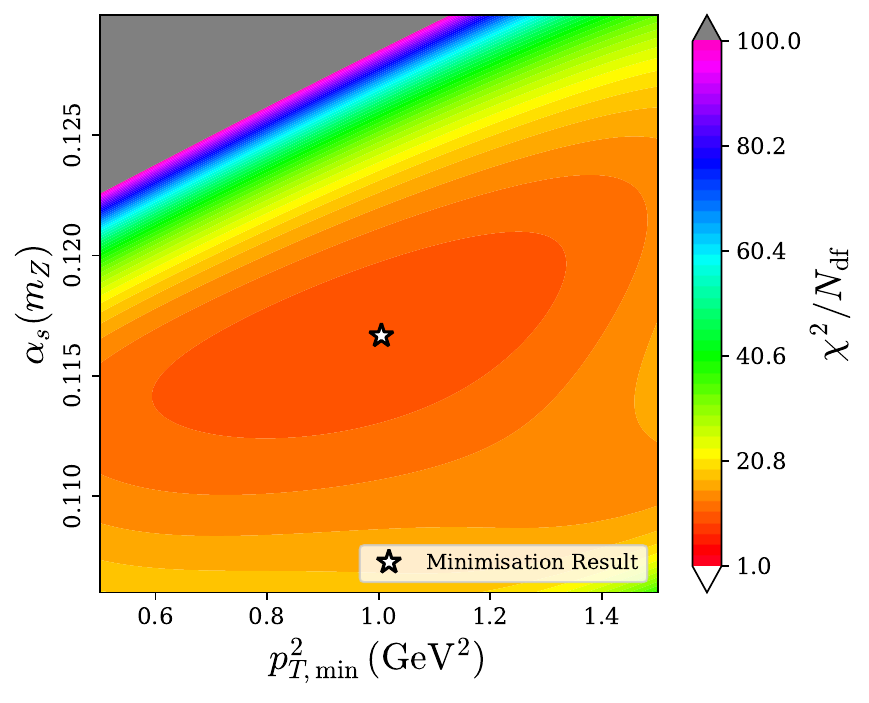}{FHP}
\labeledfig{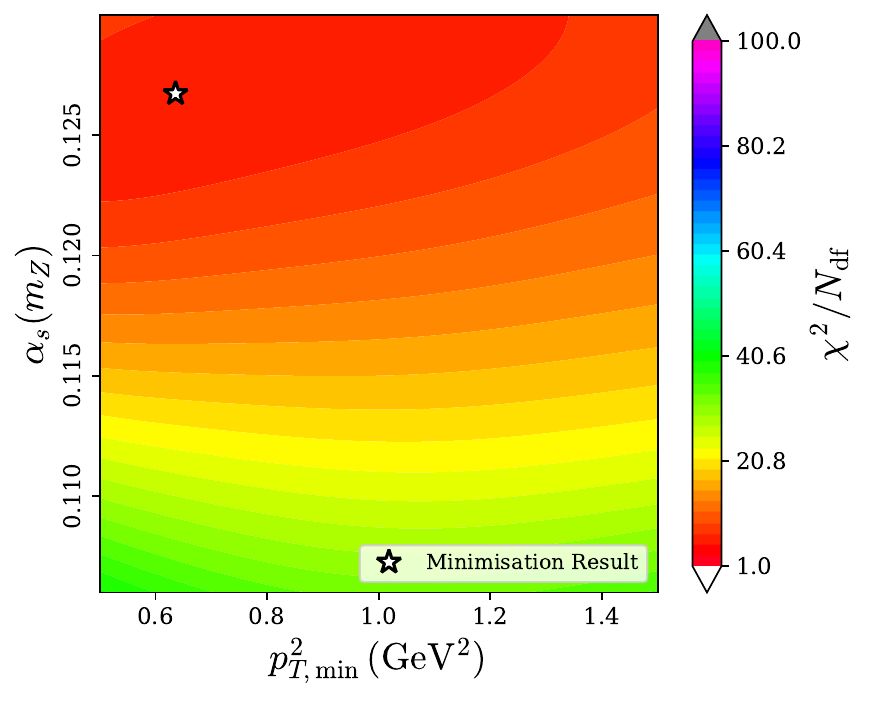}{\PGAOB}
\labeledfig{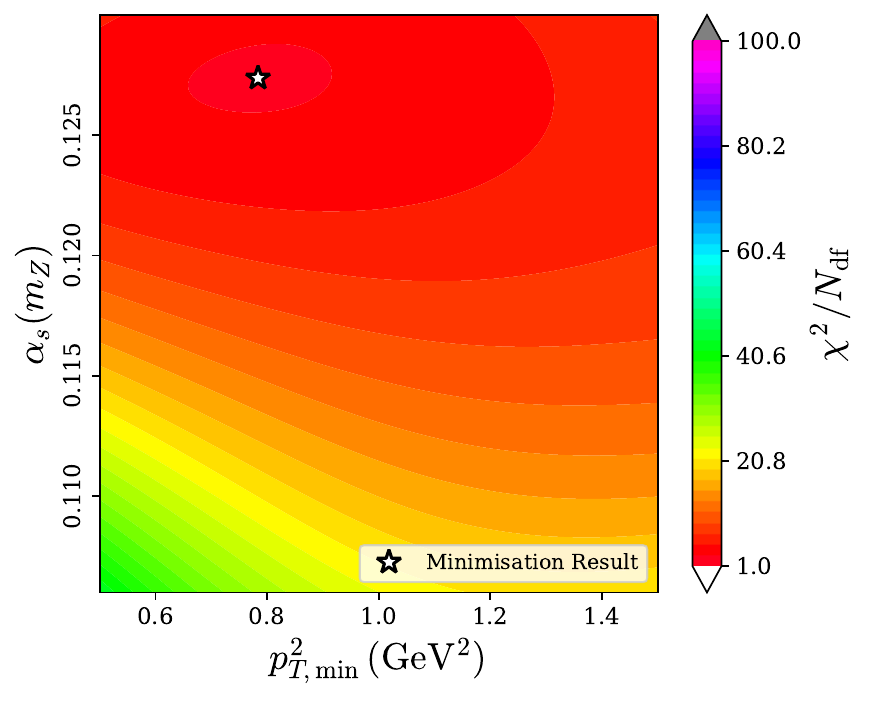}{\PGAOB}
\labeledfig{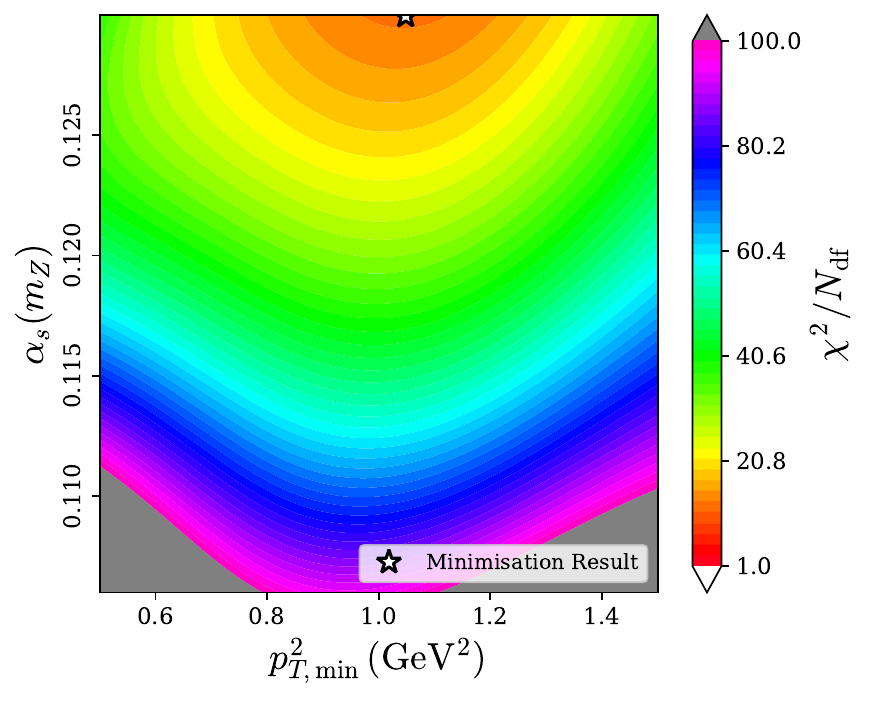}{\PGAOB}
\labeledfig{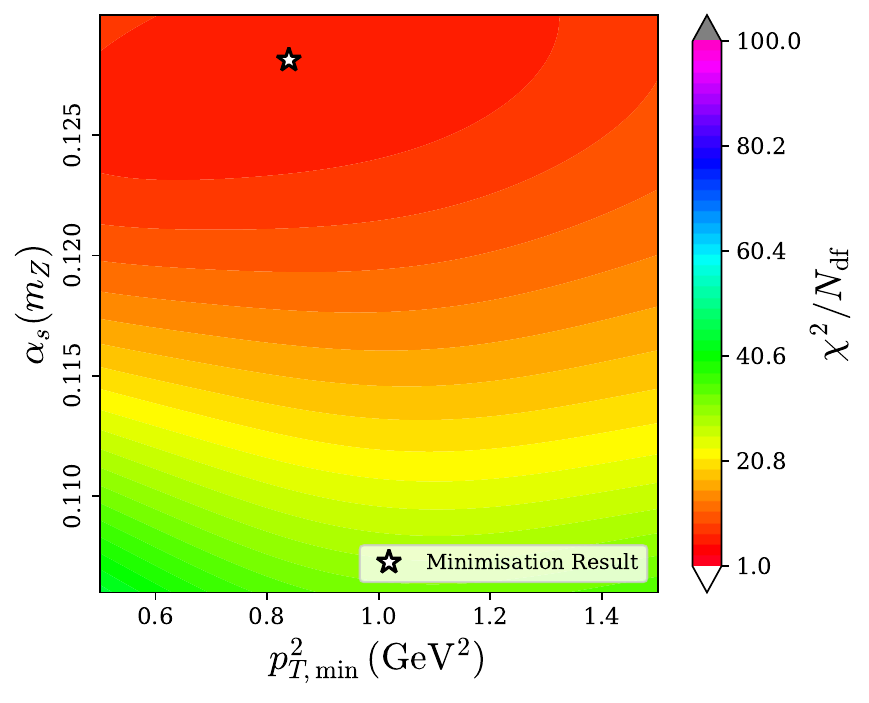}{\PGAOB}
\labeledfig{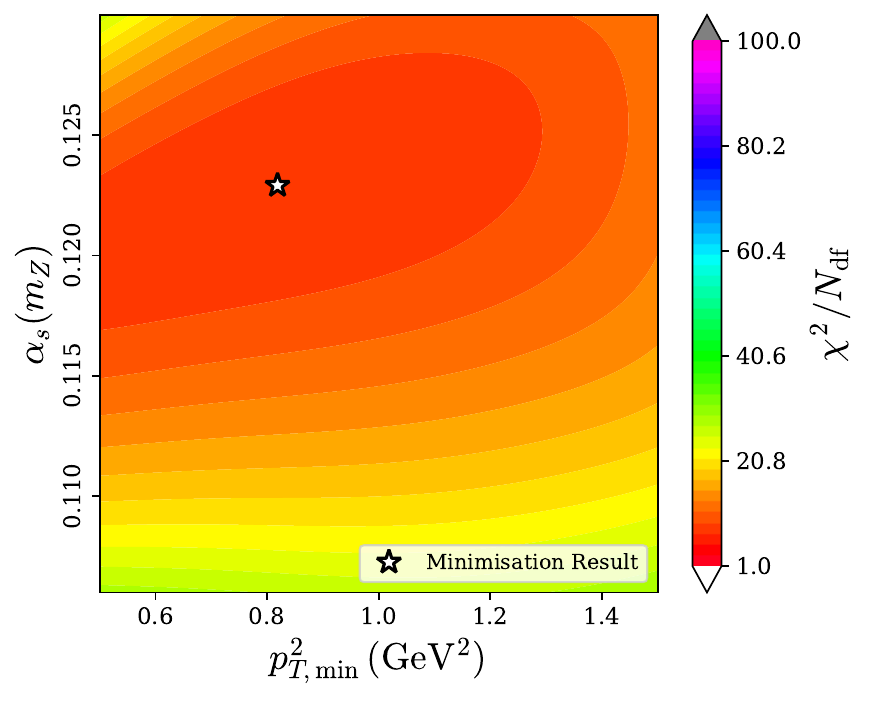}{\FHPAOB}
\labeledfig{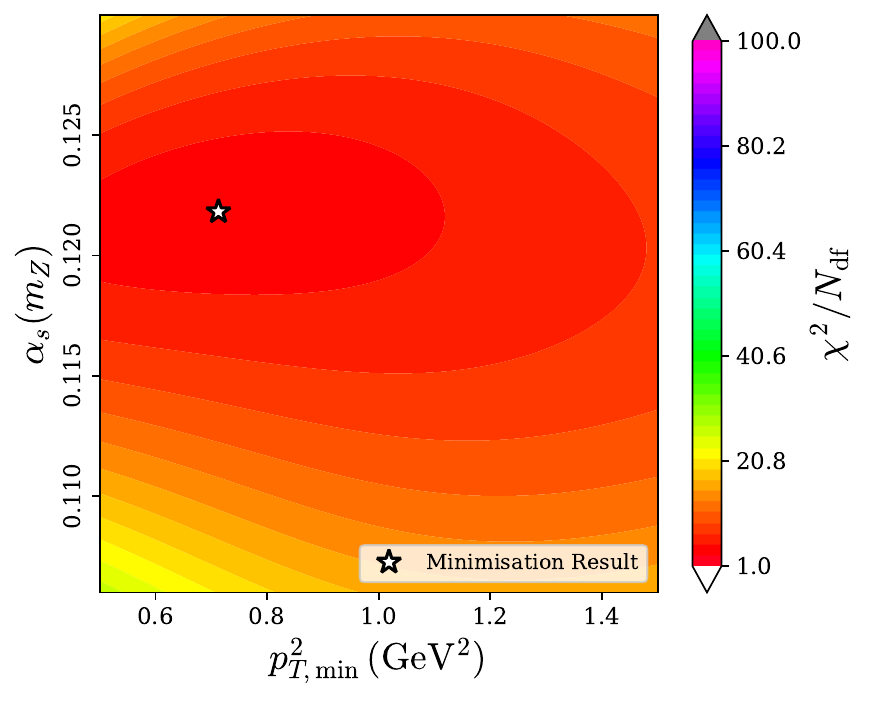}{\FHPAOB}
\labeledfig{figures/heatmaps/ChargedMult-as-pt-FHP-AOB.pdf}{\FHPAOB}
\labeledfig{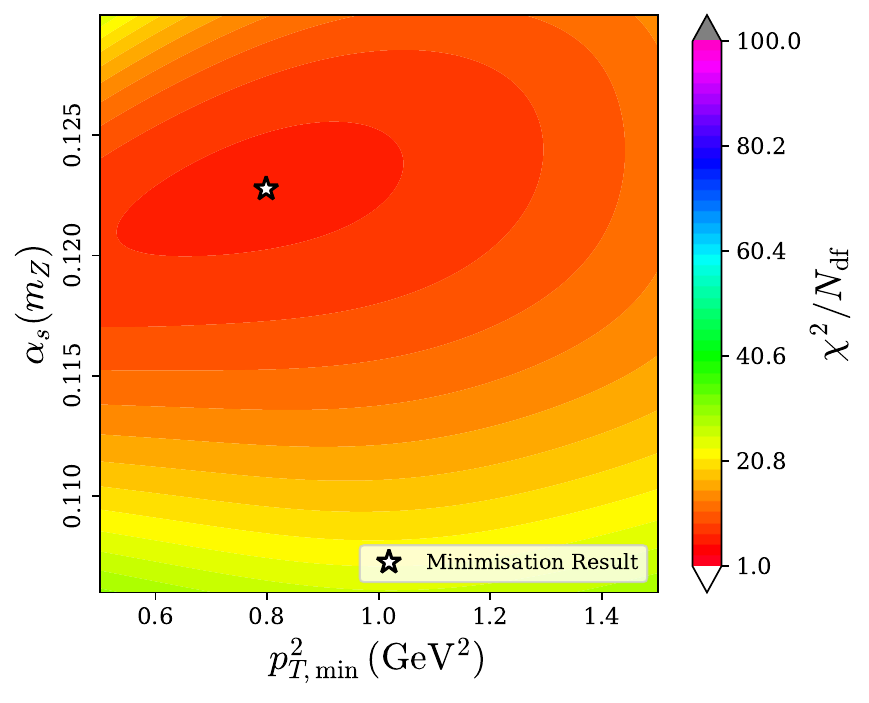}{\FHPAOB}
\caption{
$\chi^2/N_{\text{df}}$ heatmaps, varying $\alpha_s(m_Z)$ and $p_{T,\min}^2$. 
Each row shows a different shower model (in order, CS, \PG, FHP, \PGAOB,  \FHPAOB), and each column shows a different observable category (in order, Event Shapes, Jet Rates, Charged Multiplicity, All).
All heatmaps have the same key, with regions of $\chi^2/N_{\text{df}} > 100$ set to grey.
The AO-like boundary significantly improves the charged multiplicity $\chi^2/N_{\text{df}}$, and allows a good result for a large region of the phase space, as compared to the result without the AO-like boundary.
These heatmaps also explain the different tuned values of $\alpha_s(m_Z)$ and $p_{T,\min}^2$ observed in Table~\ref{tab:shower-comparison}.
Another interesting property for the CS, \PG and FHP showers is that the event shape and jet rate distributions are fairly flat in the region of good fit, and the charged multiplicity ``confines'' the fit, whose effect can be seen on the heatmap of all observables.
For \PGAOB and \FHPAOB, this effect is much milder.
}
\label{fig:heatmaps-as-pt}
\end{figure}
\clearpage

\subsection*{$\chi^2/N_{\text{df}}$ heatmaps -- $p_{T,\min}^2$ and \texttt{Clmax}}
\begin{figure}[!htbp]
\centering
\labeledfig{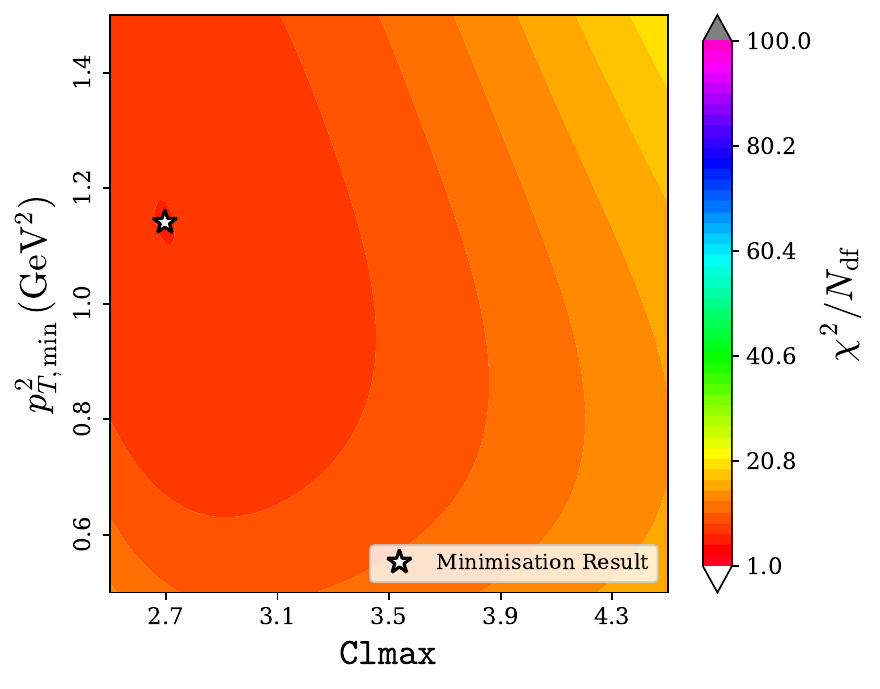}{CS}
\labeledfig{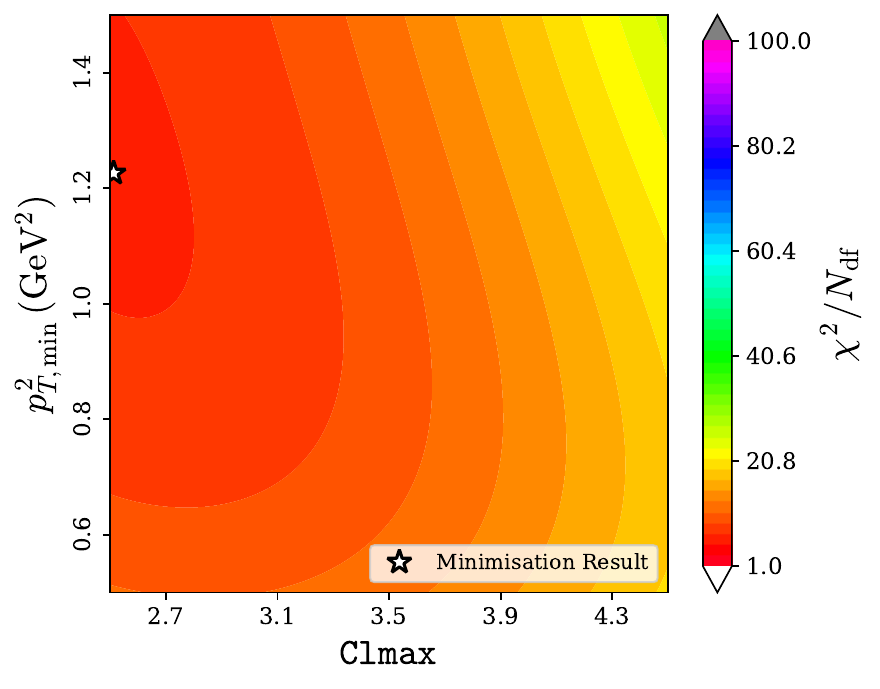}{CS}
\labeledfig{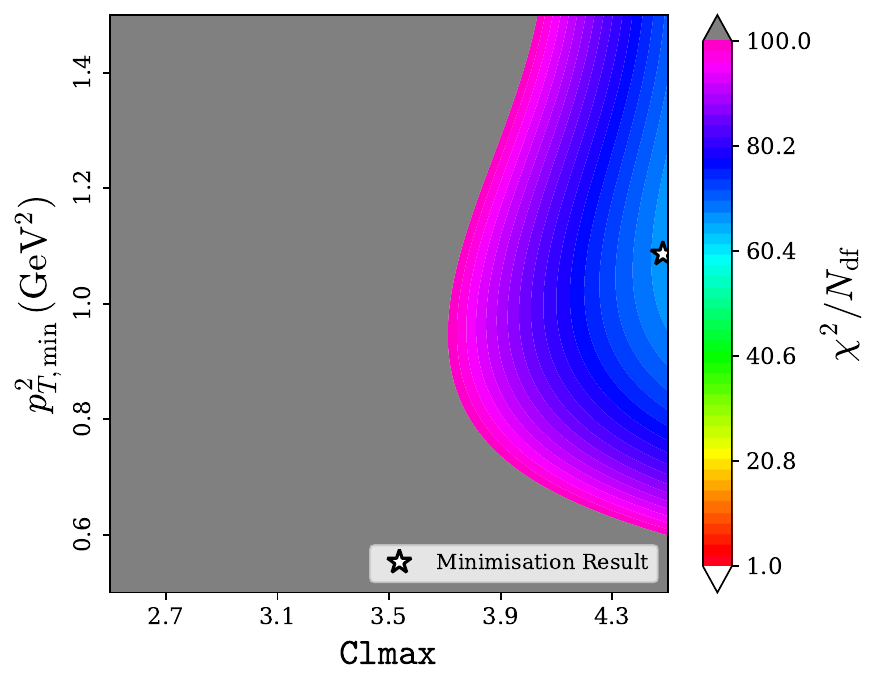}{CS}
\labeledfig{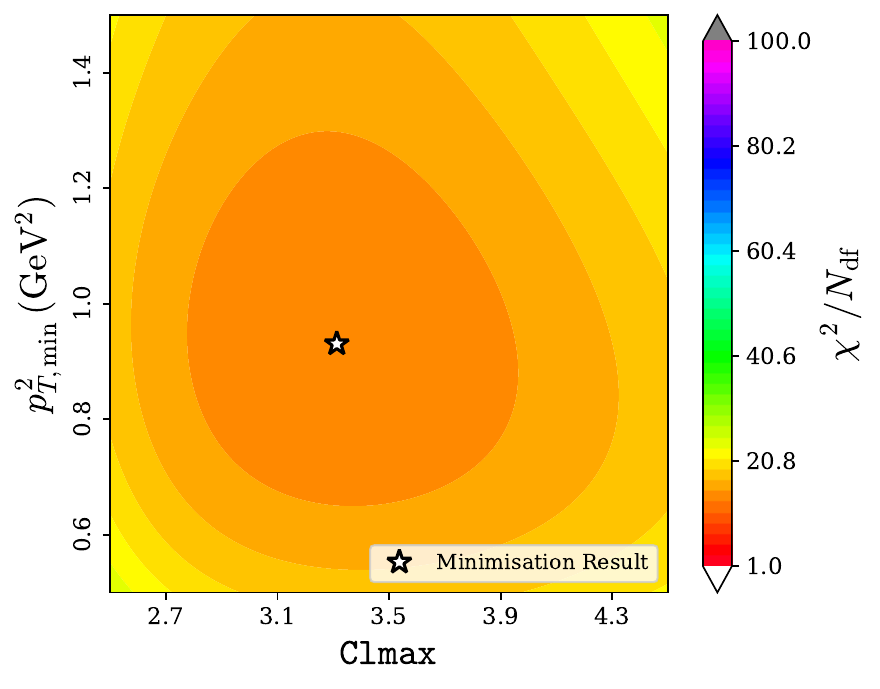}{CS}
\labeledfig{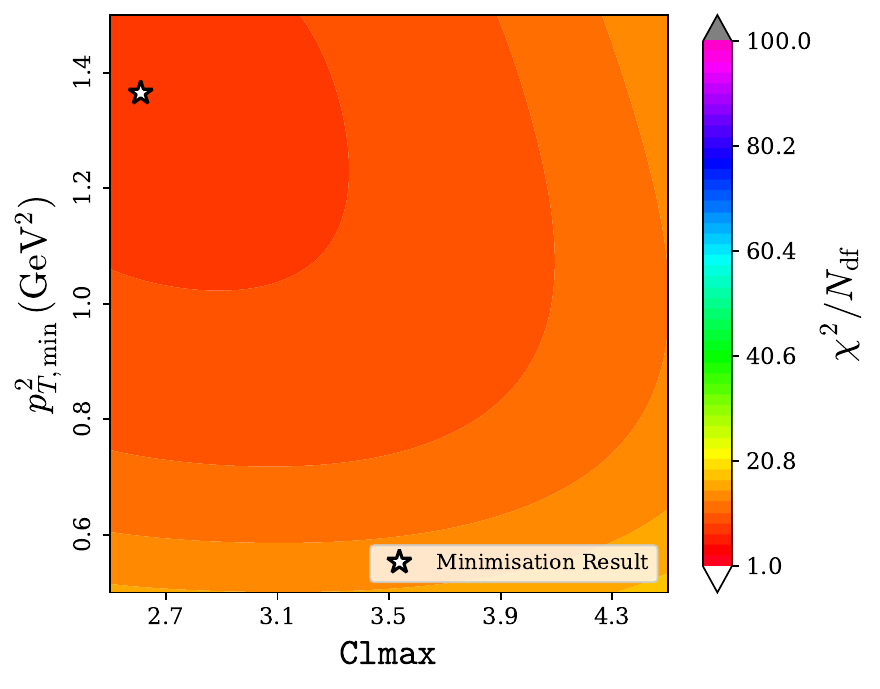}{\PG}
\labeledfig{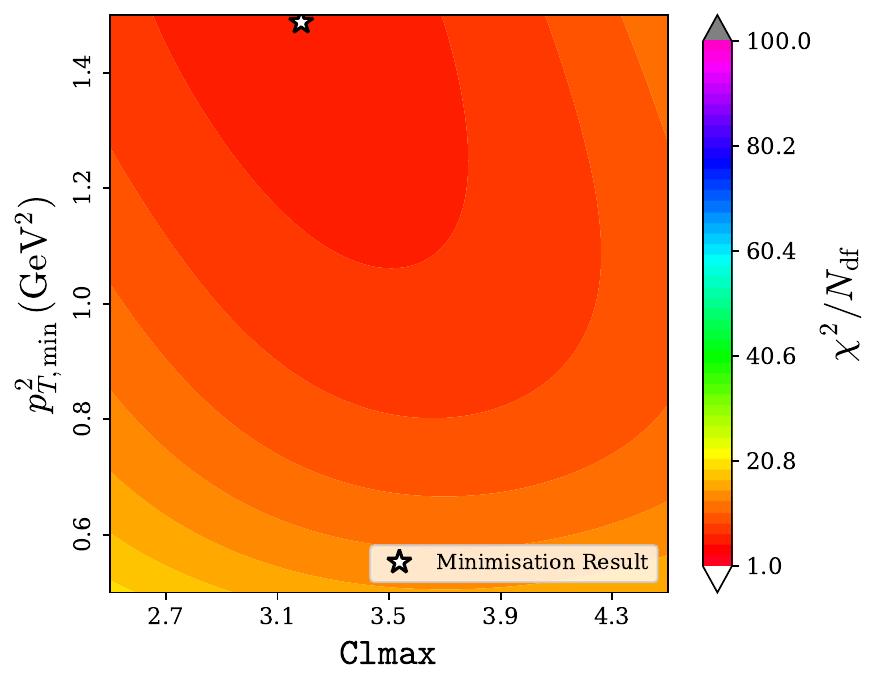}{\PG}
\labeledfig{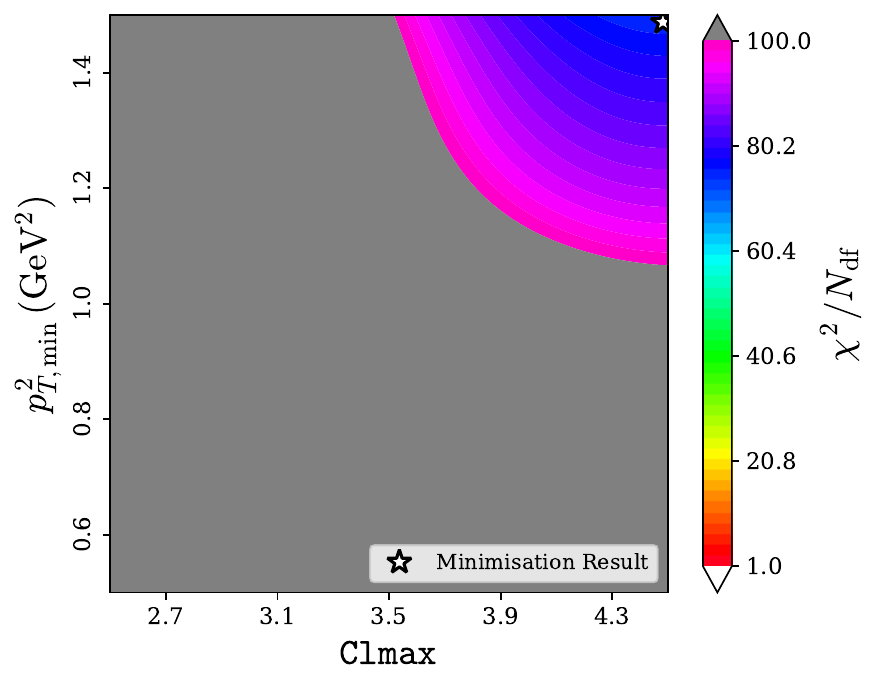}{\PG}
\labeledfig{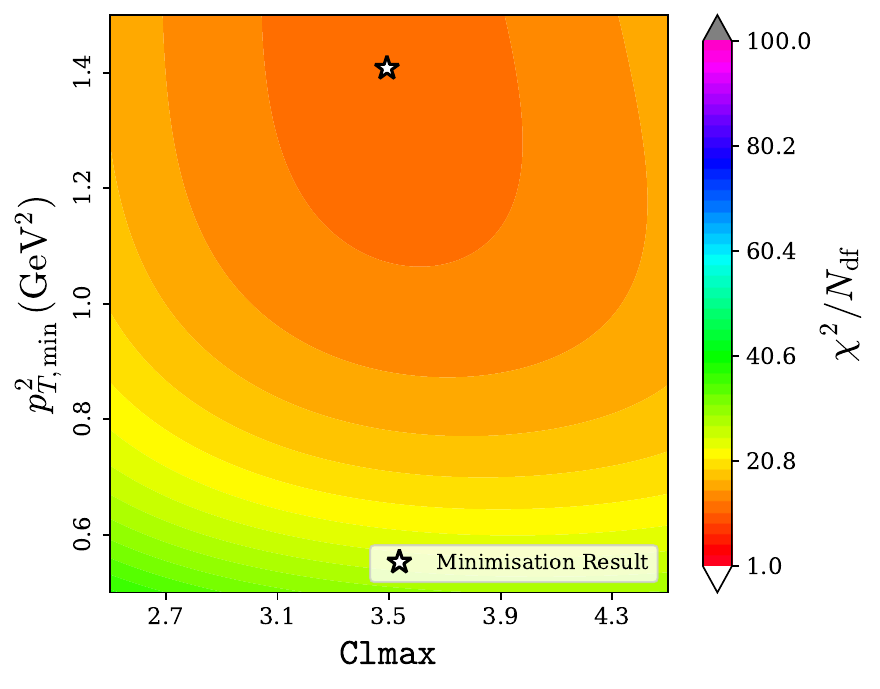}{\PG}
\labeledfig{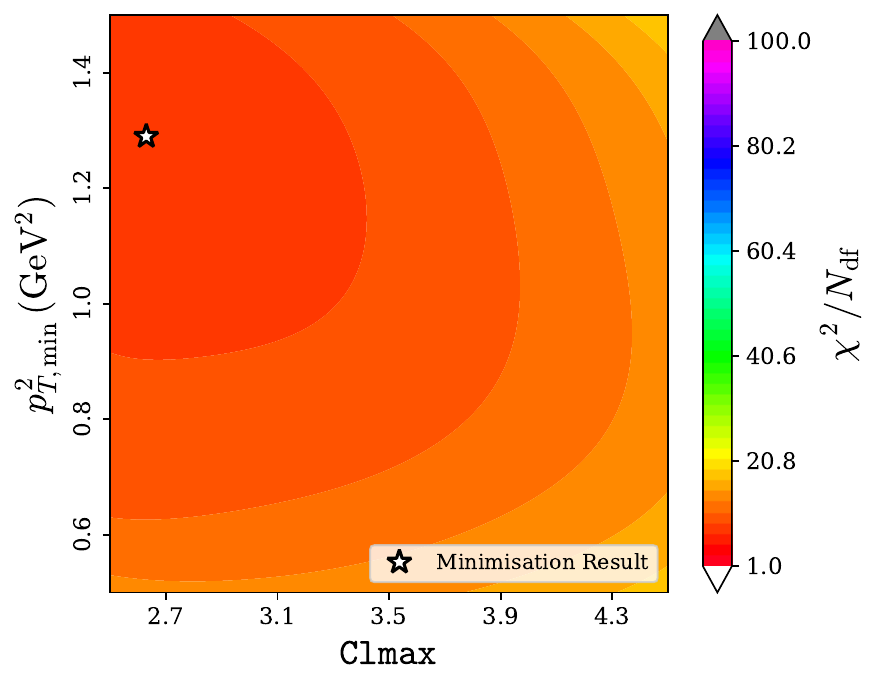}{FHP}
\labeledfig{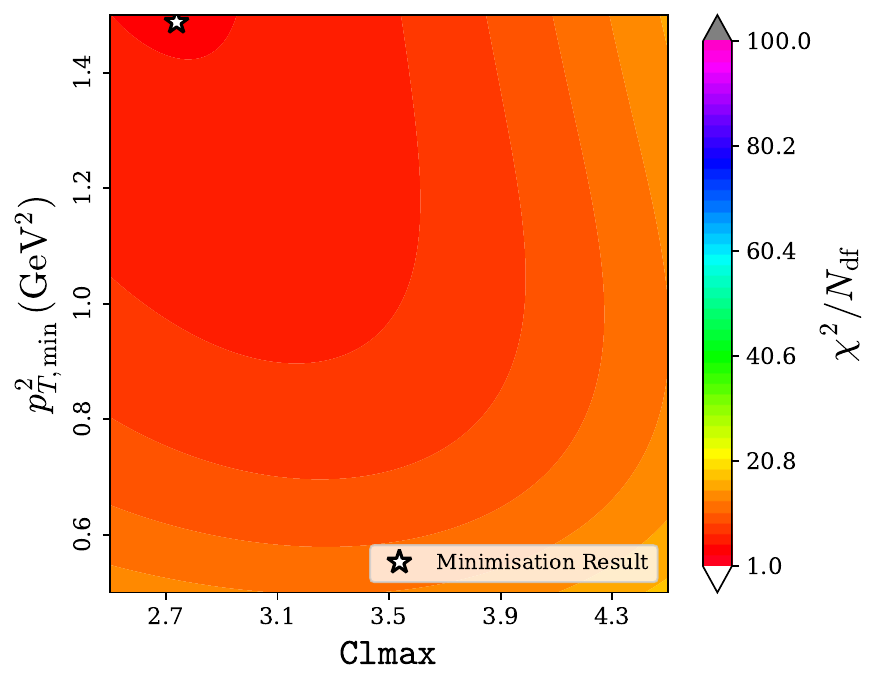}{FHP}
\labeledfig{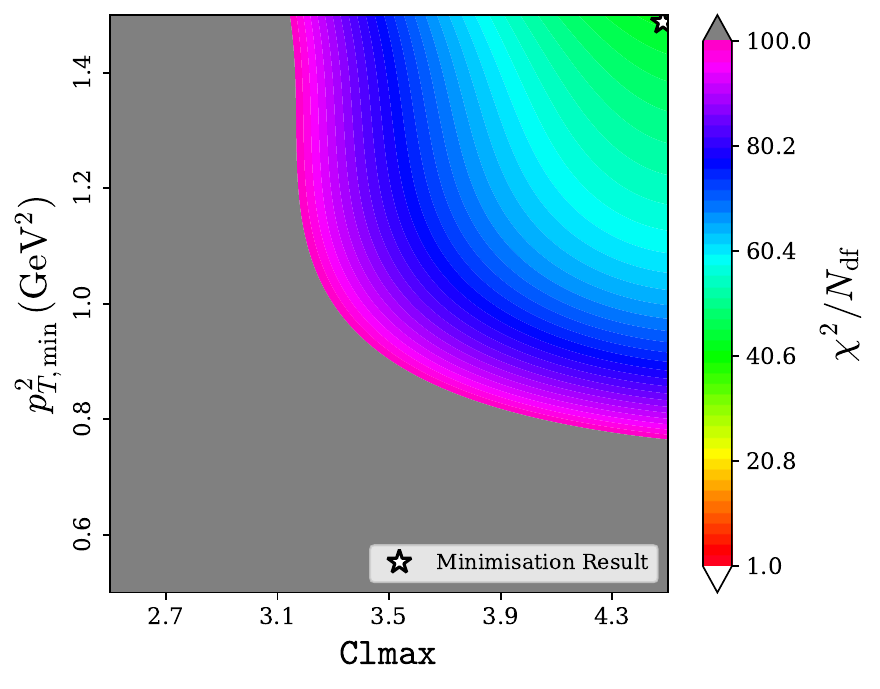}{FHP}
\labeledfig{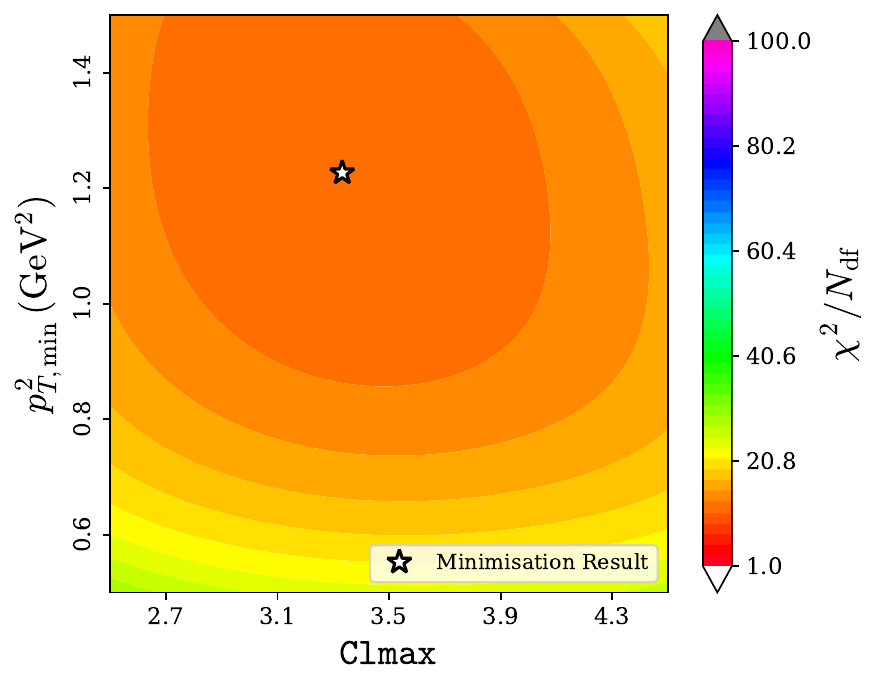}{FHP}
\labeledfig{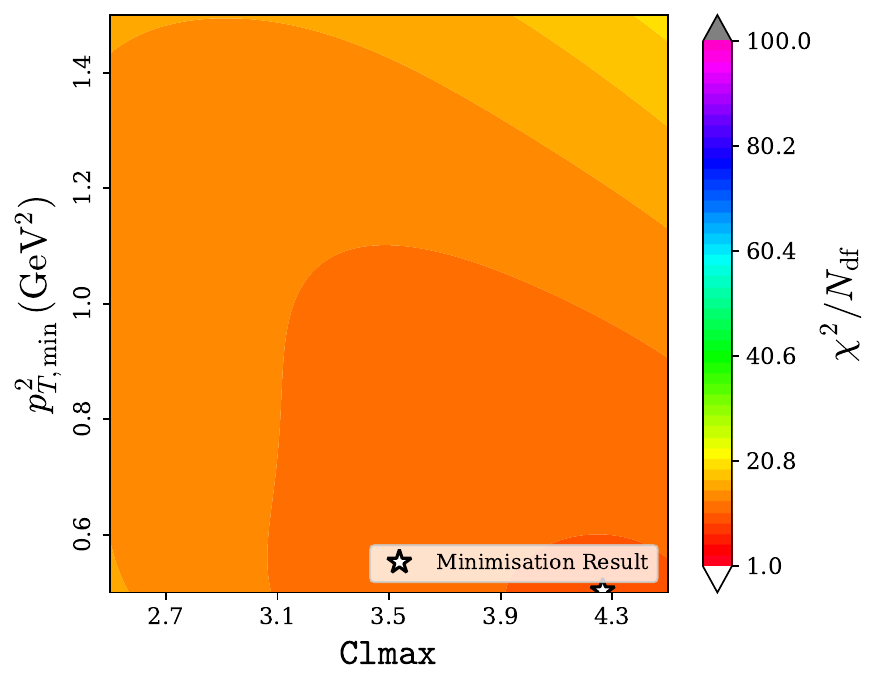}{\PGAOB}
\labeledfig{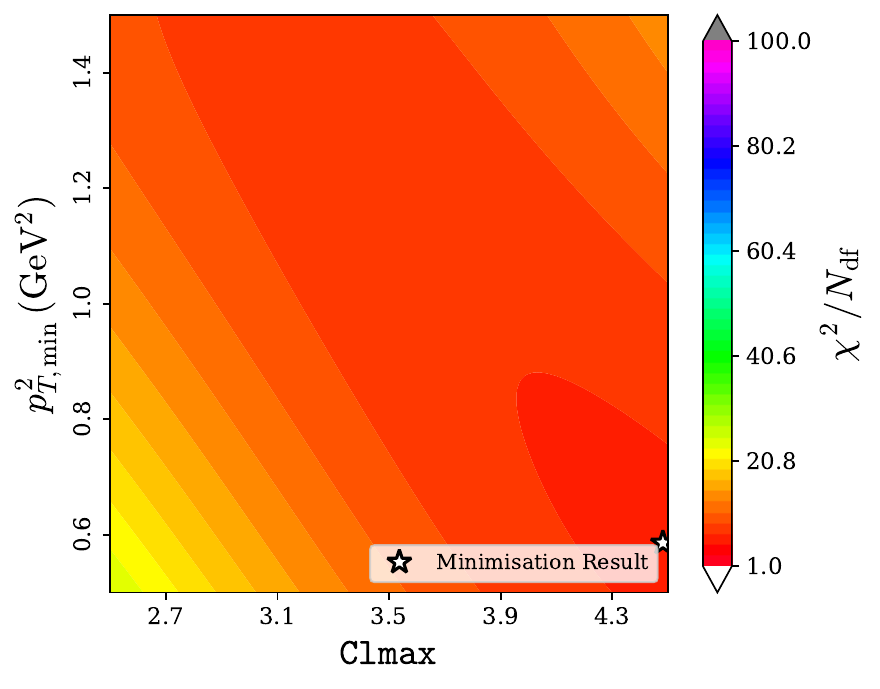}{\PGAOB}
\labeledfig{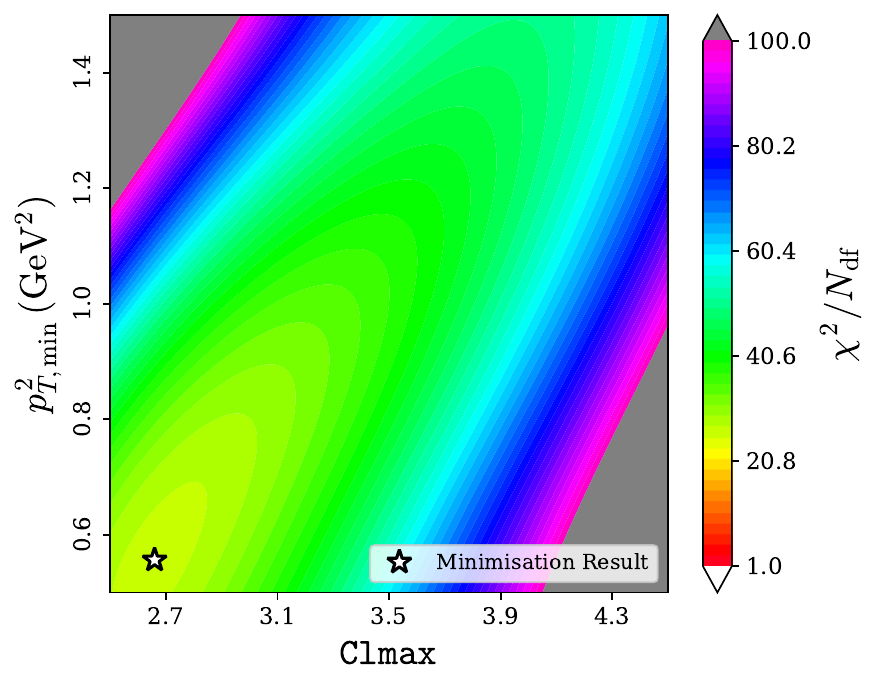}{\PGAOB}
\labeledfig{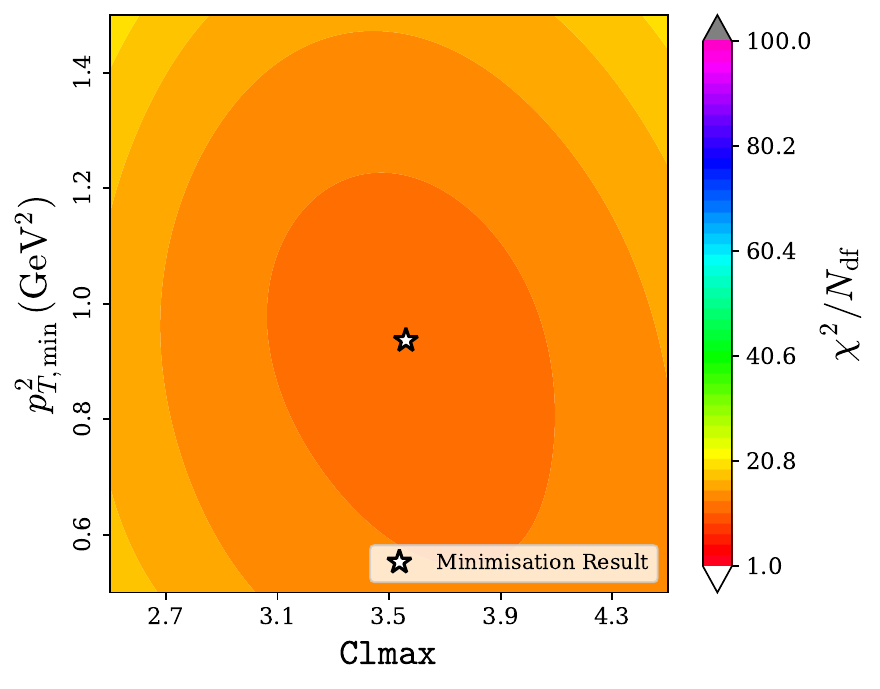}{\PGAOB}
\labeledfig{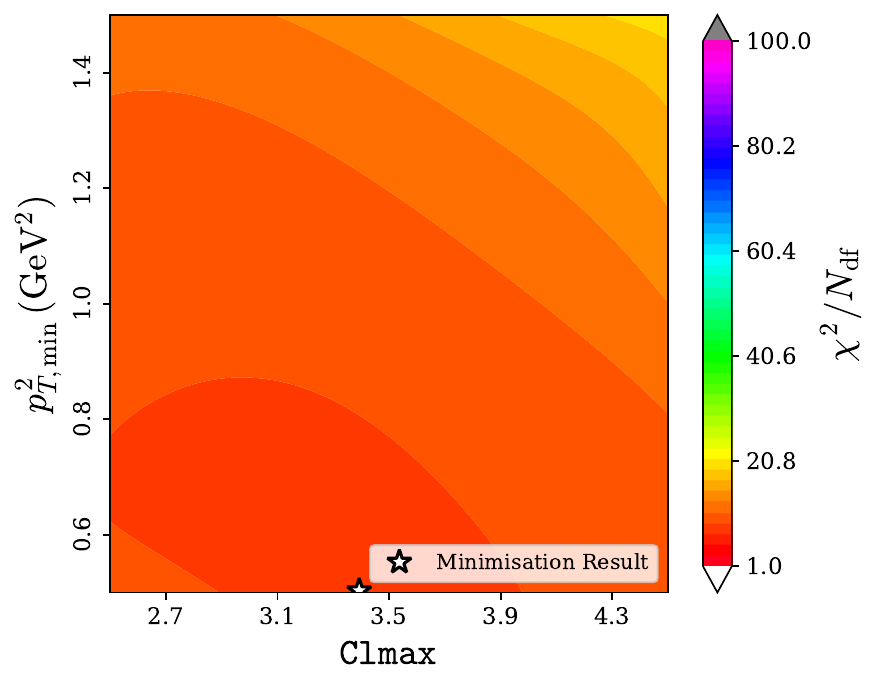}{\FHPAOB}
\labeledfig{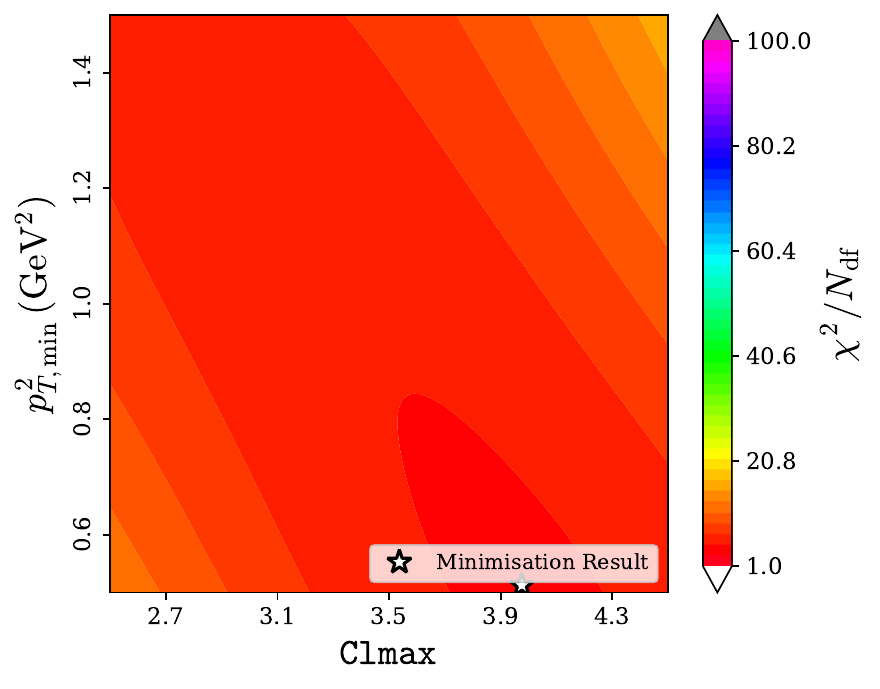}{\FHPAOB}
\labeledfig{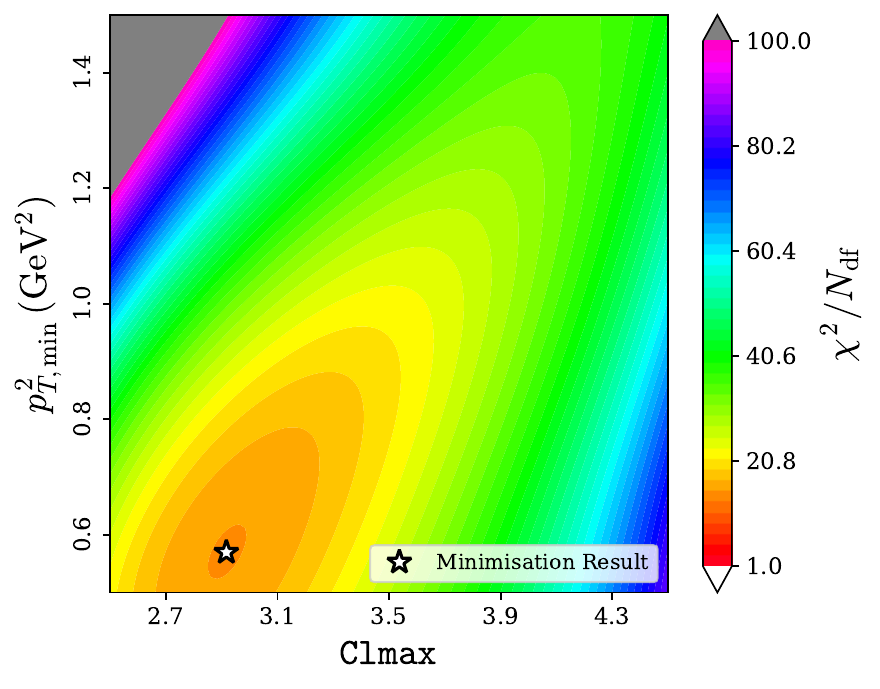}{\FHPAOB}
\labeledfig{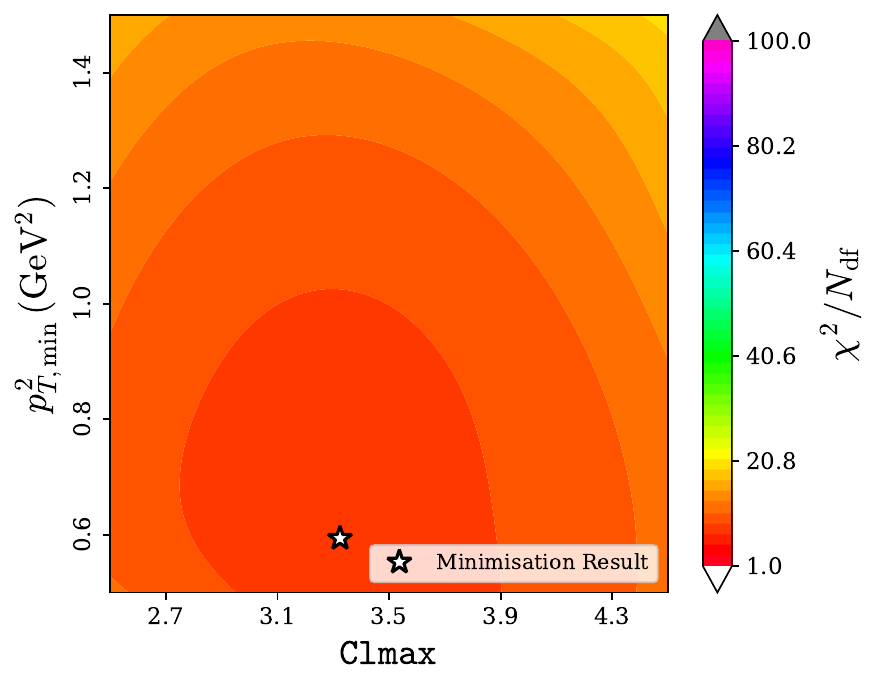}{\FHPAOB}
\caption{
$\chi^2/N_{\text{df}}$ heatmaps, varying $p_{T,\min}^2$ and \texttt{Clmax}.
This figure follows the same convention as before.
Similar properties are observed in this batch of heatmaps as in the previous figure.
However, the effect of changing these two parameters on the event shapes and jet rates is even milder; the variation in $\chi^2/N_{\text{df}}$ is of order 1.
This implies that $\alpha_s(m_Z)$ is the parameter that drives the fit in these observables.
For the charged multiplicity, as before, a very different heatmap is obtained through the AO-like boundary.
The contradiction in the contours of the heatmaps for \PGAOB and \FHPAOB yield a centrally placed contour in the heatmap of all observables.
}
\label{fig:heatmaps-pt-cl}
\end{figure}
\clearpage

\subsection*{$\chi^2/N_{\text{df}}$ heatmaps -- $\alpha_s(m_Z)$ and \texttt{Clmax}}
\begin{figure}[!htbp]
\centering
\labeledfig{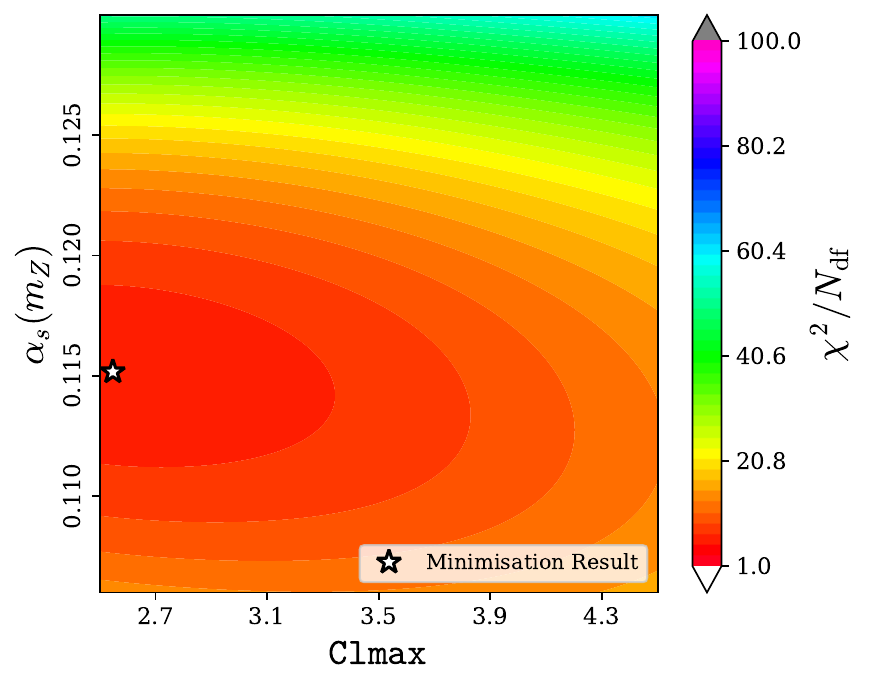}{CS}
\labeledfig{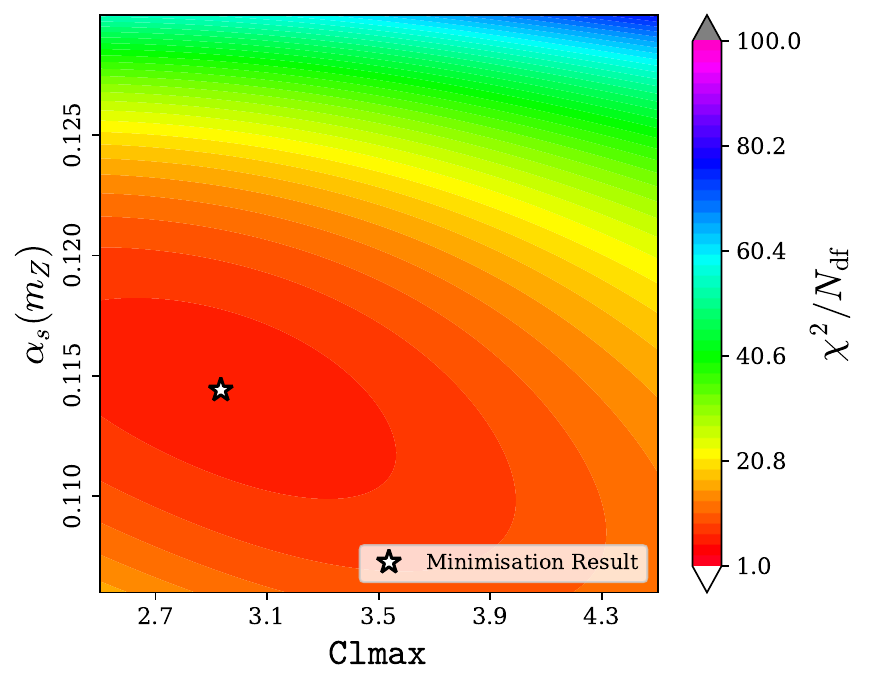}{CS}
\labeledfig{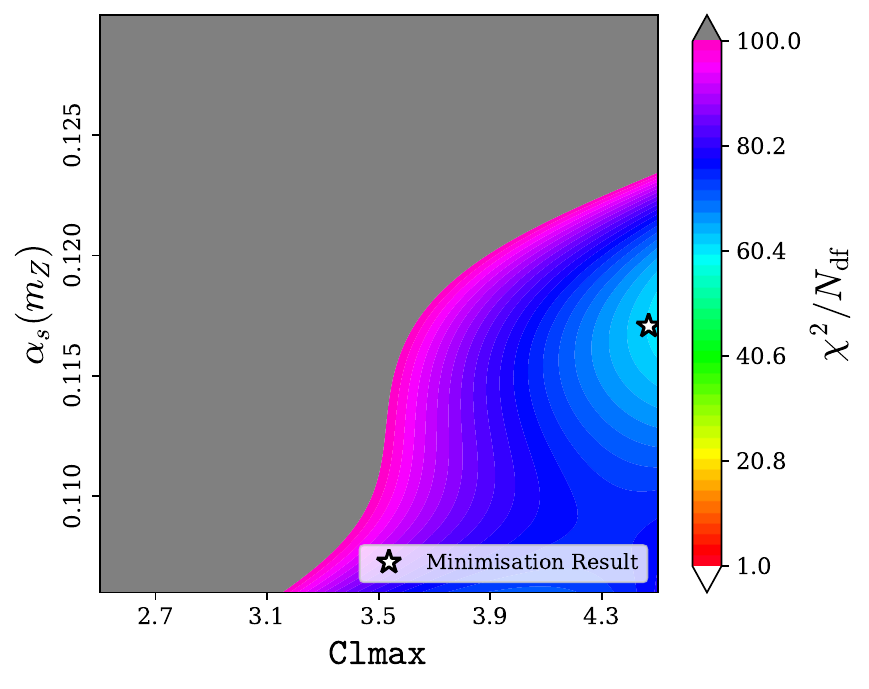}{CS}
\labeledfig{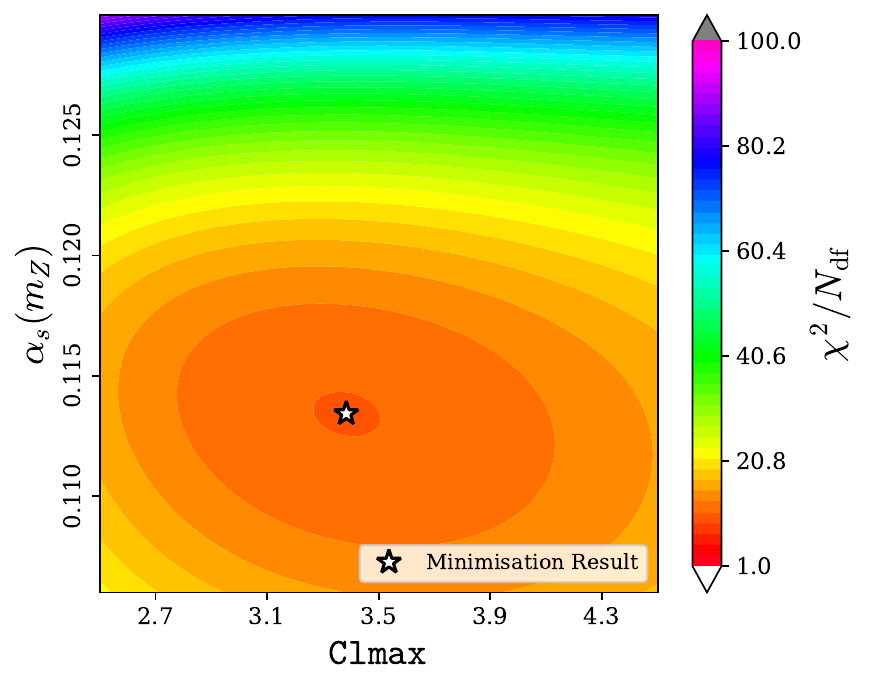}{CS}
\labeledfig{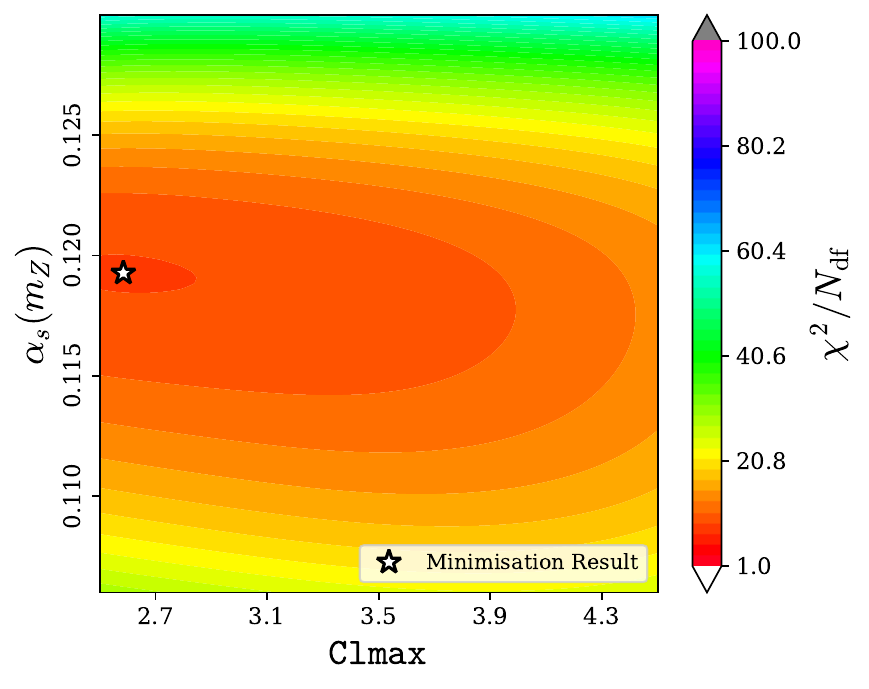}{\PG}
\labeledfig{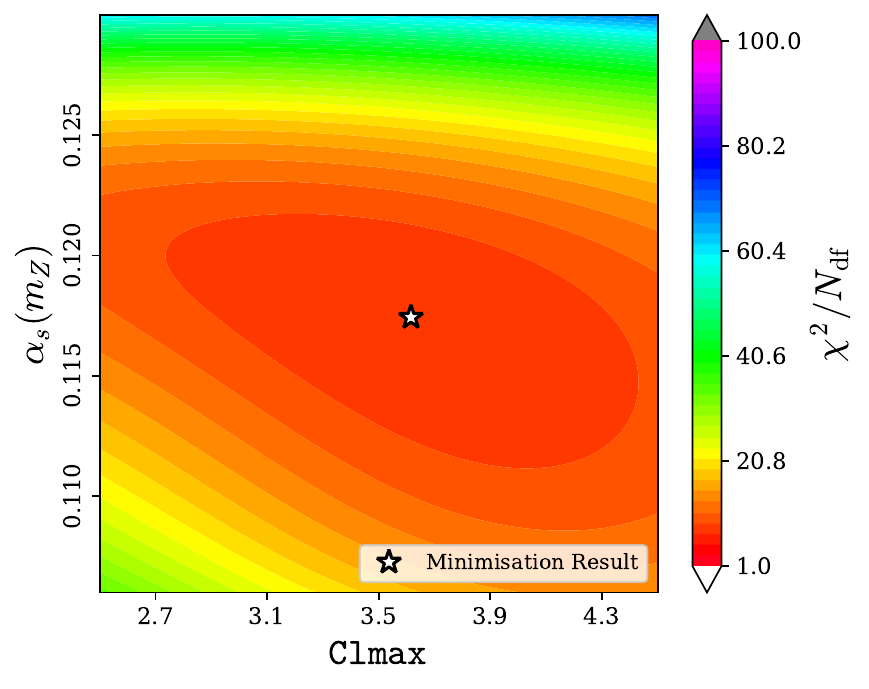}{\PG}
\labeledfig{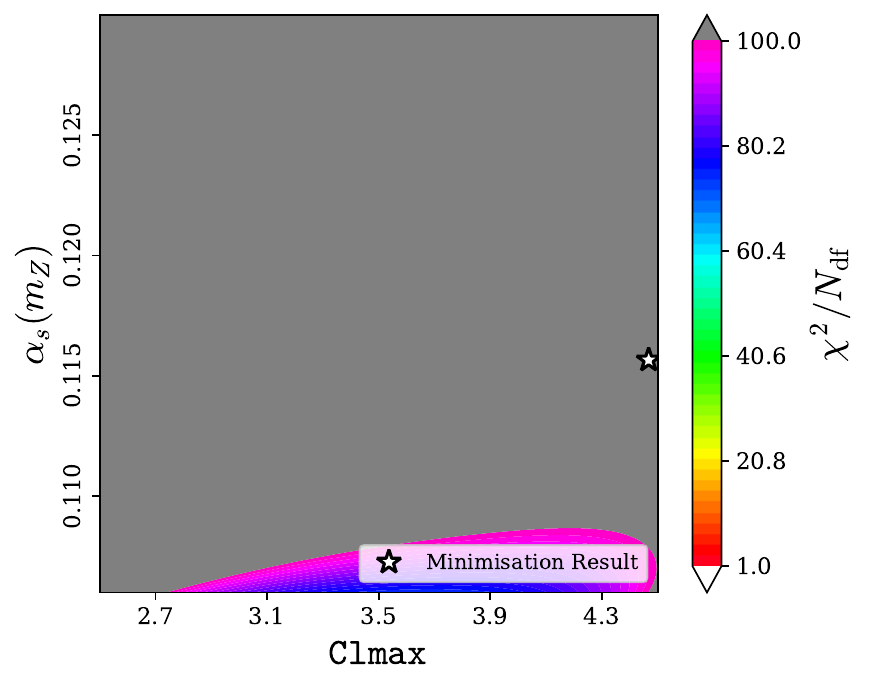}{\PG}
\labeledfig{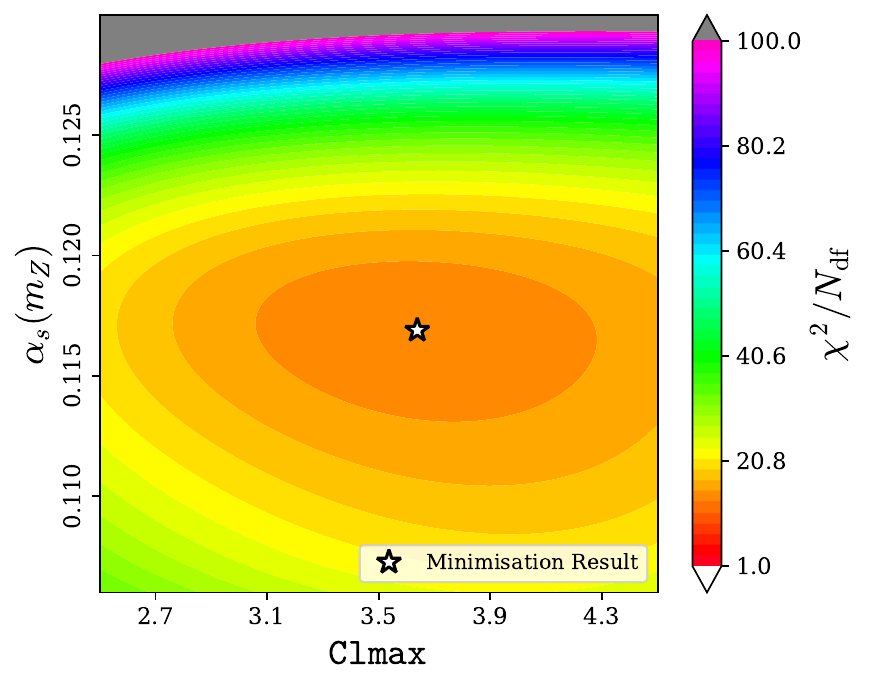}{\PG}
\labeledfig{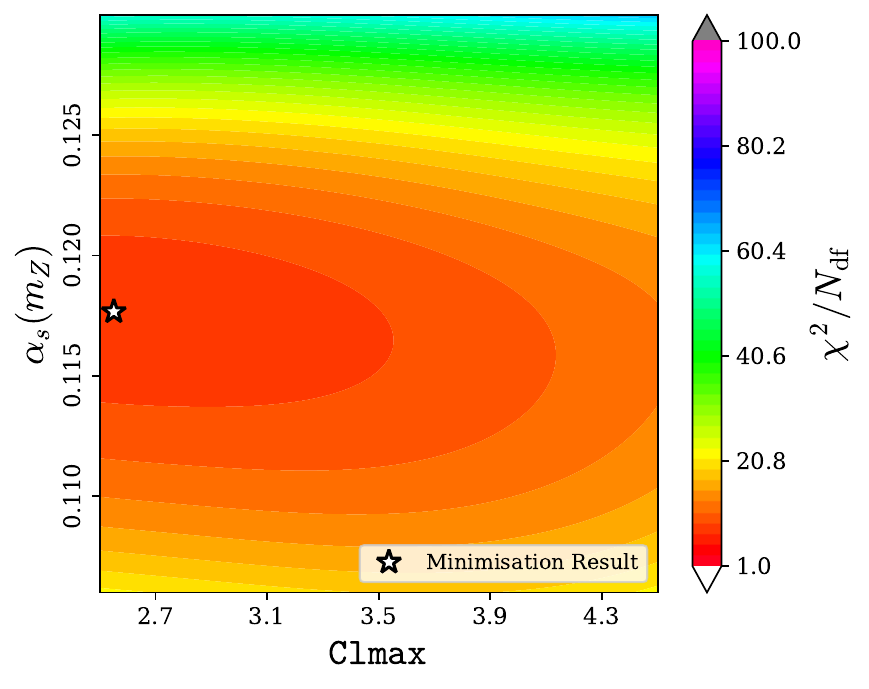}{FHP}
\labeledfig{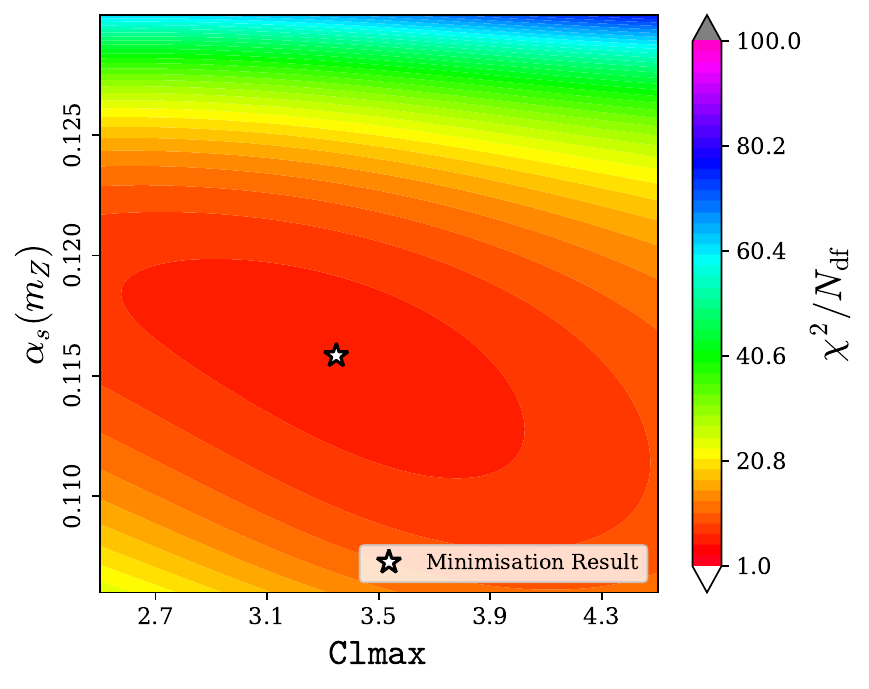}{FHP}
\labeledfig{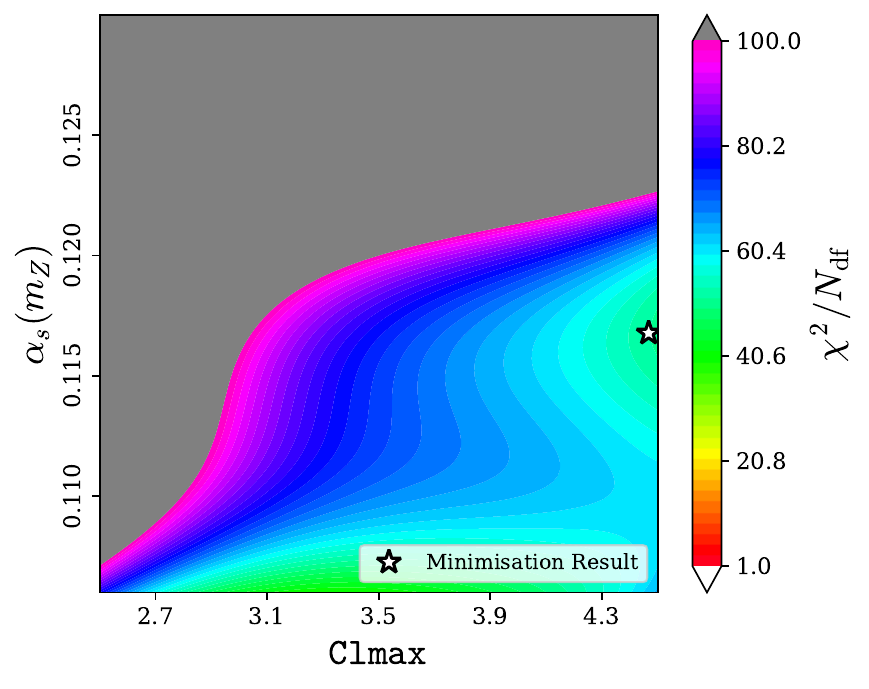}{FHP}
\labeledfig{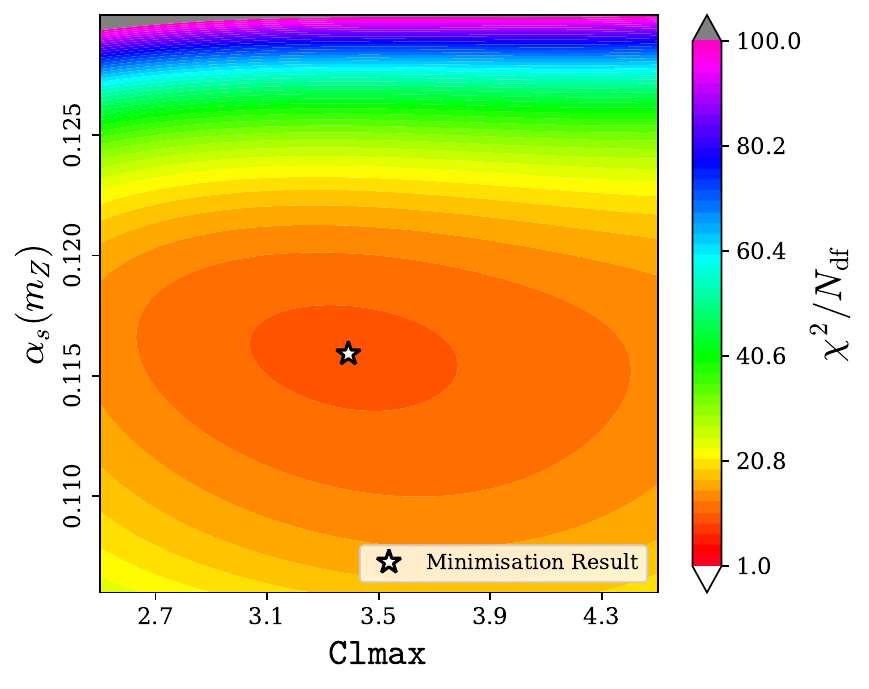}{FHP}
\labeledfig{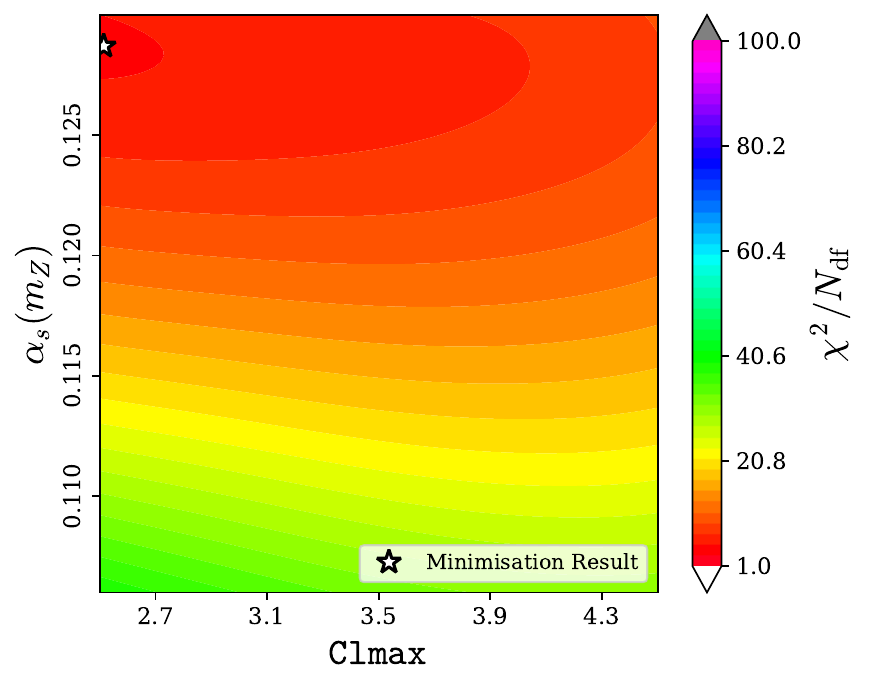}{\PGAOB}
\labeledfig{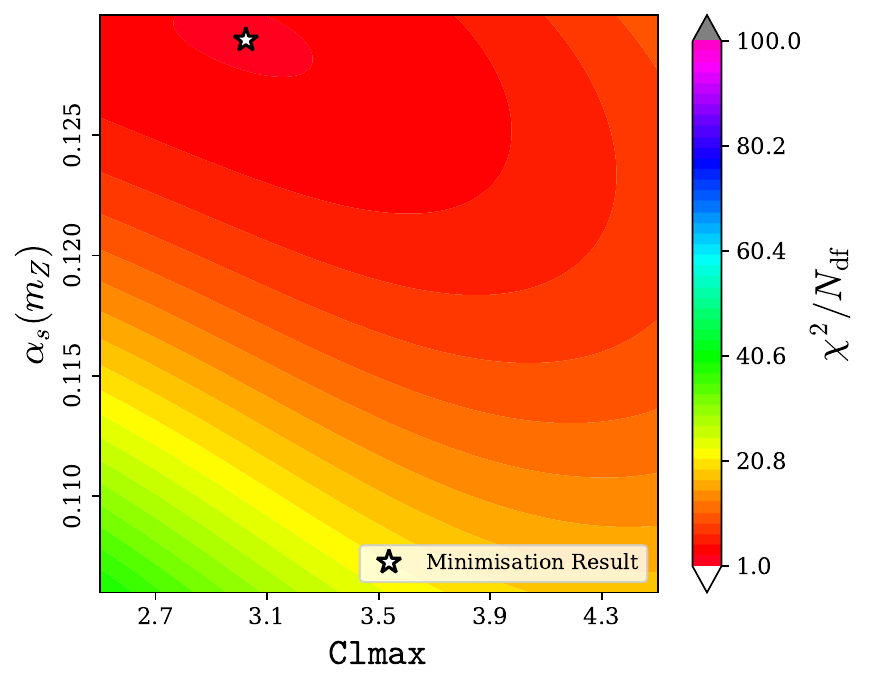}{\PGAOB}
\labeledfig{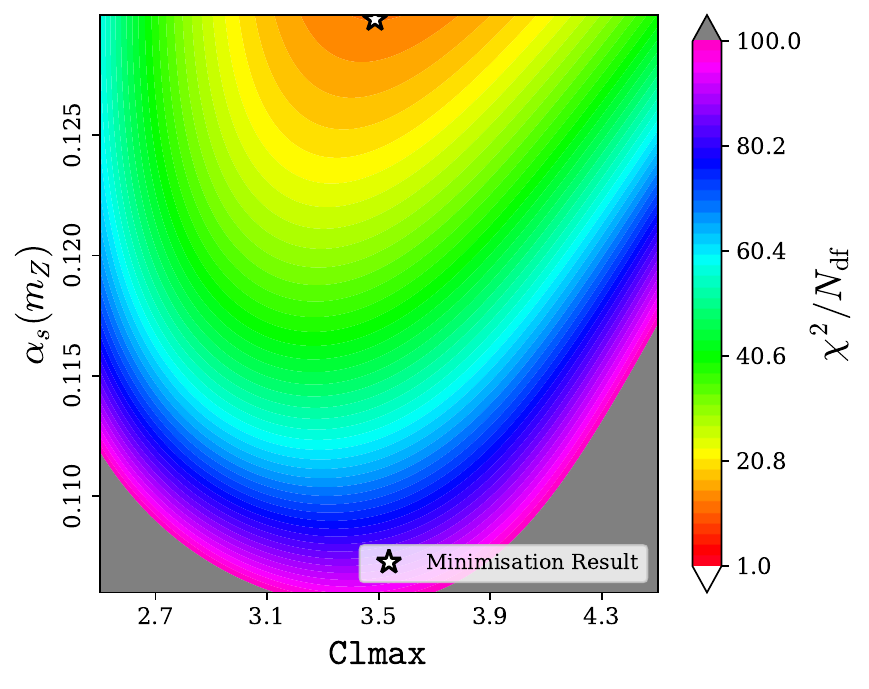}{\PGAOB}
\labeledfig{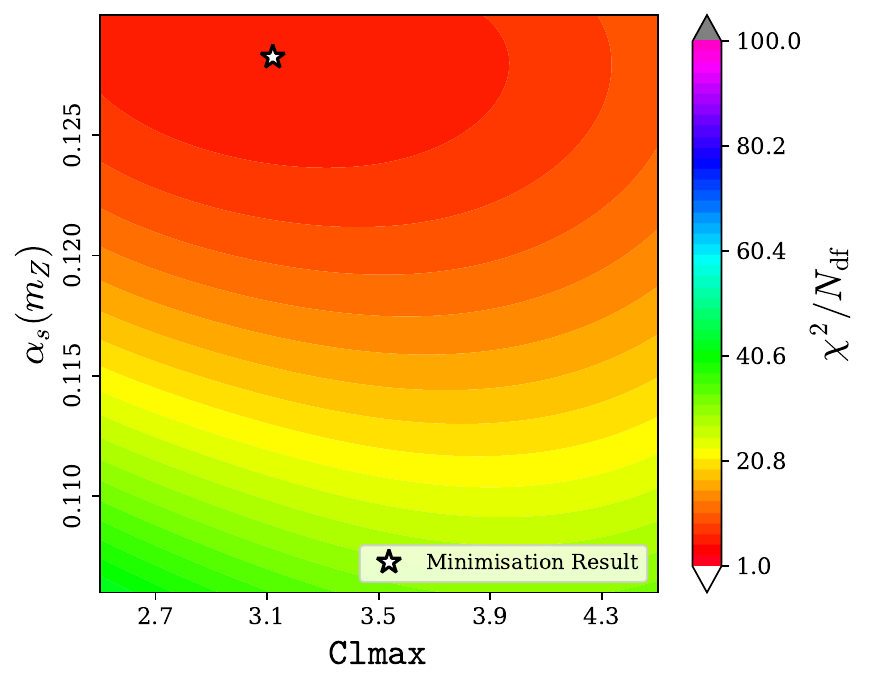}{\PGAOB}
\labeledfig{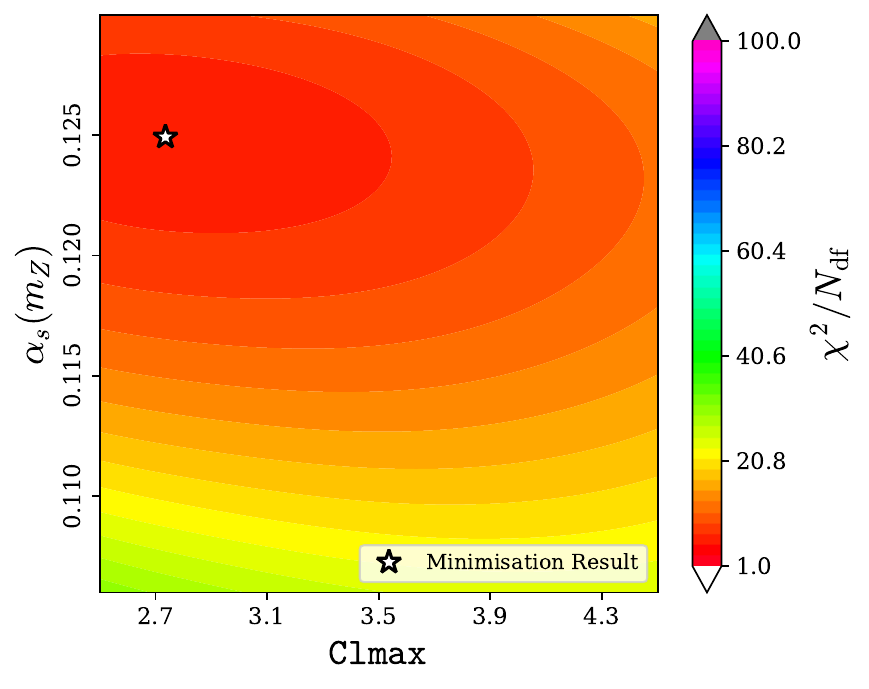}{\FHPAOB}
\labeledfig{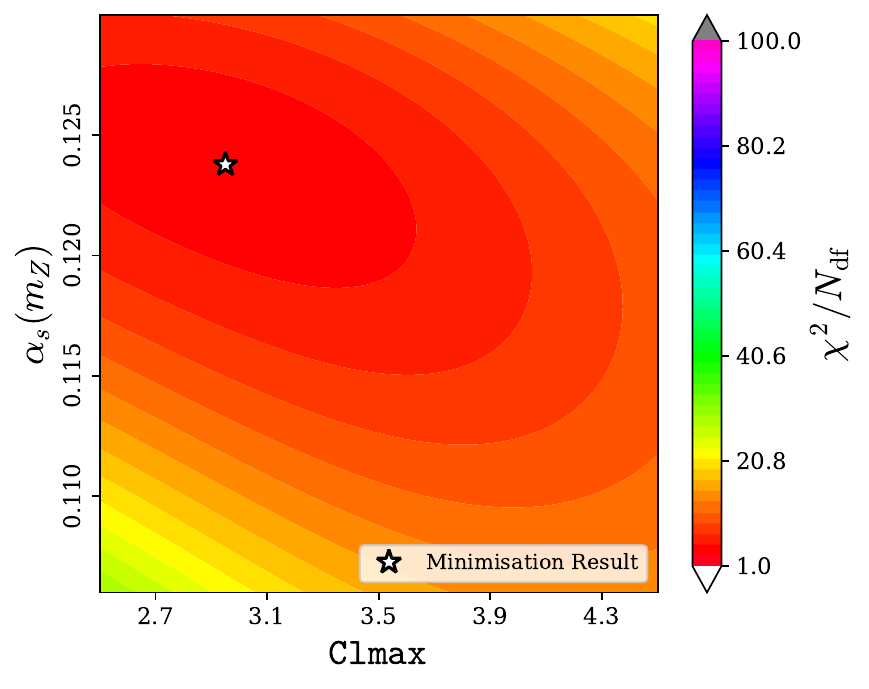}{\FHPAOB}
\labeledfig{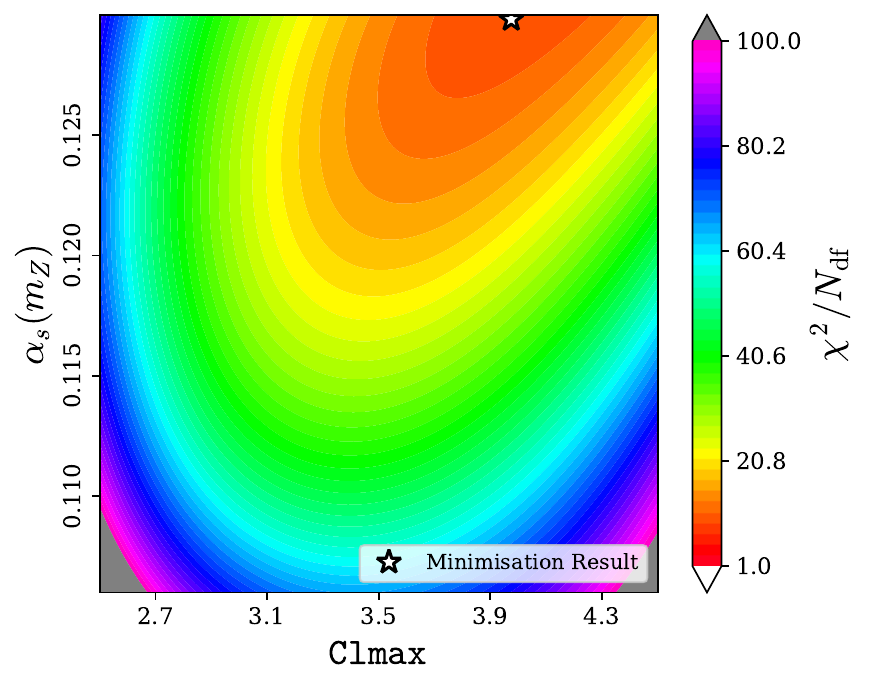}{\FHPAOB}
\labeledfig{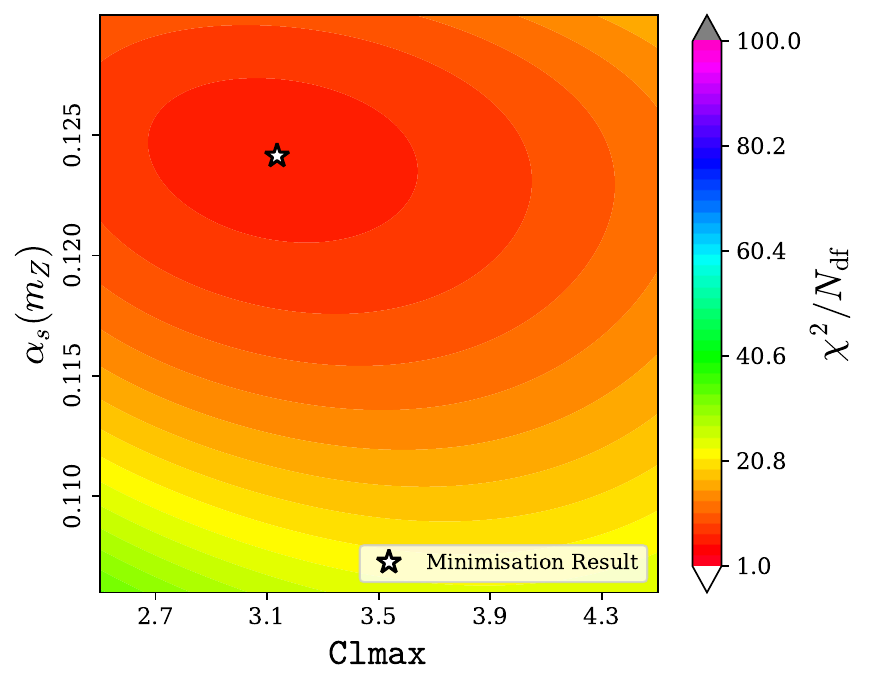}{\FHPAOB}
\caption{
$\chi^2/N_{\text{df}}$ heatmaps, varying $\alpha_s(m_Z)$ and \texttt{Clmax}. 
Comparing this set of heatmaps with the previous one reinforces the idea that $\alpha_s(m_Z)$ is the driving parameter for event shapes and jet rates.
}
\label{fig:heatmaps-as-cl}
\end{figure}
\clearpage

\subsection*{$\chi^2/N_{\text{df}}$ heatmaps -- $\alpha_s(m_Z)$ and \texttt{Clmax}, $p_{T,\min}^2=0.75 \, \mathrm{ GeV}^2$}
\begin{figure}[!htbp]
\centering
\labeledfig{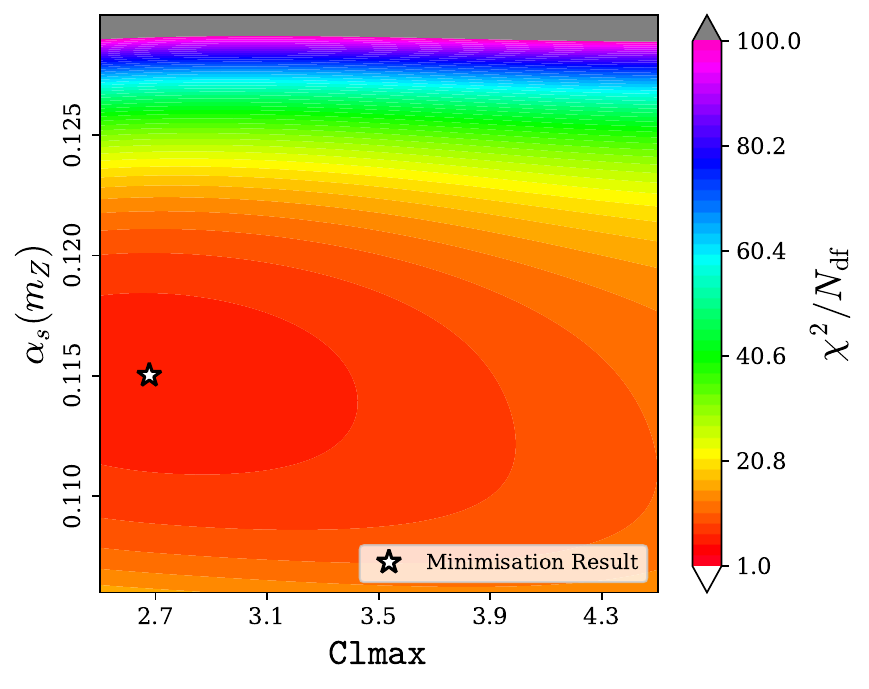}{CS}
\labeledfig{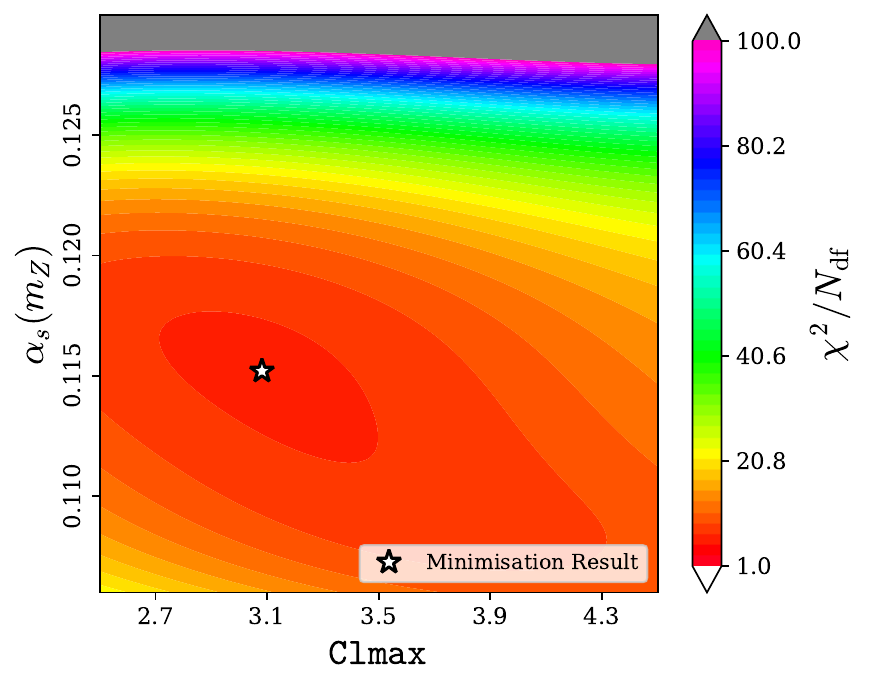}{CS}
\labeledfig{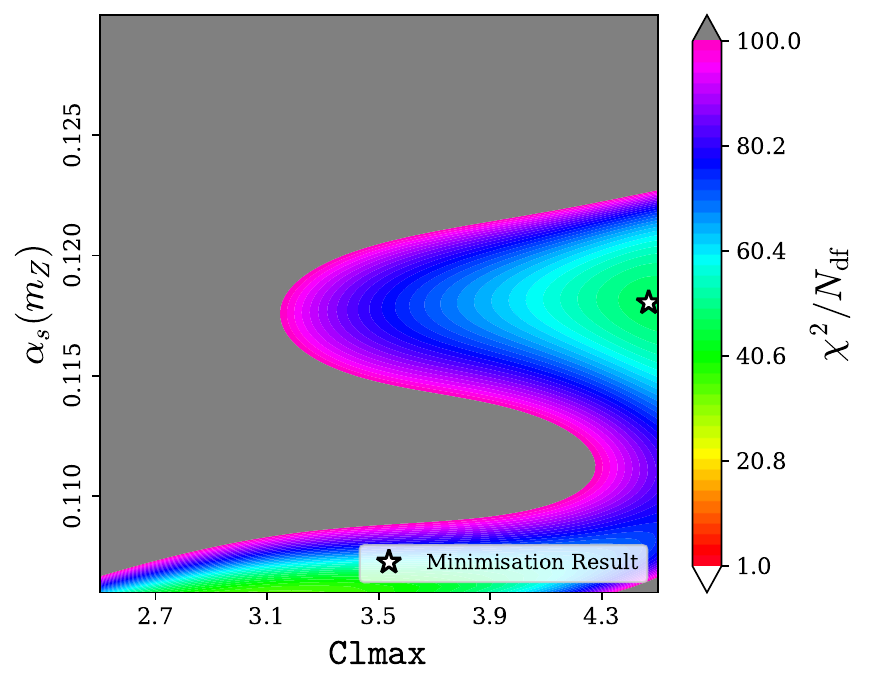}{CS}
\labeledfig{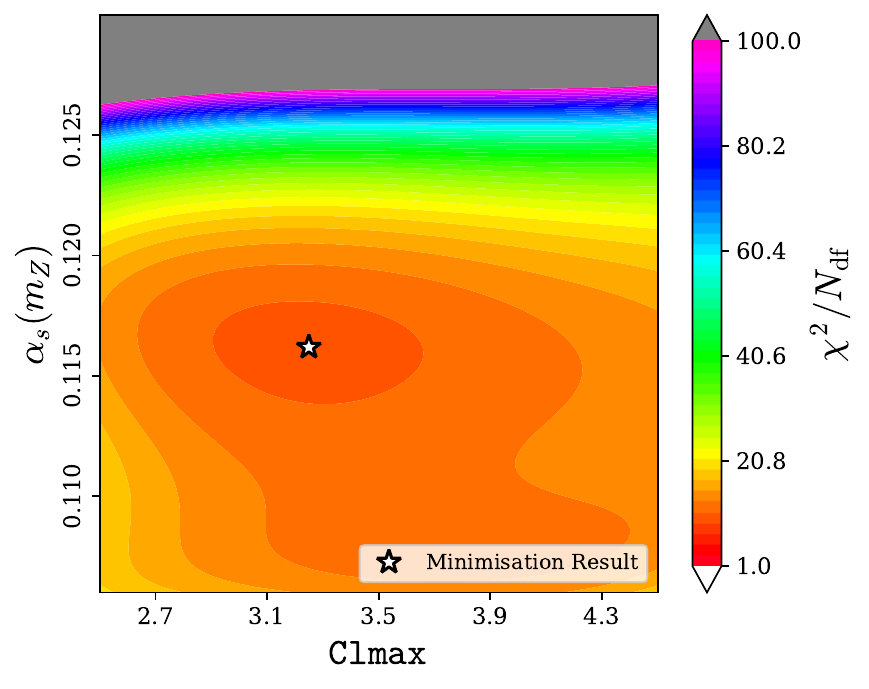}{CS}
\labeledfig{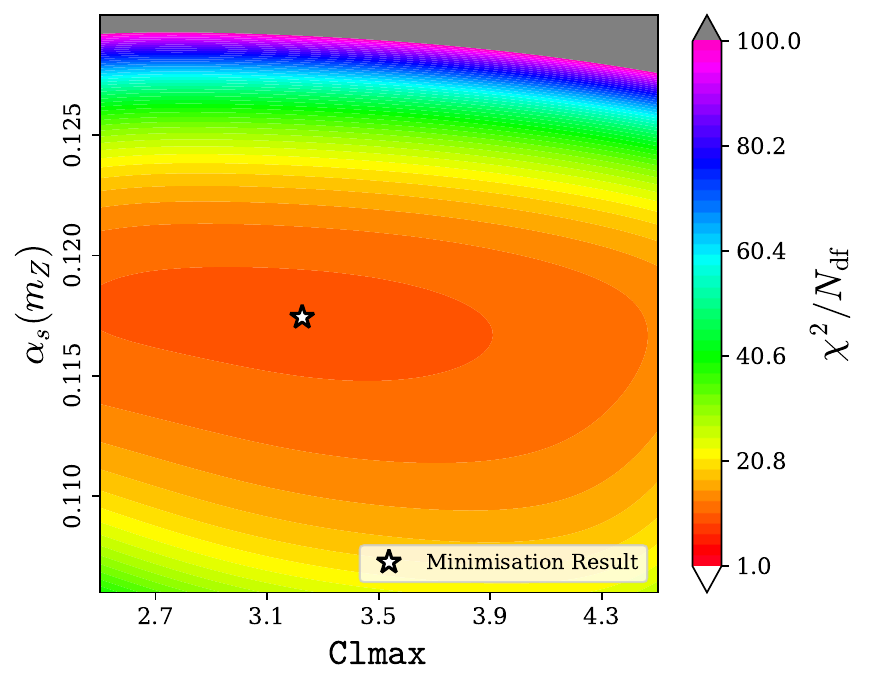}{\PG}
\labeledfig{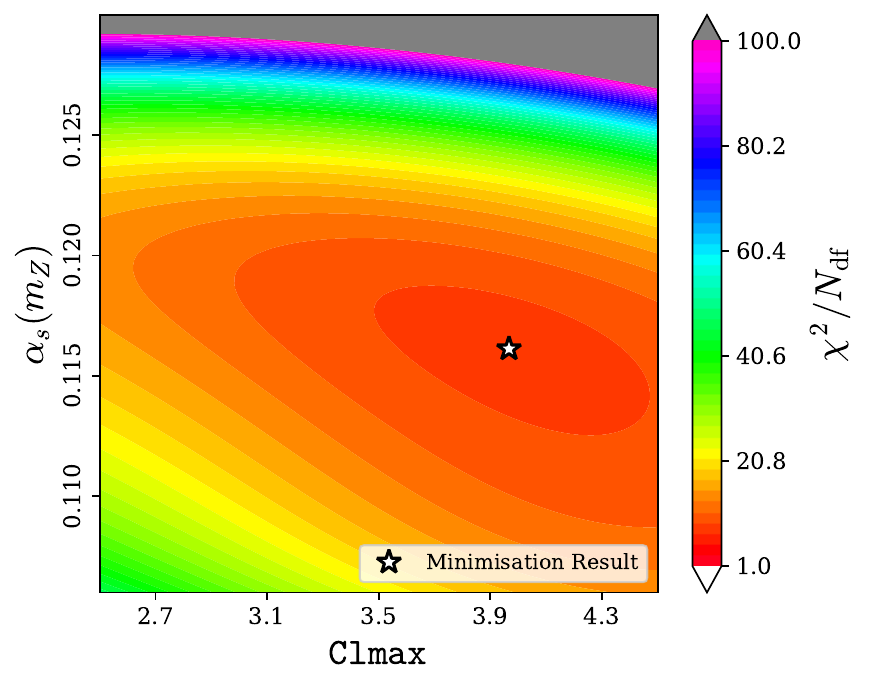}{\PG}
\labeledfig{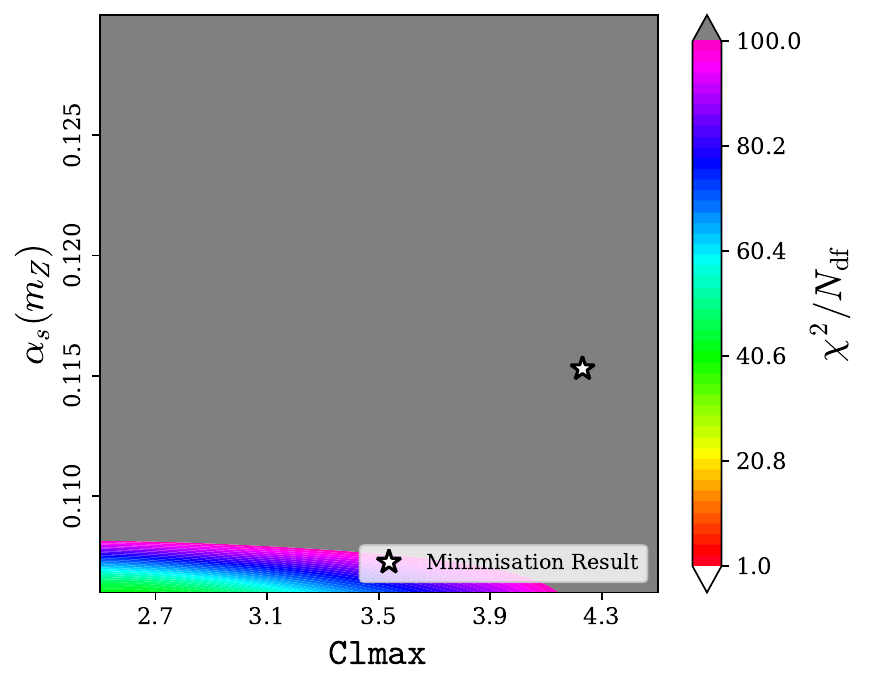}{\PG}
\labeledfig{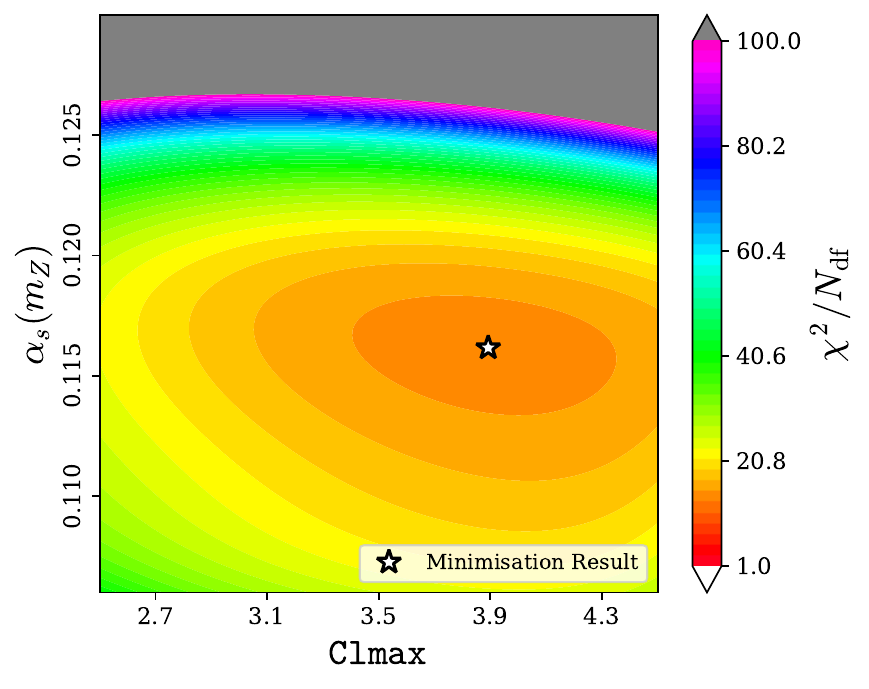}{\PG}
\labeledfig{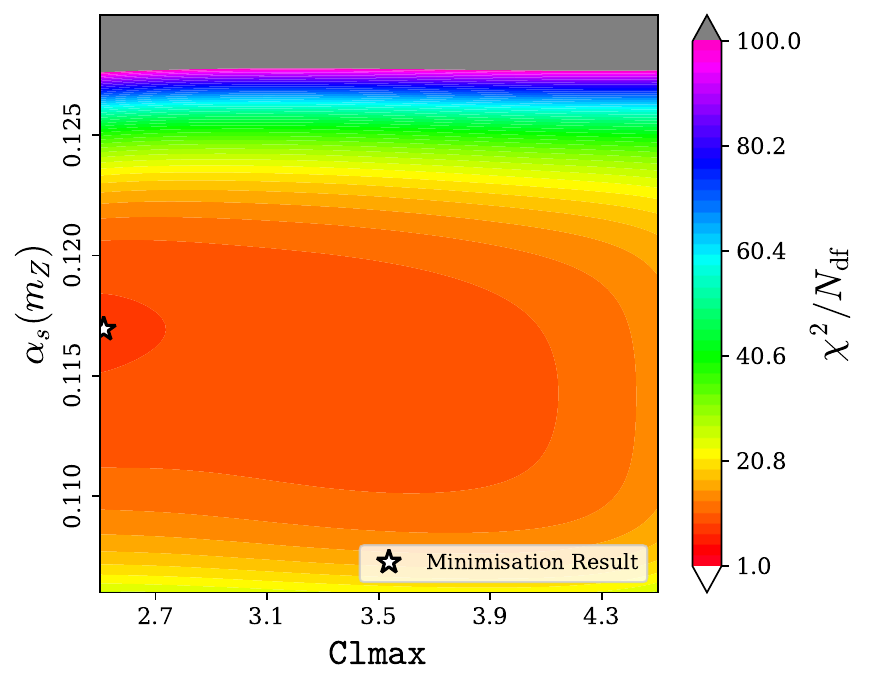}{FHP}
\labeledfig{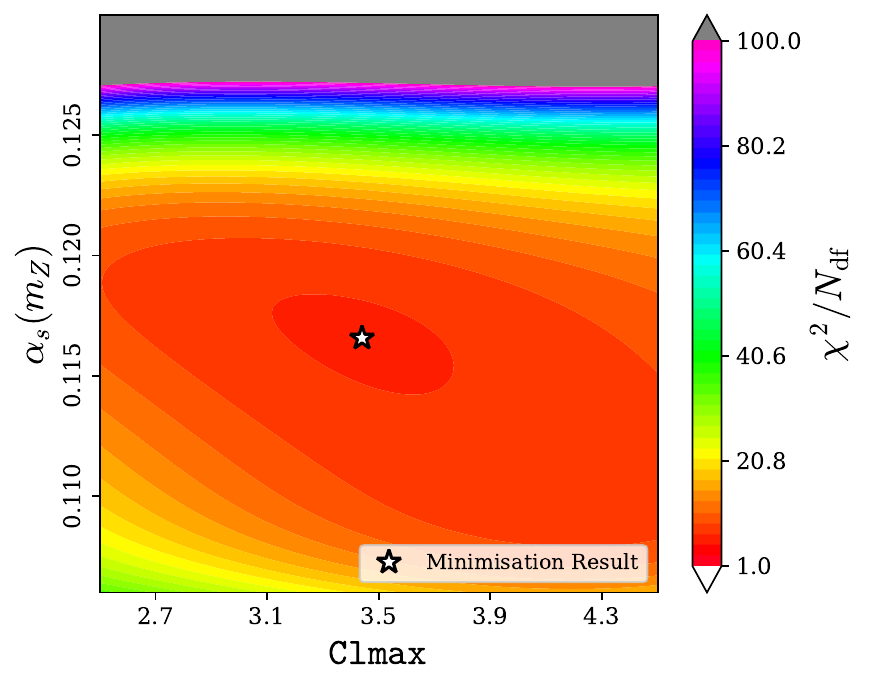}{FHP}
\labeledfig{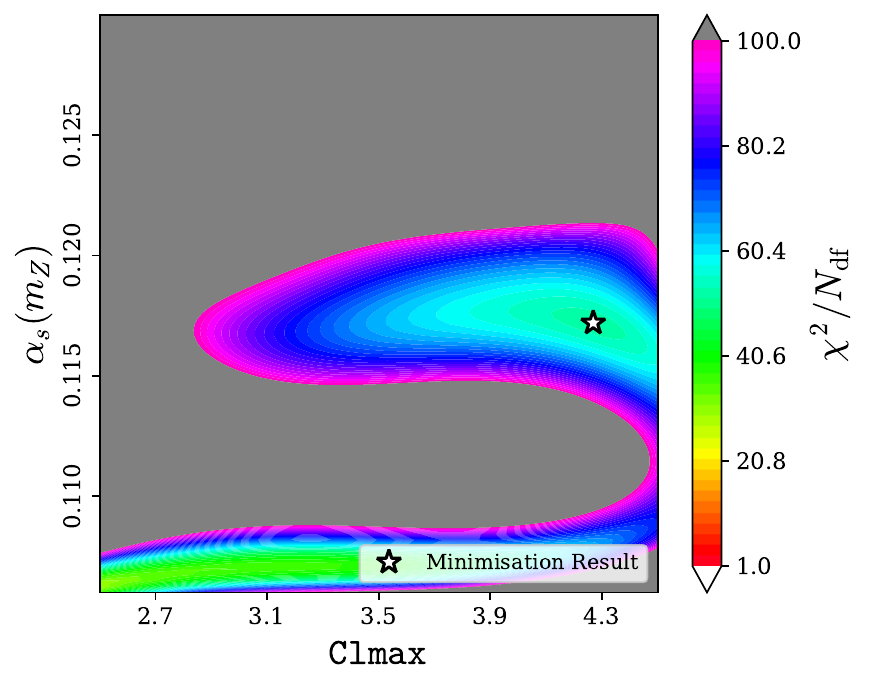}{FHP}
\labeledfig{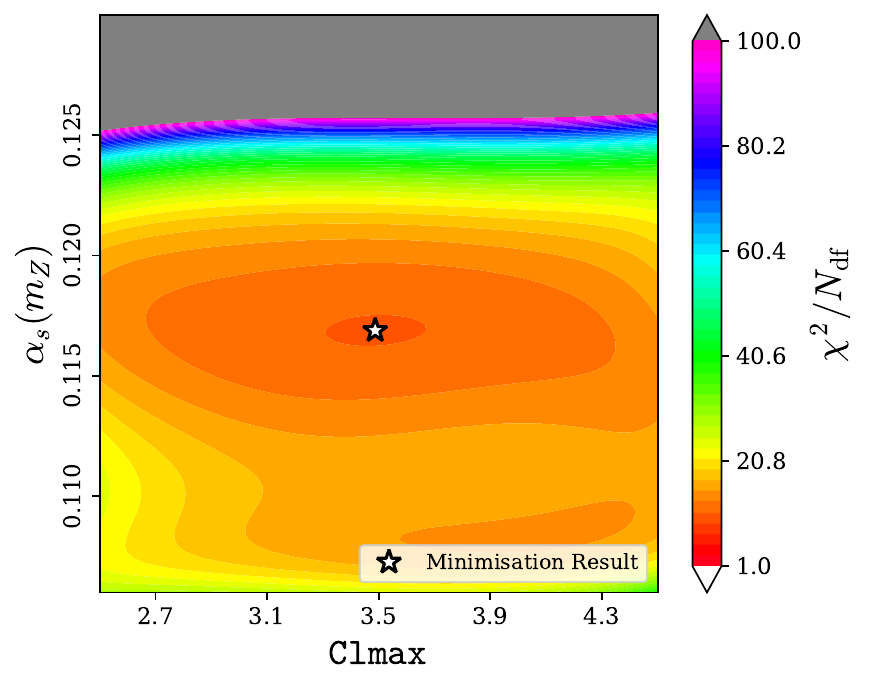}{FHP}
\labeledfig{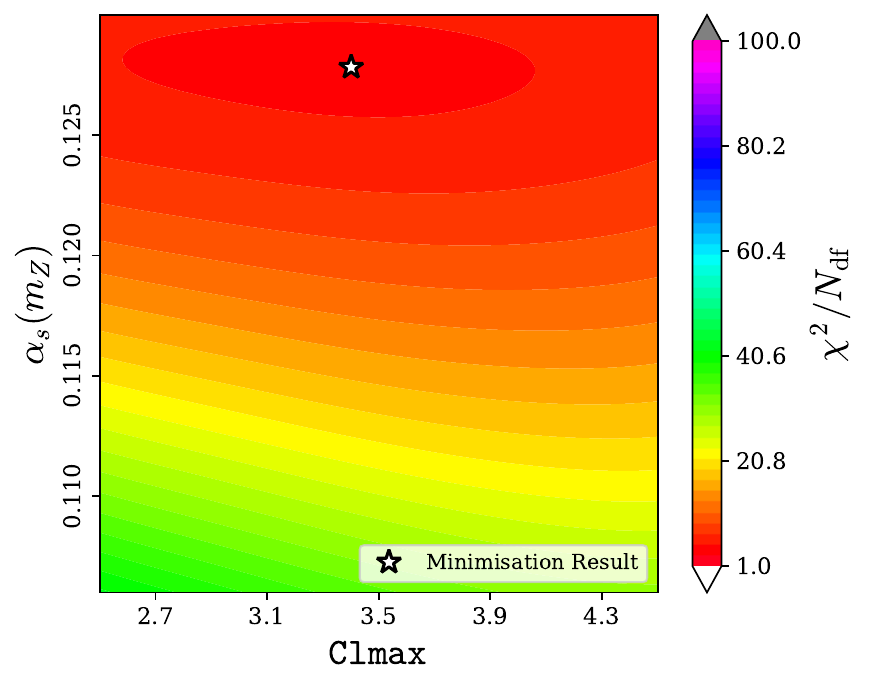}{\PGAOB}
\labeledfig{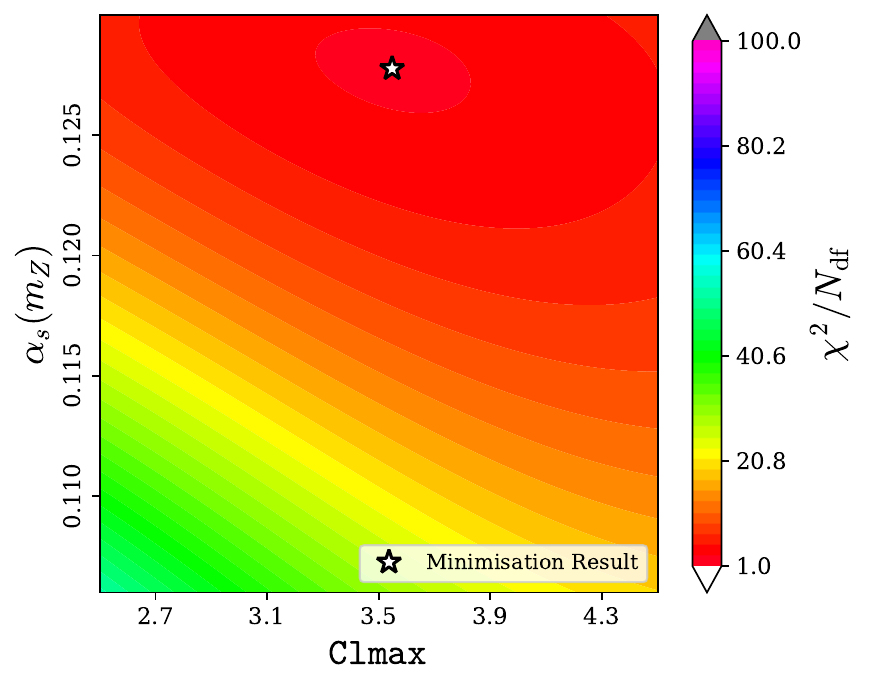}{\PGAOB}
\labeledfig{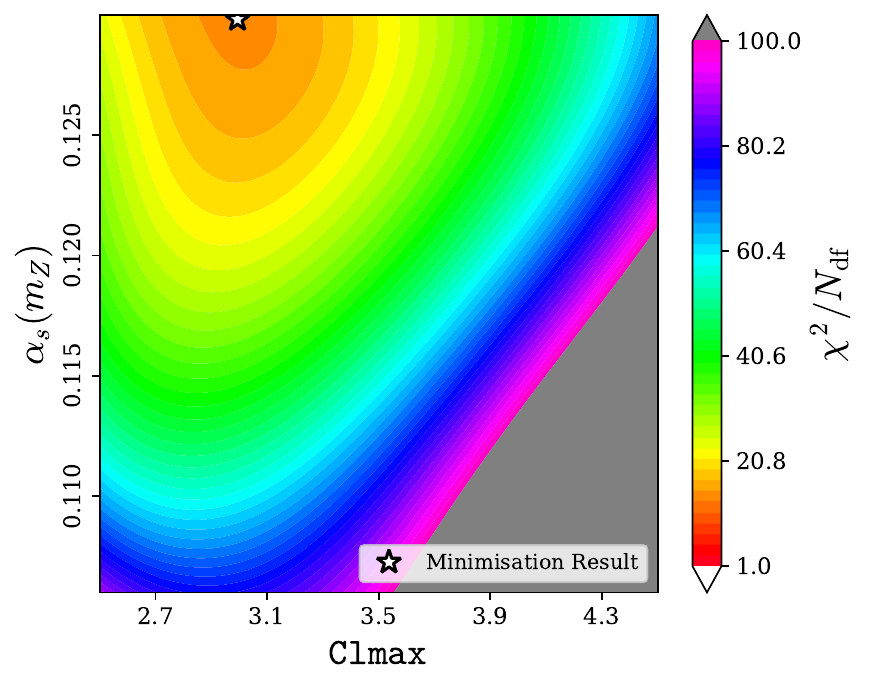}{\PGAOB}
\labeledfig{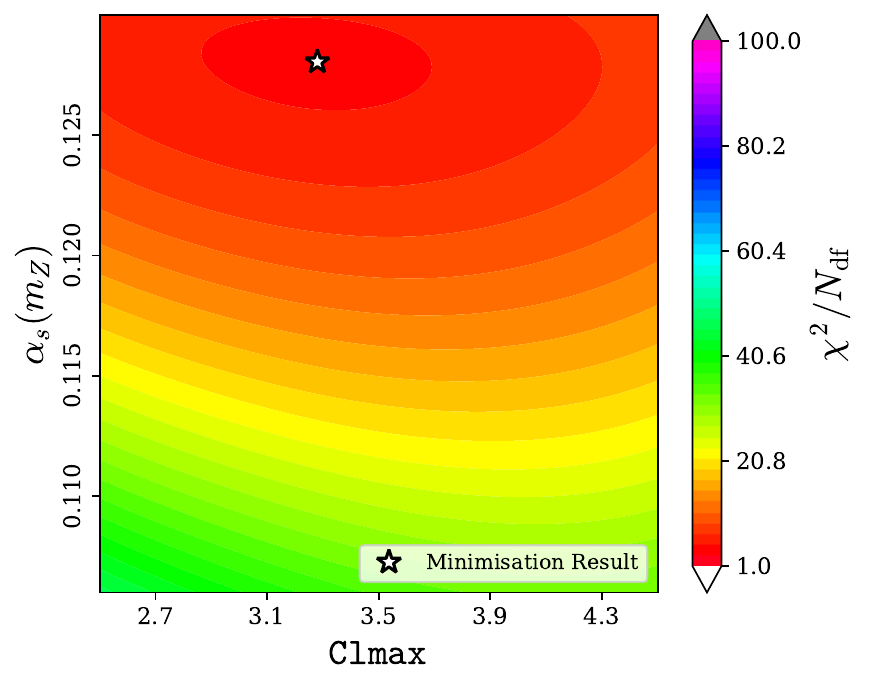}{\PGAOB}
\labeledfig{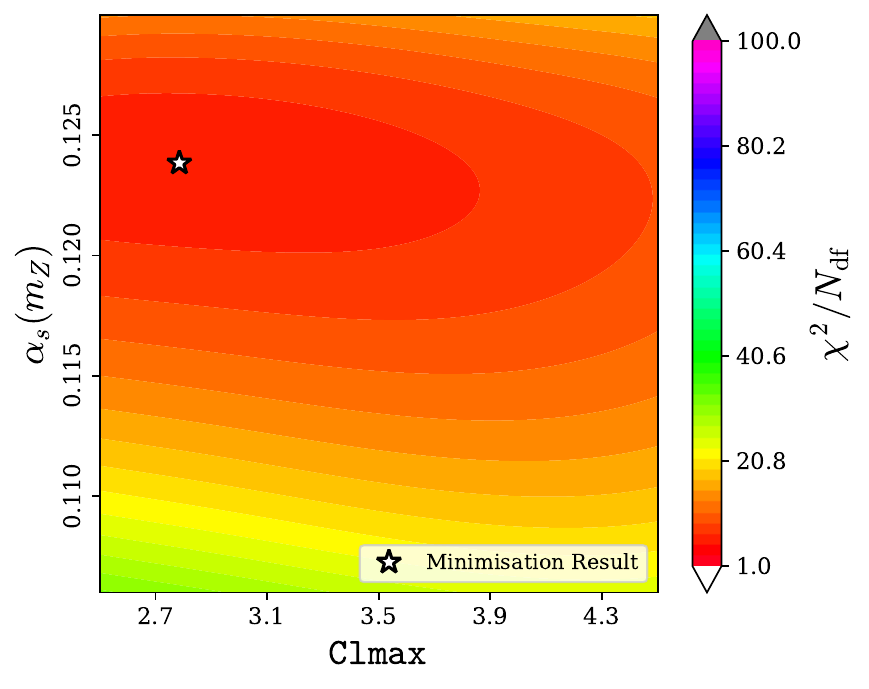}{\FHPAOB}
\labeledfig{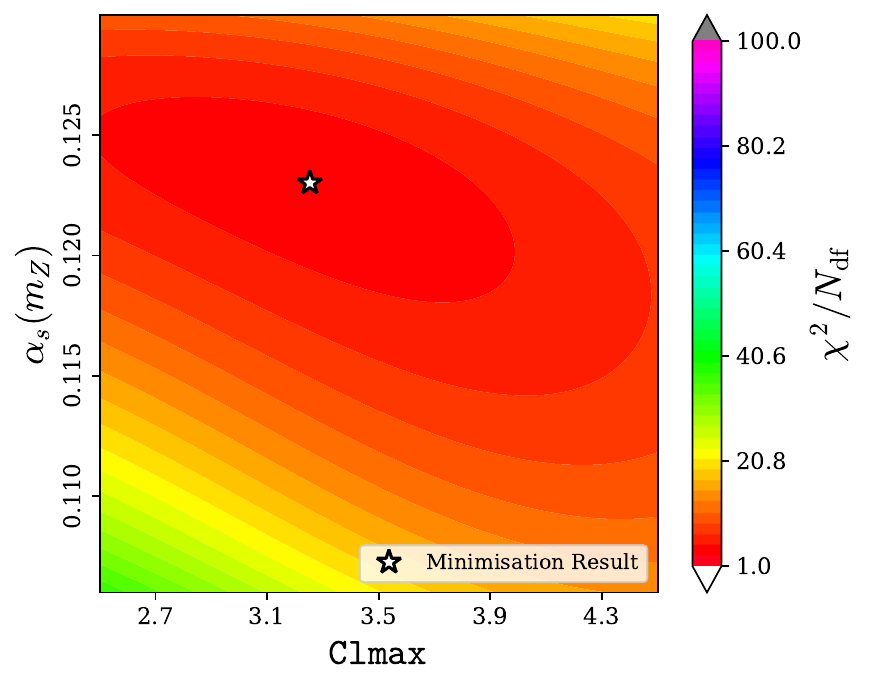}{\FHPAOB}
\labeledfig{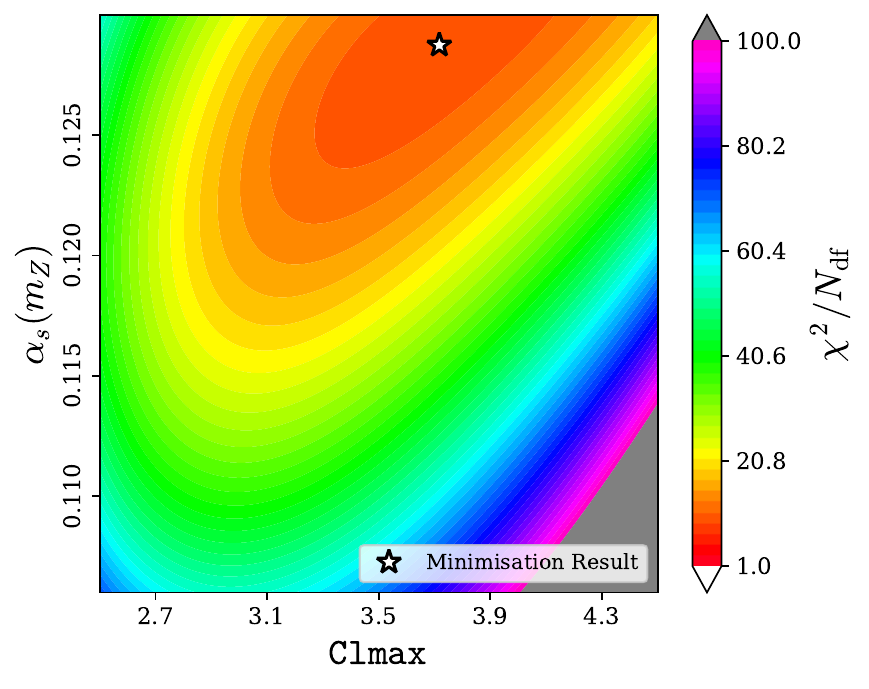}{\FHPAOB}
\labeledfig{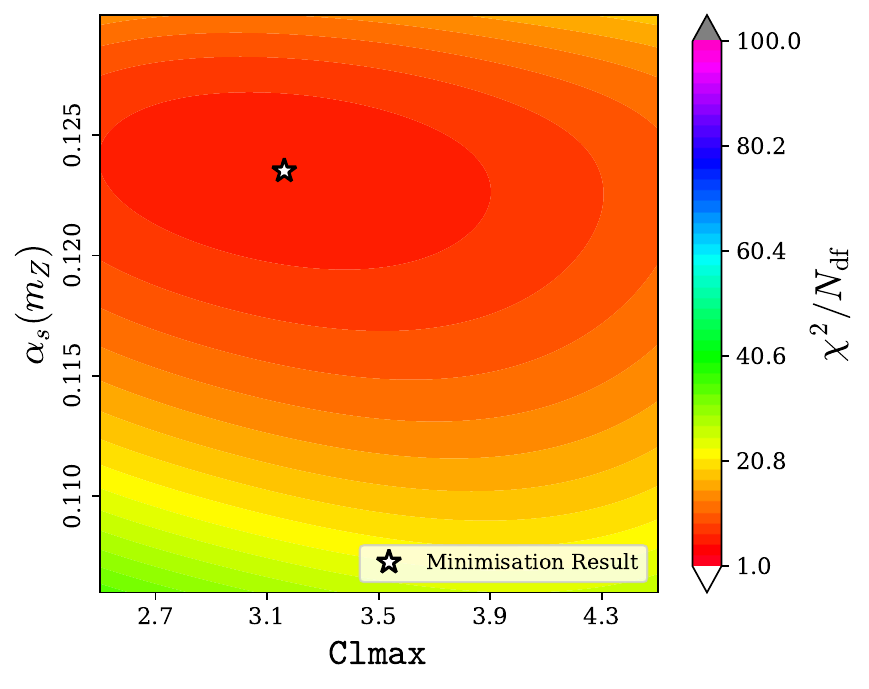}{\FHPAOB}
\caption{
$\chi^2/N_{\text{df}}$ heatmaps, varying $\alpha_s(m_Z)$ and \texttt{Clmax}, with $p_{T,\min}^2$ fixed to $0.75 \text{ GeV}^2$.
Compared with the previous figure, the charged multiplicity heatmaps exhibit large fluctuations.
}
\label{fig:heatmaps-as-cl-pt2min0p75}
\end{figure}
\clearpage

\subsection*{$\chi^2/N_{\text{df}}$ heatmaps -- $\alpha_s(m_Z)$ and \texttt{Clmax}, $p_{T,\min}^2=1.25 \, \mathrm{ GeV}^2$}
\begin{figure}[!htbp]
\centering
\labeledfig{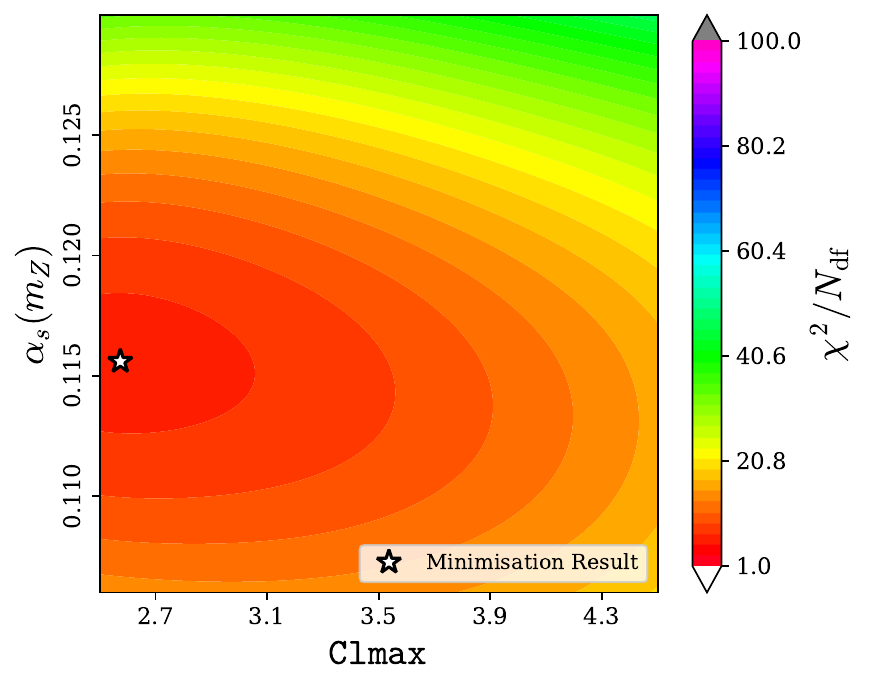}{CS}
\labeledfig{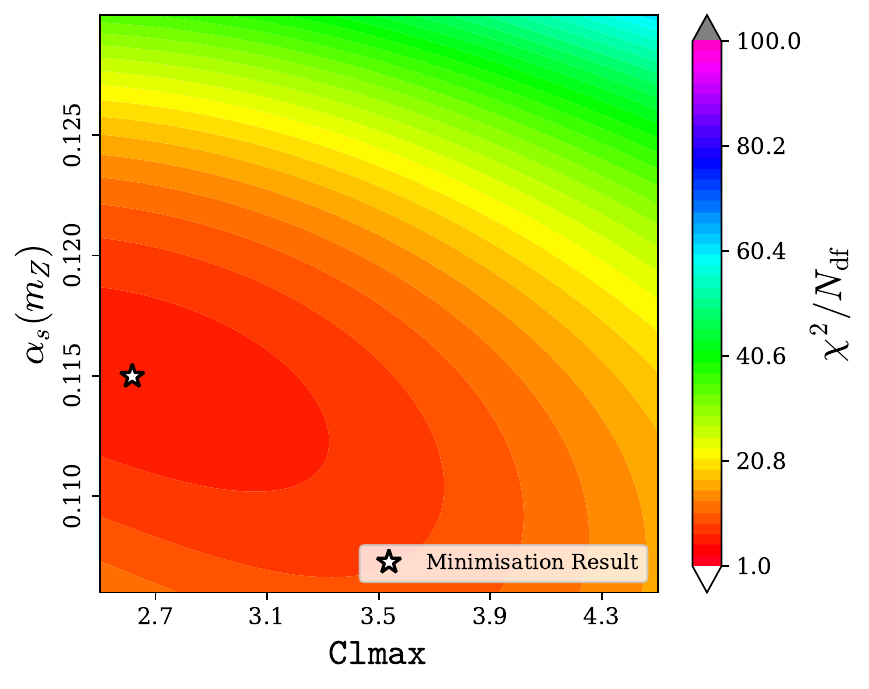}{CS}
\labeledfig{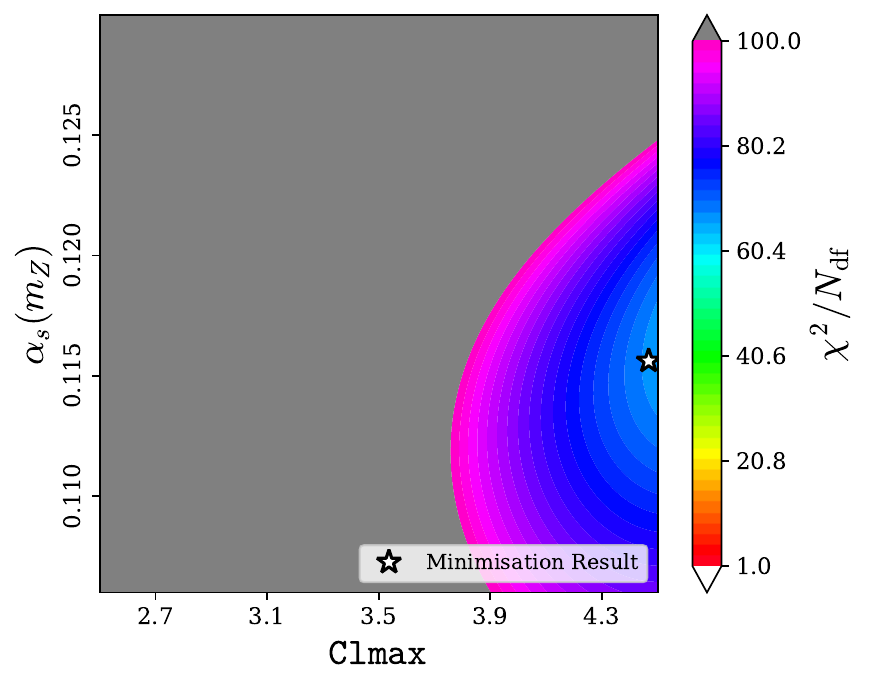}{CS}
\labeledfig{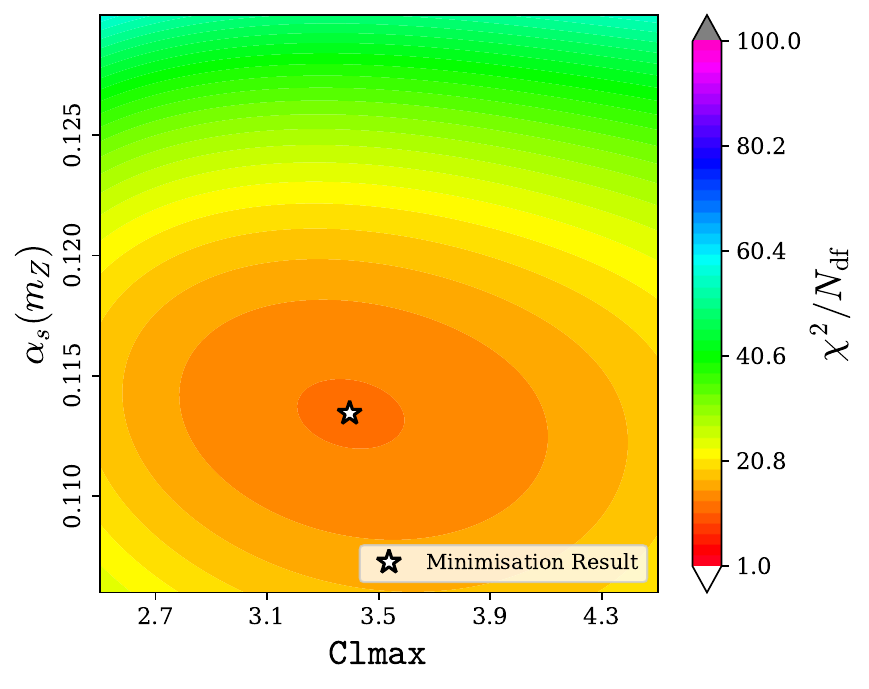}{CS}
\labeledfig{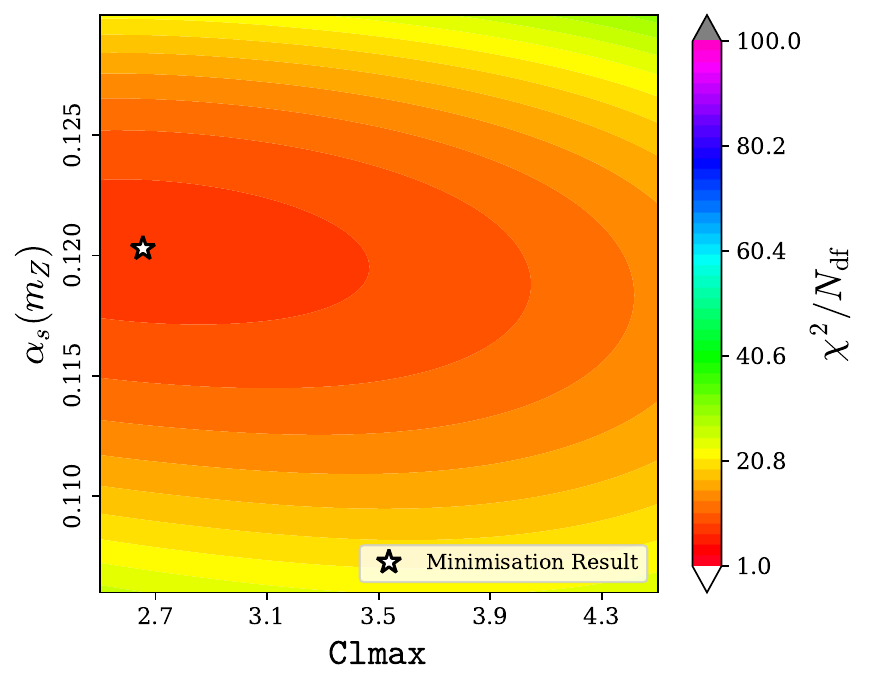}{\PG}
\labeledfig{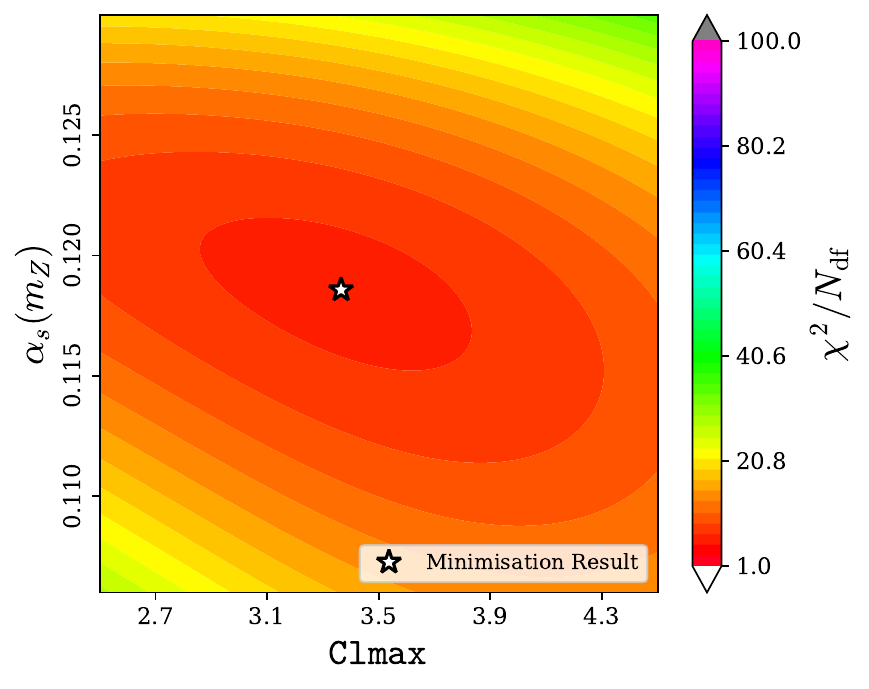}{\PG}
\labeledfig{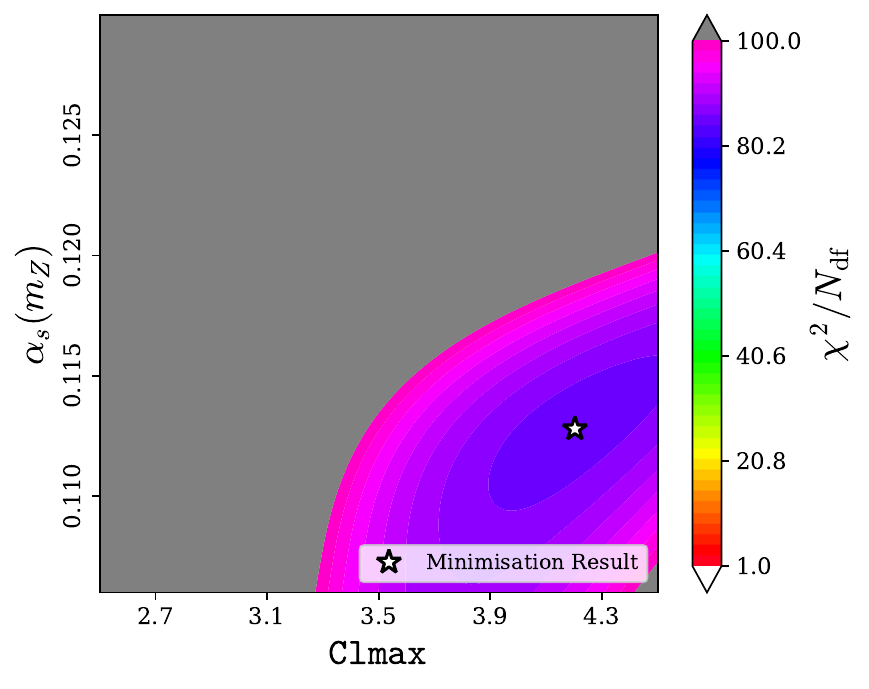}{\PG}
\labeledfig{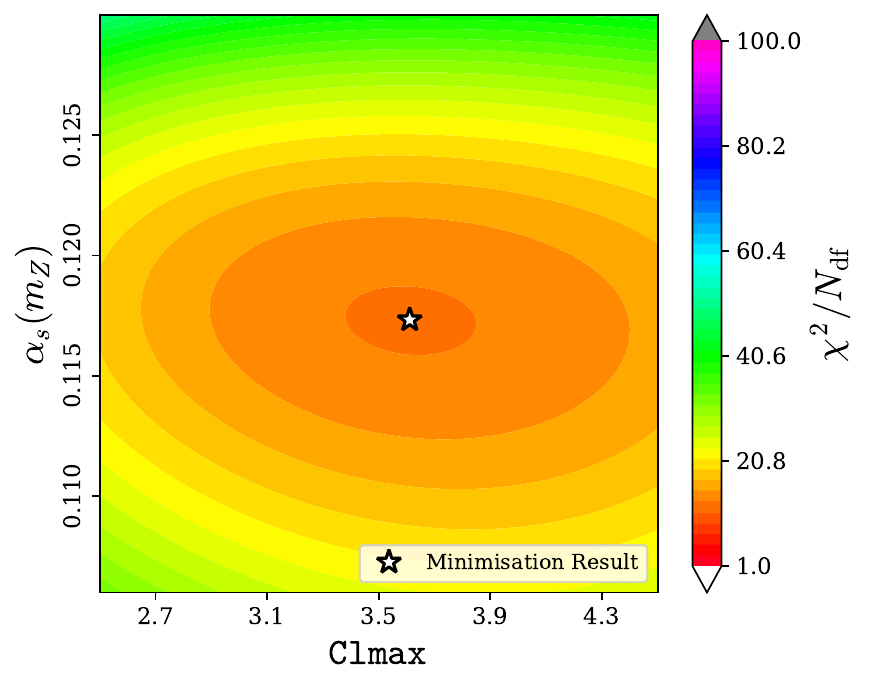}{\PG}
\labeledfig{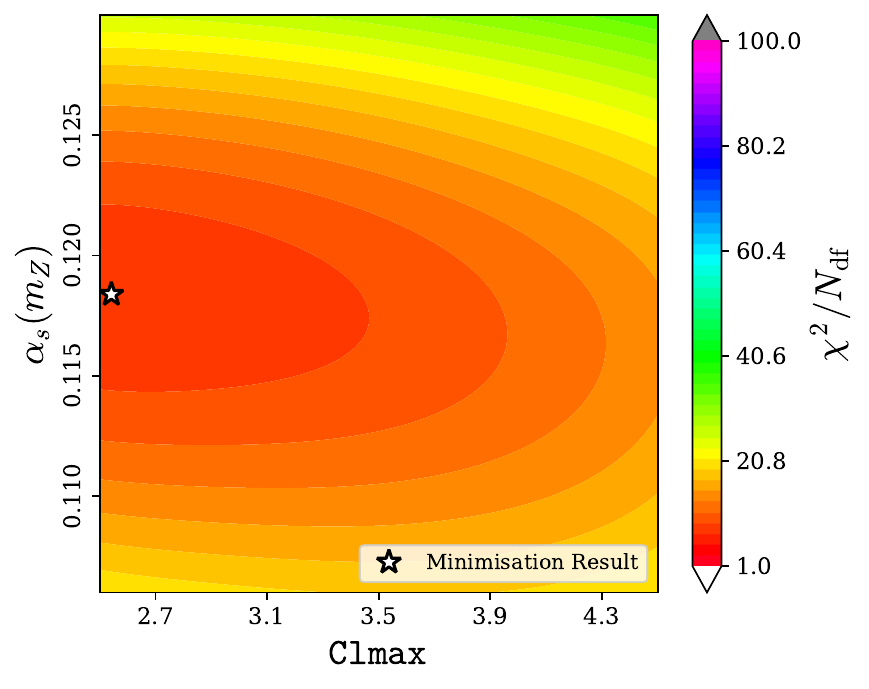}{FHP}
\labeledfig{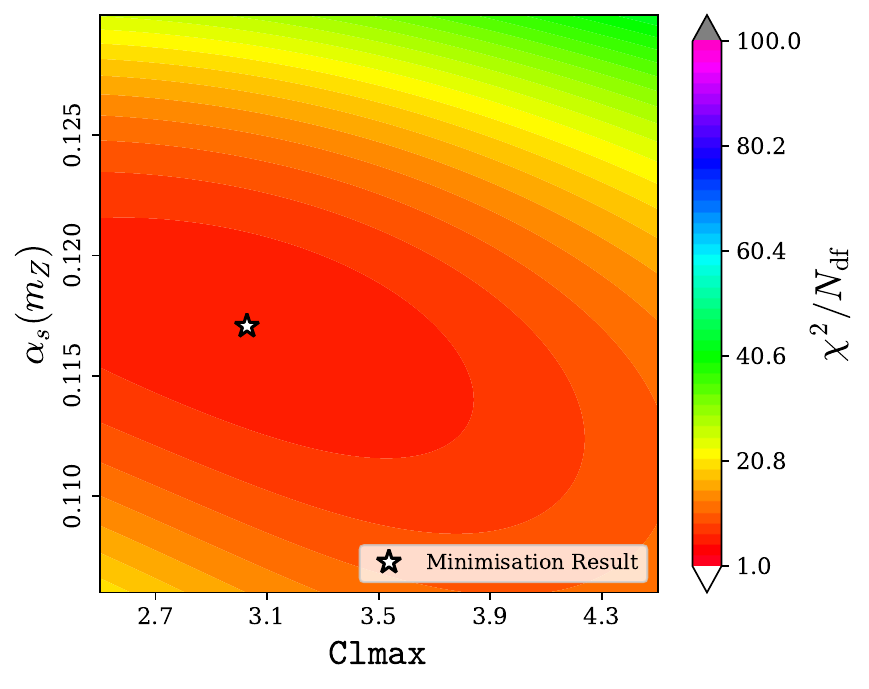}{FHP}
\labeledfig{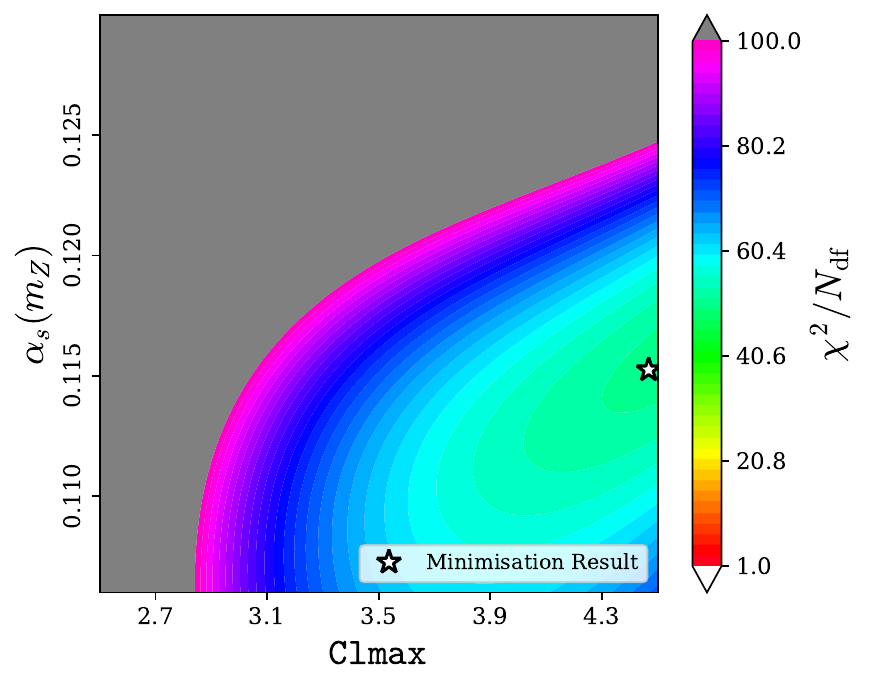}{FHP}
\labeledfig{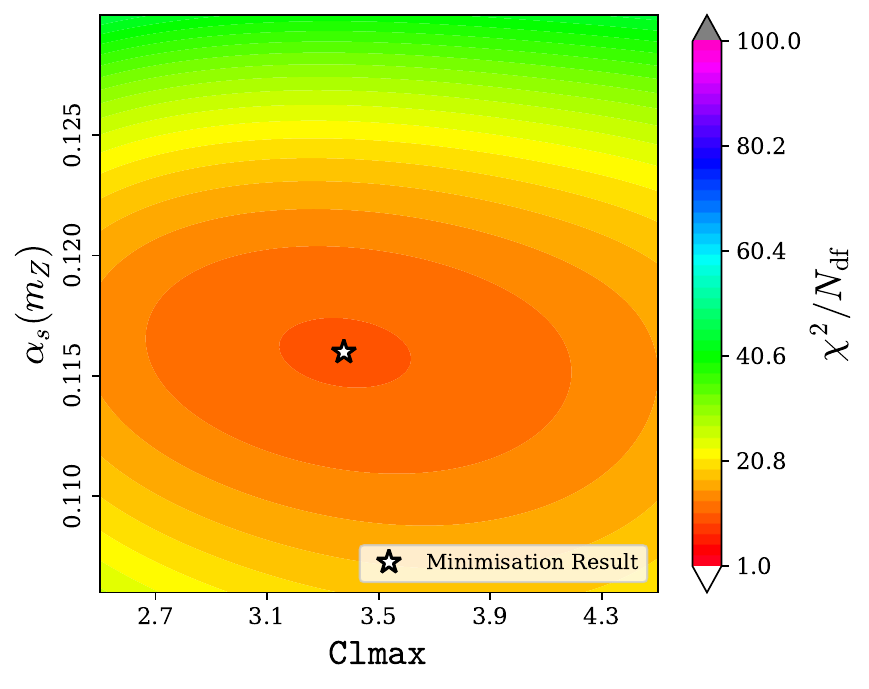}{FHP}
\labeledfig{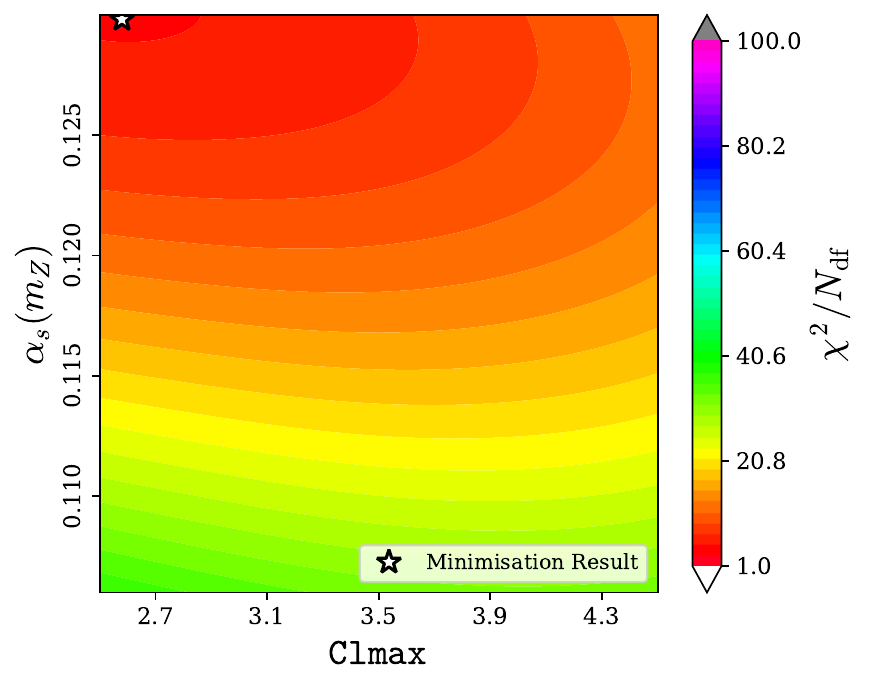}{\PGAOB}
\labeledfig{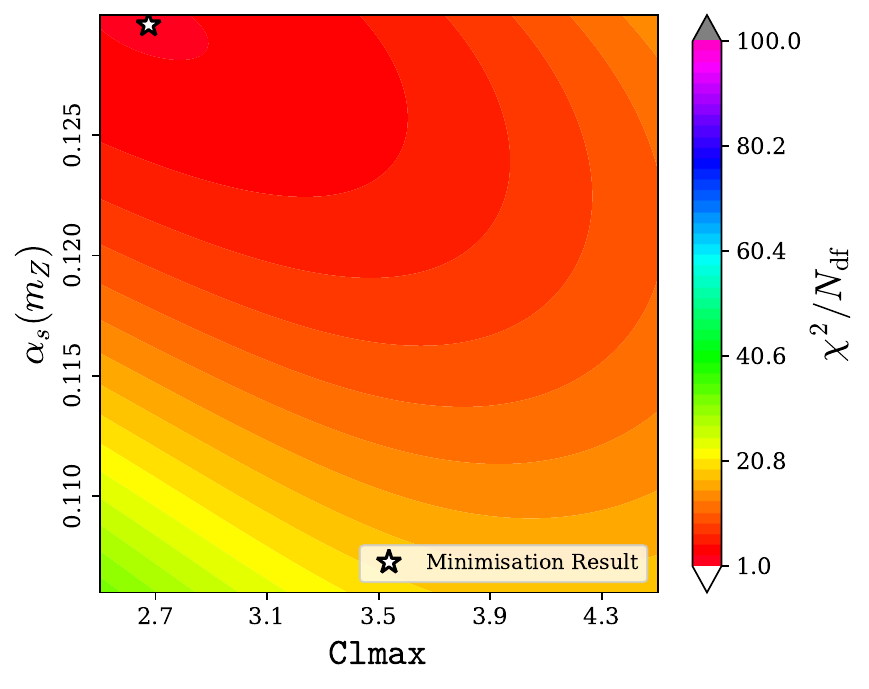}{\PGAOB}
\labeledfig{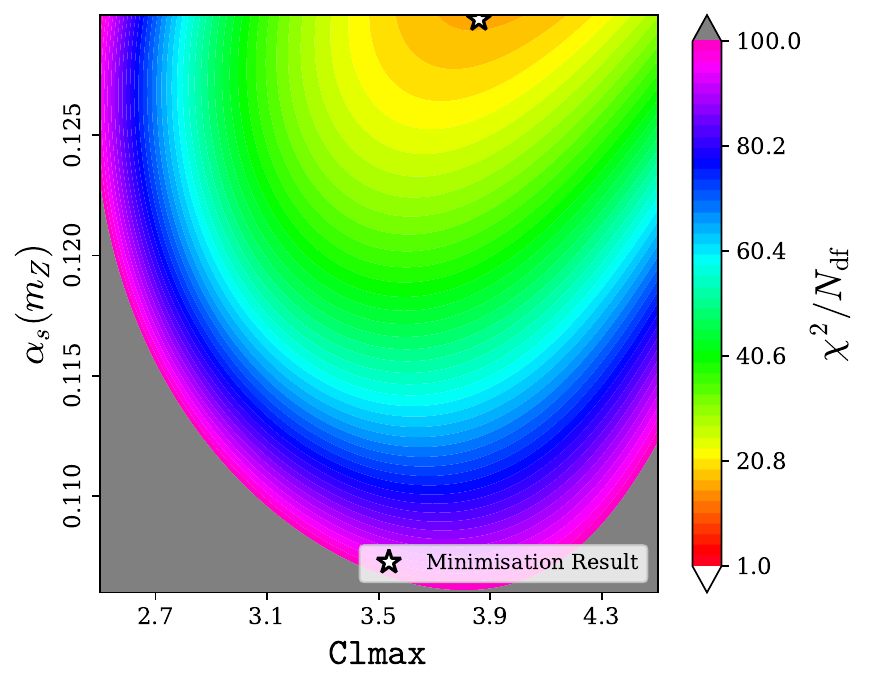}{\PGAOB}
\labeledfig{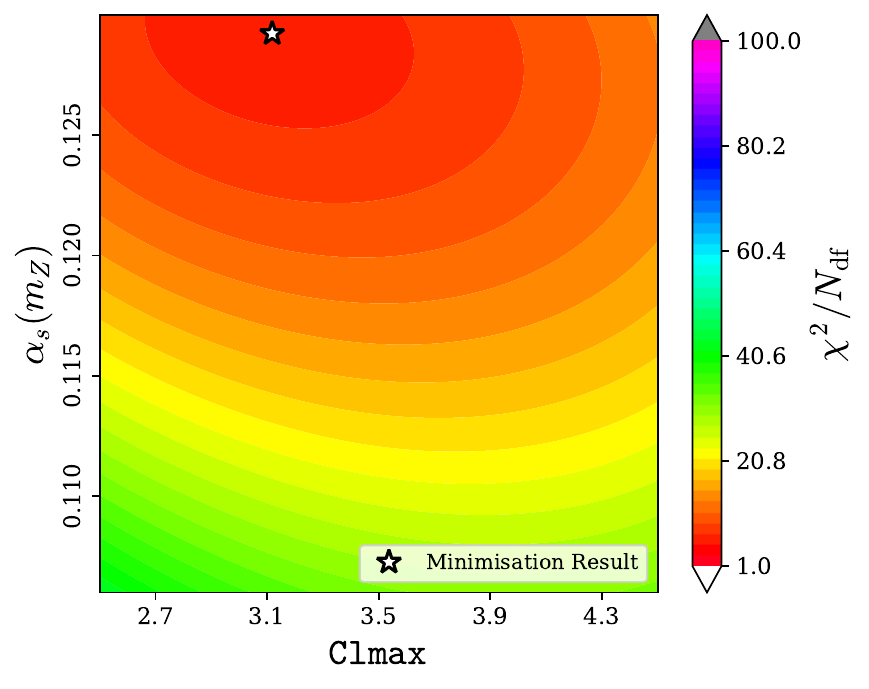}{\PGAOB}
\labeledfig{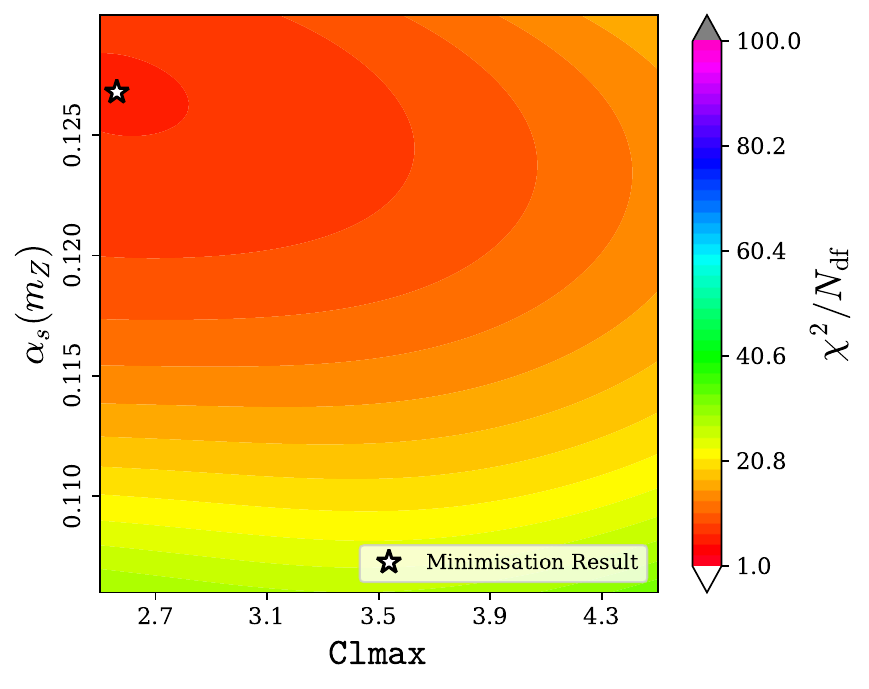}{\FHPAOB}
\labeledfig{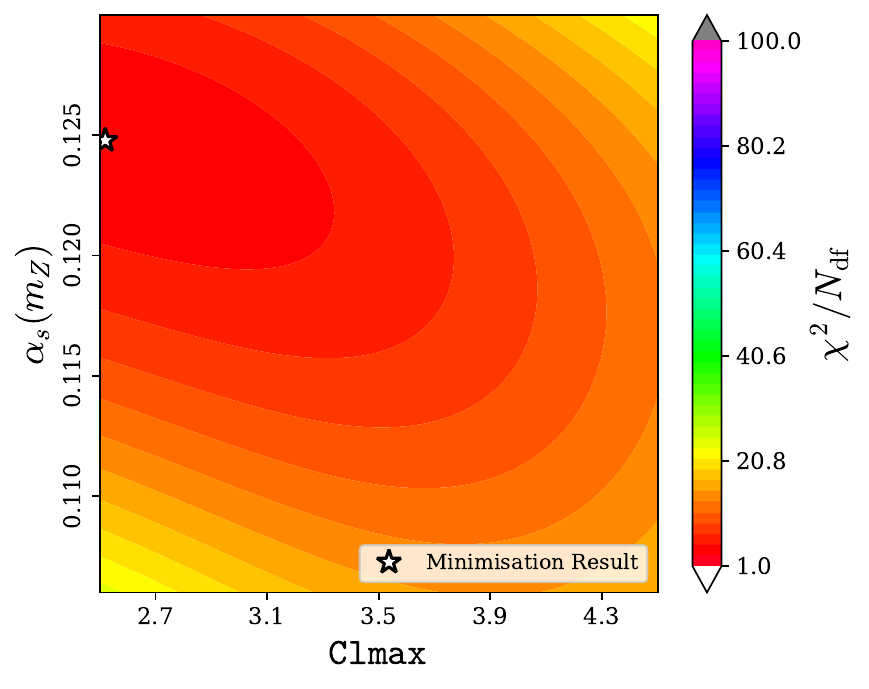}{\FHPAOB}
\labeledfig{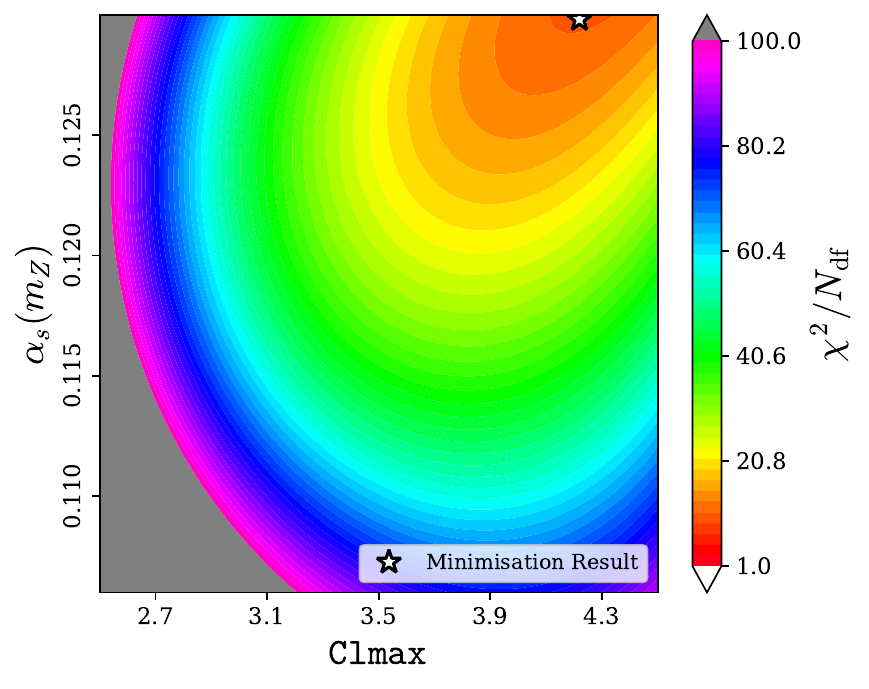}{\FHPAOB}
\labeledfig{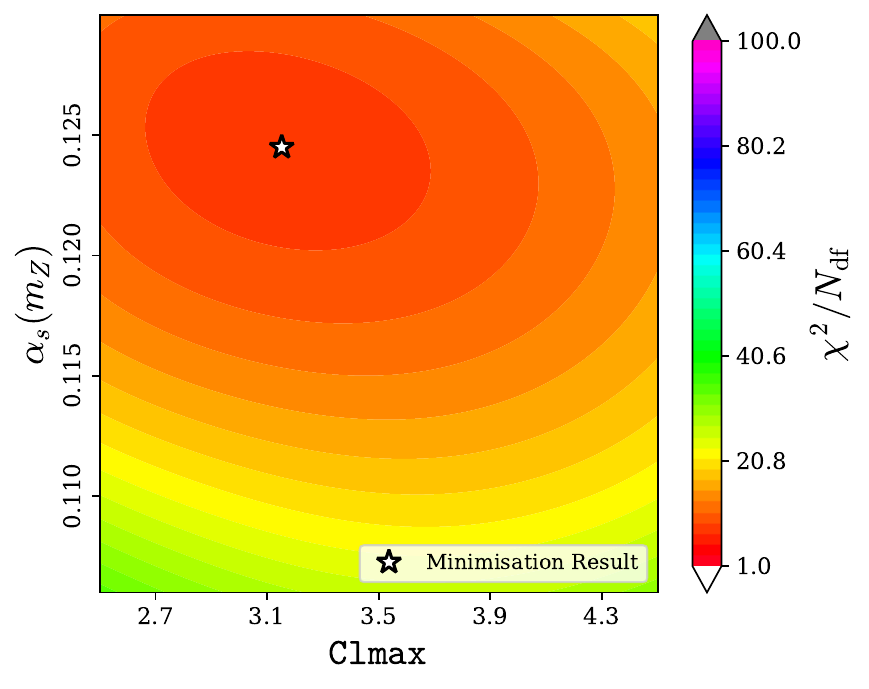}{\FHPAOB}
\caption{
$\chi^2/N_{\text{df}}$ heatmaps, varying $\alpha_s(m_Z)$ and \texttt{Clmax}, with $p_{T,\min}^2$ fixed to $1.25 \text{ GeV}^2$.
Compared with the two previous figures, the charged multiplicity heatmaps show the opposite effect, with the heatmaps' sharp features mellowing out.
This is in agreement with the idea that $p_{T,\min}^2$ shields the simulation from the phase space where perturbative physics breaks down.
}
\label{fig:heatmaps-as-cl-pt2min1p25}
\end{figure}
\clearpage